%% file: main.tex
\documentclass[twocolumn, nofootinbib, superscriptaddress, aps, a4paper]{revtex4-1}
\pdfoutput=1

	\usepackage{bbm,bm}
	\usepackage{amsmath,amssymb}
	\usepackage{graphicx}
        \usepackage{diagbox}
	\usepackage{color}
	\usepackage{xcolor}
	\usepackage{xspace}
        \usepackage{setspace}
	\usepackage{enumerate} 
        \usepackage{rotating}
	\usepackage{hyperref}
	\usepackage{tikz}
	\usetikzlibrary{decorations.pathreplacing}
	\usetikzlibrary{automata,positioning}
	\usepackage{physics}
	\usepackage{notes2bib}
\usepackage{url}

	\definecolor{darkblue}{RGB}{50,50,150}
	\hypersetup{
		colorlinks=true,
		citecolor={darkblue},
		linkcolor={darkblue},
		urlcolor={darkblue},
		pdfhighlight=/I
	}

		\renewcommand{\phi}{\varphi}
		\renewcommand{\epsilon}{\varepsilon}

		\newcommand{\cC}{\mathcal{C}}
		\newcommand{\cB}{\mathcal{B}}
		
		\newcommand{\cS}{\mathcal{S}}
		
		\newcommand{\cL}{\mathcal{L}}
		\newcommand{\bZ}{\mathbbm{Z}}
		
		\newcommand{\one}{\mathbbm{1}}
		\DeclareMathOperator{\Aut}{Aut}

		\renewcommand{\r}{{\texttt{r}}\xspace}
		\newcommand{\g}{{\texttt{g}}\xspace}
		\renewcommand{\b}{{\texttt{b}}\xspace}
		\newcommand{\x}{{\texttt{x}}\xspace}
		\newcommand{\y}{{\texttt{y}}\xspace}
		\newcommand{\z}{{\texttt{z}}\xspace}
		
		\renewcommand{\one}{{\texttt{1}}\xspace}
		\newcommand{\e}{{\texttt{e}}\xspace}
		\newcommand{\m}{{\texttt{m}}\xspace}
		\newcommand{\f}{{\texttt{f}}\xspace} 
		
		\newcommand{\rx}{{\texttt{rx}}\xspace}
		\newcommand{\ry}{{\texttt{ry}}\xspace}
		\newcommand{\rz}{{\texttt{rz}}\xspace}
		\newcommand{\gx}{{\texttt{gx}}\xspace}
		\newcommand{\gy}{{\texttt{gy}}\xspace}
		\newcommand{\gz}{{\texttt{gz}}\xspace}
		\newcommand{\bx}{{\texttt{bx}}\xspace}
		\newcommand{\by}{{\texttt{by}}\xspace}
		\newcommand{\bz}{{\texttt{bz}}\xspace}
		
		\newcommand{\fone}{{\texttt{f$_1$}}\xspace}
		\newcommand{\ftwo}{{\texttt{f$_2$}}\xspace}
		\newcommand{\fthree}{{\texttt{f$_3$}}\xspace}
		\newcommand{\ffour}{{\texttt{f$_4$}}\xspace}
		\newcommand{\ffive}{{\texttt{f$_5$}}\xspace}
		\newcommand{\fsix}{{\texttt{f$_6$}}\xspace}
		
		\renewcommand{\a}{{\texttt{a}}\xspace}
		\renewcommand{\b}{{\texttt{b}}\xspace}
		\renewcommand{\c}{{\texttt{c}}\xspace}

	    \definecolor{redanyon}{RGB}{204,0,0}
	    \definecolor{greenanyon}{RGB}{102,153,0}
	    \definecolor{blueanyon}{RGB}{51,102,204}
	    \definecolor{orangeanyon}{RGB}{255,153,0}
	    \definecolor{cyananyon}{RGB}{153,230,237}

	    \newcommand{\blackcirc}{\raisebox{2.5pt}{$\mathbin{\tikz[baseline]\draw[black,fill=black] (0,0) circle (1ex) ;}$}}
	    \newcommand{\orangecirc}{\raisebox{2.5pt}{$\mathbin{\tikz[baseline]\draw[orangeanyon,fill=orangeanyon] (0,0) circle (1ex) ;}$}}
	    \newcommand{\cyancirc}{\raisebox{2.5pt}{$\mathbin{\tikz[baseline]\draw[cyananyon,fill=cyananyon] (0,0) circle (1ex) ;}$}}
	    \renewcommand{\cross}{\raisebox{-1.6pt}{$\mathbin{\tikz [x=1.6ex,y=1.6ex,line width=.2ex] \draw (0,0) -- (1.1,1.1) (0,1.1) -- (1.1,0);}$}}%

\begin{document}


\title{Anyon condensation and the color code}

	\author{Markus~S.~Kesselring}
	\email{markus.kesselring@fu-berlin.de}
	\affiliation{Dahlem Center for Complex Quantum Systems, Freie Universit\"at Berlin, 14195 Berlin, Germany}
	\author{Julio~C.~Magdalena~de~la~Fuente}
	\affiliation{Dahlem Center for Complex Quantum Systems, Freie Universit\"at Berlin, 14195 Berlin, Germany}
	\author{Felix~Thomsen}
	\affiliation{Centre for Engineered Quantum Systems, School of Physics, University of Sydney, Sydney, New South Wales 2006, Australia}
	\author{Jens~Eisert}
	\affiliation{Dahlem Center for Complex Quantum Systems, Freie Universit\"at Berlin, 14195 Berlin, Germany}
	\affiliation{Helmholtz-Zentrum Berlin f{\"u}r Materialien und Energie, 14109 Berlin, Germany}
	\author{Stephen~D.~Bartlett}
	\affiliation{Centre for Engineered Quantum Systems, School of Physics, University of Sydney, Sydney, New South Wales 2006, Australia}
	\author{Benjamin~J.~Brown}
	\affiliation{Centre for Engineered Quantum Systems, School of Physics, University of Sydney, Sydney, New South Wales 2006, Australia}

    \begin{abstract}
	    The manipulation of topologically-ordered phases of matter to encode and process quantum information forms the cornerstone of many approaches to fault-tolerant quantum computing.
Here we demonstrate that fault-tolerant logical operations in these approaches can be interpreted as instances of anyon condensation.
	    We present a constructive theory for anyon condensation and, in tandem, illustrate our theory explicitly using the color-code model.
	    We show that different condensation processes are associated with a general class of domain walls, which can exist in both space- and time-like directions.
        This class includes semi-transparent domain walls that condense certain subsets of anyons.
        We use our theory to classify topological objects and design novel fault-tolerant logic gates for the color code.
        As a final example, we also argue that dynamical `Floquet codes' can be viewed as a series of condensation operations.
        We propose a general construction for realising planar dynamically driven codes based on condensation operations on the color code. 
	    We use our construction to introduce a new Calderbank-Shor Steane-type Floquet code that we call the Floquet color code.
    \end{abstract}

\maketitle

    \setcounter{tocdepth}{2}
    \begin{spacing}{0.9}
        \tableofcontents
    \end{spacing}

\newpage
\section{Introduction}
	\label{Sec:Introduction}

    Topological quantum error-correcting codes~\cite{Dennis02, Kitaev03, Freedman00, Levin05, Bombin06,  Nayak08} have provided the basis of many promising approaches to realise a fault-tolerant quantum computer.
    These codes are based on topological phases that robustly store quantum states in non-local degrees of freedom~\cite{Wen03, Brown16}.
    Additionally, there exist a number of distinct ways of performing logical operations on topologically protected quantum states, using unitary dynamics~\cite{Kitaev03, Freedman00}, measurement-based methods~\cite{Raussendorf06, Raussendorf07fault, Bonderson08, Bombin09CodeDefo,Fowler11twoDimensional, HorsemanSurgery, Brown17}, or combinations thereof~\cite{Bombin18, Brown2020a,Zhu2022topologicalOrder}.
    Performing these logical operations, however, is generally very resource-intensive, and so it remains a significant technical challenge to realise these current designs for quantum computing architectures.
    It is therefore important to develop novel methods of implementing robust operations that are available with topological phases in order to find more practical ways of performing the logical operations we need for universal fault-tolerant quantum computing.

    The \emph{color code}~\cite{Bombin06} is a topological quantum error-correcting code with a rich structure that can be harnessed for topological quantum computing.
    Its study was first motivated by its multitude of available transversal logic gates.
    In addition to these local constant-depth unitary logical operations, which are inherently fault-tolerant, the color code can also demonstrate various fault-tolerant measurement-based code deformations~\cite{Raussendorf07fault, Bombin09CodeDefo,Fowler11twoDimensional, HorsemanSurgery, Brown17}. All together, these operations can be combined to give universal low-overhead fault-tolerant quantum computing~\cite{Bombin09CodeDefo,Fowler11twoDimensional, Landahl14, Thomsen22}, with resource requirements that are favourable over those of the surface code.

    The versatility of the color code for performing fault-tolerant logic gates can be attributed to the underlying symmetries among its quasi-particle excitations when viewed as a topological model~\cite{Yoshida15,Bridgeman2017tensor, Kesselring18}.
    Furthermore, the color code can be decomposed into copies of more elementary phases~\cite{Bombin12, Bhagoji2015equivalence, Kubica15, Kesselring18, Roberts2020}.
    All together, these properties mean that the color code offers an excellent test-bed both to design practical ways of performing fault-tolerant quantum computation, as well as to investigate the fundamental phenomena of topological phases that enable robust logical operations.

    In this work we develop a theory of topological quantum computing in terms of \emph{anyon condensation}~\cite{Burnell18, PhysRevB.79.045316, PhysRevB.95.235119, PhysRevB.97.195124, Bombin08} for stabiliser codes, where we use the color code as a guiding example.
    Anyon condensation implements a special type of topological phase transition by identifying a subset of anyons of the phase with the vacuum particle of a condensate.
    We find that anyon condensation offers a natural way of describing many aspects of quantum computation with topological stabiliser codes.
    In particular, it offers a concise and unified language for the logical operations available in the color code, as well as other code deformation and code switching schemes.
    Furthermore, the scheme we present to perform anyon condensation is constructive in microscopic lattice models, i.e., it allows us to construct topological stabiliser codes exemplifying and implementing the various features and operations described abstractly by anyon condensation.    
    In Table~\ref{tab:PaperStructure} and the following paragraphs, we give an overview of the results obtained from implementing anyon condensation in the color code in various ways.

    As as first example, we consider condensing the anyons in a disk-shaped region of the color-code lattice.
    This allows us to recover the punctures with different types of boundaries that have been discovered earlier in the literature~\cite{Bombin06, Kesselring18}.
    The boundaries we produce can be viewed as a non-local degree of freedom, the state of which is determined by the type of anyons it has condensed, i.e., absorbed through local processes.
    Punctures provide the standard mechanism to encode quantum information in a topological code, and given these properties, we can create and braid different punctures to robustly encode and manipulate quantum information.

    In addition to known punctures, the theory of anyon condensation allows us to generalise the types of punctures that can be produced in the color code.
    Specifically, we discover what we coin a semi-puncture, where only a single boson of the color code is condensed within a disk-shaped region.
    We regard these new objects as semi-punctures in the sense that a pair of semi-punctures, of the appropriate type, can be combined to give known types of punctures.
    Indeed, a standard puncture is obtained by condensing a larger subset of color code bosons in some region.
    As we will show, new types of code deformation become available by manipulating these generalised punctures, leading to new approaches to fault-tolerant logic.

    \begin{table*}[bth]
	\centering
	\begin{tabular}{|c||c|c|c|} \hline
        & maximal \ref{sec:CondMax} & partial \ref{sec:CondPart} & trivial \ref{sec:CondTriv} \\ \hline \hline
        partial bulk & boundaries and punctures & condensates and semi-punctures & invertible domain walls \\
        \ref{sec:Cond2d} & \ref{sec:Cond2dMax} & \ref{sec:Cond2dPart}& \ref{sec:Cond2dTriv} \\
        \raisebox{-\totalheight+8pt}{\includegraphics[width=0.12\textwidth]{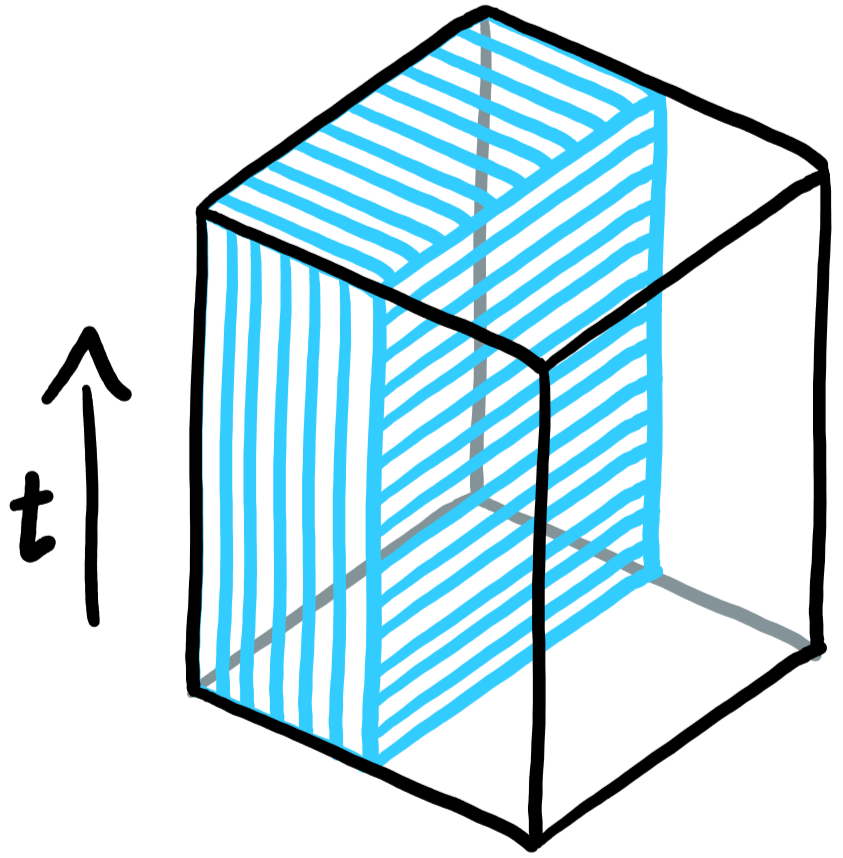}} & \raisebox{-\totalheight-1pt}{\includegraphics[width=0.13\textwidth]{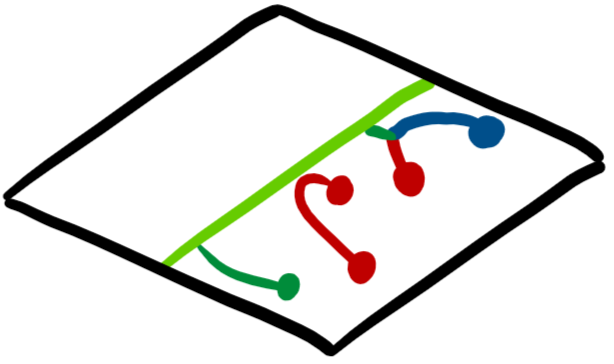}} & \raisebox{-\totalheight-1pt}{\includegraphics[width=0.13\textwidth]{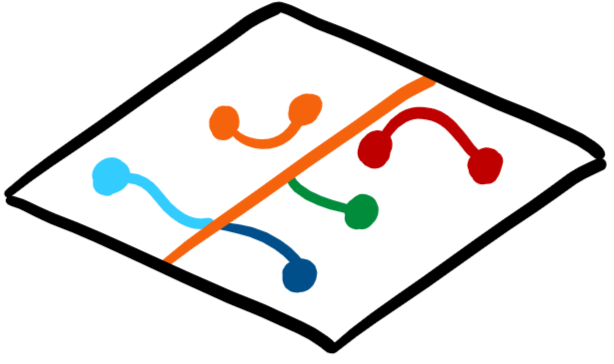}} & \raisebox{-\totalheight-1pt}{\includegraphics[width=0.13\textwidth]{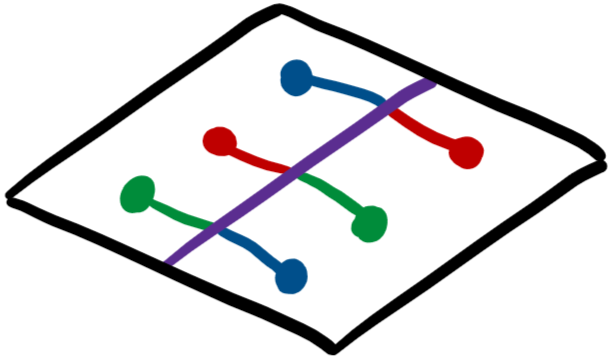}} \\
        \hline
        full bulk & readout / initialisation & partial readout / initialisation & transversal gates \\
        \ref{sec:CondTime} & \ref{sec:CondTimeMax}, \ref{sec:MicroTempBdry} & \ref{sec:CondTimePart} & \ref{sec:CondTimeTriv} \\
        \raisebox{-\totalheight+8pt}{\includegraphics[width=0.12\textwidth]{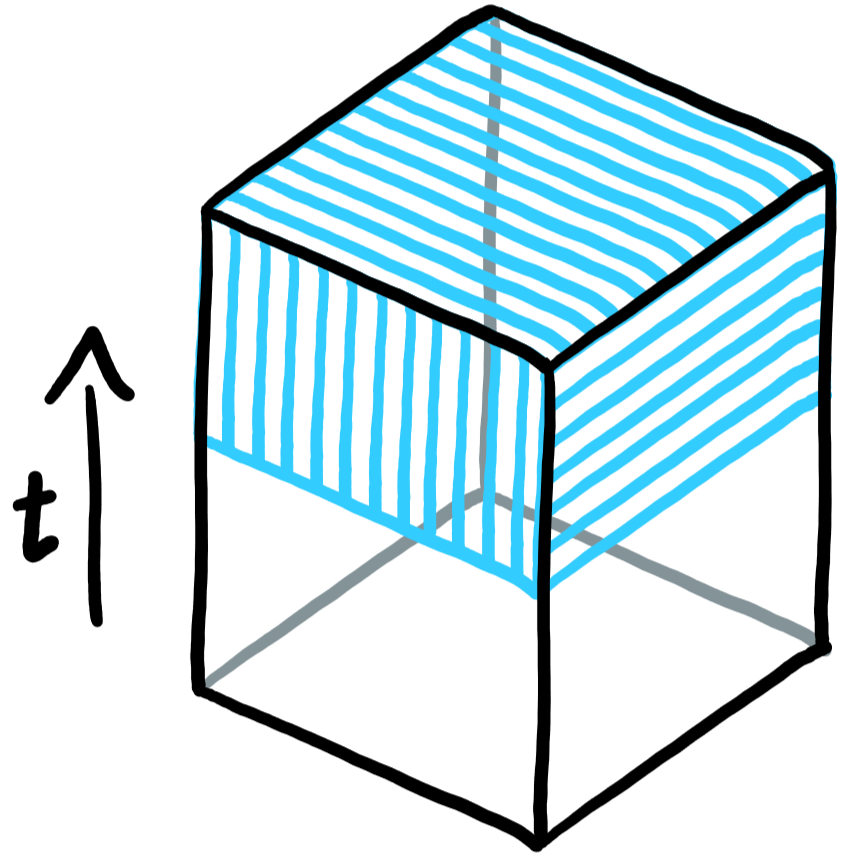}} & \raisebox{-\totalheight+8pt}{\includegraphics[width=0.07\textwidth]{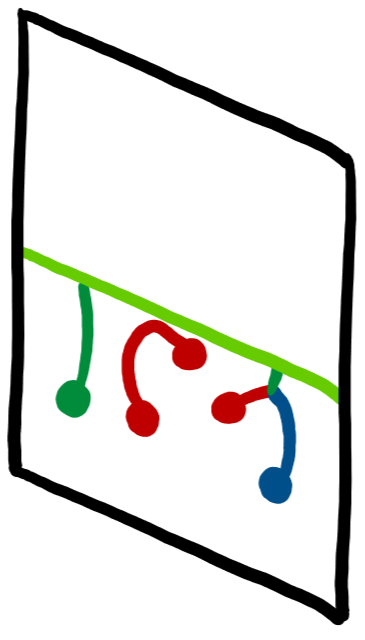}} ~ \raisebox{10pt}{\rotatebox{180}{\includegraphics[width=0.07\textwidth]{Figures/CondAnyonsTimeMax.png}}} & \raisebox{-\totalheight+8pt}{\includegraphics[width=0.07\textwidth]{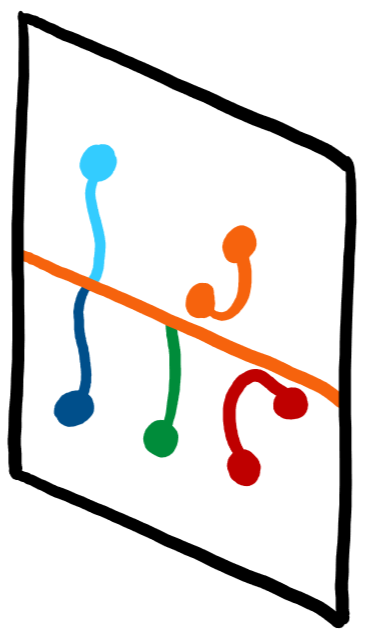}} ~ \raisebox{10pt}{\rotatebox{180}{\includegraphics[width=0.07\textwidth]{Figures/CondAnyonsTimePart.png}}} & \raisebox{-\totalheight+8pt}{\includegraphics[width=0.07\textwidth]{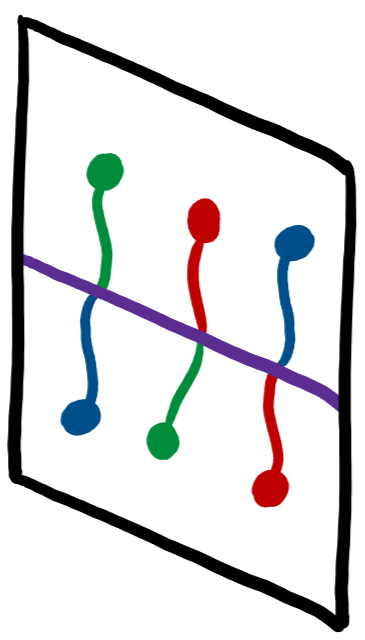}} \\
        \hline
        1d line & corners & semi-transparent domain walls & twist defects \\
        \ref{sec:Cond1d} & \ref{sec:Cond1dMax} & \ref{sec:Cond1dPart} & \ref{sec:Cond1dTriv} \\
        \raisebox{-\totalheight+8pt}{\includegraphics[width=0.12\textwidth]{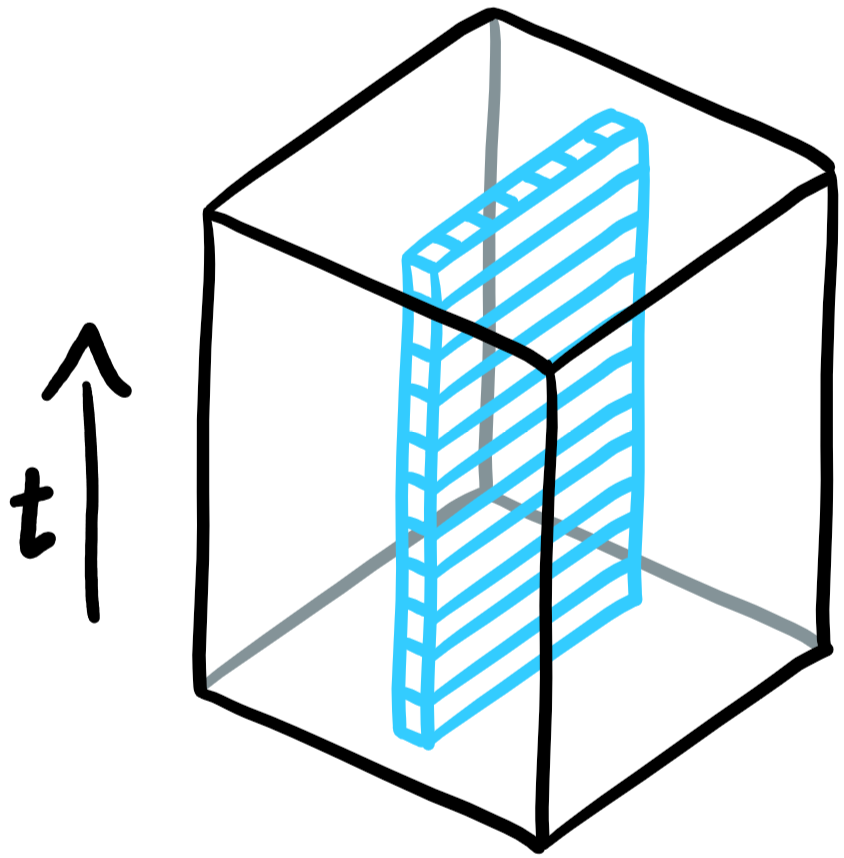}} & \raisebox{-\totalheight-1pt}{\includegraphics[width=0.13\textwidth]{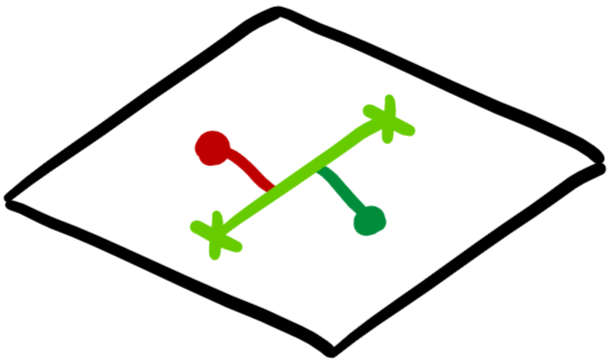}} & \raisebox{-\totalheight-1pt}{\includegraphics[width=0.13\textwidth]{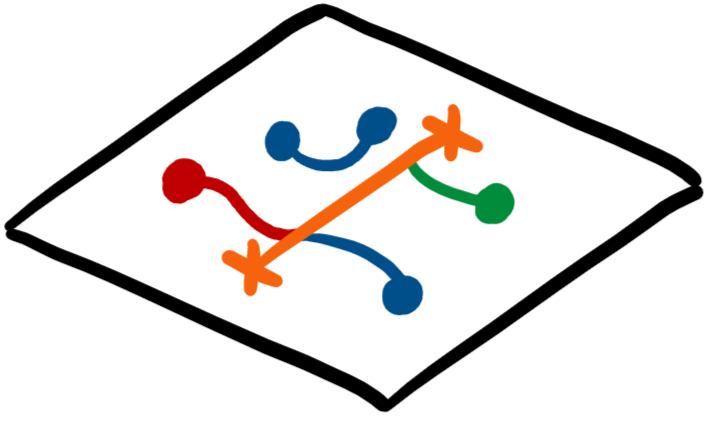}} & \raisebox{-\totalheight-1pt}{\includegraphics[width=0.13\textwidth]{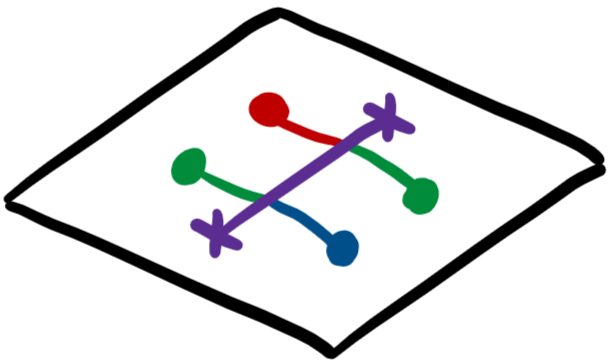}} \\
        \hline \hline
        dynamical & & Floquet codes & \\
        \ref{sec:CondDynamic} & & \ref{sec:CondDynamic} & \\
        \raisebox{-\totalheight+8pt}{\includegraphics[width=0.12\textwidth]{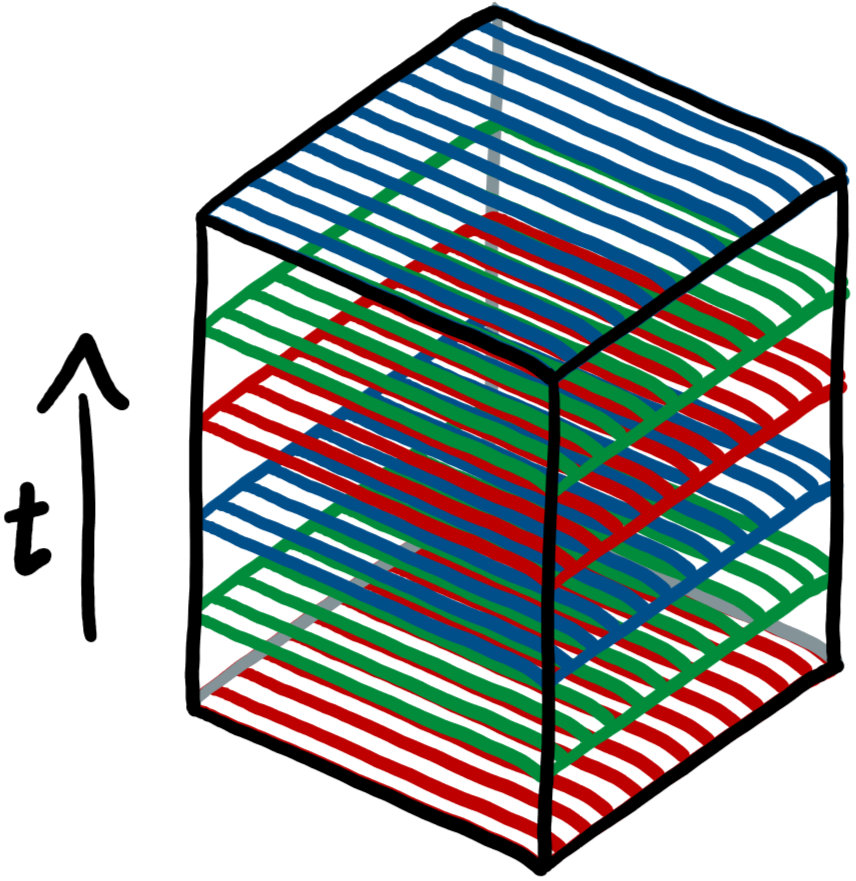}} &  &
        \raisebox{-\totalheight+8pt}{\includegraphics[width=0.07\textwidth]{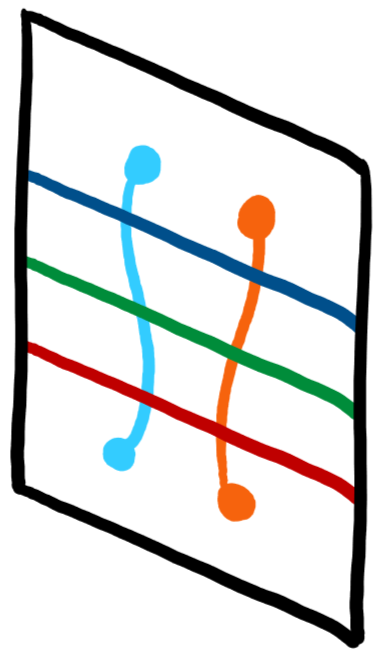}} &  \\
        \hline
    \end{tabular}
	\caption{
	    Anyon condensation on topologically ordered phases in various regions of space-time.
	    We distinguish three types of condensation -- maximal, partial or trivial -- depending on the number of anyons that are condensed.
        The different types of condensation are shown in the columns in the table, and are discussed in Secs.~\ref{sec:CondMax}~-~\ref{sec:CondTriv}.
	    We apply condensation to different regions of the space-time lattice, represented by the shaded regions in the figures of the left most column, over different sections of the work.
	    In the first row, we condense anyons in the spatial bulk.
        This introduces features such as punctures, semi-punctures and invertible domain walls~(Sec.~\ref{sec:Cond2d}).
	    In the second row, we introduce temporal domain walls using anyon condensation over the entire bulk at an instant in time.
	    This gives us protocols to initialise, manipulate and read out logical qubits in the color code~(Sec.~\ref{sec:CondTime}).
        In the third row, we condense anyons across a narrow sub-region of the bulk, in order to obtain end-points of domain walls~(Sec.~\ref{sec:Cond1d}).
        Finally, we perform a dynamic condensation, where different anyons are condensed at different time-steps.
        Using the partial condensation in the color code, we obtain a general construction for dynamically driven codes through this procedure~(Sec.~\ref{sec:CondDynamic}).
    }
	\label{tab:PaperStructure}
	\end{table*}

    We also view bulk condensation operations from a different perspective by examining how a topological phase transforms over time.
    In general, we observe a domain wall in a $(2+1)$-dimensional space-time picture as the system undergoes a phase transition between two distinct phases.
    In the example of the color code, we can condense a maximal subset of bosons, formally known as a Lagrangian subgroup of the anyon model, which transforms the phase onto the vacuum phase.
    Alternatively we can also condense a smaller subset of bosons, such that we transform the system onto the toric code phase.
    These operations all have important applications in fault-tolerant quantum computation, for logical state readout or, inversely, state preparation.
    As we will show, we find that the anyon condensation perspective shows us how to address individual color-code logical qubits for readout operations, by condensing a smaller subsets of bosons to perform logical measurements.
    We touch on the interplay between spatial and temporal boundaries and how different boundary configurations relate to the fault tolerance of a given protocol.
    As an example, we show how the theory of anyon condensation can be used to design more general stability experiments~\cite{Gidney2022stability} to evaluate the error-correction capabilities of the color code as it undergoes fault-tolerant logical operations.

    Anyon condensation is also invaluable in the classification of twist defects.
    Twist defects are obtained by terminating a domain wall that connects a topological phase to itself.
    In earlier work~\cite{Yoshida15, Bridgeman2017tensor, Kesselring18}, 72 twist defects were shown to exist when the color code phase is connected trivially to itself.
    More generally though, we find that additional domain walls can be obtained by merging a phase to itself via a non-trivial condensate~\cite{KitaevKong12,kong2014anyoncondensation}.
    This gives rise to a semi-transparent domain wall where certain charges can pass through the domain, whereas others either condense or confine.
    We demonstrate the importance of these more general types of domain wall for fault-tolerant quantum computation by investigating how semi-transparent domain walls appear in lattice surgery operations with the color code~\cite{Thomsen22}.

    As a final example, we rederive and generalise recently proposed constructions of dynamically driven `Floquet' codes~\cite{HastingsHaah21a} from the perspective of color-code anyon condensation.
    We argue that the transformations used to read out the stabilisers of dynamically driven codes can be viewed as a sequence of condensation operations, where at each step we condense a different color-code anyon.
    Our construction enables us to discover more general classes of dynamically driven codes.
    We propose one such example that we call the Floquet color code; a Calderbank-Shor Steane(CSS)-type Floquet code on the honeycomb lattice.
    We find that since our construction is based on the well-studied color code, we obtain a constructive way of designing the boundary stabilisers of dynamically driven codes~\cite{HaahHastings21b}, by appealing to the physics of the parent color-code theory.
    We, furthermore, present numerical results showing that the Floquet color code has a threshold that is very competitive with other known Floquet codes~\cite{Gidney22}.
    We remark that the Floquet color code has independently been discovered in 
    other very recent work~\cite{Davydova22,Bombin22PsiQ}. Furthermore, we note that a recent experiment has demonstrated an error detection measurement for the Floquet color code~\cite{Wootton2022measurements}.

\subsection{A guide for the reader}
    We develop the theory of anyon condensation for Abelian anyon models in the earlier sections of this work, and we investigate various instances of anyon condensation with the color code, and its applications to fault-tolerant quantum computing, in the following sections of the manuscript. We have therefore written the latter sections of the paper in a self-contained way, assuming the reader is familiar with the theory we present in the former sections. We summarise this structure in Table \ref{tab:PaperStructure}, where the different examples of anyon condensation in the latter sections of the paper are presented in the rows of the table, with respect to the different types of anyon condensation that are represented by the columns of the table. Furthermore, we offer the following guide for the reader to navigate through the various sections of the manuscript, where we emphasise the dependence of the latter sections on requisite material from former sections.

    In Sec.~\ref{sec:Preliminaries}, we review the requisite background for the color-code model, and we identify its keys properties that enable it to give rise to a number of non-trivial condensation operations. We then give a general theory for anyon condensation in Sec.~\ref{sec:AnyonCondensation}.     Specifically, we distinguish between maximal, partial, and trivial condensation.
    These different types of anyon condensation are distinguished by how excitations are transmitted across the domain wall that is produced by the condensation operation.
    Again, we show the different types of condensation are represented by the columns of Table~\ref{tab:PaperStructure}.

    Assuming the reader is familiar with the material presented in Sec.~\ref{sec:Preliminaries} and Sec.~\ref{sec:AnyonCondensation}, the microscopic examples of condensation in the following sections can be read independently. Indeed, Secs.~\ref{sec:Cond2d} and~\ref{sec:CondTime} are self contained. Sec.~\ref{sec:Cond2d} investigates domain walls that spatially separate the color code from one of its condensates, either the vacuum phase, the toric code, or the color code itself.
    This is represented by the first row of Table~\ref{tab:PaperStructure}.
    Then, in Sec.~\ref{sec:CondTime}, we investigate the different types of condensation over time-like domain walls, as shown in the second row of Table~\ref{tab:PaperStructure}.

    Sec.~\ref{sec:Cond1d} builds on the ideas we begin to develop in Secs.~\ref{sec:Cond2d} and~\ref{sec:CondTime}. In this section we describe new topological features that are produced by interfacing two phases with an intermediate condensate.
    Specifically, at the microscopic level, we show that we can make non-trivial domain walls between the color code and itself to produce different types of twist defects, where the color code phase is interfaced by a non-trvial condensate. This construction is represented on the third row of Table~\ref{tab:PaperStructure}.

    Finally, we discuss our Floquet code construction from the perspective of anyon condensation in Section \ref{sec:CondDynamic}.
    This picture we present for Floquet codes in terms of condensation operations is represented by the fourth and final row of Table \ref{tab:PaperStructure}. This section can also be read independently of Secs.~\ref{sec:Cond2d},~\ref{sec:CondTime} and~\ref{sec:Cond1d}.

\section{Preliminaries}
	\label{sec:Preliminaries}    

    The color code is a topologically-ordered phase of matter that gives rise to anyonic quasi-particle excitations.
    We start this section by introducing the theory of Abelian anyon models (Sec.~\ref{sec:PrelimAnyons}) before turning our attention to the color code.
    We introduce the color-code anyons and a microscopic color-code lattice model (Sec.~\ref{sec:PrelimCC}).
    We relate the color code to another well-known topological phase, the toric code (Sec.~\ref{sec:PrelimTCandMC}).
    In particular, we discuss first how the color code can be unfolded into two decoupled layers of toric codes (Sec.~\ref{sec:PrelimUnfolding}).
    Lastly, we introduce a space-time interpretation of topological error correcting codes (Sec.~\ref{sec:Prelimspace-time}).
        
    \subsection{Anyons}
        \label{sec:PrelimAnyons}
        
        Anyons are quasi-particles that exist in two spatial dimensions~\cite{Kitaev03, Kitaev06}.
        We denote the set of all anyons of a phase, and the data describing their behaviour, by the anyon model $\cC$.
        We label single anyons as lower case letters $\a,\b,\c \in \cC$.
        The trivial anyon, or the \emph{vacuum}, which is part of every anyon model, is denoted as $\one$.
        Let us now discuss how fusion, exchange and braiding of anyons is described.
        
        \textit{Fusion} is the process of bringing two anyons close together such that they behave as a third anyon within the same anyon model.
        We denote fusion by the $\times$ operation.
        Here, we concentrate on Abelian anyon models where fusion outcomes are unique. 
        The fusion rules of Abelian anyon models are of the form $\a \times \b = \c$ for $\a,\b,\c \in \cC$.
        Fusion with the vacuum anyon is trivial, $\a \times \one = \a$.
        Furthermore, each anyon has an anti-particle with which it fuses to the vacuum.
        In the topological phases of interest here, namely the color code and toric code, all anyons are their own anti-particles, such that $\a \times \a = \one$.
        
        \textit{Exchanging} two identical Abelian anyons results in a complex phase.
        If we exchange the position of a pair of $\a$ anyons, we denote the obtained phase as $\theta_\a$.
        This phase is called \emph{spin} and for qubit stabiliser codes takes values $\pm 1$.
        Anyons for which the self-exchange results in a $+1$ ($-1$) phase are called \emph{bosons} (\emph{fermions}).

        \textit{Braiding} is the process of moving one anyon around another before returning it to its initial position.
        Let us say we braid anyon $\a$ around a second stationary anyon $\b$, this process results in a phase called \emph{monodromy}, denoted as $M_{\a,\b}$.
        For the color code and toric code phases, the monodromy can only take values $M_{\a,\b} = \pm 1$, allowing us to use a short-hand formulation calling braiding either trivial ($M_{\a,\b} = + 1$) or non-trivial ($M_{\a,\b} = - 1$).

        In fact, it is worth pointing out that the self consistency conditions governing Abelian anyon models lead to a number of redundancies in the data.
        The identity $M_{\a,\b} = \theta_\a\theta_\b/\theta_{\a \times \b}$, for example, lets 
        us determine the self-exchange statistics of any anyon by decomposing it into two different anyons whose spin and relative braid statistics are known.
        
        We can define a microscopic, local Hamiltonian, composed of commuting Pauli interaction terms acting on qubits such that the ground space is the common $+1$ eigenspace of all the terms that give rise to a topological phase.
        We say that violated interaction terms occupy quasi-particle excitations.
        These excitations may behave like anyons.
        Unitary rotations create and transport anyons, allowing us to study their fusion, self-exchange and braiding explicitly~\cite{Bombin14}.
        In the next two sections, we study two specific examples of Hamiltonian models, the color code and the toric code.
        
    \subsection{The color code}
        \label{sec:PrelimCC}
	
        In what follows, we introduce a lattice model realising the color code~\cite{Bombin06} before turning our attention to the anyonic excitation it hosts.
        %
        %
        We employ the language of stabiliser codes~\cite{Gottesman97} to describe the microscopic lattice models realising topologically ordered phases.
        Stabiliser codes are defined by an Abelian subgroup of the Pauli group that we call the stabiliser group.
        Importantly, the common $+1$ eigenspace of the elements of the stabiliser group specifies the code space of a code.
        The group therefore does not include $-1$ as this operator has only negative eigenvalues.
        We measure stabiliser operators to detect errors.
        
\begin{figure}[tb]
	        \centering
	        \includegraphics[width=1.0\linewidth]{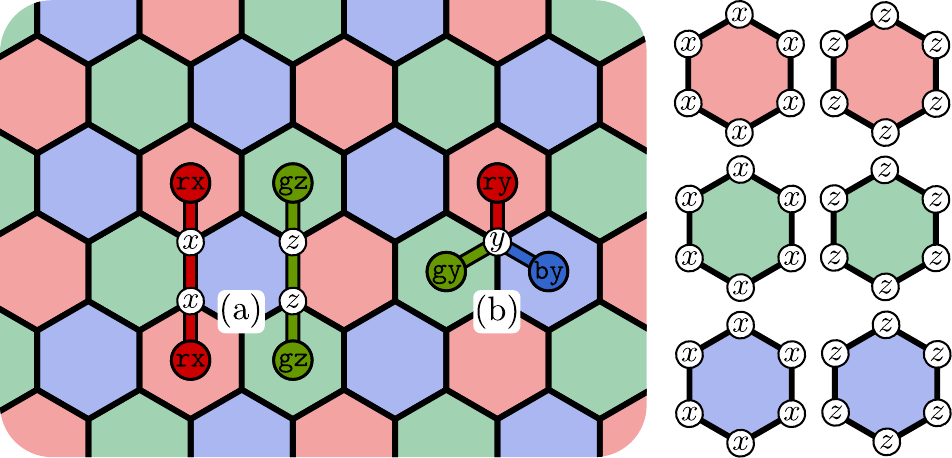}
	        \caption{
	            The lattice model of a color code.
	            A physical qubit is placed on every vertex.
	            As shown on the right, each plaquette hosts two stabilisers, one acting in the $X$-basis and one in the $Z$-basis.
	            \textbf{(a)}~Pauli rotations on the two qubits of an edge creates a pair of anyons on the plaquettes at the endpoints of said edge. In the displayed examples we show a pair of $\rx$ ($\gz$) anyons created at the end points of a string of Pauli-$X$ (Pauli-$Z$) rotations supported on a red (green) edge.
	            \textbf{(b)}~A single qubit Pauli rotation creates a triplet of anyons. This process can also be understood as fusing two anyons into the third anyon of the triplet.
            }
	        \label{fig:CCLattice}
	    \end{figure}

        The color code can be defined on any lattice which is three-colourable with respect to its faces.
        We use the colours red, green and blue.
        It is also helpful to assign colors to the edges of the lattice.
        The colour of an edge is given by the colour of the faces it connects.
        In this work, we focus on the hexagonal lattice, as shown in Fig.~\ref{fig:CCLattice}.
        
        To specify the color code we assign physical qubits to the vertices of the three-colourable lattice, and stabilisers are associated to the plaquettes of the lattice. We index the plaquettes with the symbol $p$.
        Each plaquette hosts two stabiliser generators, one of which acts in the $X$-basis on all qubits supported on the plaquette, $s_X^p$, the other in the $Z$-basis, denoted $s_Z^p$. 
        We will refer to them as $X$-type or $Z$-type stabilisers respectively.
        Note, by multiplying the two stabiliser generators on any plaquette, we obtain a $Y$-type stabiliser at plaquette $p$, i.e., $s_Y^p = s_Z^p s_X^p$.
	    
	    Given the stabilisers of a code, we can introduce a commuting Hamiltonian consisting of the negative sum of a set of stabiliser generators.
	    For topological stabiliser codes, these Hamiltonian terms commute and can be chosen to be geometrically local.
	    In the color code, we usually pick the plaquette terms $s_X^p$ and $s_Z^p$ acting in the Pauli-$X$ and Pauli-$Z$ basis on all qubits surrounding a plaquette $p$, as depicted in Fig.~\ref{fig:CCLattice}.
	    This yields the following Hamiltonian,
	    \begin{equation}
	        H_{CC} = - \sum_p s_X^p - \sum_p s_Z^p.
	    \end{equation}
	    The ground state space of this Hamiltonian coincides with the code-space of the stabiliser code.
	    Excited eigenstates are reached when some plaquette terms are violated.
	    These states differ from states in the ground state space by Pauli-rotations on single qubits.
	    We associate the violated plaquette terms with anyonic excitations and say they are created by said Pauli rotations.

        Let us now discuss color code anyons and their properties.
        The color code phase contains 16 anyons.
        Apart from the vacuum excitation, there are 9 non-trivial bosons.
        Each boson has one of three color labels, \r, \g, \b, as well as one of three Pauli labels, \x, \y, \z.
        The labels are given by the colour of the violated plaquette and the basis of the Pauli rotation which create and move the anyons.
        As an example, we say that a red plaquette whose stabiliser(s) are violated by an applied Pauli-$X$ rotation is associated with an anyon labelled \rx.
        See Fig.~\ref{fig:CCLattice}~(a), for example.
        
        Throughout this work, we find it instructive to order the $9$ color-code bosons in a $3\times3$-table~\cite{Kesselring18}, such that bosons in any given row (column) share their Pauli- (colour-) label.
        This table is referred to as the color-code boson table
        \begin{equation}
        	\begin{tabular}{ c | c | c }
                 \rx & \gx & \bx \\ \hline    
                 \ry & \gy & \by \\ \hline    
                 \rz & \gz & \bz 
            \end{tabular}
            ~ .
        	\label{eq:BosonTable}
        \end{equation}
        Let us review how the data of the color-code anyons are captured by the boson table~\eqref{eq:BosonTable}.
        All color-code anyons are their own anti particles, hence two identical anyons fuse to the trivial anyon.
        For example we get $\rx \times \rx = \one$.
        Bosons which lie in the same row or column, fuse to the third boson in said row or column.
        Examples of such fusions are $\bx \times \by = \bz$ or $\gy \times \by = \ry$. 
        Braiding between two bosons from the same row or column is trivial, e.g.,~$M_{\rx,\ry} = M_{\gz,\bz} = +1$, whereas bosons from differing rows and columns braid non-trivially, e.g.,~$M_{\gx,\rz} = -1$.

        Fusing two bosons which differ in both Pauli- and colour-label, results in one of six fermions
        \begin{align}
            \fone &= \rx \times \bz = \ry \times \gz = \gx \times \by , \\
            \ftwo &= \rz \times \bx = \ry \times \gx = \gz \times \by , \\
            \fthree &= \bz \times \gy = \gx \times \rz = \bx \times \ry , \\
            \ffour &= \rz \times \gy = \gx \times \bz = \rx \times \by , \\
            \ffive &= \rx \times \gy = \bx \times \gz = \by \times \rz , \\
            \fsix &= \bx \times \gy = \rx \times \gz = \ry \times \bz .
            \label{eq:CCFermions}
        \end{align}
        Writing color-code fermions in terms of their composite bosons lets us infer all of their relevant data.
        For example, using rules we have defined in the previous section, (Sec.~\ref{sec:PrelimAnyons}) we find that fermions labelled with an even number label braid trivially with fermions labelled by an odd number, e.g.,~$M_{\fone,\ffour} = +1$, whereas two distinct even (odd) fermions braid non-trivially, e.g.,~$M_{\fthree,\ffive} = -1$.

\begin{figure}[tb]
	        \centering
	        \includegraphics[width=1.0\linewidth]{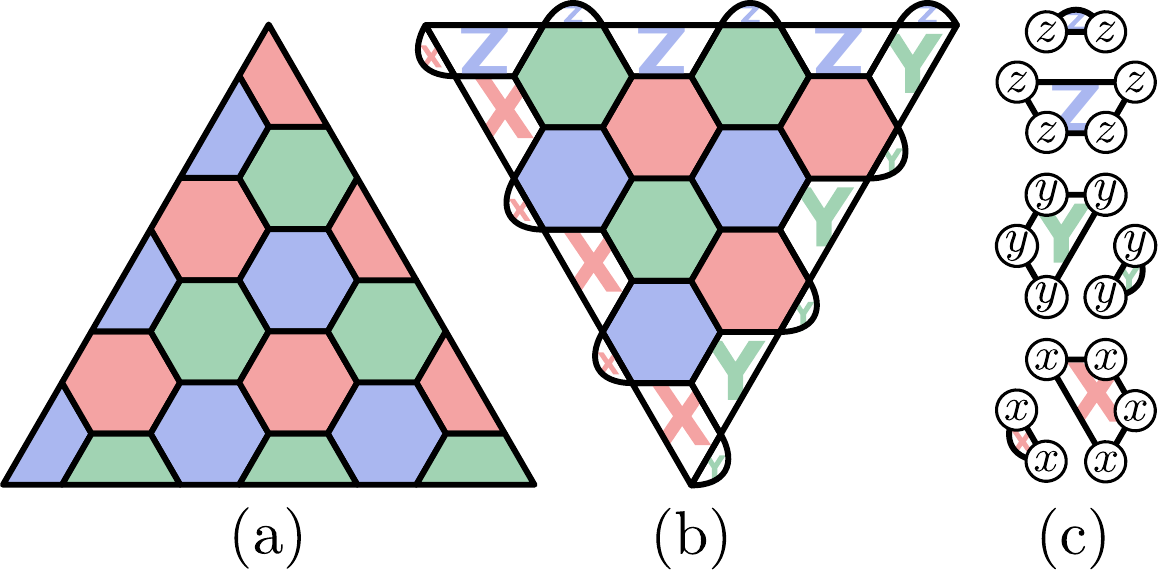}
	        \caption{
	            The triangular color code with \textbf{(a)} three distinct coloured boundaries and \textbf{(b)} with three Pauli-type boundaries, where coloured boundaries are defined in Ref.~\cite{Bombin06} and Pauli boundaries are defined in Ref.~\cite{Kesselring18}. We re-derive all of these boundary in subsection~\ref{sec:Cond2dMax} in terms of anyon condensation.
	            Each code patch encodes one logical qubit with a code distance of $d=7$ and $d=6$, respectively.
	            Plaquettes with a single coloured letter host only one stabiliser generator acting in the basis indicated by the letter, see~\textbf{(c)}.
            }
	        \label{fig:TriangularCCs}
	    \end{figure}

        To store quantum information in the color code, we can either place the code on a topologically non-trivial manifold, or we can introduce boundaries. We define the different boundary types of the color code in terms of anyon condensation in subsection~\ref{sec:Cond2dMax}.
        The prototypical example of an quantum error-correcting code in the color-code phase is the triangular color code, depicted in Fig.~\ref{fig:TriangularCCs}~(a).
        We associate a code distance $d$ with the code, given by the weight of its least-weight logical operator.
        The code depicted in Fig.~\ref{fig:TriangularCCs}~(a), for example, has a code distance $d=7$. Importantly, we have a family of codes that can be parameterised by their code distance that diverges.
        Assuming we have access to a sensible decoder~\cite{Wang10, Landahl11, Bombin12, Sarvepalli12, Delfosse17, Aloshious18, Li18, Tuckett19, Li20, Chubb21, Sabo21, Sahay2022}, we can correct all error configurations that affect fewer than some number of qubits that diverges in the code distance $d$.
        As we discuss in Sec.~\ref{sec:Cond2dMax}, logical operators in the color code appear as strings along non-trivial paths.
        In codes with boundaries, such paths may connect distinct boundaries or enclose punctures.
        This means that by increasing the system size, i.e., by separating boundaries further apart or enlarging punctures, we can increase the code distance.
        This in turn means that we can tolerate more errors. Assuming errors happen sufficiently rarely, we can decrease the probability of a logical error occurring arbitrarily close to zero by increasing the code distance.
        
	\subsection{The toric code}
        \label{sec:PrelimTCandMC}
	    
	    In this section, we discuss the well known toric code phase; a topologically ordered phase that is closely related to the color code.
        The toric code, introduced by Kitaev in Ref.~\cite{Kitaev03}, is widely regarded as the prototypical example of a topological stabiliser code.
        Its associated phase is referred to as the toric code phase, or simply the toric code.
        Here, we will briefly review its anyonic excitations and introduce an example of a microscopic lattice model in the toric code phase.

        \begin{figure}[tb]
	        \centering
	        \includegraphics[width=0.85\linewidth]{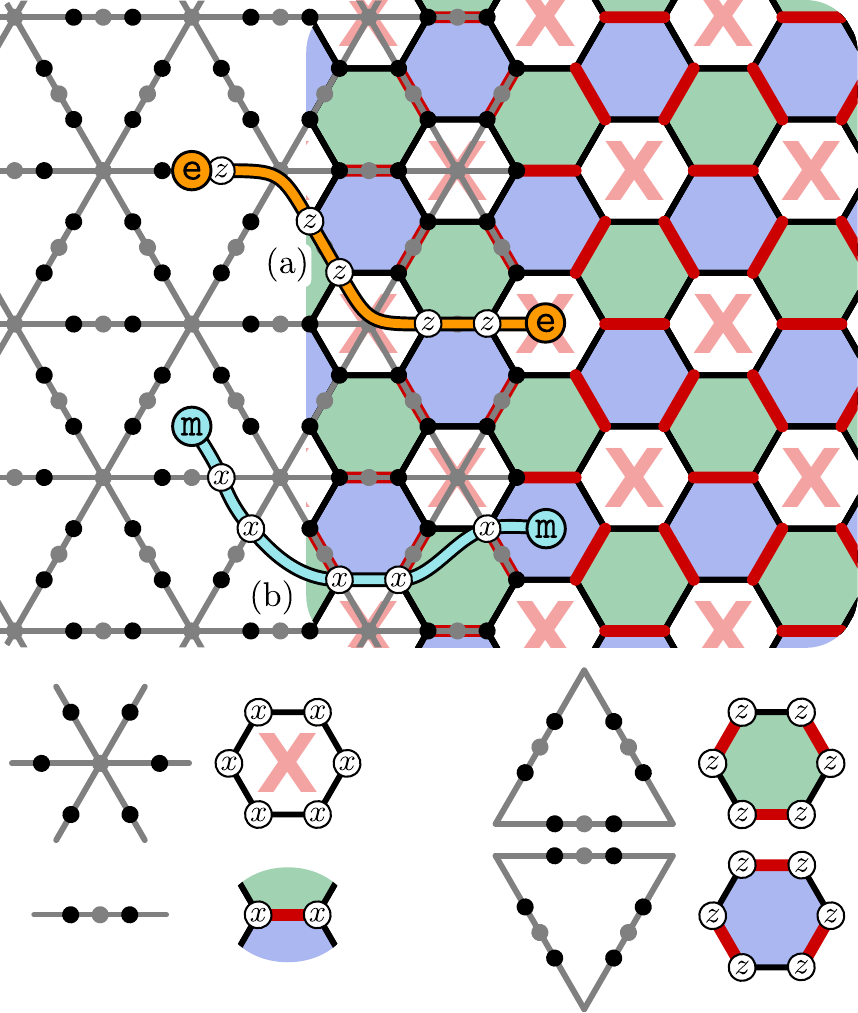}
	        \caption{
	            The lattice model of the toric code.
	            The left-hand side of the figure shows the triangular lattice with additional two-valent vertices, the right-hand side shows an equivalent representation of the same model on a three-colourable hexagonal lattice. To show their equivalence, the lattice on the left overlays the lattice on the right in the middle of the figure.
	            On the right-hand side of the figure, a physical qubit is placed on every vertex of the hexagonal lattice to the right of the figure.
	            Red edges host a two-body $XX$-stabiliser.
	            As indicated by the coloured letters, red plaquettes are stabilised by an $X$-type stabiliser.
	            Blue and green plaquettes host a $Z$-type stabiliser.
	            Note how the $XX$ term on red edges surrounding a blue or a green plaquette multiply to the $X$-type stabiliser on said plaquette.
	            Since they host both types of stabilisers, we colour the green and blue plaquettes fully. The \e and \m excitations are created and moved by $Z$- and $X$-rotations, as is shown in \textbf{(a)} and \textbf{(b)}. On the left-hand side, we show the same model defined on the standard toric code lattice where qubits are placed on the lattice edges~\cite{Kitaev03}. We mark qubits on the grey lattice with large black vertices. Pauli-X (star) stabilisers associated to the vertices of the grey lattice, and Pauli-Z (plaquette) stabilisers to the faces of the model.
                We give a key for the different types of stabiliser operators at the bottom of the figure.
            }
	        \label{fig:TC_lattice_excitations}
	    \end{figure}
        %

	    There are four anyons in the anyon model of the low-energy theory of the toric code.
	    The particle $\one$ represents the vacuum, or the trivial anyon.
	    Particles $\e$ and $\m$ are bosonic anyons with $\theta_\e = \theta_\m = +1$, and $\f$ is a fermion with $\theta_\f = -1$.
	    The anyons fuse as follows, $\e \times \e = \m \times \m = \f \times \f = \one$ and $\e \times \m = \f$.
	    Any two non-identical, non-trivial anyons braid non-trivially, e.g.,~$M_{\e,\m} = -1$.
	    
Fig.~\ref{fig:TC_lattice_excitations} shows a construction for the toric code which is particularly helpful for our discussion throughout this work.
	    We start with a hexagonal lattice, place a physical qubit on each vertex and colour the plaquettes as in Fig.~\ref{fig:TC_lattice_excitations}.
	    Each plaquette hosts a stabiliser generator acting on the surrounding qubits in a Pauli-basis defined by the colour of the plaquette.
	    Green and blue plaquettes host weight-6 $Z$-type stabilisers.
	    Red plaquettes and red edges host weight-6 and weight-2 $X$-type stabilisers, respectively.
	    The same stabilisers are obtained when following Kitaev's original toric code construction~\cite{Kitaev03} on a triangular lattice decorated with additional 2-valent vertices on every edge, as shown in Fig.~\ref{fig:TC_lattice_excitations}. In Kitaev's original description of the toric-code model, qubits are placed on the edges of some arbitrary lattice. Then, Pauli-X stabilisers are associated to the vertices of the lattice and Pauli-Z stabilisers are associated to the faces of the lattice. Specifically, Pauli-X vertex stabilizers (Pauli-Z plaquette stabilisers) are the product of Pauli-X (Pauli-Z) terms on qubits associated to edges adjacent to their respective lattice vertex (plaquette).
	    See also the matching code construction~\cite{Wootton15} which describes a construction for microscopic stabiliser models in the toric code phase based on Kitaev's honeycomb model~\cite{Kitaev06}, including the model described here.
        The excitations on red plaquettes and edges are \e anyons.
        The blue and green plaquettes host the \m anyon.

    \subsection{Unfolding the color code}
        \label{sec:PrelimUnfolding}
    
        The color code is equivalent to two decoupled layers of the toric code~\cite{Bombin12,Kubica15,Criger16}, meaning that the anyon model from two decoupled layers of toric codes is equivalent to the anyon model of the color code.
        One way of mapping the color code anyon labels to the labels given by two layers of toric codes is
        \begin{equation}
            \begin{tabular}{ c c | c | c }
                \vspace{-5pt} & \r & \g & \b  \\ 
                & & & \\
                \x ~ & $\e\one$ & $\e\e$ & $\one\e$ \\ \hline    
                \y ~ & $\e\m$ & $\f\f$ & $\m\e$ \\ \hline    
                \z ~ & $\one\m$ & $\m\m$ & $\m\one$ 
            \end{tabular} ~ ,
            \label{eq:AnyonUnfolding}
        \end{equation}
        where $\one, \e, \m, \f$ are the toric code anyons and their position in the tuple \a\b represents on which of the two layers they live.
        We refer to \eqref{eq:AnyonUnfolding} as the standard unfolding.
        There are, however, $72$ valid ways to perform the unfolding, which can be obtained by applying one of the $72$ anyon permuting symmetries of the color code~\cite{Yoshida15,Kesselring18} to the
        standard mapping~\eqref{eq:AnyonUnfolding}.
        Furthermore, we can identify the six fermions of the color code as follows,
        \begin{align}
            \fone &= \f\one, \qquad
            \fthree = \e\f, \qquad
            \ffive = \m\f, \\
            \ftwo &= \one\f, \qquad
            \ffour = \f\e, \qquad
            \fsix = \f\m.
            \label{eq:FermionUnfolding}
        \end{align}

    \subsection{Error correction in \texorpdfstring{$(2+1)$}{(2+1)}D space-time}
        \label{sec:Prelimspace-time}
        
        In two-dimensional topological stabiliser codes, we detect errors by measuring stabiliser generators to obtain a list of violated stabilisers~\cite{Kitaev03, Dennis02}.
        In reality, however, these stabiliser measurements may be imperfect and give incorrect outcomes.
        To achieve tolerance to noise in the presence of measurement errors, we must repeat the measurements multiple times~\cite{Dennis02, Wang2003confinement}.
        This transforms the space in which the stabiliser violations live into a $(2+1)$-dimensional space-time.
        
        As we have described violated stabilisers can be associated with anyonic quasi-particles.
        This association carries over to the case where we consider the full space-time picture of the stabiliser readouts.
        In fact, from a condensed matter perspective, this is very natural, as all two-dimensional topologically ordered phases and processes happening therein are described in as a $(2+1)$-dimensional space-time.
        In what follows, we make this connection explicit.
        
\begin{figure}[tb]
	        \centering
	        \includegraphics[width=1.00\linewidth]{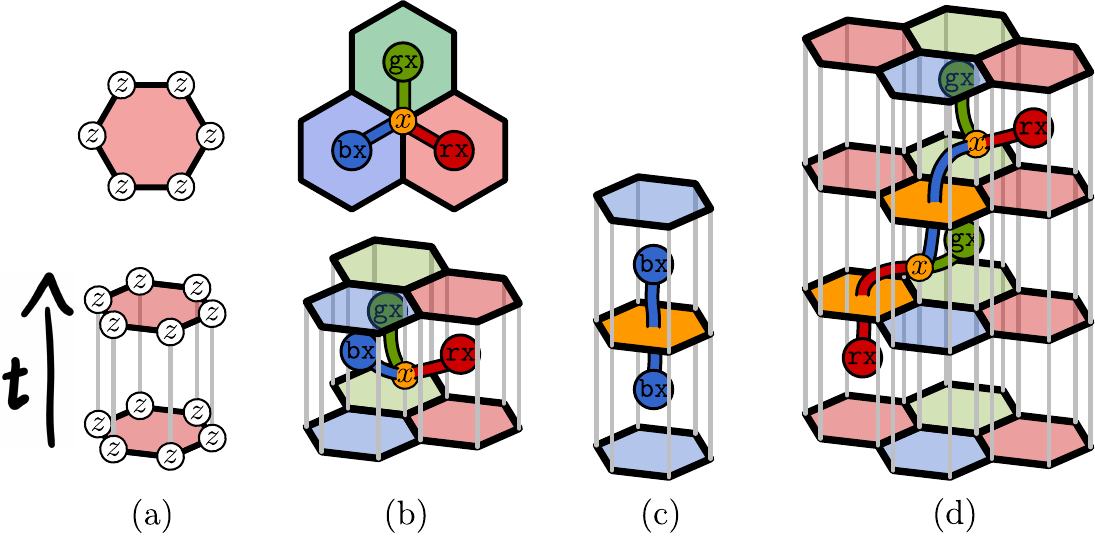}
	        \caption{
	            When performing multiple rounds of stabiliser measurements on a two-dimensional topological error correcting code, one has to consider the $(2+1)$-dimensional space-time of the process.
	            \textbf{(a)}~stabilisers (top) turn into detection cells (bottom) consisting of measurements in two consecutive time steps.
	            A detection cell heralds a defect if the two measurements do not agree.
	            \textbf{(b)}~if an error (orange) occurs in between two consecutive rounds of measurements, surrounding detection cells light up.
	            \textbf{(c)}~measurement errors (orange) light up the detection cells below and above.
	            \textbf{(d)}~general error configurations move anyons through the $(2+1)$-dimensional space-time.
	            In the depicted process, two measurement errors and two errors on physical qubits create a total of four anyonic charges.
            }
	        \label{fig:space-timeStabiliserViewPoint}
	    \end{figure}

        We begin by replacing the stabiliser generators associated with plaquettes with \textit{detection cells}~\cite{Dennis02, Wang2003confinement,  Gidney2021stim}.
        For simplicity, we assume that all stabiliser readouts are performed in parallel. Each detection cell is associated with one position in space-time $(s,t)$, given by the location of a stabiliser generator $s$ and a time-step $t$.
        A detection cell compares the outcome of the measurement of stabiliser $s$ during the $t$-th round with the result obtained for $s$ in the $(t-1)$-th round,
        see Fig.~\ref{fig:space-timeStabiliserViewPoint}~(a).
        Thus, they detect changes in measurement outcomes which might be caused by errors on the physical qubits they support or when a faulty measurement result is obtained.
	    
	    Now, we associate anyonic quasi particles with (single or sets of) violated stabiliser cells.
	    A detection cell $(s,t)$ detects errors occurring on physical qubits in the support of $s$ between time steps $t-1$ and $t$, see Fig.~\ref{fig:space-timeStabiliserViewPoint}~(b).
	    This is exactly analogous to the purely two-dimensional viewpoint, like we show it in Fig.~\ref{fig:CCLattice}, where we interpret errors as creating sets of anyons with neutral total charge.
	    Importantly, a detection cell $(s,t)$ also detects measurement errors affecting $s$ in round $t-1$ or in round $t$.
	    Thus, a single measurement error occurring when measuring the stabiliser at $s$ in round $t$ thus violates two detection cells, $(s,t-1)$ and $(s,t)$, see Fig.~\ref{fig:space-timeStabiliserViewPoint}~(b).
	    Such a process can be though of as a pair-creation in the time direction of an anyon associated with the violation of $s$.
	    In general, strings are composed of both, measurement errors as well as errors on physical data qubits, and create point-like anyons in space-time.
        As an example, we shown a world-line of a color code anyon in Fig.~\ref{fig:space-timeStabiliserViewPoint}~(d).
        
        Thus, a two-dimensional topological error-correcting code where measurements are repeated over time correspond to a microscopic model realised in $(2+1)$-dimensional space-time.
        This allows us on the one hand to explicitly study topological processes taking place in time using topological stabiliser codes as toy models.
        And on the other hand, we can draw on the wealth of results obtained in the mathematical study of topologically ordered $(2+1)$-dimensional phases of matter and apply them in real-world error correcting schemes.
        With this in mind, let us turn our attention to anyon condensation.
        
        At the end of Sec.~\ref{sec:PrelimCC}, we touched on the fault tolerance of the color code.
        Let us here make this definition more precise while including errors affecting the measurements of stabiliser terms.
        As we have seen, errors in the $(2+1)$-dimensional space-time can be interpreted as segments of strings with anyons at their end-points.
        Crucially, due to measurement errors, these error strings can also travel along the temporal direction.
        
        Logical errors occur when sufficiently many errors appear along non-trivial paths, connecting distinct boundaries~\cite{Dennis02, Strikis21}.
        Importantly, these can be spatial boundaries (see Sec.~\ref{sec:Cond2dMax}) as well as temporal boundaries (see Sec.~\ref{sec:CondTimeMax}).
        This means that we need to include errors affecting measurement outcomes into our definition of the code distance $d$.
        The code distance $d$ of a code is the lowest number of errors -- including measurement errors -- which results in a non-detectable and non-trivial logical error.
        In this work, we call a protocol fault tolerant, if $d$ grows extensively with the system size.
        In other words, by increasing the system size in the two spatial and in the temporal direction, we can reach an arbitrarily low logical error rate assuming a suitable decoder and physical error rate below threshold.

\section{Anyon condensation}
\label{sec:AnyonCondensation}

In this section, we will walk through the theory of anyon condensation in Abelian anyon models~\cite{Burnell18,Levin13}, focusing on the color-code phase as a concrete example.
In particular, we show that anyons in the condensed phase of an Abelian anyon model can be identified with cosets of a \textit{bosonic subgroup} of the anyon model, see Sec.~\ref{sec:CondTheory}.
Based on properties of the condensed subgroup, we sort condensation into three types, Sec.~\ref{sec:CondTypes}. 
We specifically describe domain walls in the context of anyon condensation in Sec.~\ref{sec:CondDWs}.
In later sections, we show that the theory of condensation we develop here underlies the physics of topologically protected operations on the encoded logical information before and after some code transformation.
To describe the details of code deformations, in Sec.~\ref{sec:CondStabs} we show how to implement anyon condensation at the microscopic level for topological stabiliser models, where the string operators of the anyons are Pauli operators.
We find that, in this class of models, condensation can be implemented by measuring low-weight string operators on a code state of what we call the \textit{parent code}.

\subsection{Condensation in Abelian anyon models}
\label{sec:CondTheory}

    Anyon condensation is a mechanism to relate certain anyon models to each other.
    Given a parent Abelian anyon theory $\cC$, a \textit{$\cC$-condensate} $\cC_\cB$ is obtained by identifying a subgroup of bosons $\cB \subset \cC$ with the trivial charge.
    An anyon in the condensed theory $\cC_\cB$ is related to a coset of anyons $\a\cB \subset \cC$.

    In particular, we start by choosing a bosonic subgroup, meaning a subgroup of bosons that is closed under fusion and that contains only bosons with trivial mutual braid statistics.
    Next, we identify this subgroup $\{\b_1,\b_2,\ldots\} = \cB\subset\cC$ with the trivial charge,
    \begin{align}\label{eq:CondCondensationCond}
        \b \equiv \one ~ \forall ~ \b \in \cB.
    \end{align}
    Hence, a pair of anyons $\a, \b \in \cC$ that differ by fusion with some elements of $\cB$ become identified in the condensate.
    In the remainder if the paper, we use the notation
    \begin{align}\label{eq:eq:CondIdentCond}
        \a \simeq \b \qq{iff} \a\cB = \b\cB.
    \end{align}
    
    Let us look at the different ways anyons of the parent theory  can be affected by condensation.
    The anyons of the parent theory $\cC$ fall into two classes.
    \begin{itemize}
        \item \textbf{Deconfined:} Any anyon $\a$ that braids trivially with all elements in $\cB$, is \textit{deconfined}
        \begin{align}\label{eq:CondConfinemnetCond}
            \a \equiv \a\cB \qq{iff} M_{\a,\b} = 1 ~ \forall ~ \b \in \cB.
        \end{align}
        The anyon \a from the parent theory then defines the anyon $\a\cB$ in the condensed theory.
        
        \item \textbf{Confined:} If $\a\in\cC$ braids non-trivially with at least one element in $\cB$, the topological spin of $\a\cB$ becomes ill-defined and $\a$ therefore becomes \textit{confined}.
        The object $\a\cB$ is not an anyon in the condensed theory.
    \end{itemize}
    
    Anyons of a condensate $\cC_\cB$, defined by condensing a bosonic subgroup $\cB\subset\cC$, are in one-to-one correspondence with a subset of cosets $\{\a\cB\,|\,M_{\a,\b}=1\,\forall~\b\in\cB\}$.
    The modular data of the condensate is given by the parent theory
    \begin{align} \label{eq:ConsCond}
        \theta_{\a\cB} = \theta_{\a} \qcomma ~ M_{\a\cB,\c\cB} = M_{\a,\c}.
    \end{align}
    The constraint on $\cB$ being a set of bosons, all of which braid trivially with $\a$ and $\b$, ensures that the topological numbers of $\a$ and $\b$ are the same.
    Furthermore, it is apparent that $\cB$ must be closed under fusion.
    
    Throughout this work, we will make use of symmetries in a given parent theory to relate different condensates.
    A symmetry of anyon model $\cC$ is a permutation of the anyon labels that leaves the anyonic data invariant.
    The group formed by all these permutations is called the automorphism group of the anyon model, $\Aut(\cC)$.
    In general, a symmetry in the parent maps a bosonic subgroup $\cB$ to a (potentially different) bosonic subgroup $\cB'$.
    The fact that it is a symmetry of the parent indicates that $\cC_\cB$ is in the same phase as $\cC_{\cB'}$.
    Symmetries that preserve $\cB$ turn into symmetries in the condensate.
   
\subsection{Types of condensation}
\label{sec:CondTypes}
    In this section, we define different types of anyon condensation.
    We use the color-code anyon model to exemplify different condensation mechanisms.

\subsubsection{Maximal condensation}
\label{sec:CondMax}
    We maximally condense a parent theory if we condense a \textit{Lagrangian subgroup} $\cL$ \cite{Levin13, davydov2017149} of bosons. A Lagrangian subgroup is a maximal bosonic subgroup, i.e., there exist no anyon of $\cC$ not included in $\cL$ that braids trivially with all elements in $\cL$.
    When a Lagrangian subgroup is condensed, all non-trivial anyons get confined and the condensate is equivalent to the trivial phase.

    The color code has six Lagrangian subgroups, each possessing three non-trivial bosons~\cite{Kesselring18}.
    They can be associated with either one of the three colour labels, or one of the three Pauli labels
    \begin{subequations}\label{eq:CC_Lagrangian}
    \begin{align}
        \cL_\r^{CC} &= \{\one,\rx,\ry,\rz\}, \\
        \cL_\g^{CC} &= \{\one,\gx,\gy,\gz\}, \\
        \cL_\b^{CC} &= \{\one,\bx,\by,\bz\}, \\
        \cL_\x^{CC} &= \{\one,\rx,\gx,\bx\}, \\
        \cL_\y^{CC} &= \{\one,\ry,\gy,\by\}, \\
        \cL_\z^{CC} &= \{\one,\rz,\gz,\bz\}.
    \end{align}
    \end{subequations}
    
    The Lagrangian subgroups are expressed using boson tables~\eqref{eq:BosonTable}.
    Condensed anyons are marked by a 
    black circle $\blackcirc$, confined charges with a $\cross$.
    The top three rows show $\cL_\r^{CC}$, $\cL_\g^{CC}$ and $\cL_\b^{CC}$, and the bottom three rows show $\cL_\x^{CC}$, $\cL_\y^{CC}$ and $\cL_\z^{CC}$, respectively.
        \begin{alignat*}{5}
                &\begin{tabular}{ c | c | c }
            \blackcirc & \cross & \cross \\ \hline
            \blackcirc & \cross & \cross \\ \hline 
            \blackcirc & \cross & \cross 
        \end{tabular}~,
        ~~
        &&\begin{tabular}{ c | c | c }
            \cross & \blackcirc & \cross \\ \hline
            \cross & \blackcirc & \cross \\ \hline 
            \cross & \blackcirc & \cross 
        \end{tabular}~,
        ~~
        && \begin{tabular}{ c | c | c }
            \cross & \cross & \blackcirc \\ \hline
            \cross & \cross & \blackcirc \\ \hline 
            \cross & \cross & \blackcirc 
        \end{tabular}~,
        \\
        &\begin{tabular}{ c | c | c }
            \blackcirc & \blackcirc & \blackcirc \\ \hline
            \cross & \cross & \cross \\ \hline    
            \cross & \cross & \cross 
        \end{tabular}~,
        ~~
        &&\begin{tabular}{ c | c | c }
            \cross & \cross & \cross \\ \hline
            \blackcirc & \blackcirc & \blackcirc \\ \hline 
            \cross & \cross & \cross 
        \end{tabular}~,
        ~~
        &&\begin{tabular}{ c | c | c }
            \cross & \cross & \cross \\ \hline
            \cross & \cross & \cross \\ \hline
            \blackcirc & \blackcirc & \blackcirc
        \end{tabular}~.
    \end{alignat*}

\subsubsection{Partial condensation}
\label{sec:CondPart}

    If the bosonic subgroup $\cB$ is not maximal, we have partial condensation.
    In this case, some of the anyons confine while some remain deconfined, depending on their braiding properties with the bosons in $\cB$.
    
    In the color code, we can chose any one of the nine non-trivial bosons to be condensed.
    In all cases, the resulting condensed phase is the toric code.
    We can depict this in the boson table~\eqref{eq:BosonTable}.
    The condensed anyon is marked with a black circle $\blackcirc$.
    We now have four deconfined color code charges; those that braid trivially with $\blackcirc$.
    Two of the deconfined charged are identified with the toric code's electric charge anyon $\e$ and marked with a $\orangecirc$.
    The other two deconfined color code anyons get identified with the toric code's magnetic flux $\m$ and marked with $\cyancirc$.
    The remaining four bosons are confined, we mark this with a $\cross$.
    There are in total 18 ways to condense the color code to the toric code.
    For each of the nine choices of condensed boson we have an additional binary choice of how to identify the deconfined charges with the toric code's \e and \m anyons.
    Some examples are shown here.
    
    \begin{align*}
        \begin{tabular}{ c | c | c }
             \blackcirc & \cyancirc & \cyancirc \\ \hline
             \orangecirc & \cross & \cross \\ \hline    
             \orangecirc & \cross & \cross 
        \end{tabular}~,
        & ~~
        \begin{tabular}{ c | c | c }
             \blackcirc & \orangecirc & \orangecirc \\ \hline
             \cyancirc & \cross & \cross \\ \hline    
             \cyancirc & \cross & \cross 
        \end{tabular}~,
        & ~~
        \begin{tabular}{ c | c | c }
             \cyancirc & \blackcirc & \cyancirc \\ \hline
             \cross & \orangecirc & \cross \\ \hline    
             \cross & \orangecirc & \cross 
        \end{tabular}~,
        \\
        \begin{tabular}{ c | c | c }
             \orangecirc & \blackcirc & \orangecirc \\ \hline
             \cross & \cyancirc & \cross \\ \hline    
             \cross & \cyancirc & \cross 
        \end{tabular}~,
        & ~~ \qquad ~~
        \hdots
        & ~~
        \begin{tabular}{ c | c | c }
             \cross & \cross & \cyancirc \\ \hline
             \cross & \cross & \cyancirc \\ \hline    
             \orangecirc & \orangecirc & \blackcirc 
        \end{tabular}~.
    \end{align*}
    
    The different ways of obtaining the toric code through condensation in the color code can be related to the color-code symmetries~\cite{Yoshida15,Kesselring18}, which we will discuss in the context of trivial condensation.

\subsubsection{Trivial condensation}
\label{sec:CondTriv}

    \textit{Trivial condensation} is where no non-trivial boson is condensed.
    The resulting phase after trivial condensation is the same as the initial phase.
    All types of trivial condensation are in one-to-one correspondence with symmetries of the parent theory which are given by the automorphism group of the anyon model, $\Aut(\cC)$.
    These are all the possible ways of relabelling the anyons such that the anyon data is preserved. We will discuss the consistency conditions that such a relabelling has to fulfil in the upcoming section in more detail.

    The color code symmetries can be read directly from the boson table, as discussed in detail in Ref.~\cite{Kesselring18}.
    Permuting any rows or columns changes the anyon labels but leaves all anyonic data the invariant.
    Additionally, thanks to the duality between Pauli- and colour-labels, reflections on the diagonals map the anyon model back to itself.
    This leads to the automorphism group $(S_3 \times S_3) \ltimes Z_2$, which contains $72$ elements.
    An example color code symmetry is the following:
    \begin{align*}
        \begin{tabular}{ c | c | c }
             \rx & \gx & \bx \\ \hline
             \ry & \gy & \by \\ \hline    
             \rz & \gz & \bz 
        \end{tabular}
        ~ \rightarrow ~
        \begin{tabular}{ c | c | c }
             \gy & \gx & \gz \\ \hline
             \by & \bx & \bz \\ \hline    
             \ry & \rx & \rz 
        \end{tabular}~.
    \end{align*}

\subsection{Domain walls and anyon condensation}
\label{sec:CondDWs}
    
    A domain wall is a one-dimensional subregion along which two phases interface.
    The type of interfaces we consider in this work correspond to \textit{gapped domain walls} between two (Abelian) topologically ordered phases $\cC$ and $\cC'$~\cite{Beigi11, KitaevKong12}.
    Anyon condensation proves to be a useful tool to study this class of domain walls.
    Furthermore, it is worth pointing out that this description is agnostic towards the space-time direction along which the phases are interfaced.
    This means that our following prescription will hold for domain walls that cut any plane of the $(2+1)$-dimensional space-time.
    
    We can interface two anyon models $\cC$ and $\cC'$ iff they share a common condensate $\cC_\cB \simeq \cC'_{\cB'}$.
    \begin{figure}[b]
        \centering
        \includegraphics[width=0.8\linewidth]{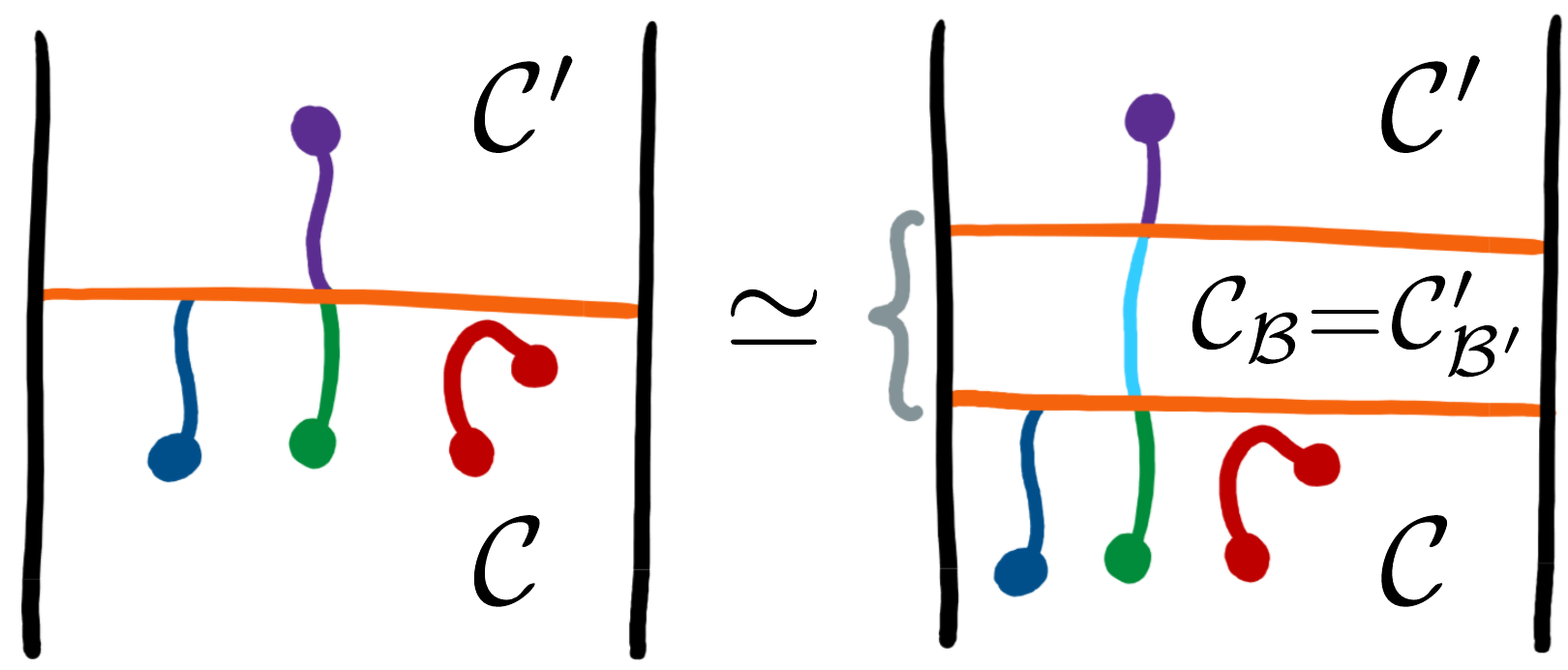}
        \caption{
            Two phases $\cC$ and $\cC'$ can be interfaced via a domain wall iff they share a common condensate $\cC_\cB = \cC'_{\cB'}$.
            A domain wall between them can be interpreted as a thin strip of the condensate phase.
        }
        \label{fig:DWCondensateEquiv}
	\end{figure}
    The domain wall can then be interpreted as a thin strip of the condensate between the two phases, as shown in Fig.~\ref{fig:DWCondensateEquiv}.
    Anyons in $\cB$ ($\cB'$) condense at the domain wall.
    Anyons that braid non-trivially with any of the condensed anyons get confined to one side of the domain wall, meaning that they cannot move without creating additional excitations.
    Anyons that braid trivially with all bosons in $\cB$ ($\cB$') are deconfined and can move through the domain wall.
    For this reason, in this context we call them mobile.

    We can derive consistency conditions on how the anyonic data on one side of the domain wall relates to the other following Fig.~\ref{fig:DWConsistencyCond}.
    \begin{figure}[tb]
        \centering
        \includegraphics[width=0.75\linewidth]{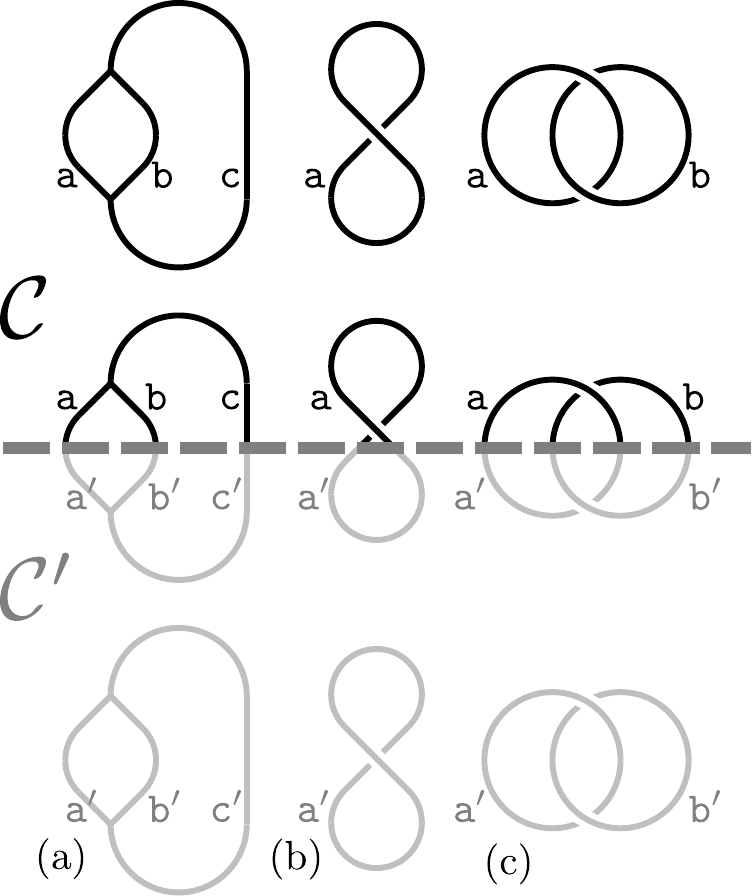}
        \caption{
            We consider two topologically ordered phases, $\cC$ and $\cC'$ and their anyons $\a, \b, \c \in \cC$ and $\a', \b', \c' \in \cC'$.
            A sequence of equivalent world-lines of anyons undergoing \textbf{(a)}~fusion, \textbf{(b)}~self-exchange and \textbf{(c)}~braiding processes on either side of a domain wall which separates $\cC$ and $\cC'$, shown by the dashed line.
            We can deform the world-lines smoothly to lie on either side of the domain wall.
            This allows us to derive consistency conditions relating the data describing the anyons on either side of the domain wall to each other.
            The second row shows the processes taking place partially on either side of the domain wall.
        }
        \label{fig:DWConsistencyCond}
    \end{figure}
    Let $\a, \b, \c$ be anyons in $\cC$ which, when moved through a domain wall get mapped to $\a', \b', \c' \in \cC'$.
    World-lines realising the three processes of fusion, self-exchange and braiding can smoothly be deformed to lie on either side of the domain wall.
    Hence, they must result in consistent outcomes. This implies that the anyonic data on either side of the domain wall must be the same, i.e., we have that
    \begin{align}
        \a\times\b=\c & \quad\Rightarrow\quad \a'\times\b'=\c'\qq{with}
        \nonumber
        \\
        \theta_{\a} = \theta_{\a'} & \qq{and} M_{\a,\b} = M_{\a',\b'} .
    \end{align}
    Furthermore, we can obtain the conditions on anyons which condense at a boundary in this fashion.
    To this end, we equate $\cC'$ to the trivial phase, which we can think of as only hosting the trivial anyon.
    This means all processes happening on this side of the domain wall must be trivial.
    From this we get closure under fusion and the triviality of self-exchange and braiding of condensible anyons, see Eq.~\eqref{eq:ConsCond}.
    
    Domain walls can be classified by the number of bosons they can condense.
    For the color code this means that the domain walls fall into three classes opaque, semi-transparent and invertible.
    All types will appear with different applications in Secs.~\ref{sec:Cond2d},~\ref{sec:CondTime} and~\ref{sec:Cond1d}.

    \textit{Opaque domain walls} are obtained when a full Lagrangian subgroup of bosons can condense from both sides.
    The corresponding condensate is the trivial phase and they have no mobile anyons.
    
    We call a domain wall \textit{semi-transparent} if certain anyons remain mobile while others condense at the domain wall.
    Semi-transparent domain walls within the same phase can be terminated within the bulk.
    
    \textit{Invertible domain walls} can only be realised within the same phase using trivial condensation.
    Anyons crossing the domain wall need to preserve their data, hence invertible domain walls are in one-to-one correspondence with the elements of the automorphism group of the anyon model, $\Aut(\cC)$.
    The endpoints of invertible domain walls are called \textit{twist defects} and behave similar to non-Abelian anyons in terms of fusion and braiding~\cite{Bombin10, Brown17, Barkeshli14, Kesselring18, bridgeman2020}.

\subsection{Anyon condensation in stabiliser models}
\label{sec:CondStabs}

    So far, we have discussed anyon condensation on an abstract level in terms of anyons.
    In what follows, we turn our attention to microscopic realisations of topologically ordered phases using stabiliser models.
    In particular, we describe how to derive a stabiliser model for a condensate from a stabiliser model of a parent phase $\cS_{\cC}$.
    Microscopically, we achieve this by adding hopping terms of the condensed bosons to the stabiliser, effectively modelling the process of anyon condensation.
    While we only consider qubit stabilisers explicitly here, the procedure can be generalised to stabilisers on qudits~\cite{Ellison21} and even for non-Pauli models~\cite{Bombin08}, if one knows the microscopic description of the string operators of the code.
    Finally, we show how to make the construction explicit by constructing the toric-code stabiliser model (see Sec.~\ref{sec:PrelimTCandMC}) from the color-code stabiliser model (see Sec.~\ref{sec:PrelimAnyons}).

    To bridge from the abstract notion of anyon condensation to microscopic models, let us reframe stabiliser operators of topological codes in terms of anyons and their string operators.
    Both the color code and toric code, as introduced in Sec.~\ref{sec:Preliminaries}, are topological lattice models whose code space corresponds to the ground space of a topologically ordered Hamiltonian.
    In any topological lattice model elementary excitations, anyons, are created at the endpoints of string operators which are supported on one-dimensional subregions, see for example Fig.~\ref{fig:CCLattice}~(a).
    This means that any closed string operator does not create any excitation and thereby leaves the ground space invariant.
    In this reading, stabilisers correspond to contractible loops of the aforementioned string operators.
    Measuring a stabiliser is equivalent to performing an interferometric charge measurement~\cite{Bombin14}.

    We now make the abstract notion of anyon condensation from Sec.~\ref{sec:AnyonCondensation} explicit in microscopic lattice models.
    In a condensate, we require that string operators that transport condensed bosons $\cB$ do not change the state, as we identify them with the trivial charge.
    We achieve this by adding all open string operators transporting bosons in $\cB$ to the stabiliser group.
    This can be achieved, for instance, by adding the set of shortest hopping terms as generators of the stabiliser group. 
    
    These string operators violate some of the original stabiliser terms. 
    To recover a commuting stabiliser group that describes the condensate, the terms of the original stabiliser group that do not commute with the new hopping terms are removed.
    This step corresponds to removing confined charges from the anyon model of the condensed phase.

    In the following, we give a recipe to construct the stabiliser group of a condensate from a parent theory $\cC$ and its stabiliser group $\cS_{\cC}$ and a bosonic subgroup $\cB \subset \cC$.
    \begin{enumerate}
        \item Define the group $\cS_{\cB}$ of open string operators for $\cB$ generated hopping terms.
        \item Remove the stabilisers from $\cS_{\cC}$ that do not commute with $\cS_{\cB}$.
        These correspond to the closed string operators of the anyons that braid non-trivially with at least one anyon in $\cB$.
        We denote the reduced stabiliser group as $\widetilde{\cS}_{\cC}$.
        \item The stabiliser group of the condensed phase $\cC_{\cB}$ is given by $\widetilde{\cS}_{\cC} \cup \cS_{\cB}$.
    \end{enumerate}
    
    This construction shows how the toric code is a condensate of the color code.
    We can view the $XX$ stabilisers on the red links, characteristic for the microscopic realisation of the toric code introduced in Sec.~\ref{sec:PrelimTCandMC}, as the hopping terms of the $\rx$ anyon in the color code.
    Following the procedure laid out above, where $\cS_\cC$ is the color code and $\cS_\cB$ is generated by the $XX$ terms on the red links of the color code lattice, the stabiliser group  $\widetilde{\cS}_{\cC} \cup \cS_{\cB}$ corresponds to closed string operators of the anyons that braid trivially with $\rx$.
    Explicitly, the color code anyons get mapped to the toric-code anyons as follows:
    $\rx \equiv \one$, $\ry \simeq \rz \equiv \e $, $\gx \simeq \bx \equiv \m$, and $\fone \simeq \fthree \equiv \f$.
    Condensing in the entire bulk results in a stabiliser model in the condensed phase.

\section{Domain walls in the color code}
    \label{sec:Cond2d}

\begin{figure}[b]
        \centering
        \includegraphics[width=0.4\linewidth]{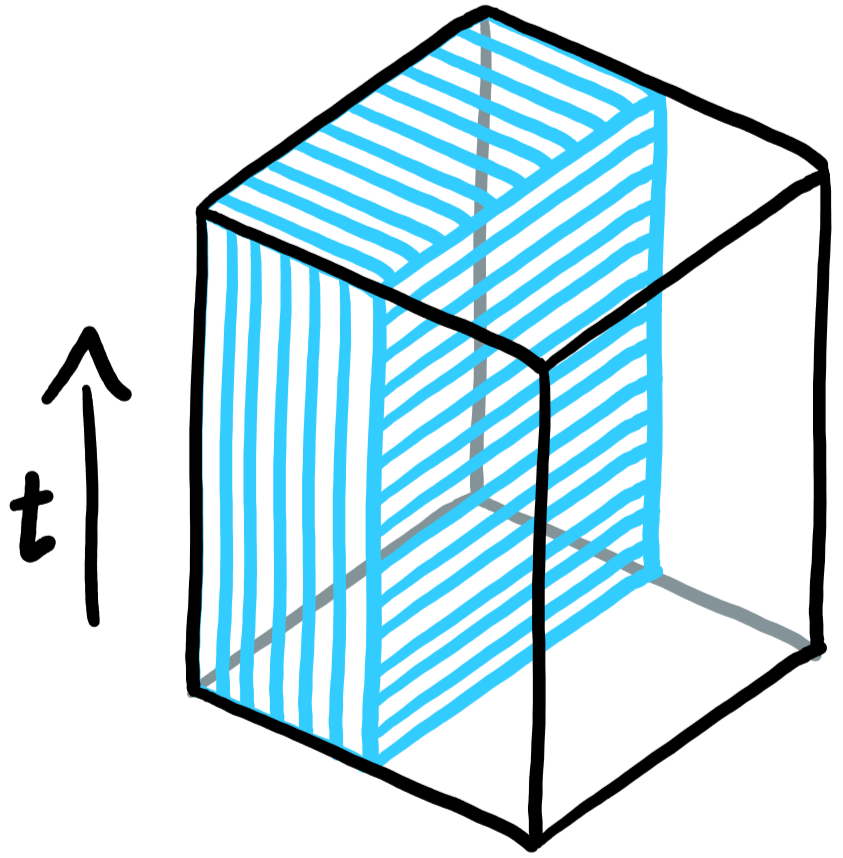}
        \caption{
            In this section, Sec.~\ref{sec:Cond2d}, we consider the condensation of anyons in given subregion (shaded in cyan).
            We keep the subregion constant in time in order to create space-like domain walls between the parent phase and the condensed phase.
        }
        \label{fig:sec4intro}
    \end{figure}

    In the coming sections, we investigate explicit microscopic examples of anyon condensation in the color-code model.
    In this section we first consider a condensed color-code phase that is spatially distinct from the color-code phase itself, see Fig.~\ref{fig:sec4intro}. 

    We consider three types of anyon condensation in this setting, where the different types of condensation are discussed in Sec.~\ref{sec:CondTypes}.
    The anyons are condensed on a sub-region of the lattice labelled $R$ where, for simplicity, we assume $R$ is a disk-shaped region, and that the color code is embedded on a manifold with a topology equivalent to that of a sphere. Upon completing the condensation operation, we obtain a domain wall at the boundary of $R$, denoted $\partial R$. 
    Applying maximal, partial or trivial condensation transforms $R$ to lie within the trivial phase, the toric code phase or the color code phase, respectively.

In Sec.~\ref{sec:Cond2dMax} we investigate condensing region $R$ to the trivial phase. This enables us to re-derive qubit encodings with punctures~\cite{Bombin09CodeDefo, Fowler11twoDimensional, Kesselring18} in terms of anyon condensation. In Sec.~\ref{sec:Cond2dPart} we consider condensing region $R$ to obtain a toric code phase. This allows us to introduce new types of domain walls between these distinct topological phases. We also introduce the notion of a semi puncture, which enables us to demonstrate new code deformations between different puncture encodings. To give a clearer perspective on semi punctures, we also reinterpret these objects in terms of the unfolded picture~\cite{Bombin12, Kubica15, Kesselring18}. For completeness, in Sec.~\ref{sec:Cond2dTriv}, consider condensing region $R$ onto the color code inself. This enables us to incorporate known transparent domain walls and twist defects for the color code~\cite{Yoshida15, Kesselring18} into our theory of anyon condensation.

    \subsection{Boundaries to the vacuum}
        \label{sec:Cond2dMax}
        
        Here we describe the boundaries between the color code and the vacuum using the language of anyon condensation.
        We show how boundaries can be used to encode quantum information in a robust manner as logical qubits.
        Finally, we discuss the structure of the logical Pauli operators.
        We give a physical interpretation of them, both as unitary operators acting on the logical state, as well as Hermitian operators used to perform measurements.

        We obtain the trivial phase from the color-code phase by condensing a complete Lagrangian subgroup of the color-code bosons.
        Hence, when applying such a condensation to a region $R$ of the system, we create a domain wall on $\partial R$ which interfaces the color code with the trivial phase.
        Such a domain wall is referred to as a boundary~\cite{Levin05}.
        As we have discussed in Sec.~\ref{sec:CondMax}, there are six Lagrangian subgroups in the color code anyon model, translating to six boundaries that terminate the color code in the spatial direction~\cite{Kesselring18}.
        These six boundaries fall into two classes: coloured boundaries and Pauli boundaries.
        Coloured boundaries are obtained if we choose to condense all three bosons with the same colour label, and we obtain Pauli boundaries if the three bosons that are condensed share a Pauli label.
        The coloured boundaries correspond to the columns of the boson table~\eqref{eq:BosonTable}, and Pauli boundaries correspond to the rows of the table.
        
\begin{figure}[tb]
	        \centering
	        \includegraphics[width=.85\linewidth]{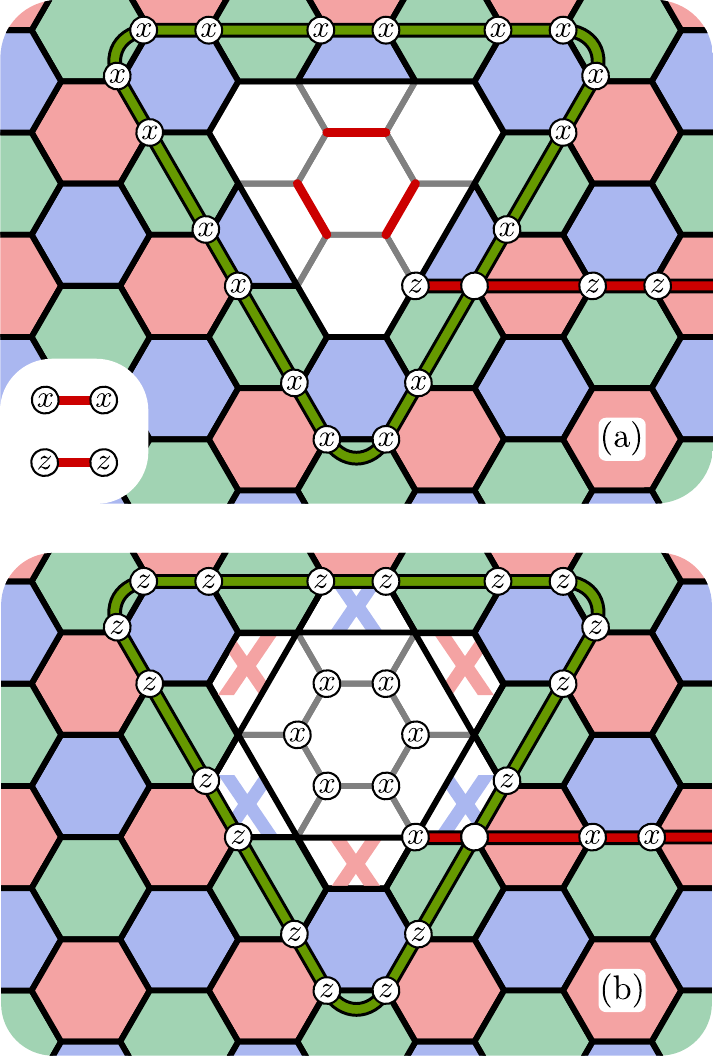}
	        \caption{
	            Punctures with different types of boundaries can be used to encode logical qubits.
	            \textbf{(a)}~shows a puncture with a coloured boundary.
	            It is introduced when all three red bosons get condensed within a disk-shaped region $R$.
	            Microscopically, condensation is achieved by adding two-body $XX$ and $ZZ$ stabilisers on red edges These edges are highlighted in the centre of the puncture.
	            \textbf{(b)}~shows a puncture terminated by a Pauli-$X$ boundary.
	            We produce the puncture by adding single qubit $X$ terms supported on qubits in the interior of $R$ to the stabiliser.
	            Some of the logical operators for these logical qubits are shown, where we assume the string exiting the figure on the right terminates on a second boundary of appropriate type.
	            Empty circles are used on qubits where the logical operators have common support, where the logical operators act on these qubits in different bases.
            }
	        \label{fig:PunctureEncoding}
	    \end{figure}

        To create a puncture microscopically, we condense anyons by adding hopping terms to the stabiliser group, as described in Sec.~\ref{sec:CondStabs}.
        To create a puncture with a coloured boundary, we consider all the edges of the chosen colour which lie within $R$ and perform Bell-pair measurements on the pairs of physical qubits supported on these edges~\cite{Fowler11twoDimensional}, see Fig.~\ref{fig:PunctureEncoding}~(a).
        To create a puncture with one of the three Pauli-boundaries, the prescription from Sec.~\ref{sec:CondStabs} dictates to add the two-qubit Pauli rotations in the appropriate Pauli basis on all the edges within region $R$.
        Alternatively, it is sufficient to perform single-qubit Pauli measurements in the chosen basis on all qubits within $R$.
        These single-qubit Pauli rotations act as simultaneous hopping terms of all three bosons with the chosen Pauli label.
        This can be seen in Fig.~\ref{fig:CCLattice}~(b), where we apply a single-qubit $Y$ rotation to decompose an \ry anyon into a \gy and a \by anyon while simultaneously moving the charge in the process.
        Importantly, the two bosons \gy and \by have a joint charge equivalent to the \ry boson, hence we can regard the single-qubit Pauli-$Y$ rotation as moving an \ry charge.
        Likewise, we can interpret the same single-qubit rotation as a hopping operator for the green or the blue bosons.
        The same argument holds for bosons with a Pauli-$X$ or $Z$ label and single-qubit Pauli-$X$ or $Z$ rotations.
        The creation of a Pauli-$X$  puncture using single qubit $X$ measurements is shown in Fig.~\ref{fig:PunctureEncoding}~(b).
        
        Let us now look at the physics of how a puncture can be used to encode and manipulate logical qubits.
        To do so, we give a physical interpretation of logical Pauli operators.
        As the boundary of a puncture can condense four anyons (including the trivial charge \one), the puncture can be in one of four states, corresponding to the four condensed bosons (including \one).
        A red puncture, for instance, can contain one of the following charges: $\left.\{ \one, \rx, \ry, \rz \right.\}$.
        Similarly, a Pauli-$X$ puncture as an example of a Pauli-puncture contains one of these four bosons: $\left.\{ \one, \rx, \gx, \bx \right.\}$.
Hence, a single puncture constitutes a four-dimensional Hilbert space which we can use to store quantum information in a robust manner.
        However, the dimension of the Hilbert space associated with a single puncture only describes the dimension of the logical subspace in the asymptotic limit of a large number of punctures. This is because we require that the charges that describe the internal states of multiple punctures respect global charge conservation~\cite{Kitaev03}. In general then the ground state degeneracy scales like $2^\Upsilon$ where $\Upsilon = 2 (\# \textrm{punctures} - C)$ where $C$ is some small correction around that depends on the boundary types of the different punctures. We find that for generic configurations of punctures we have $C = 2$. This correction can be lower for special cases where there are punctures with only one or two types of boundary.

\begin{figure}[tb]
	        \centering
	        \includegraphics[width=1.0\linewidth]{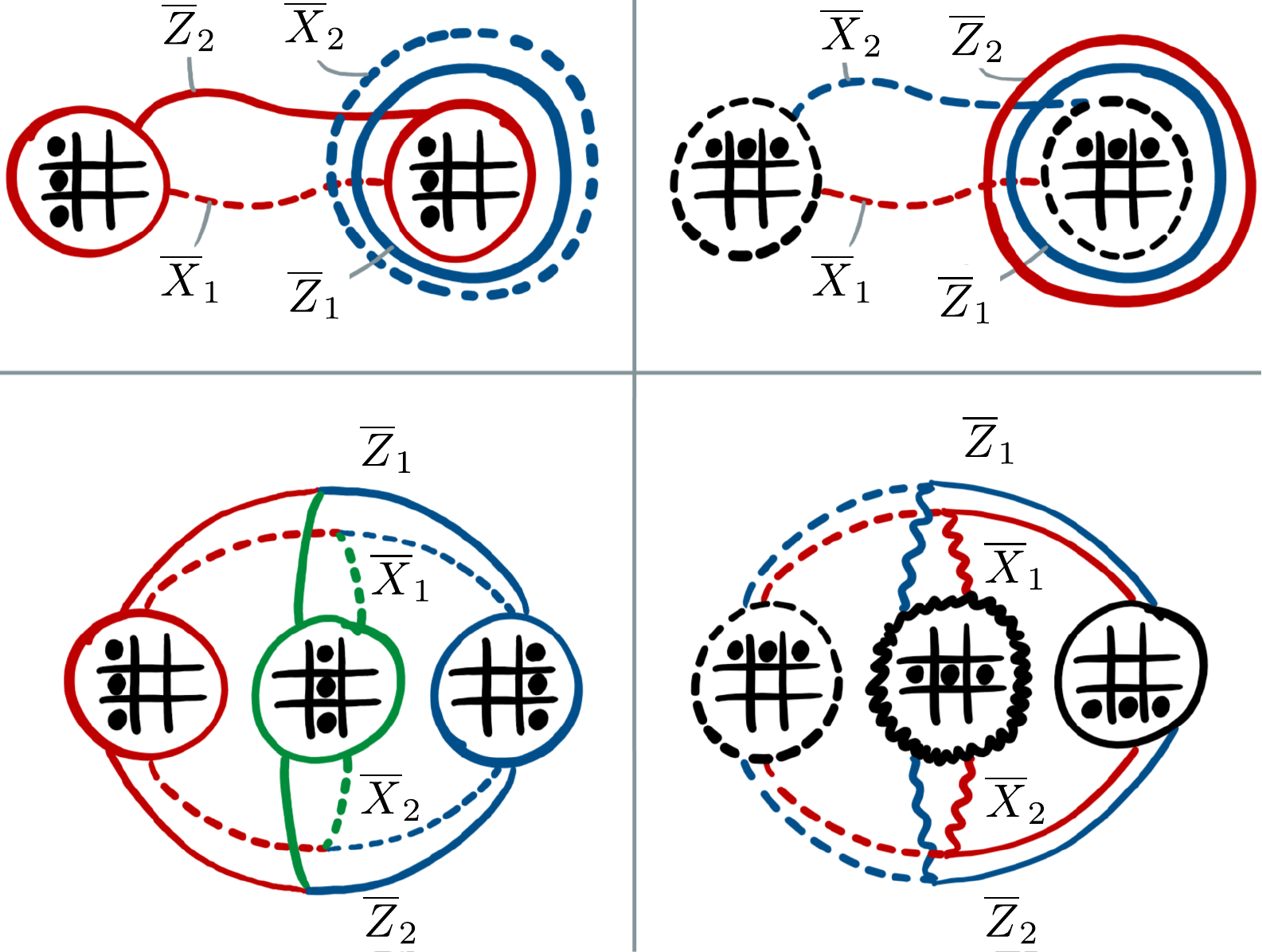}
	        \caption{
	            Four different configurations of punctures are shown.
	            The type of boundary is indicated by the bosons that are condensed inside the puncture, marked on the boson table shown inside each puncture.
	            Each configuration encodes two logical qubits, where the logical operators are depicted in the figure.
	            String operators act on edges of their their respective colour, and the basis is given by the line style:
	            Dashed lines correspond to the $X$-basis, wavy lines to $Y$ and solid lines to $Z$.
            }
	        \label{fig:PunctureEncodings}
	    \end{figure}

In general, we might prefer to encode qubits over small subsets of punctures, as this enables to perform logical operations on the encoded information.
 One simple encoding using punctures consists of a pair of punctures with the same type of boundary.
	    This puncture configuration encodes two logical qubits.
	    Fig.~\ref{fig:PunctureEncodings}~(a) shows a pair of red punctures and its logical operators, Fig.~\ref{fig:PunctureEncodings}~(b) a pair of Pauli-$X$ punctures.
	    Alternatively, a triple of coloured punctures, one of each colour, can also be used to encode logical qubits fault tolerantly.
	    This puncture configuration also encodes two logical qubits, as shown in Fig.~\ref{fig:PunctureEncodings}~(c).
	    Similarly, three Pauli punctures, again one of each type, encode two logical qubits, see Fig.~\ref{fig:PunctureEncodings}~(d).
	    Note how the left and the right halves of the figure are related by the color codes duality between the colour- and the Pauli-labels.

        Labelling different logical states by the anyon type that occupies the puncture allows us to identify the logical Pauli operators.
        Importantly, Pauli operators are both unitary and Hermitian, meaning we can interpret them as changing a state or being used as a measurement.
        Viewing the logical operators as unitaries, we require their microscopic realisations to either change the occupation of a puncture or to apply a relative phase depending on the condensed charge within a puncture.
        We achieve this by applying string operators transporting anyonic charges between punctures or by wrapping an anyonic string operator around a puncture to the encoded state.

        Specifically, we can write down an overcomplete set of logical operators as string operators that move a charge from any one puncture to another, provided the two punctures can condense a common charge that we wish to move. One can convince oneself that we can find a set of strings that do not cross, such that all of these hopping operators commute. These string operators that hop charges between punctures anti-commute with loop-like string operators that wrap around a puncture. Naturally, these loop-like string operators correspond to hopping operators for charges that are confined by the puncture that is enclosed by the loop. When viewed as a Pauli measurement operator, this set of loop-like operators can be physically interpreted as operators that measure the charge content of a given puncture.

        The weight of the least-weight logical operator, i.e., the code distance $d$ in the quantum error-correction literature, is proportional to the circumference of the punctures, and their relative separation.
        A large code distance is obtained by choosing large punctures that are well separated.

        So far we have viewed the logical Pauli operators as unitary operators.
        Now, let us consider them as Hermitian measurements.
        The naive way to perform a logical Pauli measurement is to measure the above described string operators.
        This prescription has two problems for experimental realisation, however.
        First, the number of qubits participating in the parity measurement grows with the circumference or separation of the punctures, leading to a parity measurement on extensively many qubits. A naive implementation of this measurement is not fault tolerant.
        Secondly, an single error either on a physical qubit or in the measurement apparatus might change the outcome.
        In order to obtain a fault-tolerant readout consisting of geometrically local few-qubit measurements, we can follow a procedure similar to the one introduced in Ref.~\cite{Dennis02}.
        The language of anyon condensation lends itself nicely to describe fault-tolerant protocols for readout.
        This is the topic of Sec.~\ref{sec:CondTimeMax}.

    \subsection{Domain walls to partial condensates}
        \label{sec:Cond2dPart}

        Condensing a single boson in a region $R$ transforms the color code into the toric code.
        The boundary $\partial R$ constitutes a domain wall between the two phases.
        As we discussed in Sec.~\ref{sec:CondPart}, there are $18$ possible ways to interface the two phases.
        They differ in which anyons are condensed, confined or remain mobile when approaching the domain wall, and to which anyons of the condensed phase the mobile anyons are mapped.
        
\begin{figure}[tb]
	        \centering
	        \includegraphics[width=1.0\linewidth]{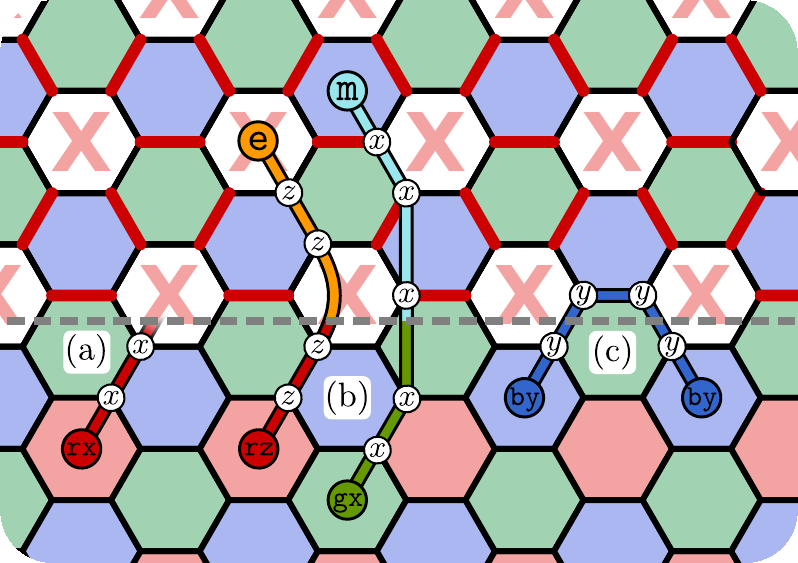}
	        \caption{
	            A domain wall (grey dashed line) between the toric code (top) and the color code (bottom).
	            Fully coloured in plaquettes host stabilisers acting in both $s_X^p$ and $S_Z^p$ stabilisers.
	            Red plaquettes marked with an $X$ host only the $X$-basis stabilisers; $s_X^p$.
	            The red-edges in the top half of the figure host $XX$ stabilisers.
	            The color code anyons either condense \textbf{(a)} or confine \textbf{(c)} at the domain wall. Otherwise they remain deconfined as they pass through the domain wall, \textbf{(b)}.
            }
	        \label{fig:CC_TC_DW}
	    \end{figure}

        Fig.~\ref{fig:CC_TC_DW} shows an example where the \rx boson gets condensed in the top half of the lattice.
        The four bosons \gy, \gz, \by, \bz that all braid non-trivially  with \rx become confined.
        The two remaining red bosons, \ry and \rz, as well as the two remaining bosons with a Pauli-$X$ label, \gx and \bx, braid trivially with \rx.
        Hence, they remain mobile and can pass through the domain wall. Upon crossing the domain wall, the mobile charges are mapped to one of the two toric code bosons.
        In the example shown, we map \ry and \rz to \e and \gx and \bx to \m.
        
        Microscopically, we introduce this domain wall by following the prescription given in Sec.~\ref{sec:CondStabs} to condense a single boson.
        Concretely, we designate a boson with a specific color and Pauli label to be condensed.
        In the example shown in Fig.~\ref{fig:CC_TC_DW} we chose to condense \rx.
        To condense the boson, we add two-body hopping terms to the stabiliser.
        Their support and the basis in which they act are given by the labels of the boson.
        To condense \rx we add $XX$ terms on all red egdes in $R$ to the stabiliser group.
        Finally, we update the stabiliser by removing terms which do not commute with the introduced hopping terms.
        In this example, we remove the $Z$-basis stabilisers on red plaquettes.
	    
	    Let us now examine the properties of the partially condensed region from the perspective of unfolding. 
	    By choosing a suitable unfolding, where we separate the color code into two disjoint copies of the toric code, we can identify the feature obtained by partial condensation in $R$ as a puncture on one of the two toric code copies.
	    Hence, we dub such a feature a semi-puncture.
        
	    With this observation, we discover the value of viewing the color-code phase from the perspective of the boson table. 
        Let us stress that we obtain a puncture on one copy of the toric code only assuming we choose a suitable unfolding map. As we have discussed in Sec.~\ref{sec:PrelimUnfolding}, there are 72 different choices of unfolding map onto copies of the toric code.
        \begin{figure}[tb]
	        \centering
	        \includegraphics[width=.65\linewidth]{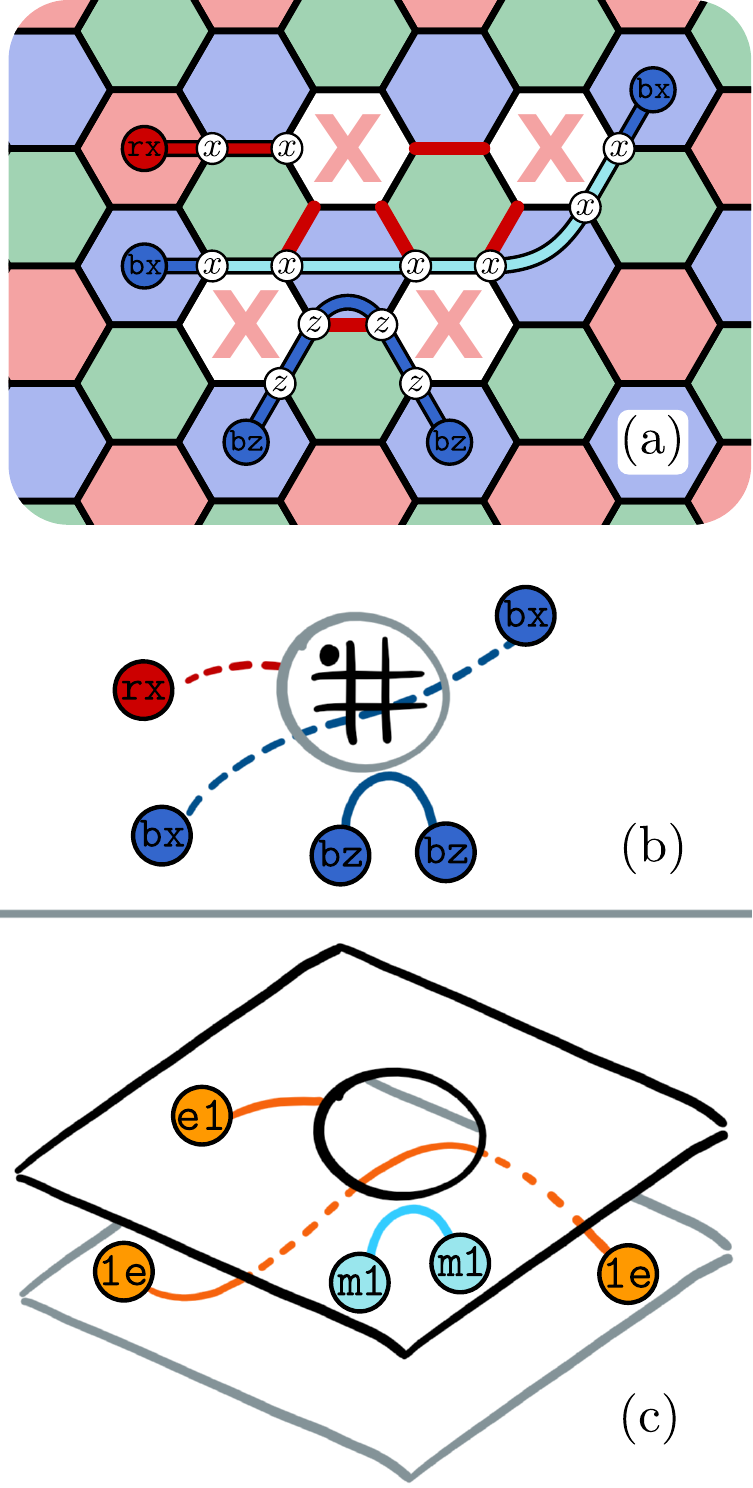}
	        \caption{
	            An \rx semi-puncture shown in both the color code picture and the unfolded toric-code picture for a suitable choice of unfolding.
                The microscopic model is shown in \textbf{(a)}, together with different types of creation operators in the vicinity of the semi-puncture.
                Specifically, the \rx anyon condenses at the semi-puncture, the \bx anyon deconfines and remains mobile, and \bz is confined, 
	            i.e., cannot enter the semi-puncture.
	            In \textbf{(b)} we give a macroscopic representation of the same semi-puncture in the color code picture. We explicitly label the semi-puncture with a boson table that marks the condensed boson in the centre of the semi-puncture.
	            We show the semi-puncture in an unfolded picture in \textbf{(c)}.
                We choose the the unfolding map shown in Eq.~\eqref{eq:AnyonUnfolding} to reveal the semi-puncture picture.
            }
	        \label{fig:SemiPunctureUnfolding}
	    \end{figure}
        However, under the same unfolding map, there are certain semi-punctures that are obtained by condensing other choices of boson, that do not immediately divide into a puncture on a single toric-code copy.
        Rather, we require the use of additional domain walls to describe all of the semi-punctures for any fixed unfolding map.	    
     
	    We therefore find the unfolded picture to be somewhat unsatisfying, because different color-code semi-punctures manifest themselves differently in the toric code picture.
        On the other hand, in the color code picture, all of the bosons are equivalent, up to symmetries among relabelling of their color and Pauli labels.
        We argue then that the color code and its corresponding boson table offers a clearer way to describe these generalised topological features we have introduced here, as the boson table symmetrises the classification of all of the different semi-punctures we can produce.
        We depict these contrasting descriptions in Fig.~\ref{fig:SemiPunctureUnfolding}.
	    
\begin{figure}[tb]
	        \centering
	        \includegraphics[width=.6\linewidth]{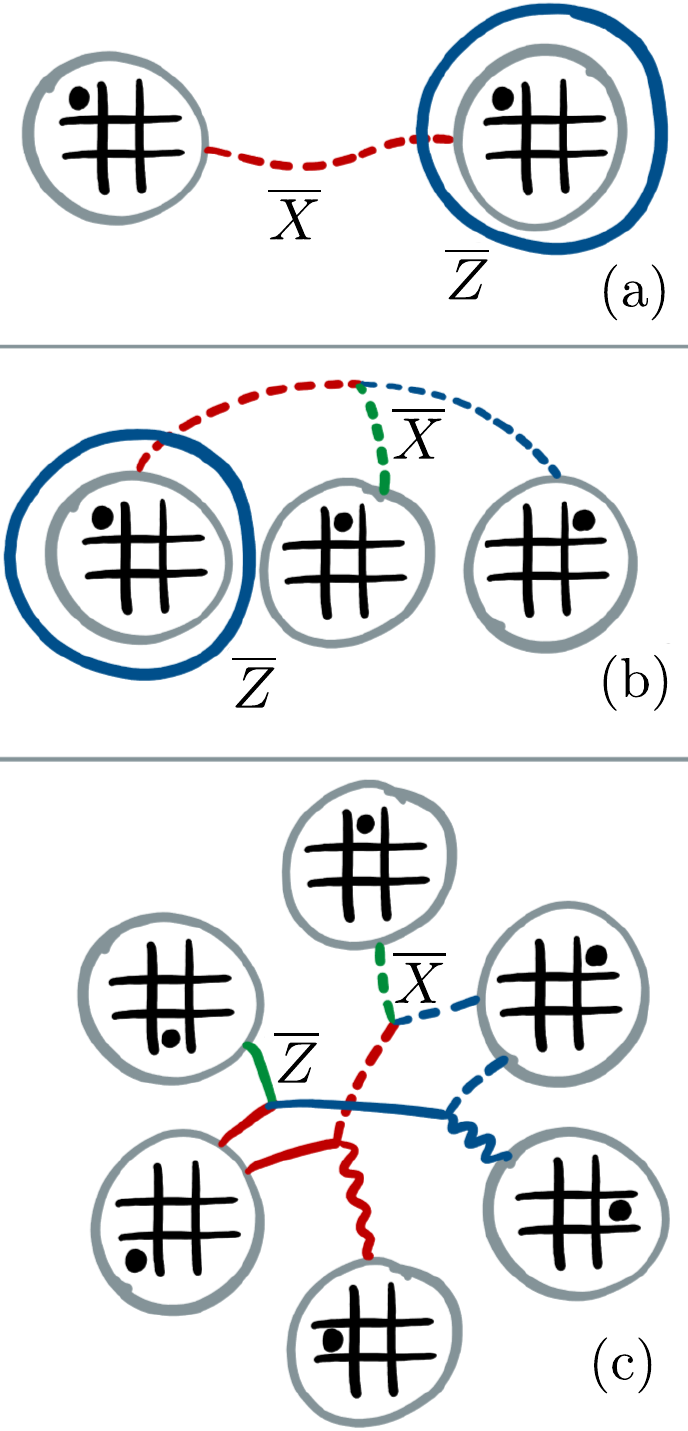}
	        \caption{
	            Encoding logical qubits using semi-punctures.
                We indicate the type of semi-puncture by marking the condensed boson on the shown boson table.
                The logical string operators are presented, where the colours of the lines indicate on which colour of edges they act.
	            Dashed, wavy and solid lines correspond to operators acting in the $X$-, $Y$- and $Z$-basis, respectively.
	            \textbf{(a)}~and \textbf{(b)}~encode one logical qubit each whereas \textbf{(c)}~encodes two logical qubits.
	            The second set of logical operators in (c) are supported to the exterior of semi-puncture configuration we have presented.
            }
	        \label{fig:SemiPunctureEncodings}
	    \end{figure}

	    Naturally, like punctures, we can use semi-punctures to encode logical information.
	    We show examples of logical encodings using semi-punctures in Fig.~\ref{fig:SemiPunctureEncodings}.
	    We can describe the physics of the associated logical operators equivalently to the logical operators encountered in Sec.~\ref{sec:Cond2dMax}.
	    They are string operators that connect semi-punctures, or string-operators that wrap around the semi-puncture.

\begin{figure}[tb]
	        \centering
	        \includegraphics[width=.95\linewidth]{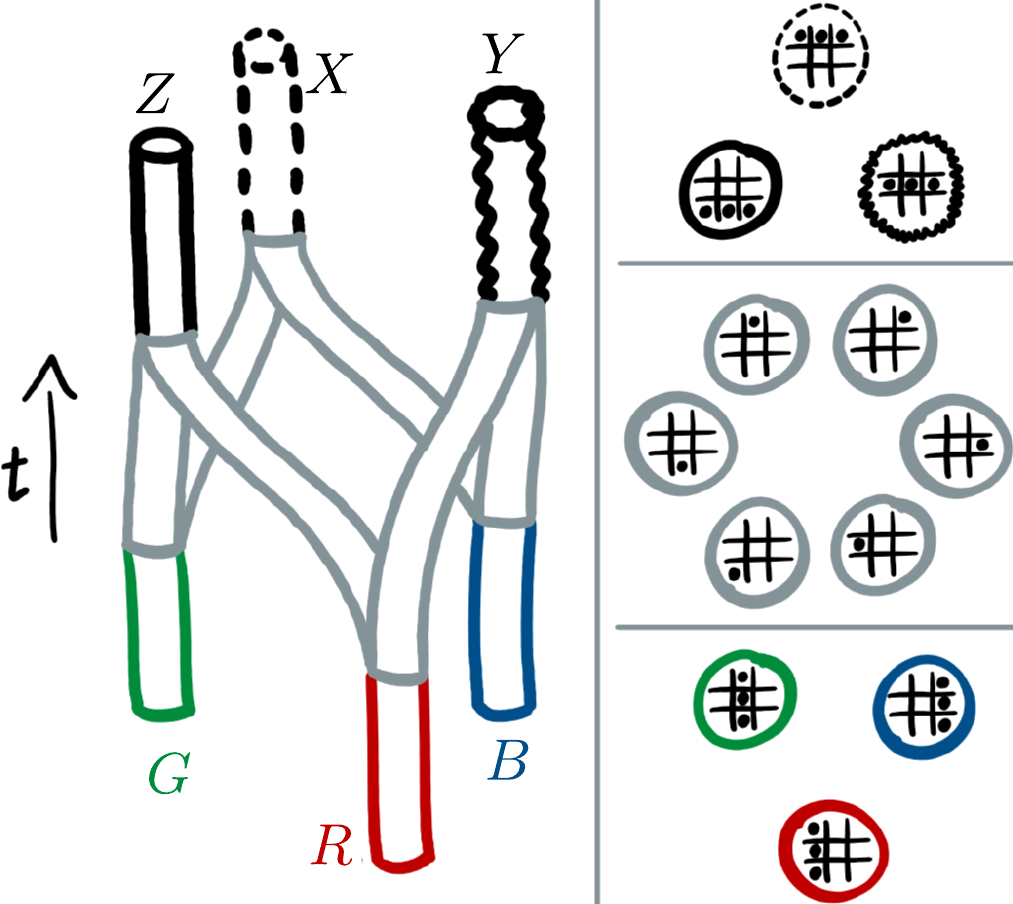}
	        \caption{
	            A code deformation protocol that is mediated by semi-punctures.
                We find that we can switch between a puncture encoding where each of the three punctures has a differently coloured boundary, onto an encoding where three punctures all have distinct types of Pauli boundary.
	            We achieve this by splitting each of the coloured punctures into two semi-punctures.
	            Next, the semi-punctures are moved and fused to form three Pauli-type punctures.
	            The left part of the figure shows the space-time diagram of the protocol while the right shows the (semi-)puncture configuration at intermediate time steps.
            }
	        \label{fig:SemiPunctureBraid}
	    \end{figure}

	    The discovery of semi-punctures also opens the door for the design of new types of code deformations.
        As an example, we show that we can transform between different configurations of punctures using semi-punctures to mediate the transition.
	    As an example, in Fig.~\ref{fig:SemiPunctureBraid} we show that we can transform between the logical qubit encoding shown in Fig.~\ref{fig:PunctureEncodings}~(b) onto the encoding in Fig.~\ref{fig:PunctureEncodings}~(c), where we  make use of the six semi-punctures at an intermediate step.
	    We add that we have already encountered the six semi-puncture encoding in Fig.~\ref{fig:SemiPunctureEncodings}~(c). In addition to this example, we find that semi-punctures also emerge when performing a readout addressing only some of the logical qubits encoded in a color code.
        We discuss this example in the following section, in Sec.~\ref{sec:CondTimePart}.

	    The examples we present show us the discovery of semi-punctures may be helpful to find new fault-tolerant logical operations.
        These may be helpful, for instance, to reduce the resource overhead of color code based quantum computation.
	    Furthermore, considering that they can be combined with corners to obtain mixed boundary semi-punctures~\cite{Delfosse16,Benhemou22}, we have presented a significant landscape to design and explore code-deformation protocols in the future.

    \subsection{Invertible domain walls}
        \label{sec:Cond2dTriv}
        
        When condensing a set of anyons, we implicitly or explicitly make use of the symmetries of the underlying topological phases.
        This is also true in the trivial case, where no non-trivial anyons are condensed.
        This leaves us interfacing the parent phase with itself while applying an anyon permuting symmetry.
        These symmetries are the automorphisms of the anyon model $Aut(\cC)$.
        In the case of the color code, there are 72 such automorphisms, as we have briefly summarised in Sec.~\ref{sec:CondMax}.
        To each automorphism, we can associate one domain wall;
        see Ref.~\cite{Kesselring18} for a detailed discussion on automorphisms of the color code and the associated spatial domain walls.

        Anyons are distinguished by their fusion and braiding properties.
        Consider applying a symmetry given by the automorphism $Aut(\cC)$, i.e., trivial condensation, in a simply connnected closed region $R$.
        This creates an invertible domain wall along the boundary $\partial R$.
        We say the domain walls acts as $Aut(\cC)$ on the anyon crossing it.
        
        However, note that there is no way of detecting the domain wall using only operations based on moving the color code anyons, i.e. fusion and braiding.
        This is because, by definition, the relative behaviour of the anyons is independent of the presence of a the domain wall, see Fig.~\ref{fig:DWConsistencyCond}.

        Domain walls can, however, be terminated.
        In doing so we create so called twist defects at the end-points.
        This is the topic of Sec.~\ref{sec:Cond1dTriv}.
        Alternatively, we can consider applying an automorphism on an entire code patch, applying a transversal logical operation on the encoded qubits.
        This is the topic of Sec.~\ref{sec:CondTimeTriv}.
        
\section{Temporal domain walls}
    \label{sec:CondTime}
    
\begin{figure}[b]
        \centering
        \includegraphics[width=0.4\linewidth]{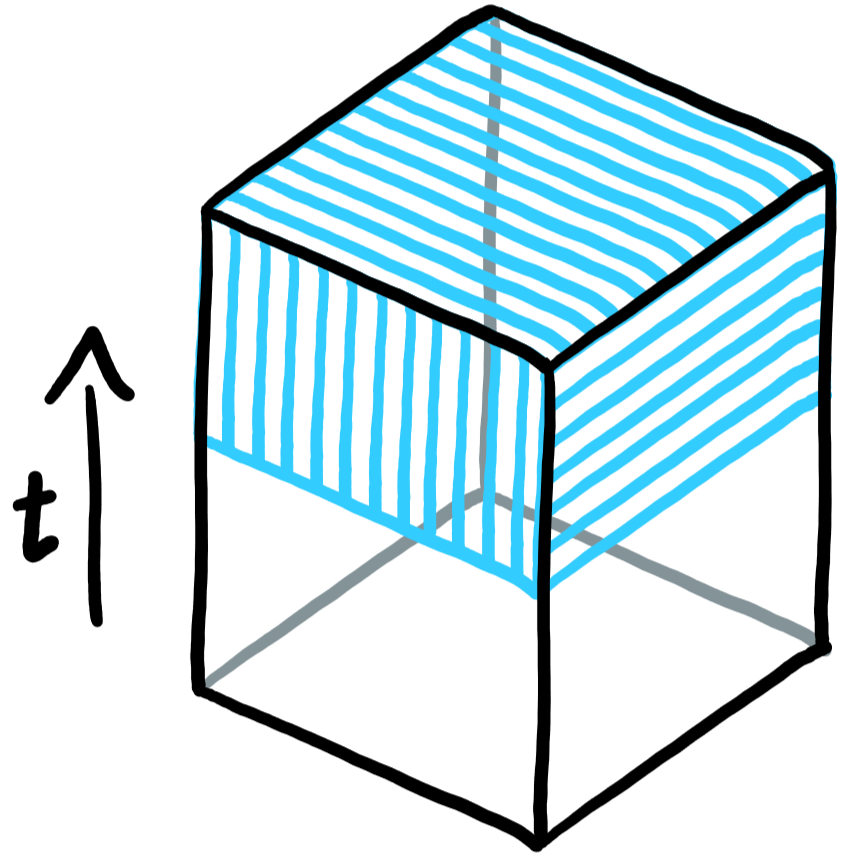}
        \caption{
            In this section we create temporal domain walls between the parent phase and the condensed phase.
            We achieve this by condensing anyons in the whole spatial bulk after (or before) a given point in time.
            The region of the space-time diagram in which the condensation takes place is shaded in cyan.
        }
        \label{fig:CondensationTimeSlice}
    \end{figure}

    In this section, we discuss temporal domain walls.
    They are commonplace in topological quantum computation as they describe initialisation or injection of a code state, the application of a locality preserving gate or the readout step at the end of a computation.
    Temporal domain walls are introduced by changing the stabiliser group over time, see Fig.~\ref{fig:CondensationTimeSlice}.
    In particular, here we argue that anyon condensation is a well suited framework, not only to describe temporal domain walls, but also to construct them microscopically as topologically protected deformations of the stabiliser group.
    In this sense, our treatment of temporal domain walls in this section is exactly analogous to that of spatial domain walls discussed in Sec.~\ref{sec:Cond2d}.

    On a high level, quantum information is processed by changing logical operators over time.
    Since logical Pauli operators in two-dimensional topological error-correcting codes are anyonic string operators, keeping track of how anyons get mapped when passing through a domain wall gives us a physical interpretation for the action of different topological operations on logical qubits.

    As we show in Sec.~\ref{sec:CondTimeTriv}, we can relate locality preserving gates with invertible domain walls. This enables us to include transversal gates~\cite{Bombin06, Eastin09} in our theory of anyon condensation.
    We also find that initialisation and readout can be interpreted in terms anyon condensation. In Sec.~\ref{sec:CondTimeMax} we go through the details of fault-tolerant state readout~\cite{Bombin09CodeDefo, Fowler11twoDimensional} in terms of the anyon condensation picture we have introduced. 
    We develop this discussion further in Sec.~\ref{sec:MicroTempBdry}, where we describe at the microscopic details of a readout operation in terms of error-correction space-time. 
    In Sec.~\ref{sec:BoundaryInterplay} we elaborate on the theory of temporal boundaries further by showing how temporal boundaries interact with spatial boundaries during certain known color-code state preparation procedures~\cite{Bombin09CodeDefo, Fowler11twoDimensional, Landahl14}.

    In Sec.~\ref{sec:StabilityExperiment} we describe new stability experiments for the color code which exemplify nicely the interplay between different types of spatial and temporal boundaries. This builds on recent work introducing the notion of a stability experiment~\cite{Gidney2022stability}.
    Finally, in Sec.~\ref{sec:CondTimePart} we introduce partial initialisation and partial readout for the color code. This gives us new readout protocols based on semi-transparent temporal domain walls. These operatiosn enables us to address specific logical qubits encoded on some region of the color-code lattice, without interacting with other qubits encoded on the same region.

    \subsection{Invertible domain walls and Clifford gates}
    \label{sec:CondTimeTriv}
    
\begin{figure}[tb]
            \centering
            \includegraphics[width=0.95\linewidth]{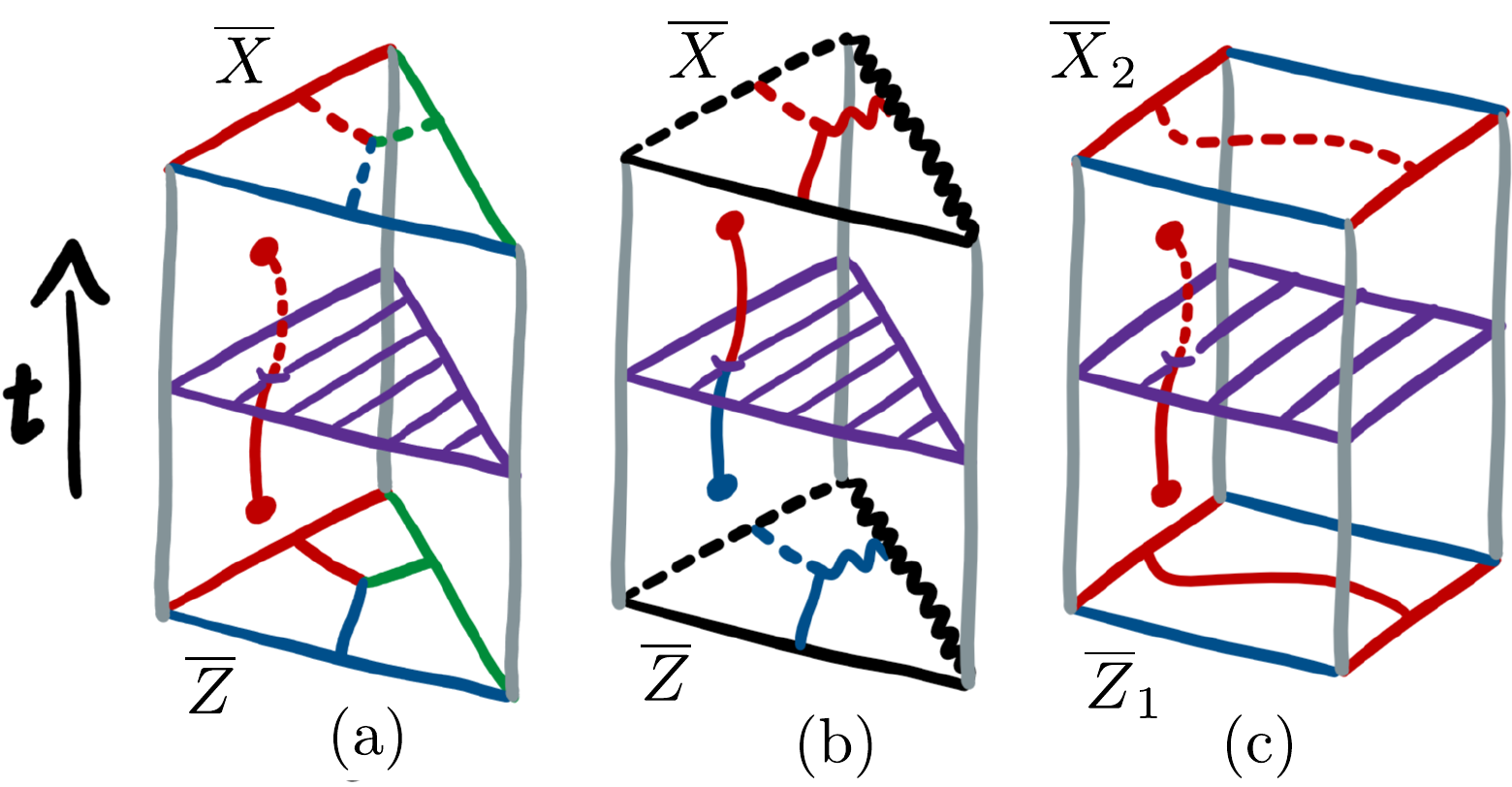}
            \caption{
                Transversal gates in the triangular color code correspond to invertible domain walls in the time direction.
                The action of the domain walls (purple) is depicted by the effect an anyon crossing it.
                Dashed (solid) world-lines crossing the domain wall correspond to anyons with a Pauli-$X$($Z$) label.
                In the triangular color code with coloured boundaries \textbf{(a)}, a domain wall which turns the Pauli-$Z$ label into the Pauli-$X$ label changes the logical $\overline{Z}$ operator into $\overline{X}$.
                For the triangular color code with Pauli boundaries \textbf{(b)}, the domain wall needs to permute the colour labels of the anyons in order to act as a logical Clifford gate.
                In a rectangular color code with RBRB boundaries, as depicted in \textbf{(c)}, a transversal Hadamard implements to the logical gate $\overline{H}_1 \circ \overline{H}_2 \circ {\rm \overline{SWAP}}_{1,2}$.
                Dashed, wavy and solid lines correspond to the Pauli-$X$, $Y$ and $Z$ basis, respectively.
            }
            \label{fig:CliffordGatesInspace-time}
        \end{figure}

        A temporal invertible domain wall corresponds to a symmetry of the anyon model, $Aut(\cC)$.
        This means all anyons can traverse the domain wall and in doing so get mapped to potentially different anyons.
        As the logical Pauli operators are associated with anyon strings, they get permuted.
        Thus, introducing an invertible temporal domain wall acts as a logical Clifford gate.
        We show examples in Fig.~\ref{fig:CliffordGatesInspace-time}.
        
        In the triangular color code, all single-qubit Clifford gates are transversal~\cite{Bombin06}.
        This leads to an apparent mismatch between the number of different invertible color-code domain walls, $72$~\cite{Yoshida15,Kesselring18}, and the size of the Clifford group acting on one qubit, $6$ if we ignore phases.
        We assume the applied symmetry to not change the boundaries of the code in order to preserve the code space.
        In the example of the triangular color code with coloured boundaries, a symmetry that permutes any of the colour labels would also change the boundary type of some of the boundaries.
        Similarly, any symmetry which applies the duality transformation exchanging the Pauli- and the colour-labels of the anyonic charges of the color code transforms coloured boundaries into Pauli boundaries.
        Thus, we exclude them here.
        This leaves us with the $|S_3| = 6$ symmetries which solely permute the Pauli labels.
        These are exactly the elements of the Clifford group without phases.
        Applying any of these gates transversally on all physical qubits applies the equivalent logical gate~\cite{Bombin06}.
        
        Interestingly, this argument holds true for any boundary configuration which contains only coloured boundaries - or equivalently, any configuration only containing Pauli boundaries.
        Meaning that any such code contains exactly $6$ gates which can be applied transversally as an invertible temporal domain wall.
        In the case of the square color code with RBRB boundaries (see Fig.~\ref{fig:CliffordGatesInspace-time}~(c) and Fig.~\ref{fig:ReadOut}), a generating set for the transversal gates that can be realised is as follows:
        A Hadamard gate $H$ applied transversally to all of the physical qubits exchanges the logical operators $\overline{Z}_1 \leftrightarrow \overline{X}_2$ and $\overline{X}_1 \leftrightarrow \overline{Z}_2$, corresponding to a logical swap gate followed by a Hadamard gate on each logical qubit, $\overline{H}_1 \circ \overline{H}_2 \circ {\rm \overline{SWAP}}_{1,2}$.
        This transversal gate is shown in Fig.~\ref{fig:CliffordGatesInspace-time}~(c).
        The gate exchanging the Pauli $Y$ and $Z$ basis when applied transversally to all physical qubits is, up to phases, equivalent to the application of $\overline{H}_1 \circ {\rm \overline{CNOT}}_{1,2} \circ \overline{H}_1$.
        The other three non-trivial single qubit Pauli permuting gates can be generated from the two given examples.
        On the logical level, they can all be composed of ${\rm CNOT}$s and Hadamard gates.
        
        Similarly, through the duality between the colour-labels and the Pauli-labels, we can argue that any color codes that is terminated exclusively by Pauli-boundaries also has $6$ transversal Clifford gates.
        It may be interesting to relax the assumtion that the boundaries before and after the transversal gates have to match.
        This might also combine in a non-trivial way with the code deformations we have introduced in Sec.~\ref{sec:Cond2dPart} that smoothly transform between different boundary types.
    
        Lastly we point out that invertible temporal domain walls have applications beyond the implementation of logical gates.
        Certain Floquet codes~\cite{HastingsHaah21a, Vuillot21, Gidney21, HaahHastings21b, Gidney22, Paetznick22} can be interpreted as so called automorphism codes, where invertible domain walls on different subregions are periodically introduced~\cite{aasen22automorphism}.
        In particular, in this reading of the honeycomb code~\cite{HastingsHaah21a} every time step introduces a domain wall around $\frac{1}{3}$ of the plaquettes, such that after $3$ steps an automorphism has been applied to the whole code. 
        In this work we argue that anyon condensation is a well suited tool to study and construct Floquet codes, see Sec.~\ref{sec:CondDynamic}.

    \subsection{Initialisation and readout}
        \label{sec:CondTimeMax}
    
        In this section, we study the readout, and implicitly the initialisation, of logical qubits encoded in color codes.
        These processes are naturally described as maximal anyon condensation.
        We discuss the relationship between the logical operators we address in the process and the Lagrangian subgroups which are condensed.
        Condensing anyons in time introduces temporal domain walls.
        We defer a detailed discussion of temporal boundaries to Sec.~\ref{sec:MicroTempBdry}. 
    
\begin{figure}[tb]
	        \centering
	        \includegraphics[width=1.0\linewidth]{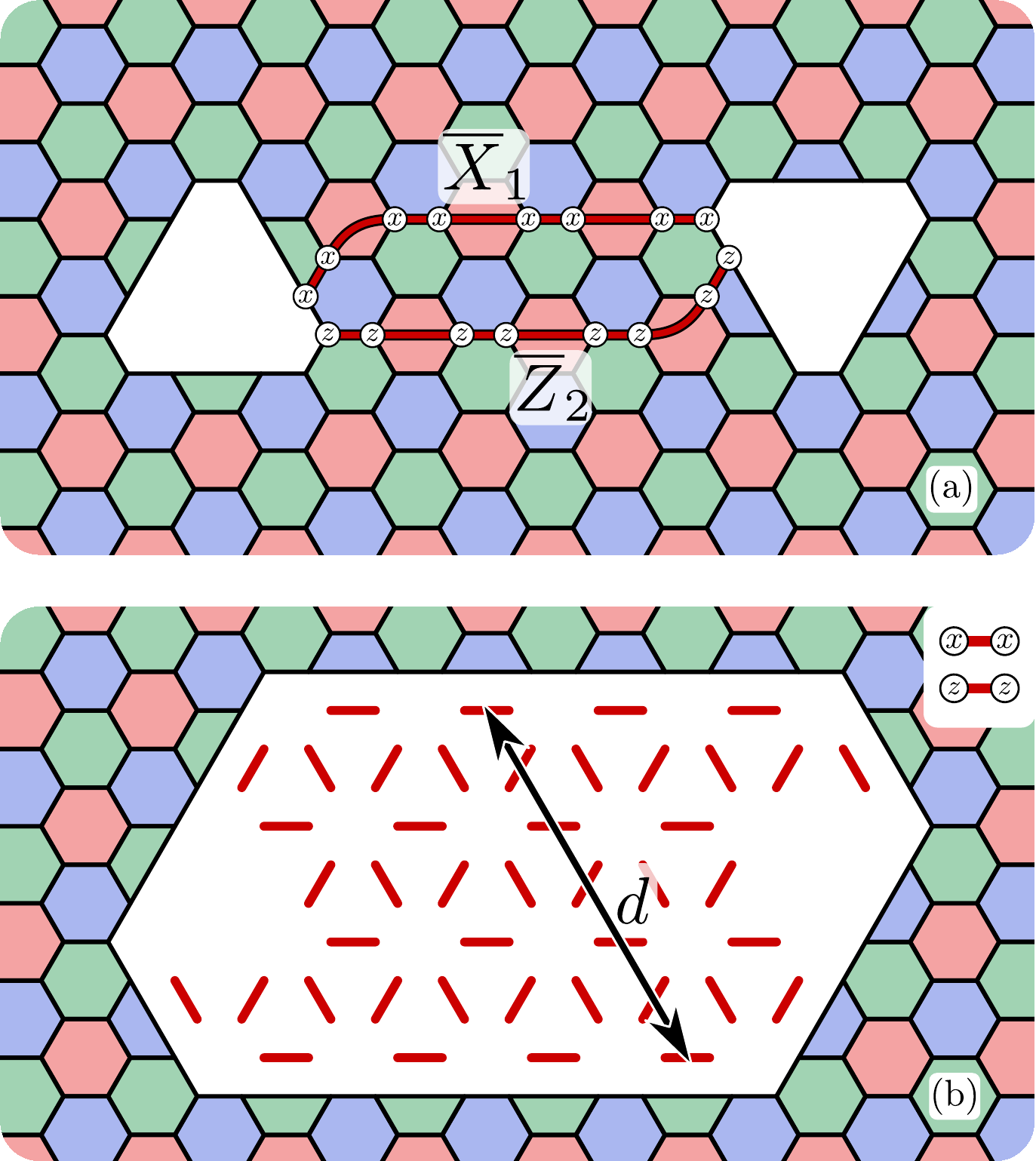}
	        \caption{
                Reading out logical qubits by condensing anyons.
	            Two red punctures, as shown in \textbf{(a)}, encode two logical qubits.
	            A red string connecting the two punctures constitutes a logical operator.
	            To read out the logical operators $\overline{X}_1$ and $\overline{Z}_2$, we can measure the $XX$ and $ZZ$ parity on all red links in a region $R$ encompassing the two punctures, as depicted in \textbf{(b)}.
	            This process condenses all red anyons and creates one large red puncture in $R$.
	            The inverse process initialises the two logical qubits in an eigenstate of $\overline{X}_1$ and $\overline{Z}_2$.
            }
	        \label{fig:ReadOutPuncturesFull}
	    \end{figure}

        We begin by revisiting the encoding scheme encountered in Sec.~\ref{sec:Cond2dMax}, where two red punctures are encoding two logical qubits. 
        Specifically, we show how to simultaneously read out logical operators $\overline{X}_1$ and $\overline{Z}_2$, see Fig.~\ref{fig:ReadOutPuncturesFull}~(a).
        Both of these logical operators, $\overline{X}_1$ and $\overline{Z}_2$, can be interpreted as string operators that transport red anyons between the punctures.
        Hence, they are composed of two-body hopping terms on red edges, see Fig.~\ref{fig:ReadOutPuncturesFull}~(a).

        To read out the logical operators, we begin by defining a region $R$ on which $\overline{X}_1$ and $\overline{Z}_2$ can be fully supported.
        Then, we measure all red hopping terms in $R$, i.e., 
        the $XX$ and $ZZ$ parities on all red edges.
        These measurements can be combined to infer the values of $\overline{X}_1$ and $\overline{Z}_2$.
        Choosing a suitable region $R$ allows to correct for errors on the value of the logicals.
        For details see Sec.~\ref{sec:BoundaryInterplay}.
        Likewise, this readout scheme is used to perform measurement-based logical gates in lattice surgery protocols presented in Ref.~\cite{Thomsen22}
        and in Sec.~\ref{sec:Cond1dLS} of this work.  

        The above example can be understood in terms of anyon condensation.
        In particular, we condense all red anyons $\cL_R$ when measuring the red hopping terms.
        In fact, it is straight forward to generalise the concept of condensing anyons to read out logical qubits to topological stabiliser codes.
        The logical Pauli operators $\{\overline{L}_i\}$ that can be read out simultaneously have to be composed of hopping terms of bosons contained in the same Lagrangian subgroup $\cL$.
        This guarantees that we read out a commuting set of logical operators, as their corresponding anyons braid trivially, by the definition of a Lagrangian subgroup.
        
        Let us look at how we read out logical Pauli operators at the physical level.
        To this end, we consider two different stabiliser groups, $\cS_{C}$ being the stabiliser group of a topological stabiliser code $C$, and $\cS_{M}$ which is generated by the microscopic hopping terms of the anyons in $\cL$, see Sec.~\ref{sec:CondStabs}.
        Note how all logical operators in question are contained in this stabiliser, $\overline{L}_i \in \cS_M$.
        Thus, condensing the Lagrangian subgroup $\cL$ by measuring the generators of $\cS_M$ measures the eigenvalues of $\overline{L}_i$.
        Thanks to the construction in terms of anyon condensation, we guarantee the geometric locality and thus the bounded weight of the operators that we measure.
        
        To initialise logical qubits in a certain eigenstate of their logical Pauli operators, we follow an equivalent procedure, reversing the time direction.
        Concretely, this means we change from an initialisation stabiliser group $\cS_I$ to the code's stabiliser group $\cS_{C}$.
        We construct $\cS_I$ to contain the hopping terms composing $\overline{L}_i$, thus initialising the system in eigenstates of $\overline{L}_i$. 
        Note that the sign the logical Pauli $\overline{L}_i$ carries depends on the choice of signs for the generators of $\cS_i$.

\begin{figure}[tb]
            \centering
            \includegraphics[width=1.0\linewidth]{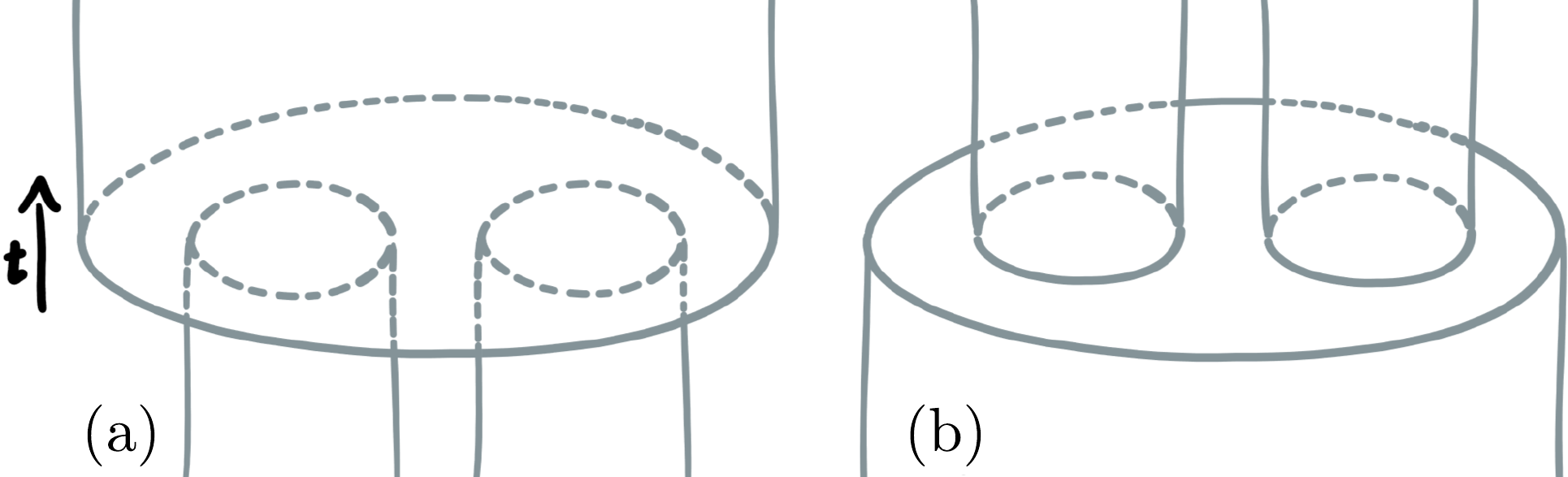}
            \caption{
                The space-time diagrams corresponding to the readout process depicted in Fig.~\ref{fig:ReadOutPuncturesFull}.
                \textbf{(a)}~During readout, the two punctures turn into one large puncture.
                \textbf{(b)}~Initialisation is achieved by flipping the time-direction, i.e., by starting with one large red puncture which then gets deformed into two smaller red punctures.
                In both cases, the horizontal boundaries are temporal boundaries.
            }
            \label{fig:InitialisationReadOutDoulbePuncturespace-time}
        \end{figure}

        The described initialisation and readout protocols condense a full Lagrangian subgroup of anyons.
        This introduces boundaries between the code and the vacuum which lie perpendicular to the time direction.
        In the color code readout example from Fig.~\ref{fig:ReadOutPuncturesFull}, for instance, we introduce a red temporal boundary between the color code and the vacuum phase.
        We can depict this using the space-time picture, see Fig.~\ref{fig:InitialisationReadOutDoulbePuncturespace-time}.
        The microscopic details of temporal color code boundaries are discussed in Sec.~\ref{sec:MicroTempBdry}.

    \subsection{Microscopics of temporal boundaries}
        \label{sec:MicroTempBdry}
        
        It is helpful to view fault-tolerant logical operations in the complementary three-dimensional space-time picture.
        In this picture we identify spatial domain walls as lying perpendicular to a spatial directions.
        On the other hand, we identify domain walls which lie perpendicular to the time direction with temporal domain walls.
        In fact, we find that the space-time picture reveals a duality between the spatial domain walls we have introduced in Sec.~\ref{sec:Cond2dMax} and the temporal domain walls.
        Here, we will describe this duality in the microscopic picture, focusing on boundaries.
        
        We conduct our investigation into the behaviour of different types of charges, identified by different types of detection cells, by focusing on two examples.
        In one example we discuss a temporal boundary where we read out logical qubits using single-qubit Pauli-$X$ measurements, and a second example where we read out the color code with Bell measurements on the green edges.
        In short, we study a temporal Pauli-boundary as well as a temporal colour boundary. 
        We remark though that focusing on these two cases is without loss of generality, due to the Pauli-label and colour-label exchange symmetries of the color code.
        Likewise, while we concentrate on the readout step by investigating the detection cells obtained by deforming $\cS_{CC}$ onto $\cS_{M}$, we can reproduce the same discussion at the initialisation step by, instead, deforming the stabiliser group $\cS_{I}$ onto $\cS_{CC}$.
        This case differs only in the sense that the direction of time is reversed, as discussed in Sec.~\ref{sec:CondTimeMax}.
        
        The characteristic features of spatial boundaries are their ability to condense a Lagrangian subgroup of bosonic excitations of the underlying anyon model.
        In the space-time picture, we replace the notion of stabilisers for identifying point-like charges with that of a detection cell, see Sec.~\ref{sec:Prelimspace-time}.
        Moreover, for each distinct anyon that we can measure in the two-dimensional picture, we have a corresponding detection cell in the space-time picture.
        We therefore find a one-to-one correspondence between our description for bosonic charges in the space-time picture with the more conventional two-dimensional idealisation of the color code.
        Taking this perspective, we find that we can view the charges in space-time to either condense or confine at a temporal boundary between the color code and the vacuum phase.
        Moreover, we find that we can obtain a temporal boundary that corresponds to any of the six Lagrangian subgroups of bosonic charges in the color-code model.
        Following the prescription given in Sec.~\ref{sec:CondStabs}, we realise the found temporal boundaries explicitly in microscopic lattice models.
        
        Let us begin our discussion by considering the temporal boundary where we read out the color code with Pauli-$X$ measurements.
        We will identify the condensed charges, before looking at confined charges.
        A signature of a boundary being able to condense a given type of charge, $\a$, is that an individual charge of that type can be created locally at the boundary, seemingly violating the fusion rules in the bulk of the system.
        Indeed, the fusion rules of an anyon model are modified close to a boundary in general.
        As such, identifying a configuration of errors that gives rise to a single charge close to a boundary is indicative that a boundary condenses (the anti particle of) that given charge type. 
        
        \begin{figure}[tb]
            \centering
            \includegraphics[width=0.75\linewidth]{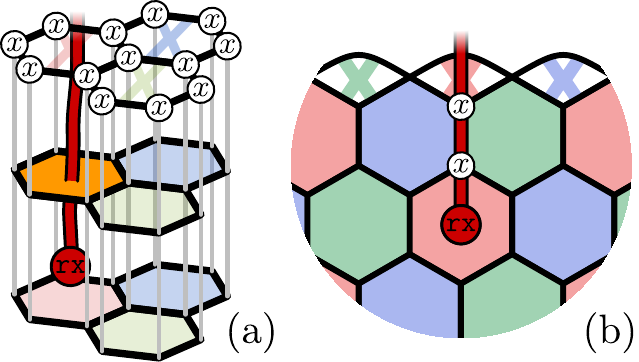}
            \caption{
                Condensed charges at Pauli-$X$ type boundaries in the temporal in spatial direction.
                \textbf{(a)}~The final reading of the stabilisers of the color code before we make a transversal Pauli-$X$ measurement on all the physical qubits, thereby condensing all charges with a Pauli-$X$ label.
                A measurement error (marked in orange) on the final reading of a Pauli-$Z$ stabiliser before the code is read out gives rise to a detector cell that identifies a single \rx charge in the space-time picture.
                \textbf{(b)}~An analogous charge configuration at a spatial Pauli-$X$ boundary.
                We show an error configuration that gives rise to a single \rx charge.
            }
            \label{fig:Pauli-X-condensed}
        \end{figure}
        
        In Fig.~\ref{fig:Pauli-X-condensed}~{(a)}, 
        we show an error configuration where a single \rx charge is created at the temporal boundary where we measure all qubits in the Pauli-$X$ basis.
        Specifically, we show a measurement error on the final reading of the six-body Pauli-$Z$ stabiliser.
        Its only corresponding detection cell indicates the detection of a \rx boson, where we recall the convention we have adopted with Pauli-$Z$ stabilisers detect bosons with Pauli-$X$ labels.
        Indeed, this is the final detection cell that we measure of this type at this readout process, as we cannot infer the values of Pauli-$Z$ stabilisers from the choice of readout operation.
        Identifying a single \rx charge signifies that \rx charges and condensed at this boundary.
        We show the charge configuration at a spatial boundary in Fig.~\ref{fig:Pauli-X-condensed}~{(b)} to emphasise the analogy between this temporal boundary and a spatial boundary.
        In both cases, we can find similar error configurations where detector cells identify individual charges of any colour, provided they have a Pauli-$X$ label.
        For a temporal $X$-boundary they are introduced by measurement errors to Pauli-$Z$ stabilisers of the appropriate colour.
        For a spatial boundary, they are created by an $XX$ error supported on an edge of the appropriate colour.

\begin{figure}[tb]
            \centering
            \includegraphics[width=1.00\linewidth]{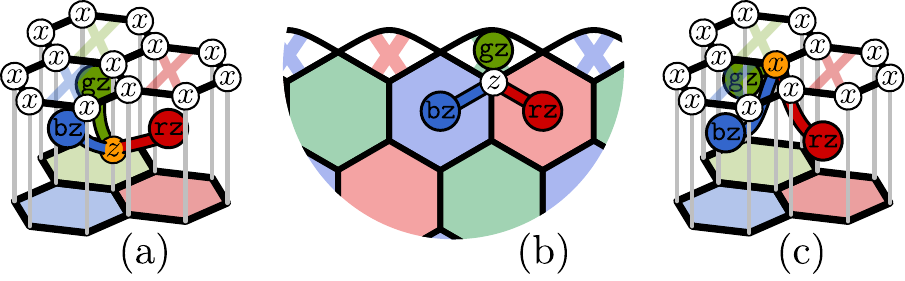}
            \caption{
                Configurations of confined charges at a Pauli-$X$ boundary.
                {\bf (a)}~A physical Pauli-$Z$ error close to the temporal domain wall created by measuring all physical qubits in the Pauli-$X$ basis creates a configuration of confined charges.
                {\bf (b)}~We compare the confined charges of the temporal Pauli-$X$ boundary to an analogous configuration of charges created at a spatial Pauli-$X$ boundary of the color code.
                We show the error and the charge configuration on the two-dimensional lattice.
                {\bf(c)}~A single measurement error, marked by the orange qubit, that occurs on one of the physical-qubit measurements at the readout step creates a configuration of confined charges equivalent to that shown in (a).
            }
            \label{Fig:Pauli-X-confined} 
        \end{figure}

        While certain charges are condensed at a temporal boundary, we find that other types of charges are confined.
        Let us posit that, since charges with Pauli-$X$ labels are condensed at Pauli-$X$ boundaries, all charges with other labels must be confined.
        In Fig.~\ref{Fig:Pauli-X-confined}~{(a)} we show a single Pauli-$Z$ error that gives rise to a configuration of color-code charges that is allowed by the bulk fusion rules that are confined at a Pauli-$X$ boundary.
        We contrast this configuration in the space-time picture with an analogous configuration of defects at a spatial boundary in the two-dimensional picture in Fig.~\ref{Fig:Pauli-X-confined}~{(b)}.
        In the space-time picture, Fig.~\ref{Fig:Pauli-X-confined}~{(a)}, these detection cells are completed by taking the last six-body reading of the Pauli-$X$ type stabilisers, and comparing them with the values of the same stabilisers that are inferred from the single-qubit measurements made at the readout step.
        Indeed, we can find no error configuration that locally annihilates any one of the individual charges of this configuration such that the fusion rules of this error configuration are violated.
        We find the same configuration of charges if we have a measurement error at the final readout step, see Fig.~\ref{Fig:Pauli-X-confined}~{(c)}.
        
        We can attribute the confinement of charges during this condensation procedure to a fault-tolerant readout step.
        Measurement errors at this final readout step also give rise to confined subsets of charges.
        In Fig.~\ref{Fig:Pauli-X-confined}~{(c)}, we show a measurement error that occurs on one of the physical qubits during the readout step.
        We find that this configuration of charges identified by the detection cells is identical to that of Fig.~\ref{Fig:Pauli-X-confined}~{(b)}, i.e., a physical error that produces a configuration of charges that is consistent with the fusion rules of the bulk phase.
        We can compare the detection of confined charges, to the detection of condensed charges at this temporal boundary, to find that it is not \textit{a priori} obvious that we should expect to detect charges that respect the color code fusion rules close to a temporal boundary.
        The confining effect during the readout enables us to employ standard decoding methods to identify errors during readout.
        
        With this example, let us finally note that the distinction between the confined charges with a Pauli-$Y$ label and those with a Pauli-$Z$ label is not well defined at the Pauli-$X$ boundary.
        At the microscopic level, this is due to the fact that at the final time step where we infer the values of Pauli-$X$ stabiliser detection cells at readout, we do not make a Pauli-$Z$ detection-cell measurement.
        Macroscopically too, this is a generic feature of a boundary that condenses charges with a Pauli-$X$ label.
        This is due to the fact that we can locally create individual charges with a Pauli-$X$ label of any colour and arbitrarily fuse them with the confined charges that are in the vicinity of this boundary.
        As such, charges with a Pauli-$Y$ label and a Pauli-$Z$ label are indistinguishable when they are close to a Pauli-$X$ boundary.
        This is another example of how anyon condensation leads to identification of distinct confined anyons, as described in Sec.~\ref{sec:CondTheory}.
        
        Let us next look at the temporal boundary created by making Bell measurements on the lattice edges for some choice of colour.
        Without loss of generality we choose to make Bell measurements on the green edges.
        Like the temporal domain we have already discussed that gave rise to a Pauli-$X$ type boundary, here we find that the temporal domain wall we produce is analogous to that of a green color-code boundary.
        
        \begin{figure}[tb]
            \centering
            \includegraphics[width=0.75\linewidth]{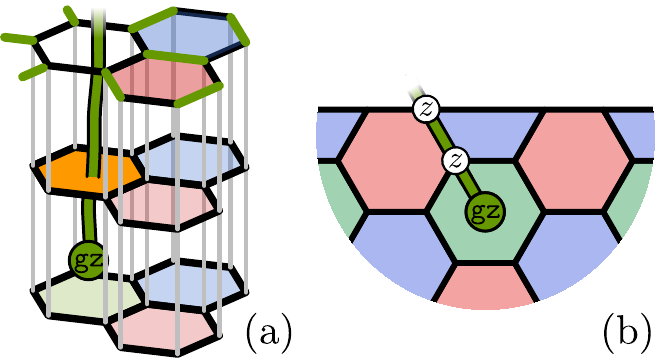}
            \caption{
                A condensed charge at green boundaries.
                \textbf{(a)}~A single measurement error on the final reading of a green six-body stabiliser immediately before the readout step is identified by a single detection cell only.
                This is consistent with the behaviour of a temporal boundary that condenses green charges in the space-time picture.
                \textbf{(b)}~A configuration of physical errors that creates a single green charge at a spatial boundary of the color code, as shown in two dimensions.                
            }
            \label{Fig:green-condensed}
        \end{figure}
        
        Again, we begin by looking at the condensed charges in this example.
        In Fig.~\ref{Fig:green-condensed}~{(a)}, we show a measurement error during a reading of a six-body Pauli-$X$ stabiliser.
        Given that we cannot infer the values of the green stabilisers from the Bell measurements we measure during readout, this is the final reading of this specific detection cell.
        This means the detection cells give rise to a syndrome configuration showing a single \gz charge.
        We show the charge configuration next to a green boundary of the color code in two dimensions in Fig.~\ref{Fig:green-condensed}~{(b)}.
        One can check that we could have equivalently made a single \gx or \gy charge with different configurations of measurement errors at the final reading of the green stabiliser.
        We therefore observe physics that is consistent with that of a green boundary at the temporal domain wall we created in the space-time picture, where we performed a measurement that condenses the green charges to read out the logical qubit.

        \begin{figure}[tb]
            \centering
            \includegraphics[width=0.8\linewidth]{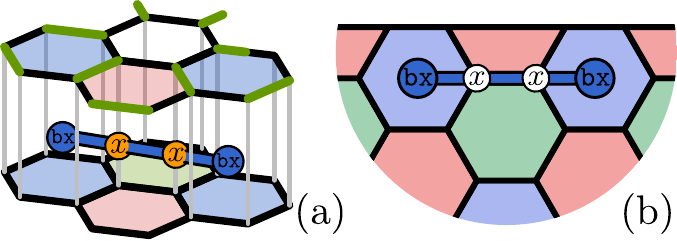}
            \caption{
                Confined charges at green boundaries.
                \textbf{(a)}~Errors supported on a blue edge creates charges at two blue detection cells.
                This is an example of a configuration for confined charges at a green boundary.
                \textbf{(b)}~An analogous configuration of charges to that of (a) shown at a green spatial boundary in two dimensions.
            }
            \label{Fig:green-confined}
        \end{figure}
        Let us lastly look at the confined charges at the green temporal boundary.
        In Fig.~\ref{Fig:green-confined}~{(a)} we show a two-qubit error supported on a blue edge.
        The figure shows detection cells identifying a pair of \bx charges.
        We compare the charge configuration found at the temporal boundary to the more familiar green spatial boundary shown in two dimensions in Fig.~\ref{Fig:green-confined}~{(b)}.
        
\begin{figure}[b]
            \centering
            \includegraphics[width=0.75\linewidth]{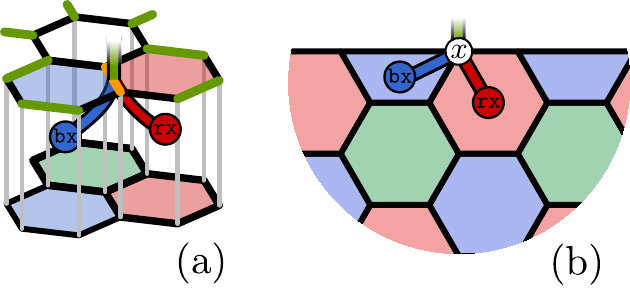}
            \caption{
                Charges at green boundaries.
                \textbf{(a)}~A measurement error on a two-body measurement that occurs during readout creates charges at two detection cells; a blue cell and a red cell.
                This is an allowed charge configuration at a green boundary, where individual green charges can be created locally.
                \textbf{(b)}~An analogous configuration of charges to that of (a) shown at a green spatial boundary in two dimensions.
                We note that the net charge in both of the displayed charge configurations is that of a green anyon.
                This is consistent with the behaviour of a boundary that condenses green charges where individual green charges can be created locally.
            }
            \label{Fig:green-intermediate}
        \end{figure}

        We can also look at charges at the temporal boundary created by making Bell measurements at the green edges.
        In Fig.~\ref{Fig:green-intermediate}~{(a)} we show a single measurement error that occurs during the Bell measurement readout step.
        We note that one can find a physical error on a single qubit, that occurs just before the readout, that is equivalent to this measurement error in the sense that the two local errors give rise to an equivalent charge configuration.
        Here we focus on a measurement error.
        The figure shows detection cells identifying a pair of charges; an \rx and a \bx charge.
        While this charge configuration is inconsistent with the fusion rules that are allowed in the bulk of the color code, this configuration is in fact allowed at a boundary that condenses green charges.
        Indeed, as we can create green charges locally at a green boundary, the colouring of the confined red- and blue-labelled charges with the same Pauli label becomes ill-defined.
        As such, the creation of any even parity of charges that take any colour other than green is allowed near to a green boundary.
        We observed similar physics at the Pauli-$X$ boundary discussed earlier in this section, except where in the previous example the Pauli labels become ill defined, here the color labels become ambiguous in this example.
        We compare the charge configuration found at the temporal boundary to the more familiar green spatial boundary shown in two dimensions in Fig.~\ref{Fig:green-confined}~{(b)}.
        
        In fact, the last example of a blue and a red charge close to a green boundary has an unusual quirk that we finally point at here.
        Specifically, we can take two different perspectives on this charge configuration.
        At a microscopic level we can view it as a pair of confined charges.
        Alternatively, from a more macroscopic perspective we can view the charge as a single green charge.
        The fusion rules of the color code are such that the union of a red and a blue charge with the same Pauli label fuse to give a green charge.
        As such, we can view the charge configurations shown in Fig.~\ref{Fig:green-intermediate} as demonstrating a net charge that has a green label from a global perspective.
        Of course, as we have mentioned, this is consistent with the physics of a green boundary, where individual green charges can be created locally.

    \subsection{Interplay between temporal and spatial boundaries and fault-tolerance}
        \label{sec:BoundaryInterplay}
        
        To investigate the fault-tolerance properties of a quantum computation in topological stabiliser codes the space-time picture is essential.
        In this section, we begin by using the established framework of anyon condensation to describe error detection at boundaries.
        We then analyse the fault tolerance of a space-time computational scheme by studying the configuration of its boundaries.

        In the previous section, Sec.~\ref{sec:MicroTempBdry}, we have developed a microscopic theory for the temporal boundaries in the space-time picture.
        This allows us to view fault-tolerant quantum computational protocols with the color code as space-time volumes enclosed by one of six boundaries in both the space and time direction, where the boundary types correspond to the Lagrangian subgroups of the color code.
        We can analyse these volumes to determine processes that give rise to logical errors. 
        
        At any given time-step of a computation, the logical state of the encoded qubits is determined by the parity of charges condensed at the boundaries (or other condensation objects, like twist defects as discussed in Sec.~\ref{sec:Cond1d}).
        A non-trivial logical Pauli error corresponds to an anyon string supported on a topologically non-trivial path.
        This means that it cannot be deformed continuously to a point.
        Note that one has to take into account how anyon strings interact with boundaries and other types of defects when deforming them through space-time.
        Fault-tolerance is achieved when the support of any non-trivial logical error grows with the system size.
        
\begin{figure}[tb]
            \centering
            \includegraphics[width=0.9\linewidth]{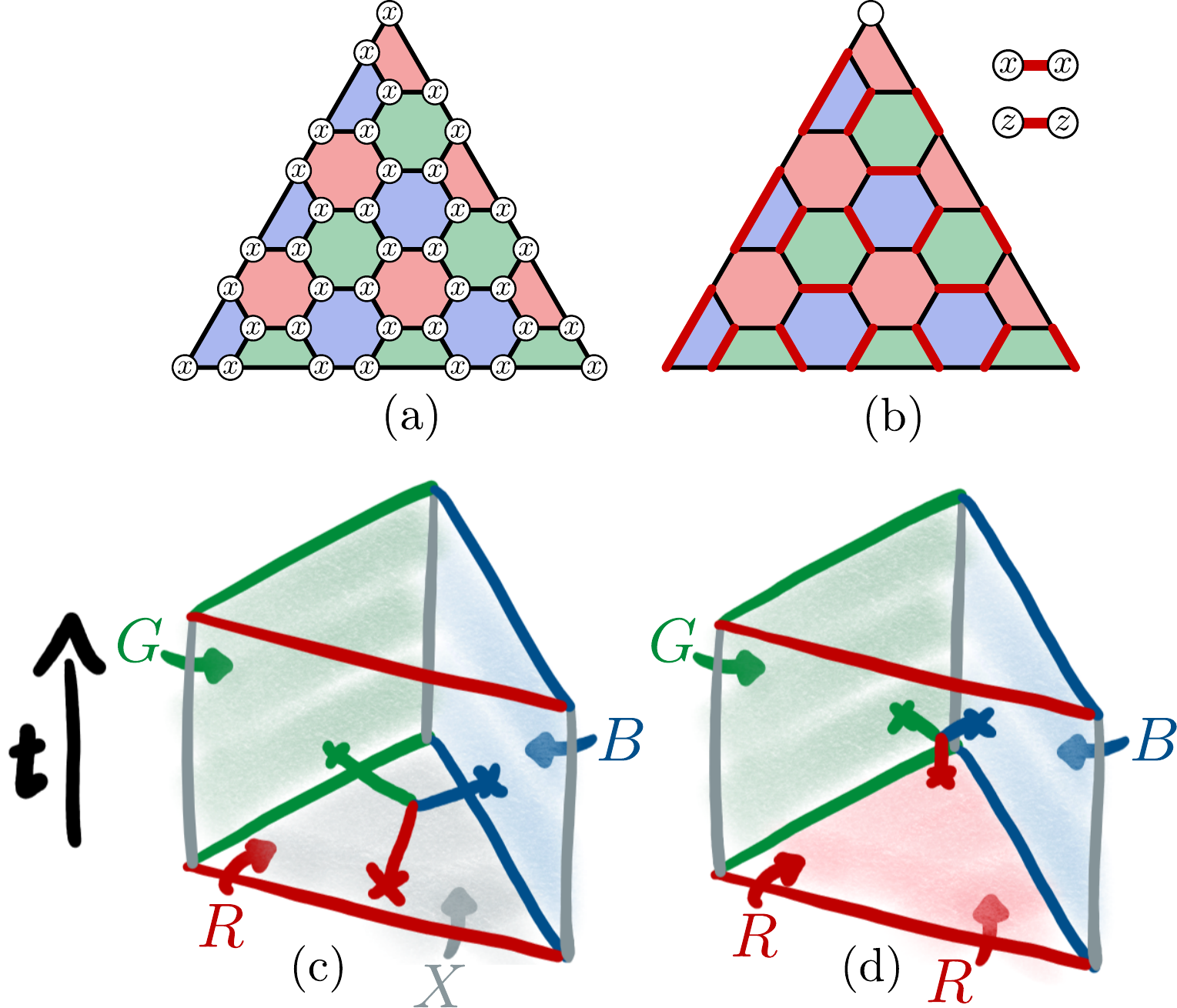}
            \caption{
                A triangular color code with coloured boundaries can be initialised (and read out) fault-tolerantly using one of the three temporal Pauli boundaries.
                In \textbf{(a)}, we show as an example how initialising a $\overline{X}$ eigenstate is achieved by preparing all physical qubits in an eigenstate of the single qubit $X$ operator, creating a temporal $X$-boundary.
                If a coloured boundary is used, as shown in \textbf{(b)} with the example of a red temporal boundary, the process is not fault-tolerant and corresponds to the state injection of the single qubit state on the qubit in the top corner of the code.
                To see why this is the case, we can check the support of the logical operators in the space-time picture of the process.
                In \textbf{(c)}, we can see that the weight $d$ is maintained throughout the process.
                On the other hand, in \textbf{(d)}, we can find a constant size logical operator on one of the corners.
            }
            \label{fig:InitialisationTrinagularCC}
        \end{figure}

        To ensure fault-tolerance, we have to pick a suitable boundary configuration in our computation such that all of the non-trivial logical errors have a macroscopic length.
        To illustrate this, let us look at different temporal boundaries to initialise the triangular color code, see Fig.~\ref{fig:InitialisationTrinagularCC}.
        We find that the $X$ boundary is a valid temporal boundary to initialise this code in a logical $\overline{X}$ eigenstate.
        This is because it confines all bosons with a Pauli-$Z$ or Pauli-$Y$ label.
        This leaves only the three spatial coloured boundaries for the logical $\overline{Z}$-error ($\overline{Y}$-error) strings to terminate, see Fig.~\ref{fig:InitialisationTrinagularCC}~(c).

        If we were to use a red boundary, on the other hand, we can now find a logical error with constant support.
        It spans between the spatial green and blue boundary as well as the temporal red boundary, see Fig.~\ref{fig:InitialisationTrinagularCC}~(d).
        Such configurations can easily be spotted by keeping the space-time diagram of a computational operation in mind.
        Finally, we would like to point out that even this non-fault-tolerant protocol has its uses in quantum computational schemes.
        In Ref.~\cite{Landahl14} a similar protocol is proposed to inject logical non-Pauli eigenstates from single qubits into patches of colour code.
        The equivalent readout protocol teleports the state of the logical qubit onto a single physical qubit.

\begin{figure}[tb]
            \centering
            \includegraphics[width=1.00\linewidth]{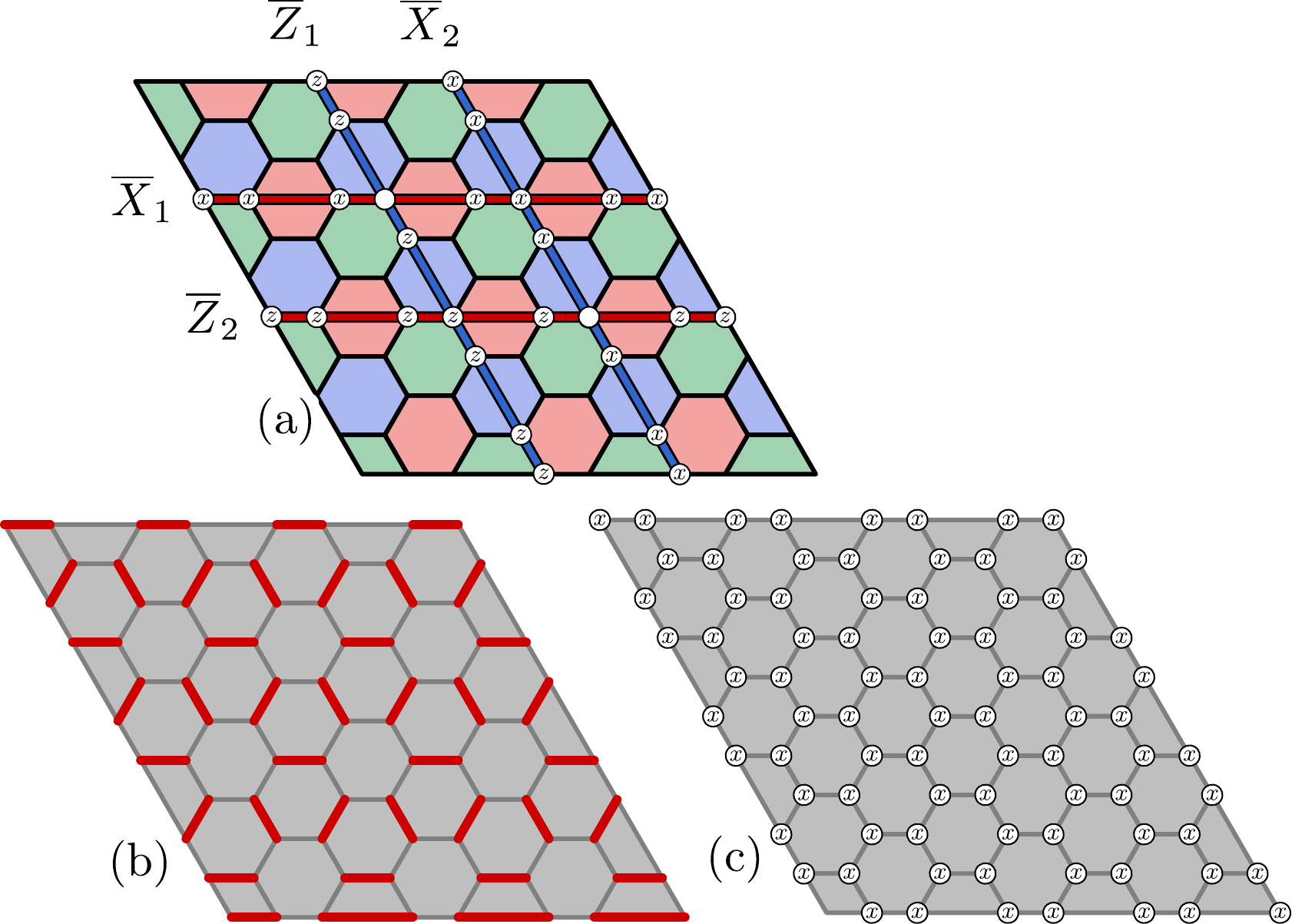}
            \caption{
                We show a rectangular patch of color code encoding two logical qubits and its logical operators, see \textbf{(a)}.
                This code can be read out or initialised in five different ways in a fault-tolerant manner.
                \textbf{(b)}~shows an example where all red edges get initialised/read out with Bell pair stabilisers, i.e., $XX$ and $ZZ$.
                This corresponds to initialising/measuring $\overline{X}_1$ and $\overline{Z}_2$.
                Choosing instead the blue coloured temporal boundary (not shown) would initialise/measure $\overline{Z}_1$ and $\overline{X}_2$.
                \textbf{(c)}~corresponds to interfacing to the vacuum with the Pauli-$X$ boundary by initialising/measuring every single qubit in the $X$-basis.
                This prepares/reads out $\overline{X}_1$ and $\overline{X}_2$.
                Similarly, choosing the temporal $Z$ boundary corresponds to the initialisation/readout of $\overline{Z}_1$ and $\overline{Z}_2$ (not shown).
                Lastly, using the temporal Pauli-$Y$ boundary, we initialise/read out the product of $\overline{Z}_1\overline{X}_2$ and $\overline{X}_1\overline{Z}_2$ (not shown).
            }
            \label{fig:ReadOut}
        \end{figure}

        As a last example, consider the square color code in Fig.~\ref{fig:ReadOut}.
        In Sec.~\ref{sec:CondTimeMax}, we discussed how the logical qubits can be initialised respectively read out using anyon condensation.
        The basis in which the logical qubits are read out/initialised is determined by the strings that can condense at the boundary which is introduced by the condensation.
        For example, the red temporal boundary in Fig.~\ref{fig:ReadOut} reads out(initialises) the parity of $\overline{X}_1$ and $\overline{Z}_2$.
        Importantly, a subset of the logical Paulis are red anyon strings connecting the two red boundaries.
        Let us turn our attention to the case when the temporal boundary has a different color than all the logical strings that can condense at any spatial boundary, e.g. a green boundary for the code in Fig.~\ref{fig:ReadOut}.
        In fact the logical Pauli-$Y$s 
        can be represented by a green string operators connecting opposing corners.
        Hence, one might assume that we can use a green temporal boundary to initialise.
        However, this process is not fault-tolerant.
        Again, we can see this by looking at the boundary configuration in the space-time.
        Here, at all four corners of the code, a green temporal boundary is interfaced with a red and a blue spatial boundary.
        This means we can find a operator of constant support who changes the parity of the charges condensed at these three boundaries with constant support, similar to the one shown in Fig.~\ref{fig:InitialisationTrinagularCC}~(d).
        Nonetheless, this protocol can be used as an injection/teleportation scheme between the rectangular color code and a $[\![4,2,2]\!]$-code, where the qubits in the corners are the physical qubits of this small code.

        The two cases where we described a injection/teleportation between a small code and a color code can be interpreted as a code switching~\cite{bombin2016dimensional} protocol.
        Indeed, here we see that we switch between a two-dimensional and a zero-dimensional code.
        Alternatively, by condensing anyons in the bulk but not along a boundary of the code, we can switch between a two-dimensional and a one-dimensional code.
        We can also extend this to the third spatial dimension and interpret the gauge fixing protocol presented in Ref.~\cite{bombin2016dimensional} to a condensation procedure.
        Specifically we can obtain a two-dimensional color code from a three-dimensional color code by condensing its charges in the bulk, up to its boundary. 
        
        So far, we have discussed the fault-tolerance in initialisation and readout protocols considering relatively simple examples where the logical qubits were encoded in the spatial boundary configuration.
        In general, all condensation objects have to be considered, like corners and twists which are the subject of Sec.~\ref{sec:Cond1d}.

        Combining anyon condensation with the space-time picture results in some no-go~\cite{PhysRevLett.102.110502,PhysRevLett.110.170503,PhysRevA.91.012305,PhysRevA.102.022403,Webster2022Universal} theorems for stabiliser-based topological quantum computation.
        In any topological stabiliser code, non-trivial anyon strings define the logical Pauli operators.
        From this we can deduce that temporal domain walls can only be used to initialise logical Pauli-eigenstates, apply logical Clifford operations or perform Pauli-basis readouts.
        To promote this set of topologically protected operations to an universal one, one needs to make use of topological codes in higher spatial dimensions~\cite{bombin2016dimensional}.
        Alternatively, one could include some operation which is not topologically protected, like the described state injection, to inject non-Pauli eigenstates.
        Combined with state distillation protocols~\cite{PhysRevA.71.022316}, or other appropriate means, allows us obtain a universal set of fault-tolerant logical operations.

    \subsection{Stability experiments in topological codes}

        \label{sec:StabilityExperiment}
        In the following we discuss memory and stability experiments in topological error correcting codes.
        We find that the theory of spatial and temporal boundaries that we have introduced is well suited to explain stability experiments as well as to devise variants thereof.
        After reviewing memory and stability experiments for general topological codes, we turn to the color code and show how it can be used to perform a combined memory-stability experiment.
        
        To test the performance of an error correction code as a quantum memory, we can perform a memory experiment.        
        These experiments consist of three parts.
        First, we begin by fault-tolerantly initialising the logical qubit(s) of the code in a certain state.
        Next, we let the code idle for a given period of time while measuring its stabiliser generators.
        And finally, we measure the logical qubit(s) and verify if they remained in the initialised state or not.
        For topological codes, such experiments check if we can tolerate the errors affecting the physical qubits and the gadgets used to perform the stabiliser measurements.
    
        In addition to strings of errors introducing unwanted transformations to logical qubits, logical errors may also occur during a computation due to strings of faults that align in the time-like direction in the space-time picture.
        This can be a problem, for instance, when we perform gates by code deformations.
        In recent work~\cite{Gidney2022stability}, a simple experiment has been proposed to check the performance of the toric code against time-like logical errors.
        The experiment is called \textit{stability experiment} and consist of a patch of toric code which does not encode any logical qubits.
        However, it is initialised and read out in a manner allowing us to check for occurrences of strings of non-correctable errors in the measurements of the stabiliser generators.
        As the direction of these measurement errors is in the time-direction, the experiment effectively checks for temporal logical errors.
        These are errors connecting distinct temporal boundaries.
        Considering the full space-time of the experiment, we can see that it is equivalent to a memory experiment ``rotated by $90^\circ$", i.e., where one of the spatial directions is exchanged with the temporal direction.
        The following discussion of said experiments in terms of spatial and temporal boundaries allows us to construct stability experiments for any topological stabiliser code.
        We turn to the color code as a concrete example.
        Interestingly, we find that we can use the color code to perform a stability and a memory experiment simultaneously.

        \begin{figure}[tb]
            \centering
            \includegraphics[width=1\linewidth]{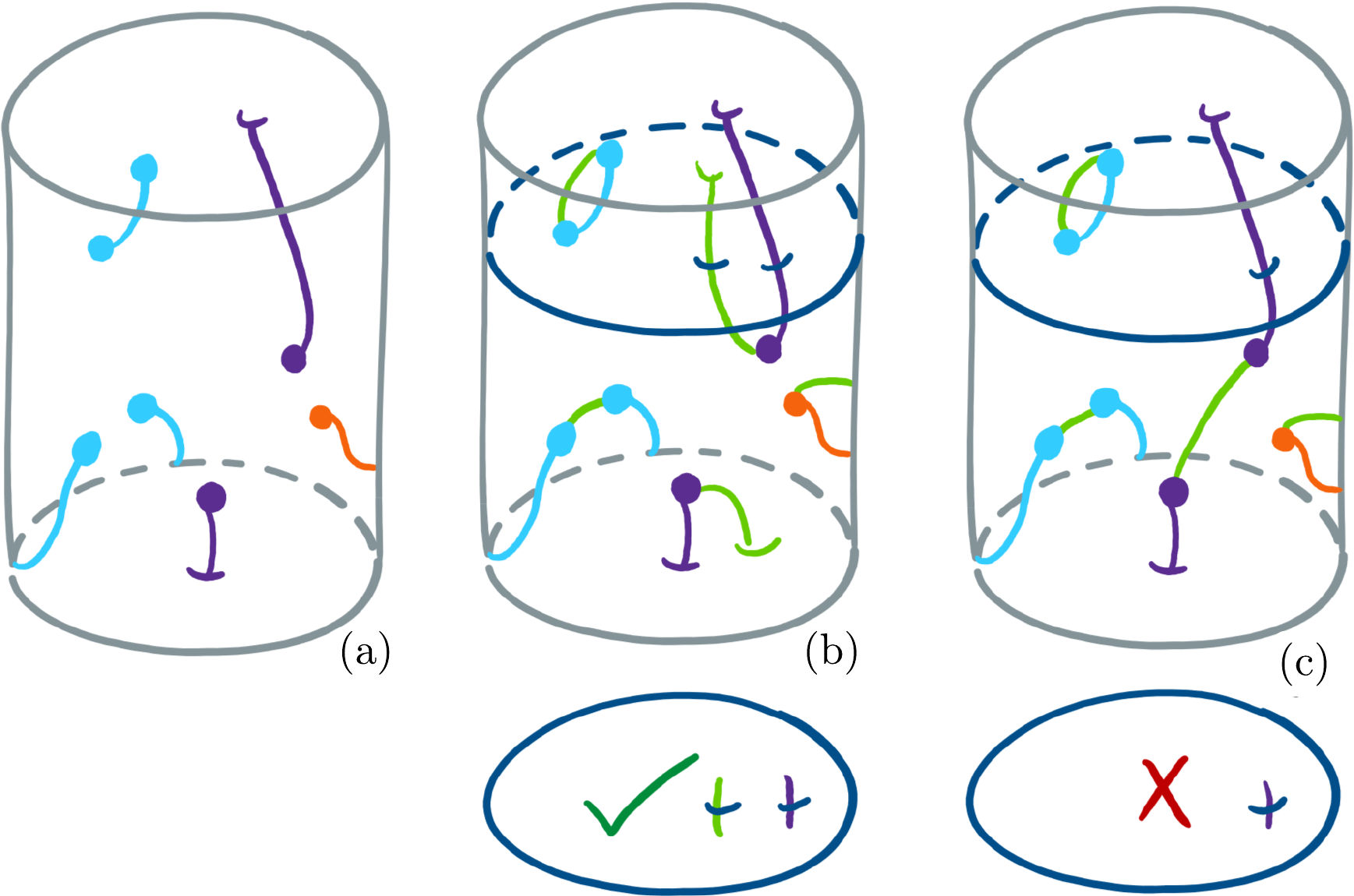}
            \caption{
                The space-time diagram of a simple stability experiment.
                The temporal boundaries at the top and the bottom are of one type, the spatial boundaries on the sides are of a different type.
                \textbf{(a)}~errors occurring during the experiment correspond to string operators creating anyons in space-time.
                We depict different types of anyons using different colours.
                The stability experiment is concerned with anyons which can condense only at the top or bottom boundary (drawn in purple) or at the corners (cyan).
                Errors creating anyons which can condense on a spatial boundary (orange) are not relevant to the outcome of the experiment.
                \textbf{(b)}~shows a successful experiment, as the errors together with the applied corrections (green) lead to an even number of string operators of the considered anyons through any time slice (dark blue).
                This implies that the evaluated conserved quantities are indeed conserved.
                \textbf{(c)}~shows a failed experiment.
                Here, the errors together with the applied correction leads to a temporal logical error.
                We see this as a violation of the conserved quantity since an odd number of string operators cross a given time slice.
            }
            \label{fig:Stabilityspace-timeSketch}
        \end{figure}
        In its simplest form, a stability experiment constitutes a cylinder in space-time with one type of spatial boundary wrapping around the cylinder and a second type of temporal boundaries capping it off at the top and bottom.
        We call the Lagrangian subgroups describing the temporal boundaries $\cL_1$ and the spatial boundary $\cL_2$.
        A space-time sketch of the experiment is shown in Fig.~\ref{fig:Stabilityspace-timeSketch}.
        We are interested in the anyonic charges which can condense at the top and bottom but not at the sides of the cylinder.
        These are charges which are not contained in the Lagrangian subgroup describing the spatial boundary $\cL_2$ but can condense at the top/bottom boundary or the corner between the two boundaries.
        More precisely, we consider anyons in $\cL_1 \times \cL_2 \backslash \cL_2$.
        Depending on the boundaries used, a different number of anyons fulfil the required property.
        The stability experiment then consists in checking whether an even or odd number of string operators corresponding to the anyons in question cross a given time-slice.
        This can be inferred from the parity of a set of stabiliser measurements.
        In Ref.~\cite{Gidney2022stability}, the product this set of stabiliser measurements is called a \emph{conserved quantity}~\cite{Kitaev03, Brown22conservation}.
        
        In the color code, a stability experiment can be performed using any combination of spatial and temporal boundaries, as long as two different boundaries are used.
        This is true since for any pair of different boundaries described by the Lagrangian subgroups $\cL_1$ and $\cL_2$ respectively, there exists at least one boson which is in $\cL_1$ but not in $\cL_2$.
        To maximise the significance of a performed experiment, we want to maximise the number of conserved quantities, i.e., maximise the number of bosons in $\cL_1$ but not in $\cL_2$.
        This is achieved by either picking two distinct coloured boundaries or two distinct Pauli boundaries.
        Supposing initialising single qubit states and performing single qubit measurements is simpler than preparing and measuring Bell-pairs, using Pauli-boundaries is experimentally simpler.

        \begin{figure}[tb]
            \centering
            \includegraphics[width=1.00\linewidth]{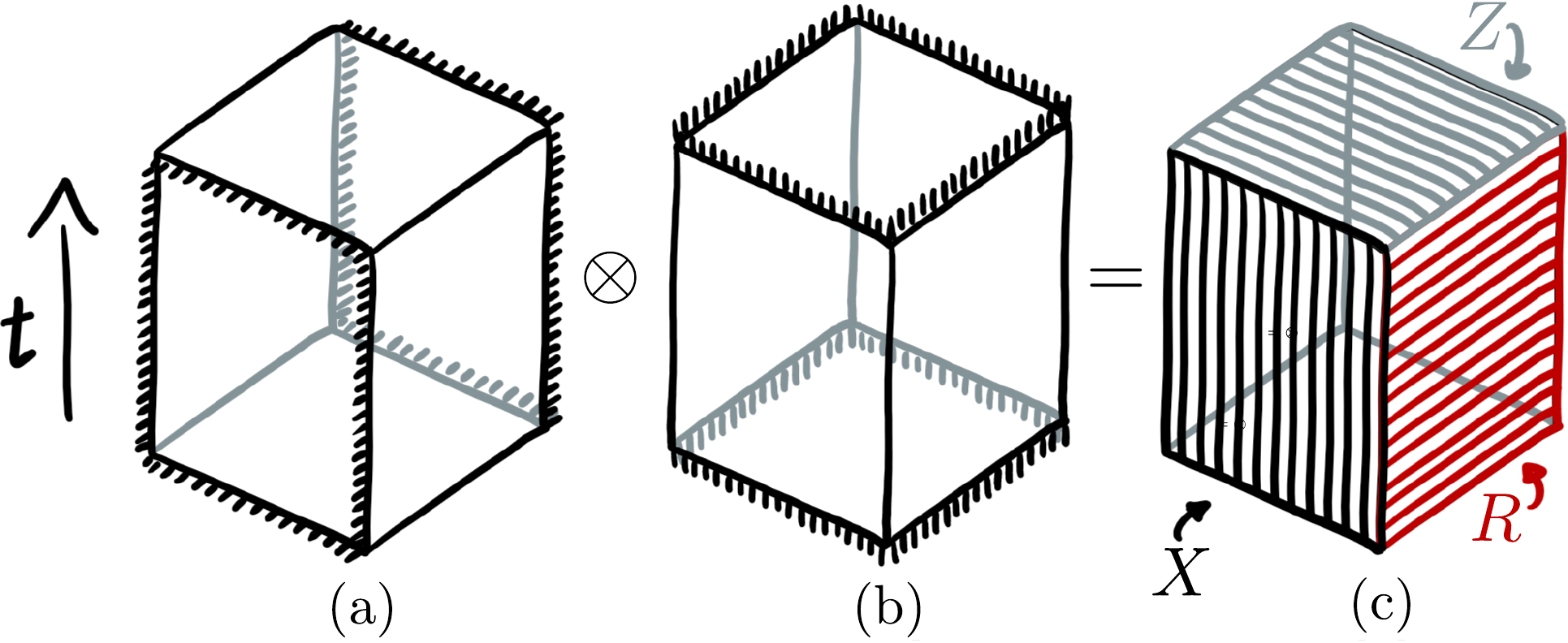}
            \caption{
                \textbf{(a)}~shows a surface code memory experiment and \textbf{(b)}~a surface code stability experiment.
                If we overlay the two we get a combined stability and memory experiment.
                As shown in \textbf{(c)}, this can be realised within a single color code.
                It looks like a cube in space-time where opposite boundaries are of the same type.
                A suitable boundary configuration has, for example, two pairs of Pauli boundaries and one pair of coloured boundaries.
            }
            \label{fig:StabMemExpCombined_space-time}
        \end{figure}
\begin{figure}[b]
            \centering        
            \includegraphics[width=.95\linewidth]{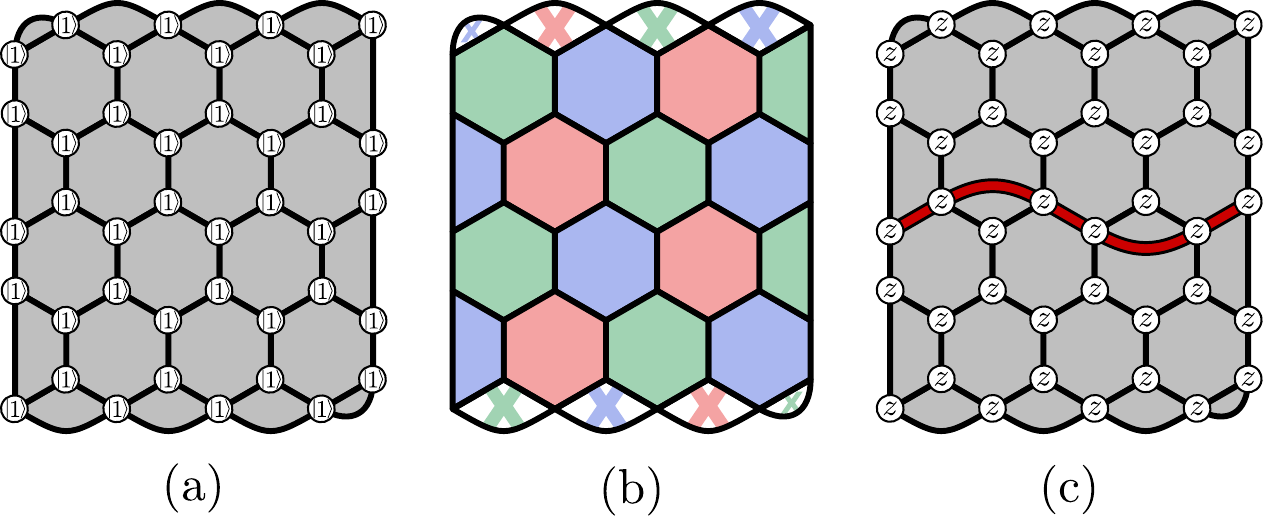}
            \caption{
                The combined memory-stability experiment using the color code.
                \textbf{(a)}~we initialise the code using a Pauli-$Z$ boundary by preparing all physical qubits in the $\ket{1}$ state.
                \textbf{(b)}~the stabilisers of a rectangular patch of color code with two red and two Pauli-$X$ boundaries are measured for $d$ rounds.
                \textbf{(c)}~we terminate the experiment by performing single-qubit Pauli-$Z$ measurements on all physical qubits.
                After error correction, a successful experiment reveals an even parity in all blue and green $X$-type stabilisers and a $+1$ outcome for depicted red string operator.
            }
            \label{fig:StabMemExpCombined_Stabilisers}
        \end{figure}

        As mentioned above, the goals of stability and memory experiments are similar.
        In both we check for the presence of a logical error by validating the parity of a space-time slice perpendicular to the direction of the logical operator.
        The difference is the direction in which the logical operator runs, a spatial direction for the memory experiment and a temporal direction for the stability experiment.
        If we overlay two code patches in the toric code phase, one used to carry out a stability experiment and a second one to perform a memory experiment, we obtain a code patch in the color code phase.
        This is sketched in Fig.~\ref{fig:StabMemExpCombined_space-time}.
        Such a code is capable of simultaneously checking for the presence of temporal and spatial logical errors within the same experiment, using certain anyon parities.
        The corresponding space-time diagram is shown in Fig.~\ref{fig:StabMemExpCombined_space-time}.
        A range of color code boundary conditions are suitable to carry out such a combined memory-stability experiment.
        We might, for example, initialise and terminate the experiment using a temporal Pauli-$Z$ boundaries and terminate the code in the spatial directions using red and Pauli-$X$ boundaries on opposite sides.
        The microscopic lattice model of this experiment together with instructions for initialisation, readout and evaluation are given in Fig.~\ref{fig:StabMemExpCombined_Stabilisers} and its caption.

    \subsection{Partial initialisation and readout}
        \label{sec:CondTimePart}
        
        A temporal domain wall introduced by partial condensation condenses a subset of bosons which do not form a Lagrangian subgroup.
        This implies that the logical Pauli operators which correspond to the condensed anyons get initialised or read out, depending on the orientation of the domain wall.
        However, since some anyons remain mobile through the introduced semi-transparent domain wall, some logical degrees of freedom remain encoded throughout this process.
        We hence call the effect of semi-transparent temporal domain walls \textit{partial initialisation} and \textit{partial readout}.
        Note that a semi-transparent domain wall can also act non-trivially on the mobile anyons and hence implement logical gates on the associated logical degrees of freedom.

\begin{figure}[tb]
	        \centering
	        \includegraphics[width=1.00\linewidth]{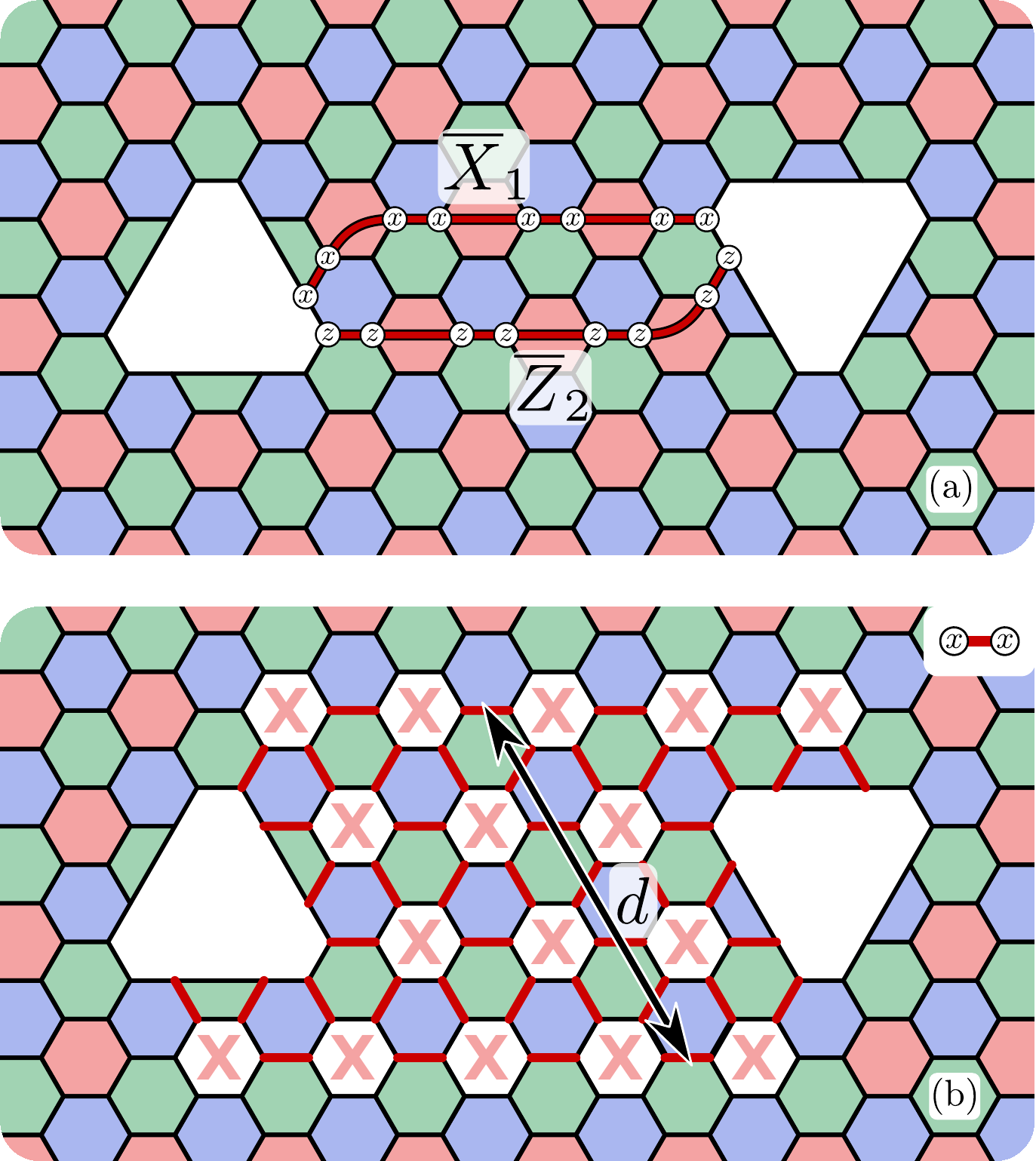}
	        \caption{
	            Two red punctures as shown here encode two logical qubits.
	            The logical operators $\overline{X_1}$ and $\overline{Z}_2$ are shown in \textbf{(a)}.
	            They can be chosen to have the exact same support.
	            However, it is still possible to read out one but not the other.
	            To measure $\overline{X_1}$, for example, we measure the $XX$ terms on the red edges in a region $R_L$, as shown in \textbf{(b)}.
	            The width $d$ of the region determines how many fault readouts can be tolerated.
            }
	        \label{fig:PartialReadOutPunctures}
	    \end{figure}

        Let us now illustrate partial condensation on exemplary instances of the color code.
        First, consider two qubits encoded in two punctures in the color code, see Fig.~\ref{fig:PartialReadOutPunctures}.        
        This is equivalent to the example shown in Sec.~\ref{sec:CondTimeMax} in Fig.~\ref{fig:ReadOutPuncturesFull}, where we condense a full Lagrangian subgroup to read out two logical qubits at once.
        Here now, we perform a partial condensation to obtain the value of one logical degree of freedom while leaving another one encoded.
        For example, we can measure the eigenstates of $\overline{X}_1$ while leaving the second qubit encoded.
        We achieve this by condensing \rx in region $R$.
        The second logical qubit remains encoded, as it's logical operators commute with all the \rx hopping terms while not being a combination thereof.
        Similarly, reversing the process lets us encode a second qubit in the punctures.
        Condensing different bosons leads to the partial readout of different logical degrees of freedom.
        For instance, if we condense \ry we measure the value of $\overline{X}_1\overline{Z}_2$.
        The logical degree of freedom which remains encoded is a logical parity qubit with $\overline{X} \equiv \overline{X}_1 \simeq \overline{Z}_2$ and $\overline{Z} \equiv \overline{Z}_1\overline{X}_2$.
	    
\begin{figure}[tb]
            \centering
            \includegraphics[width=0.95\linewidth]{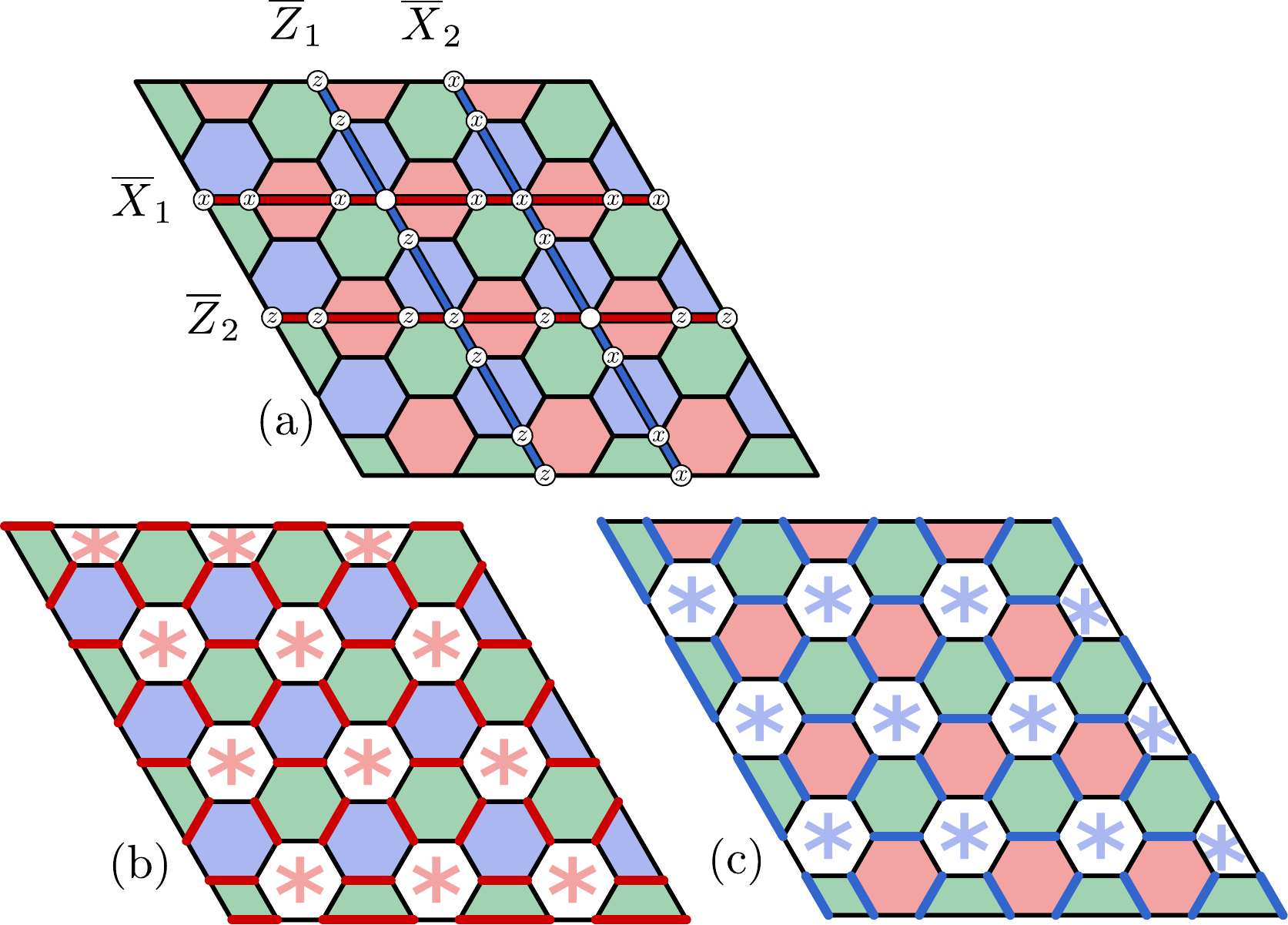}
            \caption{
                A color code encoding two logical qubits and its logical operators, see \textbf{(a)}.
                A color code boson gets condensed to obtain a toric code, encoding one logical qubit.
                Depending on which boson is chosen to be condensed, one logical operator is being read out.
                The asterisks are placeholders for the Pauli-label of the condensed boson.
                \textbf{(b)}~depicts the readout of $\overline{X}_1$, $\overline{Z}_2$ or $\overline{X}_1 \overline{Z}_2$ if the chosen boson is $\rx$, $\rz$ or $\ry$ respectively.
                Similarly, in \textbf{(c)} we condense $\bx$, $\bz$ or $\by$ to read out $\overline{X}_2$, $\overline{Z}_1$ or $\overline{Z}_1\overline{X}_2$.
                Condensing a green boson results in a partial teleportation of one of the logical qubits into a $[\![4,1,2]\!]$-code supported on the four corner qubits.
            }
            \label{fig:PartiaReadOut}
        \end{figure}

	    As a second example, let us return to the rectangular color code patch, see Fig.~\ref{fig:PartiaReadOut}.
        First, let us consider the partial condensation of the \rx anyon.
        To condense \rx, we measure the $XX$ hopping terms on all red edges.
        From these measurements we infer the value of the logical operator $\overline{X}_1$.
        The second logical qubit remains encoded in the obtained code.
        We identify this to be a surface code, see Fig.~\ref{fig:TC_lattice_excitations}.
        Similarly, condensing \rz by measuring red $ZZ$ terms measures $\overline{Z}_2$, leaving the first qubit encoded in the surface code we produce.
        If we choose to condense \ry, we measure the product $\overline{X}_1\overline{Z}_2$.
        The surface code now encodes a parity qubit with the logical operators $\overline{X} \equiv \overline{X}_1 \simeq \overline{Z}_2$ and $\overline{Z} \equiv \overline{Z}_1\overline{X}_2$.
        These new logical operators are composed of hopping terms of the deconfined bosons $\e \equiv \rx \simeq \rz$ and $\m \equiv \gy \simeq \by$ respectively.
        Condensing one of the three blue bosons, \bx, \by or \bz lets us infer the value of the logical Pauli operators $\overline{X}_2$, $\overline{Z}_1\overline{X}_2$ or $\overline{Z}_1$, respectively.
        Condensing a green boson leads leaves one logical qubit encoded in the obtained surface code.
        The other qubit, however, is not read out, but now encoded in a $[\![4,1,2]\!]$ code which is supported on the four corner qubits.

        The fault-tolerance of a computation in the color code involving partial initialisation (readout) can be assessed with the same methods as described the Sec.~\ref{sec:BoundaryInterplay}.
        Together with spatial domain walls, this completes the space-time picture of computations in the color code and provides a unified tool to design and study topologically protected computational protocols in 2D topological stabiliser codes.

\section{Terminating the color code domain walls}
    \label{sec:Cond1d}

    \begin{figure}[b]
        \centering
        \includegraphics[width=0.4\linewidth]{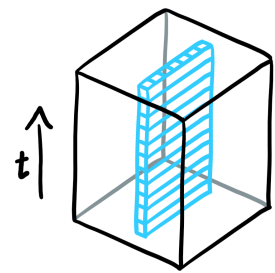}
        \caption{
            In this section, Sec.~\ref{sec:Cond1d}, we create consider domain walls between the parent phase and itself.
            We create them by condensing a one-dimensional spatial region, as depicted by the $(1+1)$-dimensional cyan region in the $(2+1)$-dimensional space-time figure.
        }
        \label{fig:CondensationOneDim}
    \end{figure}
    In this section we terminate the domain walls realisable in the color code and study the features emerging at the end-points.
    To terminate a domain wall, we can condense color code anyons in a one-dimensional spatial, open region.
    This is depicted as a $(1+1)$-dimensional object in $(2+1)$-dimensional space-time in Fig.~\ref{fig:CondensationOneDim}.
    Depending on the type of condensation we apply, we obtain different types of domain walls between the color code and itself.
    While we start the discussion by terminating invertible domain walls, we extend the theory here to include opaque and semi-transparent domain walls as well.

    In what follows, in Sec.~\ref{sec:Cond1dTriv} we review the theory of twist defects in the color code~\cite{Kesselring18} and describe how they are manifest in the space-time picture. In Sec.~\ref{sec:Cond1dMax} we investigate the corners of the color code~\cite{Kesselring18}, i.e., twist defects that divide two distinct boundary types. We reinterpret corners in terms of the condensates of their adjacent boundaries, to incorporate corners into our theory of anyon condensation with the color code. In Sec.~\ref{sec:Cond1dPart} we use the theory of anyon condensation to classify a new type of domain wall that we refer to as a semi-transparent domain wall. This elaborates on the theory that was briefly introduced in Ref.~\cite{Thomsen22} where semi-transparent domain walls were employed in fault-tolerant quantum-computing processes with the color code. Finally, in Sec.~\ref{sec:Cond1dLS} we discuss the physics of lattice surgery of the color code~\cite{Landahl14, Thomsen22}, in terms of the semi-transparent domain walls we have discussed throughout this section using the language of anyon condensation.

    \subsection{Invertible domain walls and twist defects}
        \label{sec:Cond1dTriv}
        
        Anyon models have associated domain walls that transform the anyons onto other anyon types as the domain walls are crossed~\cite{Bombin10,KitaevKong12,Barkeshli14}.
        These domain walls can be terminated, where we call their endpoints twist defects or simply twists.
        For the color code, the $72$ distinct twist defects that are described macroscopically and microscopically in Refs.~\cite{Yoshida15, Kesselring18}.
        Here, we will briefly recall some important results which we make use of in the following sections.
        In particular we show three examples of invertible domain walls in the lattice model.
        Furthermore, we talk about the twist defects that terminate at invertible domain walls, and how they can be used to store logical qubits.
        Finally, we discuss how domain walls and twist defects are manifest in the space-time picture of topological phases of matter.
        
\begin{figure}[tb]
	        \centering
	        \includegraphics[width=.8\linewidth]{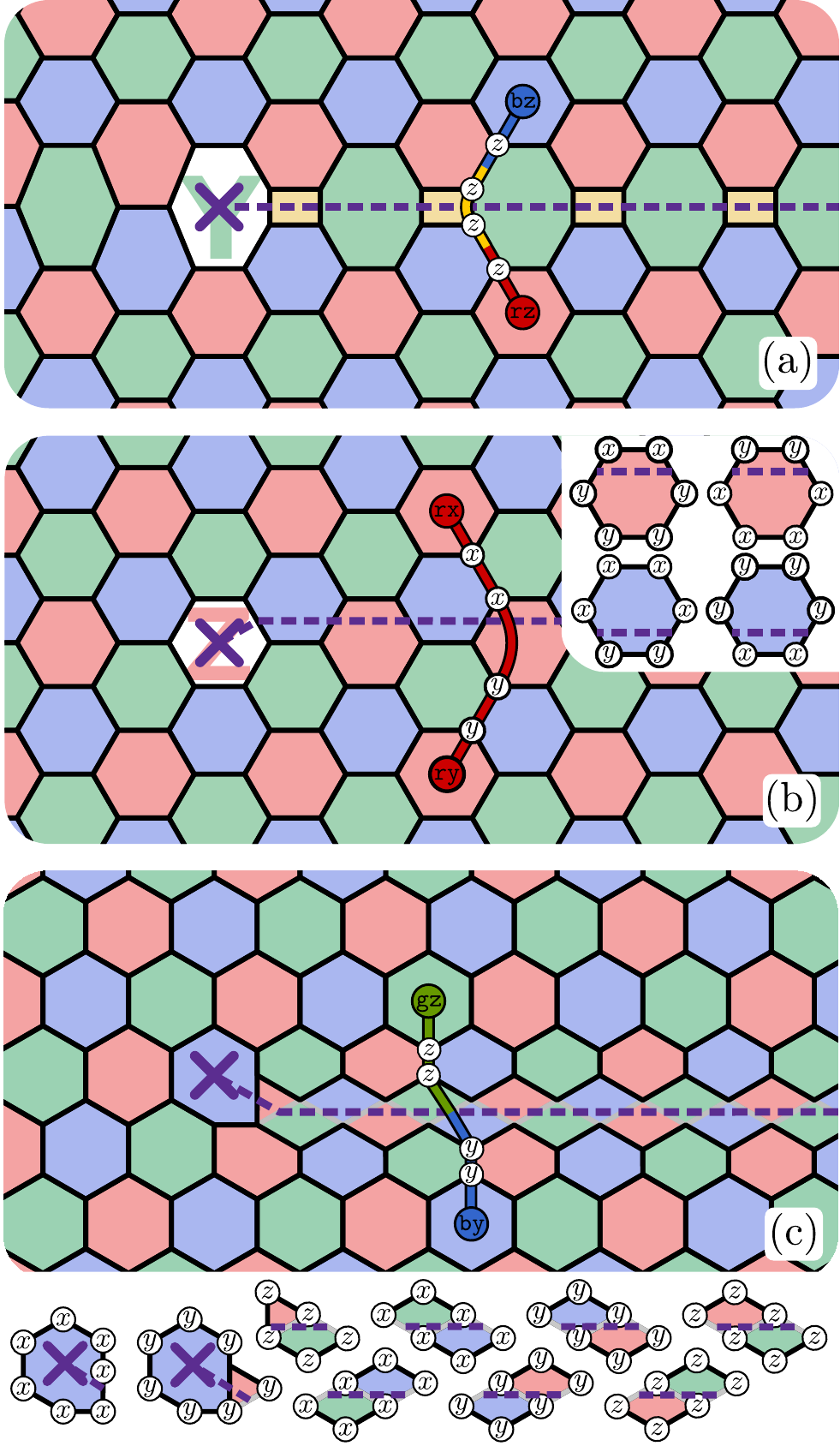}
	        \caption{
	            Three invertible domain walls (purple dashed line) in the color code and the twist defects terminating them (purple cross).
	            In each case we show one example of an anyon crossing the domain wall.
	            \textbf{(a)}~shows an example of a colour permuting domain wall.
	            Here, the red and blue colour label of anyons crossing the domain wall get exchanged.
	            The green plaquette at the endpoint hosts only one stabiliser acting in the $Y$-basis on the $7$ qubits in its support.
	            \textbf{(b)}~shows a Pauli-label permuting domain wall which exchanges the Pauli-$X$ and Pauli-$Y$ labels.
	            \textbf{(c)}~shows the duality domain wall which exchanges the colour with the Pauli label.
            }
	        \label{fig:InvertDWsAndTwists}
	    \end{figure}

        Let us start by showing microscopic realisations of three different invertible domain walls and their twist defects, see Fig.~\ref{fig:InvertDWsAndTwists}.        
        In each case, the stabilisers along the domain wall are changed.
        In the case of the colour permuting domain wall presented in Fig.~\ref{fig:InvertDWsAndTwists}~(a), the support of the stabilisers is changed according to the new lattice geometry.
        This change is such that the tricolourability of the faces is violated by the addition of twist defects to the lattice.
        This inconsistency in the colouring leads to a permutation of the colour label of anyons moving around the twist defect.
        The domain wall we show in Fig.~\ref{fig:InvertDWsAndTwists}~(b) does not change the lattice geometry, but instead we change the basis in which the stabilisers along the domain wall act.
        This leads to a permutation of the Pauli labels of anyons crossing it.
        Finally, Fig.~\ref{fig:InvertDWsAndTwists}~(c) shows a domain wall implementing the duality symmetry between the colour and the Pauli label of the color codes anyons.
        The three domain walls we show generate all $72$ invertible domain walls in the color code~\cite{Kesselring18}.
	    
	    The twist defects at the end points of the domain wall can condense certain color-code anyons.
	    As such, we can find string operators that transport appropriate choices of anyonic excitations between twist defects.
        These string-like operators correspond to logical operators from the perspective of quantum error correction.
	    By arranging configurations of twist defects on the lattice such that all the twist defects are sufficiently well separated, we encode logical qubits robustly.
	    Twist defects which condense many charges thus increase the size of the logical Hilbert space more than twists which only condense fewer anyons.
	    Formally, we capture this by assigning a quantum dimension to each type of twist defect~\cite{Dong_2008, Bombin10, Brown13TEEtwist, BONDERSON2017}.
	    For details on twist defect in the color code see Ref.~\cite{Kesselring18}.

\begin{figure}[tb]
	        \centering
	        \includegraphics[width=.4\linewidth]{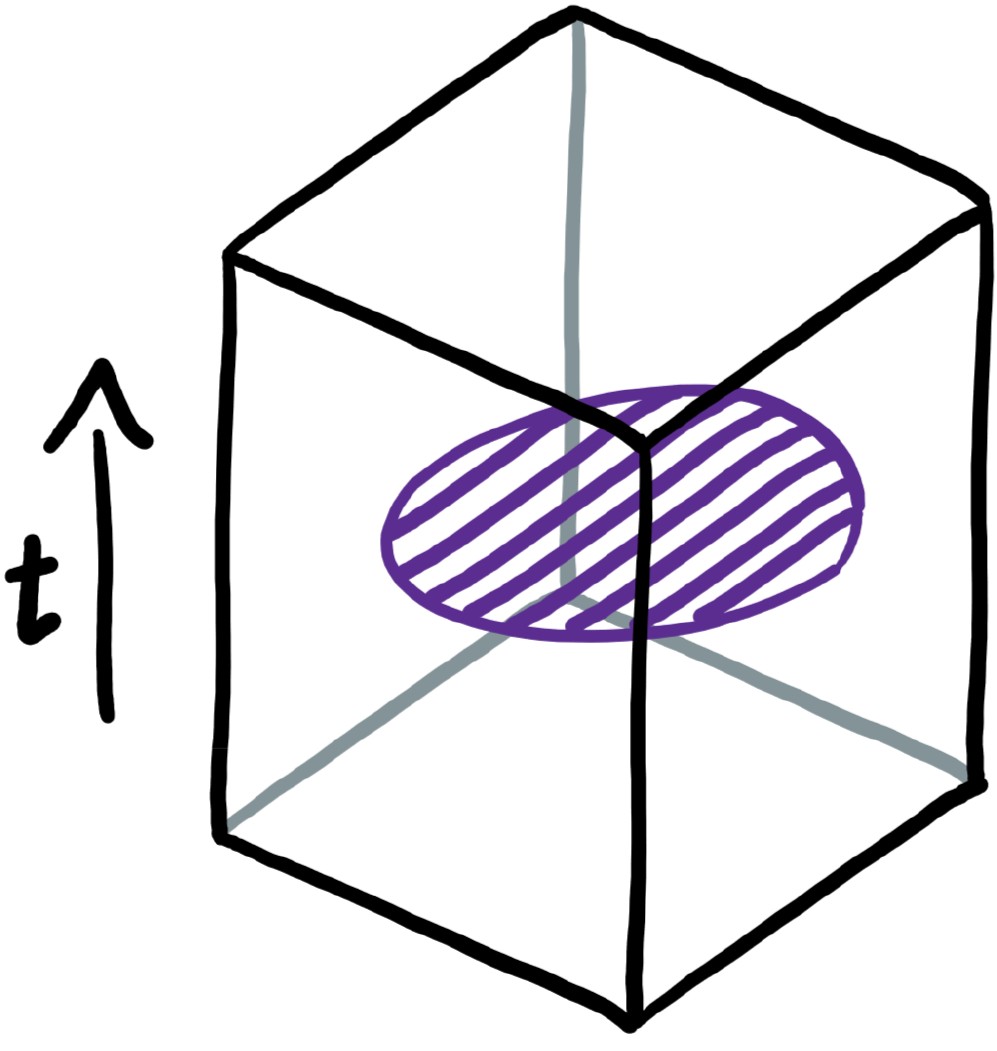}
	        \caption{
	            In space-time domain walls correspond to two-dimensional membranes and their one-dimensional boundaries are twist defects.
	            A pair of twist defects are created and moved apart before their are brought back together and annihilated.
	            }
	        \label{fig:TransvGatesRotatedTwist}
	    \end{figure}

        In a space-time picture we keep track of the position of the twist defect and the position of the physical defect line over time.
        In doing so, we obtain a two-dimensional membrane for the domain wall which is terminated by the one-dimensional world-line of the twist defect, as shown in Fig.~\ref{fig:TransvGatesRotatedTwist}.
        As the $(2+1)$-dimensional topological space-time we consider is isotropic from a macroscopic perspective, we can deform the world-lines of the twist defects arbitrarily, and the processes they undergo will be equivalent up to continuous deformations of the world-lines of the twist defects.
        On the other hand, the orientation of domain walls affects the microscopic details of the implementation of a defect braiding process in a physical system.
        These microscopic processes are distinct from an error correction point of view.
        
	    The interaction between twist defects and other topological features are also discussed in Refs.~\cite{Brown17,Kesselring18}.
        Notably, it is interesting to examine the interactions that can occur as twist defects approach boundaries.
        The process that occurs depends on two degrees of freedom;
        firstly by the automorphism associated to the domain wall that gives rise to the twist defect at its end point $Aut(\mathcal{C})$,
        and secondly by the Lagrangian subgroup that specifies its bosonic condensate $\mathcal{L}$.

        We distinguish two cases on how these degrees of freedom, or equivalently how the twist defect and the boundary, interplay with each other.
        In the first case, the associated symmetry leaves the Lagrangian subgroup of the boundary invariant, i.e., $Aut(\mathcal{L}) = \mathcal{L}$.
        In this case, an individual twist effectively `vanishes' at the boundary.
        We can regard this as twist condensation, named to reflect the analogy between this process and anyon condensation where an anyon is absorbed, or `vanishes' at the boundary.
        In contrast, if the associated anyon symmetry non-trivially alters the elements of the Lagrangian subgroup associated to the boundary, such that $Aut(\mathcal{L}) = \mathcal{L}' \not= \mathcal{L}$, then the presence of the twist non-trivially changes the physics of the boundary.
        Specifically, we find that the twist is confined at the boundary.
        This object has been coined a corner~\cite{Brown17, Kesselring18}, as they are often found at the corners of topological codes.
        We discuss corners in the following section.

    \subsection{Corners between boundaries}
        \label{sec:Cond1dMax}
        
        Corners are points on a boundary where the boundary type changes.
        They are important features of topological error-correcting codes proposed for quantum computation, such as the surface code~\cite{Dennis02} and the triangular color code~\cite{Bombin06} depicted in Fig.~\ref{fig:TriangularCCs}~(a).
        In the following, we offer two constructive interpretations that can give rise to equivalent corners.
        First, we view them as confined twist defects.
        This allows us to make statements about the computational power of different types of corners.
        Secondly, we use the language of anyon condensation to introduce two distinct boundaries such that a corner is prepared in between them.
        This allows for a general procedure to construct corners in topological error-correcting codes.
        
        Earlier in this manuscript, we have already encountered corners, see for example Fig.~\ref{fig:InitialisationTrinagularCC} which features the triangular color code.
        This code is terminated by three distinct colour boundaries and the three points where the boundary type changes are corners.
        We say an anyon can condense at a corner if it is contained in the union of the two Lagrangian subgroups which describe the boundaries interfacing at the corner.
        Note that the union of two distinct Lagrangian subgroups, each consisting of anyons of a given colour label, contains a generating set of all color code anyons.
        As such, a corner interfacing two distinct coloured boundaries can condense all $16$ color code anyons.
        Indeed, the same holds true for corners interfacing two distinct Pauli boundaries.
        We discuss corners interfacing a coloured boundary with a Pauli boundary later in this section.

        \begin{figure}[tb]
	        \centering
	        \includegraphics[width=.75\linewidth]{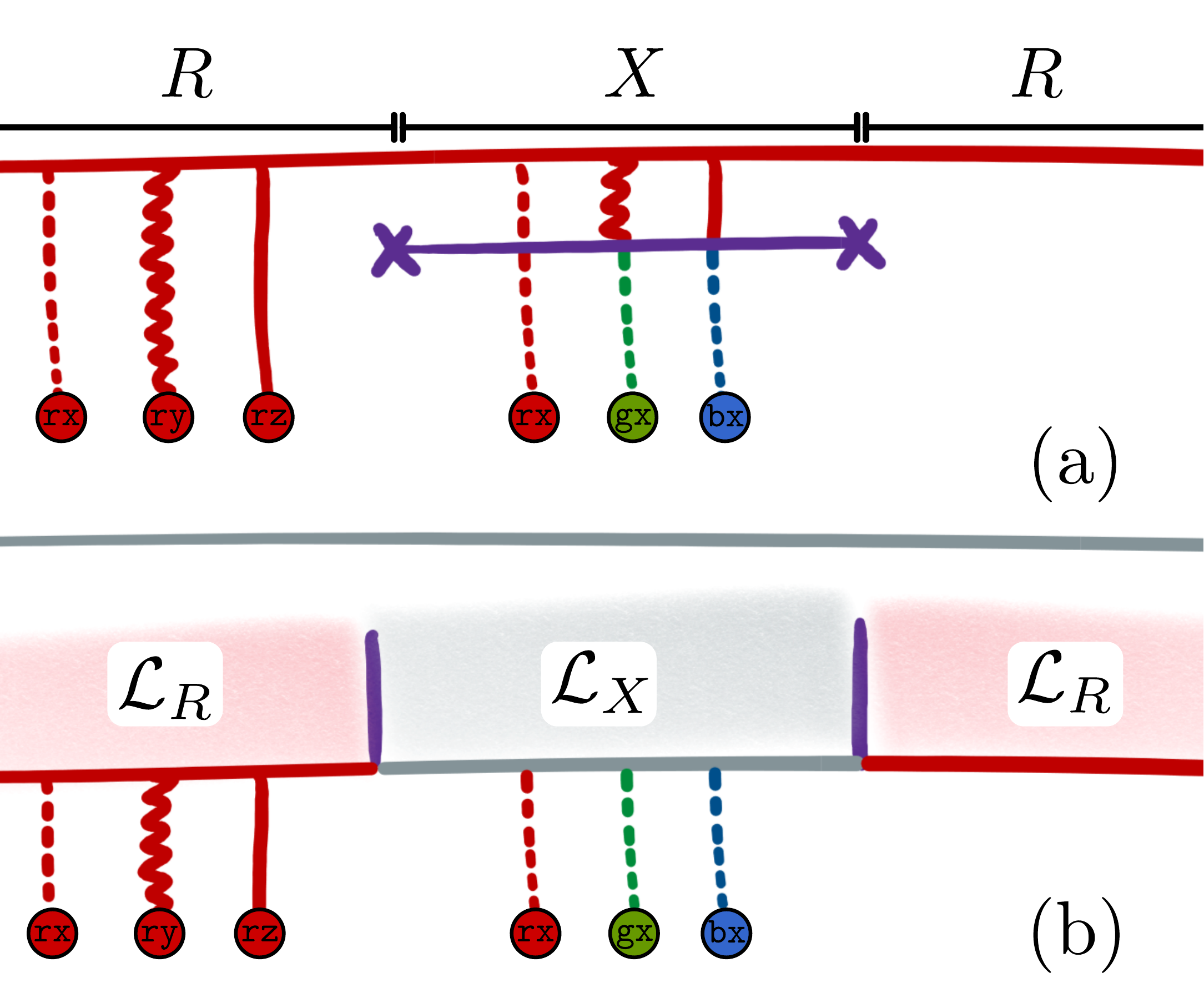}
	        \caption{
                Corners are the points where the boundary changes type.
                They can be interpreted as condensed twists defects.
                \textbf{(a)}~shows a domain wall close to the boundary to demonstrate how an invertible domain wall can change the boundary type.
                \textbf{(b)}~explains how to construct boundaries microscopically using anyon condensation.
                In the red and grey regions we condense the red and the Pauli-$X$ Lagrangian subgroups $\cL_R$ and $\cL_X$, respectively.
            }
	        \label{fig:CornersTwistsAndCondensation}
	    \end{figure}
        %
        
        Let us now interpret corners as confined twist defects, see Sec.~\ref{sec:Cond1dTriv}.
        To this end, we start with a uniform boundary between the color code and the vacuum.
        Next, we introduce a pair of twist defects and move them close to the boundary.
        This may change the type of boundary, as anyons now need to cross a domain wall before reaching the boundary.
        If the anyon permuting symmetry applied when crossing the domain wall changes the Lagrangian subgroup corresponding to the boundary, then the confined twists defects correspond to a non-trivial corner.
        If, however, the automorphism corresponding to the twist defect leaves the Lagrangian subgroup invariant, we say that the twist defect condenses and no non-trivial corner is introduced.
        We show an example of a confining twist in Fig.~\ref{fig:CornersTwistsAndCondensation}~(a).
        Here, a colour-Pauli-duality twist transforms a red boundary into a Pauli-$X$ boundary.
        Importantly, this interpretation shows us that we can pull the corners out far away from the boundaries where they can be braided as bulk twist defects to perform logical gates~\cite{Brown17}.
        Similarly, this interpretation allows us to modify the lattice realisation of topological codes which can increase the number of encoded qubits.
        This is done in Refs.~\cite{Yoder17,Kesselring18}, where the corners are moved as twists into the centre of the lattice.
        Finally, the insight that we can identify corners with twists can be of use in designing lattice surgery protocols in certain topological error correction codes.
        In Ref.~\cite{Twist}, for instance, a twist based modification of the lattice surgery protocol is presented that allows all three logical Pauli operators in a patch of surface code to be addressed.

        A second interpretation of corners is found by noticing that they can be created by condensing different Lagrangian subgroups in adjacent regions.
        This is shown in Fig.~\ref{fig:CornersTwistsAndCondensation}~(b).
        Here, we see three neighbouring regions where, from left to right, we condense the Lagrangian subgroups $\cL_R$, $\cL_X$, and $\cL_R$.
        We therefore create a red, a Pauli-$X$ and a red boundary, respectively.
        The corners appear at triple points where the boundaries of these two condensed regions meet the color code phase.
        This perspective shows us that corners can be interpreted as end-points of opaque domain walls, as shown in Table~\ref{tab:PaperStructure}.
        An opaque domain wall is essentially a narrow puncture, which, if it consists of two distinct boundaries, features non-trivial corners at its end-points.

        \begin{figure}[tb]
	        \centering
	        \includegraphics[width=.85\linewidth]{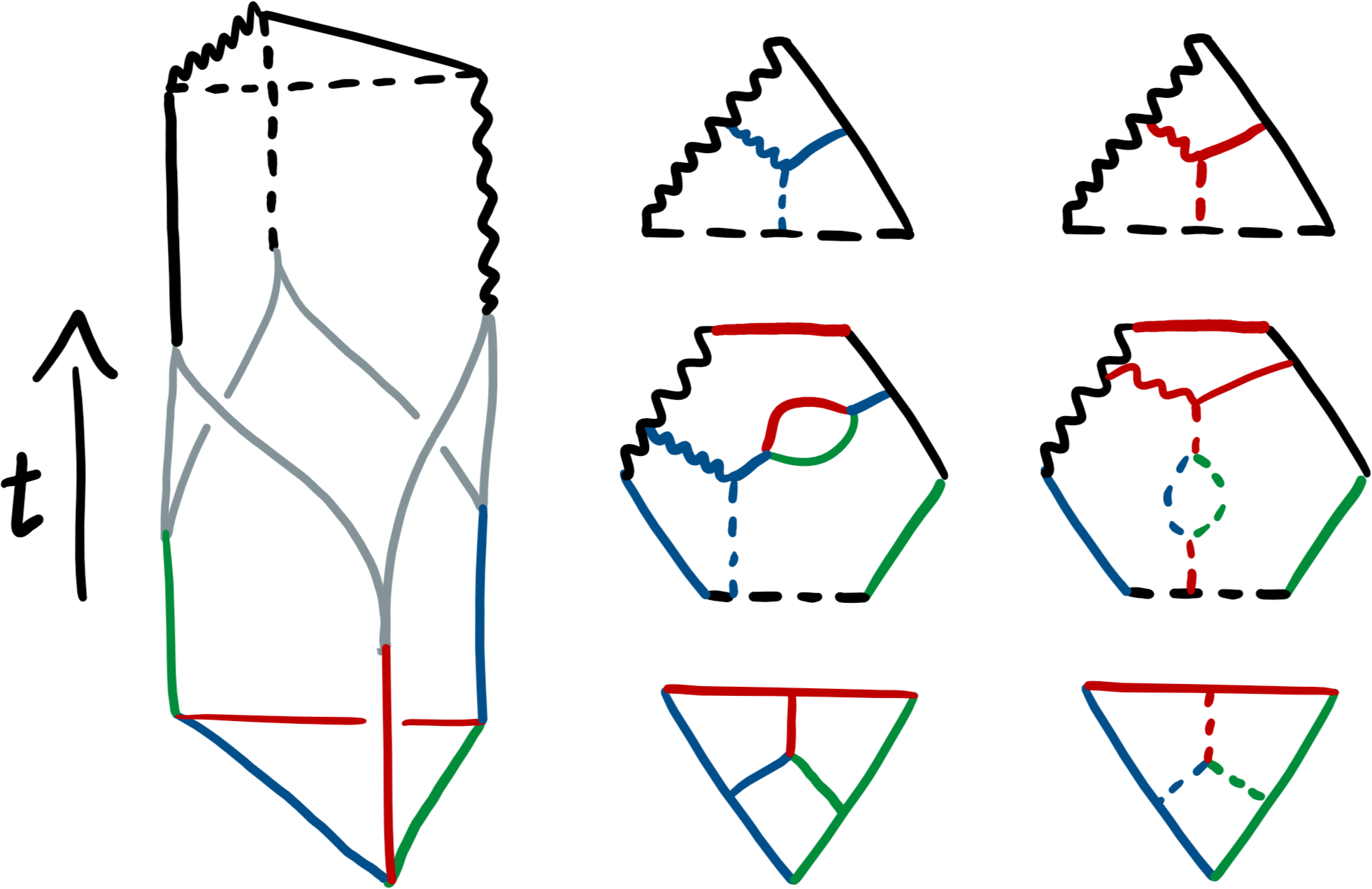}
	        \caption{
	            The space-time picture of a triangular color code undergoing a transformation that changes its boundary type.
	            We show the space-time process on the left while depicting three instances of the two-dimensional code at different times to the right of the figure.
	            In separate pictures of the lattice, we show the various deformations for both the logical $Z$ and logical $X$ operators over time.
	            We start at the bottom with a triangular color code with coloured boundaries.
	            We then split each of its corners into two semi-corners and move them apart.
                When they meet a semi-corner originating from a distinct corner, they fuse together to form a corner interfacing two Pauli boundaries.
                Dotted, wavy and solid lines correspond, in turn, to the Pauli-$X$, $Y$ and $Z$ basis.
            }
            \label{fig:SwitchingBoundaryTriangularCodeHexagonVersion}
	    \end{figure}
        %
        
        Let us now use the second interpretation we have presented for corners to generalise this class of topological features.
        Our new construction generalises the notion of corners that we have already encountered, as the two Lagrangian subgroups describing the neighbouring boundaries have non-zero overlap.
        This means that there exists a non-trivial anyon $\a$ for which $\a \in \cL$ and $\a \in \cL'$.
        We dub these corners semi-corners by, again, appealing to the unfolded picture of the color code. In an appropiate unfolded picture, we find that color code semi-corners appear as non-trivial corners on only one of the two toric-code layers.
        As opposed to the corners between two distinct coloured boundaries (or two Pauli boundaries), semi corners can not condense all of the species of color-code anyons.
        The corner between a red coloured boundary and a $X$-Pauli boundary, for example, condenses the following anyons:
        $\{\one,\rx,\ry,\rz,\gx,\bx,\ftwo,\fthree\}$.

\begin{figure}[tb]
	        \centering
	        \includegraphics[width=.65\linewidth]{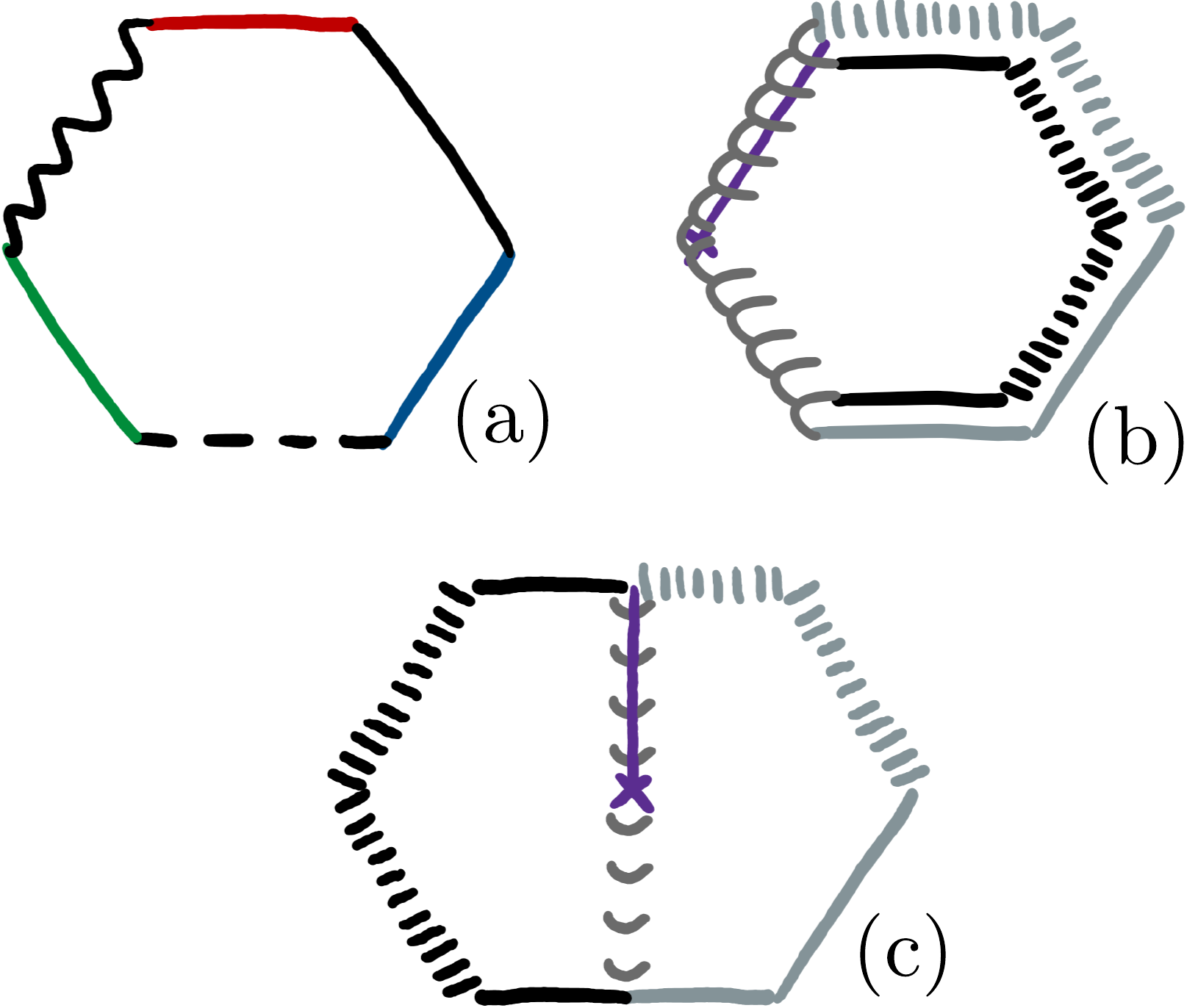}
	        \caption{
	            \textbf{(a)}~The hexagonal color code which is terminated by all six different boundaries, hosting six semi-corners.
	            \textbf{(b)}~shows the same code unfolded as two stacked toric code layers.
	            Different boundaries correspond to stacked toric code boundaries or a combination of folds and toric code domain walls, drawn in purple.
	            \textbf{(c)}~to obtain one single layer of toric code, we unfold the top layer to the left.
	            This leaves us with one central twists defect.
	            This code is known as the triangular surface code~\cite{Yoder17}.
            }
	        \label{fig:HexColorCodeUnfoldingSketch}
	    \end{figure}

        As an example of their utility, we can use semi-corners to transform color codes with coloured boundaries into color codes with Pauli boundaries.
        We show this process in Fig.~\ref{fig:SwitchingBoundaryTriangularCodeHexagonVersion} where semi-corners are depicted as grey lines.
	    As a middle stage of this transformation we encounter a color code which hosts all six distinct color code boundaries.
	    In the following, we unfold this code, see Fig.~\ref{fig:HexColorCodeUnfoldingSketch}.
	    Interestingly, we obtain the surface code with a twist~\cite{Yoder17}.
	    
    \subsection{Semi-transparent domain walls}
        \label{sec:Cond1dPart}    
    
        We have seen examples of topological features that condense charges, confine charges, and also allow deconfined charges to remain mobile.
        In general, we can find topological features that allow all three of these processes to occur over the charges of some anyon model.
        A semi-transparent domain wall permits all three of these processes to occur.
        We obtain a semi-transparent domain wall if we perform partial condensation along a one-dimensional sub-region of the lattice.
        More generally, we can obtain and classify all of the semi-transparent domain walls by composing them with other domain walls.    
    
        In what follows we classify the color codes semi-transparent domain walls using an abstraction based on the color-code boson table~\eqref{eq:BosonTable}.
        This allows us to divide all $162$ semi-transparent domain walls into $8$ classes.
        We can therefore explore how they can be used to store and manipulate logical qubits.
        We will also describe the physics of a semi-transparent domain wall for the color code by viewing it in the unfolded picture.
        We note that the exposition given in this subsection expands on the discussion given in the appendix of Ref.~\cite{Thomsen22}.
        Before we present our general classification of semi-transparent domain walls, let us first show an explicit microscopic example.
        
\begin{figure}[tb]
	        \centering
	        \includegraphics[width=1.00\linewidth]{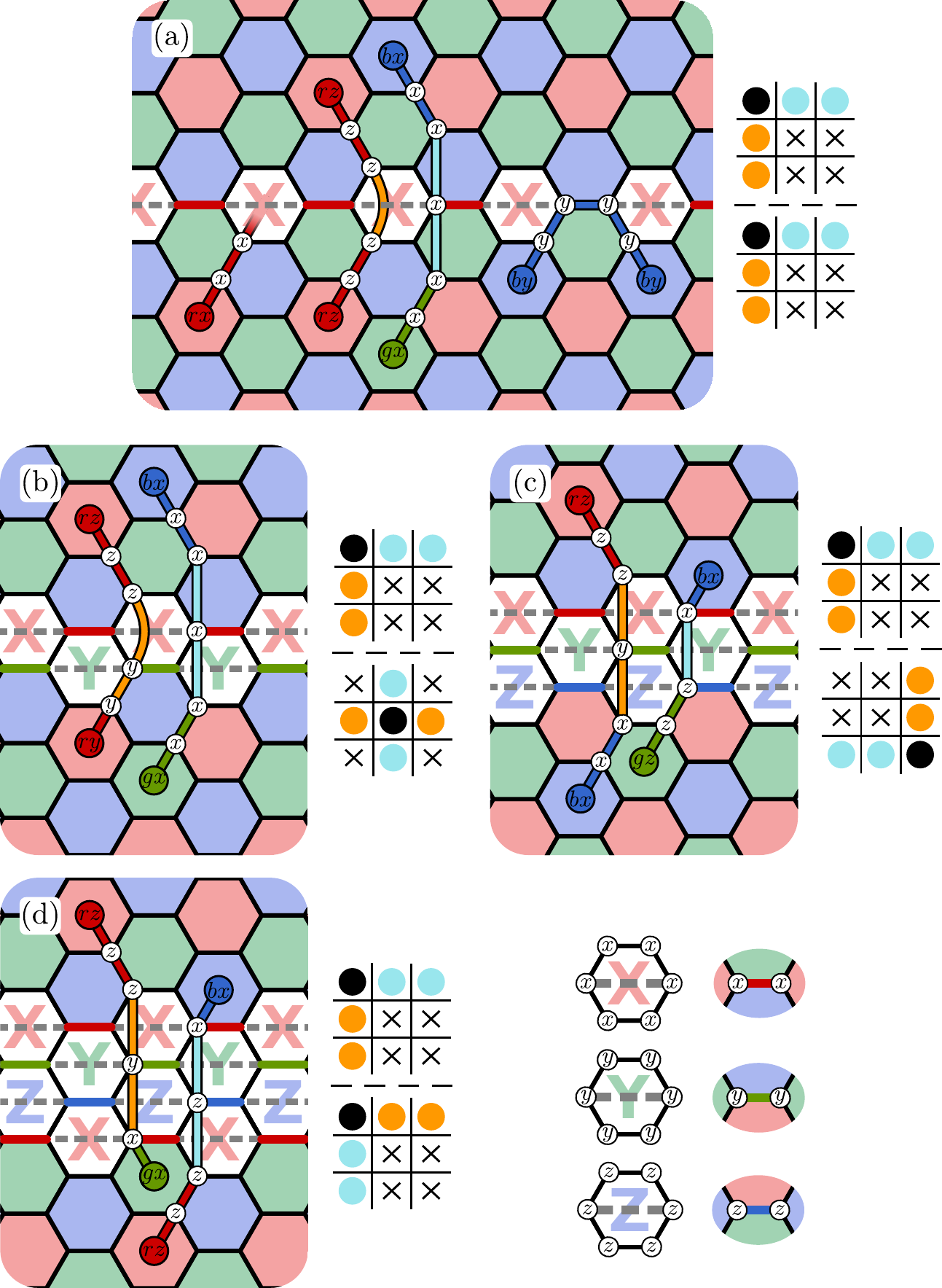}
	        \caption{
	            Four different semi-transparent domain walls realised microscopically in the color code.
	            Each coloured plaquette hosts an $X$- and a $Z$-type stabiliser.
	            All exceptions and additional stabilisers are given in the bottom right of the figure.
                In each case we present the lattice realisation of the domain wall and show different anyons crossing the domain wall.
                In \textbf{(a)} we additionally show the \rx charge condensing and a \by charge being confined to one side of the domain wall.
                Next to the lattice realisation of the domain wall we draw the effect the domain wall has on all color code anyons in terms of the color codes boson table, as discussed in the main text.
    \textbf{(b)}, \textbf{(c)} and \textbf{(d)} show other examples of semitransparent domain walls together with their corresponding boson tables. Deconfined excitations are also shown on each lattice.        }
	        \label{fig:CC-CC-DW}
	    \end{figure}

        We introduce a semi-transparent domain wall to the color code by applying a partial condensation to a one-dimensional subregion of the lattice.
        As discussed in Sec.~\ref{sec:CondPart}, one color-code boson is chosen to be identified with the trivial charge in order to perform a partial condensation.
        In Fig.~\ref{fig:CC-CC-DW}~(a) we show a domain wall where the \rx anyon is chosen to be condensed on the color-code lattice.
        One can check that this domain wall condenses the \rx charge no matter which side it approaches the domain wall.
        For a general semi-transparent domain wall, however, this is not the case and anyons approaching the domain wall from opposite sides might behave differently.
        We can construct one such example microscopically by composing two distinct bosons to be condensed along adjacent one-dimensional regions.
        Fig.~\ref{fig:CC-CC-DW}~(b) shows a horizontal semi-transparent domain wall which we construct by condensing \rx above the domain wall and \gy below the domain wall.
        We follow the procedure laid out in Sec.~\ref{sec:CondStabs} to obtain a valid stabiliser realisation.

\begin{figure}[tb]
	        \centering
	        \includegraphics[width=.65\linewidth]{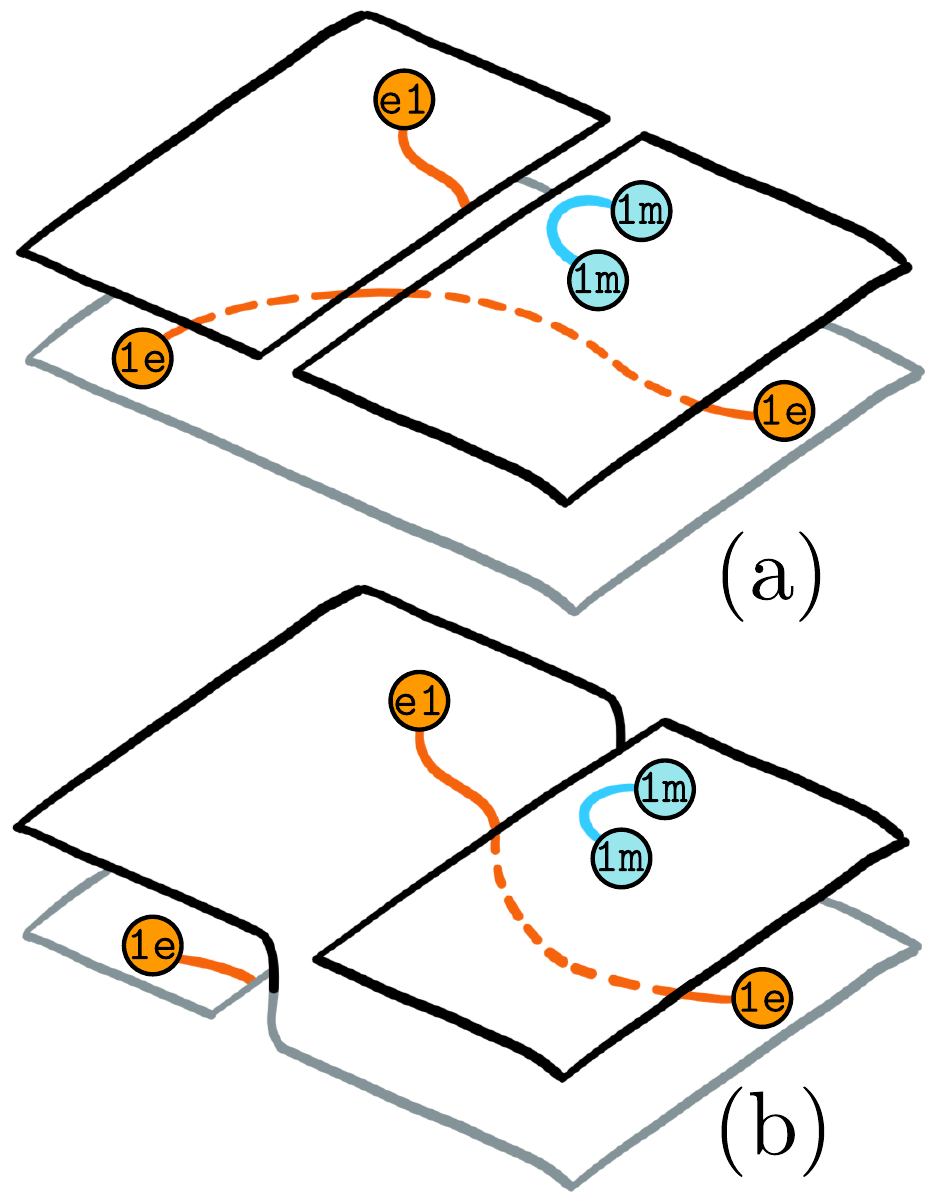}
	        \caption{
	            We show two semi-transparent color code domain walls after unfolding.          
	            \textbf{(a)}: In the simplest case, the domain wall corresponds to a narrow puncture on one layer (top) while the other layer (bottom) is continued.
	            This means that charges on the bottom layer remain mobile while charges on the top layer condense or confine.
	            From this simple example, we can construct all semi-transparent domain walls in the color code by adding a transparent domain wall.
	            One such example is shown in \textbf{(b)}, where a layer-swapping domain wall is introduced.
            }
	        \label{fig:SemiTranspDWunfolded}
	    \end{figure}

	    We obtain a clearer understanding of the physics of semi-transparent domain walls of the color code by unfolding them into two layers of the toric code~\cite{Bombin12,Kubica15}.
        Fig.~\ref{fig:SemiTranspDWunfolded}~(a) shows an example of an unfolded semi-transparent domain wall.
        Here, the top layer hosts a narrow puncture while the bottom layer is connected across the domain wall.
        Anyons on the bottom layer can pass through the domain wall, i.e., they remain mobile, while anyons on the top layer either condense or confine according to the chosen boundary type of this toric code layer.
        While this is the simplest example of an unfolded semi-transparent color code domain wall, the general case can be obtained from it readily.
        More precisely, we redundantly obtain all possible semi-transparent domain walls in the color code from this simple example by adding transparent domain walls to either side of the displayed semi-transparent domain wall.
        One such example is shown in Fig.~\ref{fig:SemiTranspDWunfolded}~(b), where we additionally introduce a layer-swapping domain wall.
        
        Clearly, one can conceive of many different types of semi-transparent domain wall for the color code.
        In order to catalogue the color codes semi-transparent domain walls, we will once again make use of the boson table~\eqref{eq:BosonTable}.
        Specifically, we take two copies of the boson table, one corresponding to the `top side' of the domain wall and the other corresponding to the `bottom side'.
        Examples are shown on the right of the domain walls in Fig.~\ref{fig:CC-CC-DW}.
        We make use of the notation introduced in Sec.~\ref{sec:CondPart} to denote anyons that condense, those that confine, and those that remain deconfined.
        On either side, one of the nine bosons is condensed, marked by a $\blackcirc$ label.
        The other 8 bosons then get marked by $\cross$, $\orangecirc$ or $\cyancirc$, depending whether they confine or deconfine at the domain wall.
        Charges marked with $\cross$ remain confined to their corresponding side of the domain wall.
        The remaining anyons are deconfined and remain mobile, i.e., can be moved across the domain wall.
        We label these remaining anyons $\orangecirc$ or $\cyancirc$, corresponding to electric and magnetic charges, respectively.
        Upon transmission through the domain wall, anyons marked with the $\orangecirc$ ($\cyancirc$) labels of the top side are mapped onto anyons marked with the $\orangecirc$ ($\cyancirc$) labels of the bottom.

        This characterisation using two boson tables suffices to find the total number of semi-transparent domain walls.
        First of all, we can choose which of the nine color-code bosons we condense, $\blackcirc$, on both the top side and bottom side of the domain wall arbitrarily.
        This fixes the confined charges on each grid.
        We have one final degree of freedom, namely, how the mobile charges from the top get mapped to the mobile charges on the bottom.
        Without loss of generality, let us arbitrarily fix the $\orangecirc$ and $\cyancirc$ labels on the top grid.
        There are now two possible choices to configure the $\orangecirc$ and $\cyancirc$ labels on bottom grid. Given these rules, let us now count the semi-transparent domain walls.
        Given the nine choices of condensing anyons on the top and bottom side grid, we obtain $9 \times 9 = 81$ semi-transparent domain walls.
        Then, together with the binary choice for how to configure the $\orangecirc$ and $\cyancirc$ labels on the bottom grid we arrive at $81 \times 2 = 162$ semi-transparent domain walls.
        
        \begin{figure}[tb]
	        \centering
	        \includegraphics[width=.7\linewidth]{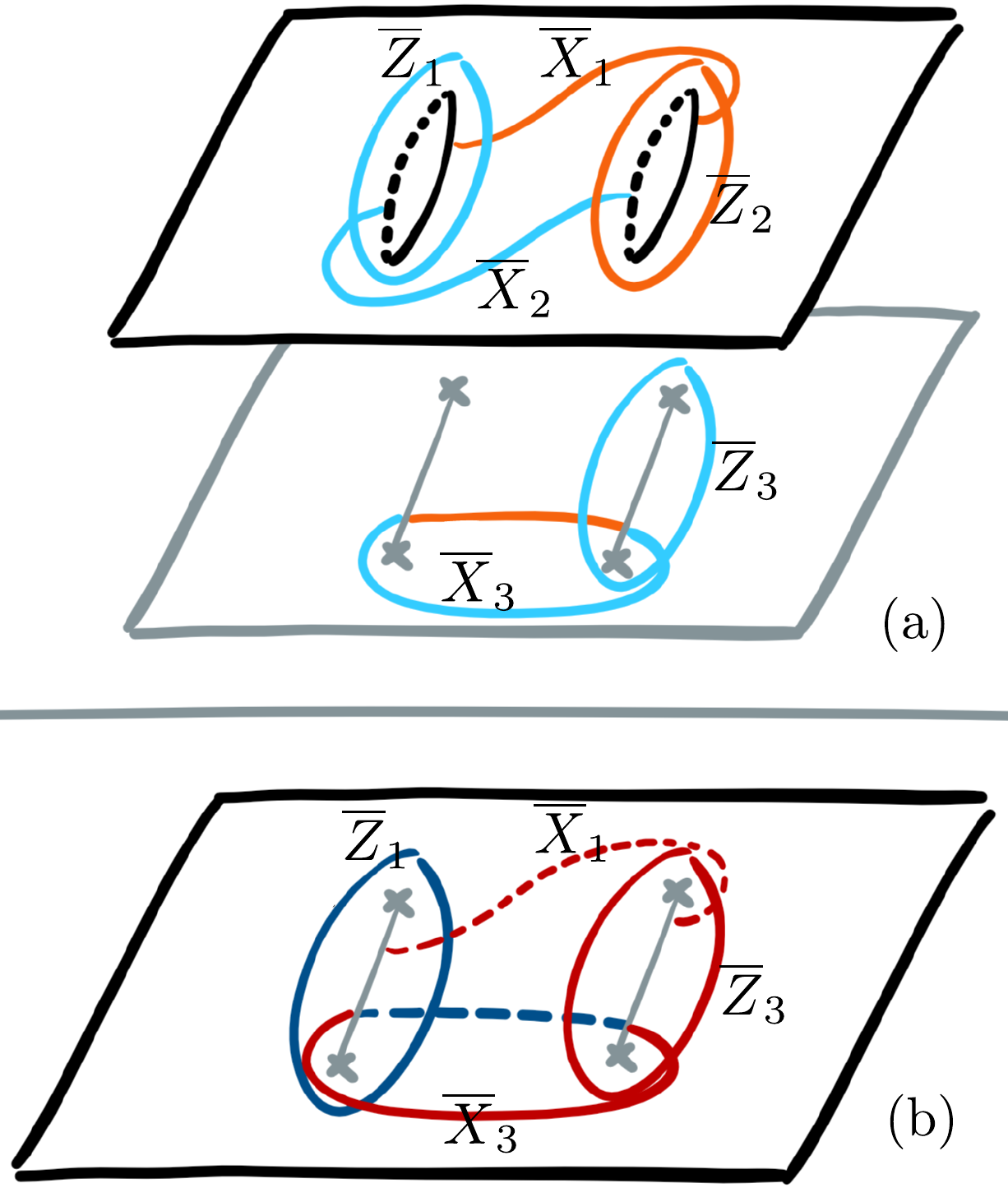}
	        \caption{
	            Two pairs of semi-twists are shown and some of the logical operators they support.
	            \textbf{(a)}~shows the unfolded version where the semi-transparent domain walls correspond to punctures on the top layer and an invertible toric code domain wall on the bottom layer.
	            \textbf{(b)}~shows the same configuration in the color code.
	            Here, we omit the logical operators $\overline{X}_2$ and $\overline{Z}_2$ for clarity.
	            In the toric code, we represent operators corresponding to electric (magnetic) anyons \e (\m) with orange (blue) lines.
	            In the color code, we draw the colour corresponding to the colour label (red or blue, here) and the Pauli-$X$($Z$) label corresponds to dashed (solid) lines.
            }
	        \label{fig:SemiTwistEncoding}
	    \end{figure}

        Let us now consider the end-points of these semi-transparent domain walls.
        We call them semi-twists. We will also briefly discuss how they can be characterised, and used to store quantum information in a robust manner.
        To interpret semi twists, we can follow the unfolding procedure above to obtain two decoupled layers of toric code.
        See Ref.~\cite{Benhemou22} where objects that can be interpreted as semi-twists on a single-layer of the toric code are discussed.
        In the case of the color code we find that for any domain wall there exists an unfolding in which it is composed of a narrow puncture (with possibly two distinct boundaries) on one layer of toric code and an invertible domain wall (possibly the trivial one) on the other.
        Hence, the semi-twists can be regarded as corners ``on top of'' twists.

        Both corners as well as twist defects can condense certain charges.
        This can be used to encode logical information in pairs of semi-twist defects.
        The associated logical operators are either string operators transporting a charge from one semi-twist to another or strings wrapping around a pair of semi-twists.
        Examples of the logical operators associated to semi twists are shown in Fig.~\ref{fig:SemiTwistEncoding}.

        Each of the 162 semi-twists can be associated with one of eight classes.
        The classes are obtained by checking if the condensing anyons on the top side and the bottom side share both their colour and Pauli label (class 1), just the Pauli label (class 2), just the colour label (class 3) or neither of the two labels (class 4).
        Within each class we introduce a subclass $A$ or $B$ depending on how the mobile anyons get mapped when crossing the domain wall.
        If ``rows get mapped to rows'' and ``columns get mapped to columns" we are in subclass $A$, if ``rows get mapped to columns'' and vice versa, we are in subclass $B$.
        Table~\ref{tab:stDWclasses} shows an example of a semi-transparent domain wall in the boson table notation for each of the $8$ classes as well as the number of elements in each of the classes.
        \begin{table}[tb]
        	\centering
        	\begin{tabular}{c|c|c|c|c}
                & 1 & 2 & 3 & 4 \\ 
                \hline
                \raisebox{-35pt}{$A$} &
                \raisebox{-\totalheight+4pt}{\includegraphics[width=0.07\textwidth]{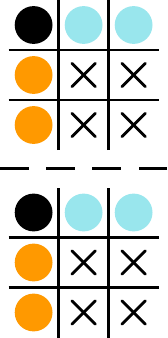}} & 
                \raisebox{-\totalheight+4pt}{\includegraphics[width=0.07\textwidth]{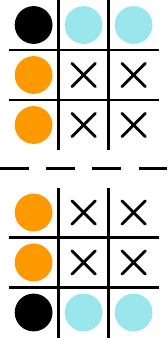}} & 
                \raisebox{-\totalheight+4pt}{\includegraphics[width=0.07\textwidth]{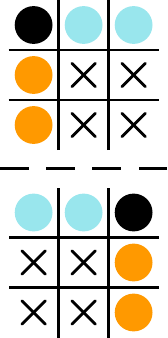}} & 
                \raisebox{-\totalheight+4pt}{\includegraphics[width=0.07\textwidth]{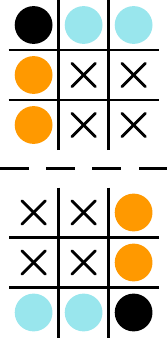}} \\
                &
                \raisebox{-1pt}{$9$} &
                \raisebox{-1pt}{$18$} &
                \raisebox{-1pt}{$18$} &
                \raisebox{-1pt}{$36$} \\
                \hline
                \raisebox{-35pt}{$B$} &
                \raisebox{-\totalheight+4pt}{\includegraphics[width=0.07\textwidth]{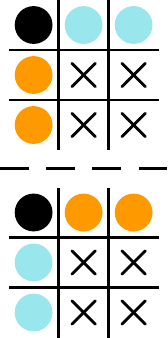}} &
                \raisebox{-\totalheight+4pt}{\includegraphics[width=0.07\textwidth]{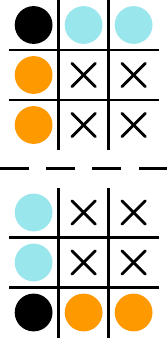}} &
                \raisebox{-\totalheight+4pt}{\includegraphics[width=0.07\textwidth]{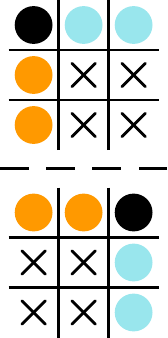}} &
                \raisebox{-\totalheight+4pt}{\includegraphics[width=0.07\textwidth]{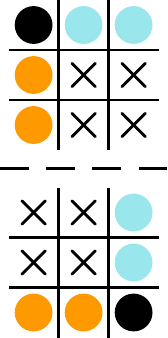}} \\
                & 
                \raisebox{-1pt}{$9$} &
                \raisebox{-1pt}{$18$} &
                \raisebox{-1pt}{$18$} &
                \raisebox{-1pt}{$36$} \\
            \end{tabular}
            \caption{
                The $162$ semi-transparent domain walls in the color code can be separated into $8$ classes.
                Here, we show an example of one member of the class as well as the number of distinct domain walls in each class.
            }
            \label{tab:stDWclasses}
        \end{table}
        
\begin{figure}[tb]
	        \centering
	        \includegraphics[width=1.00\linewidth]{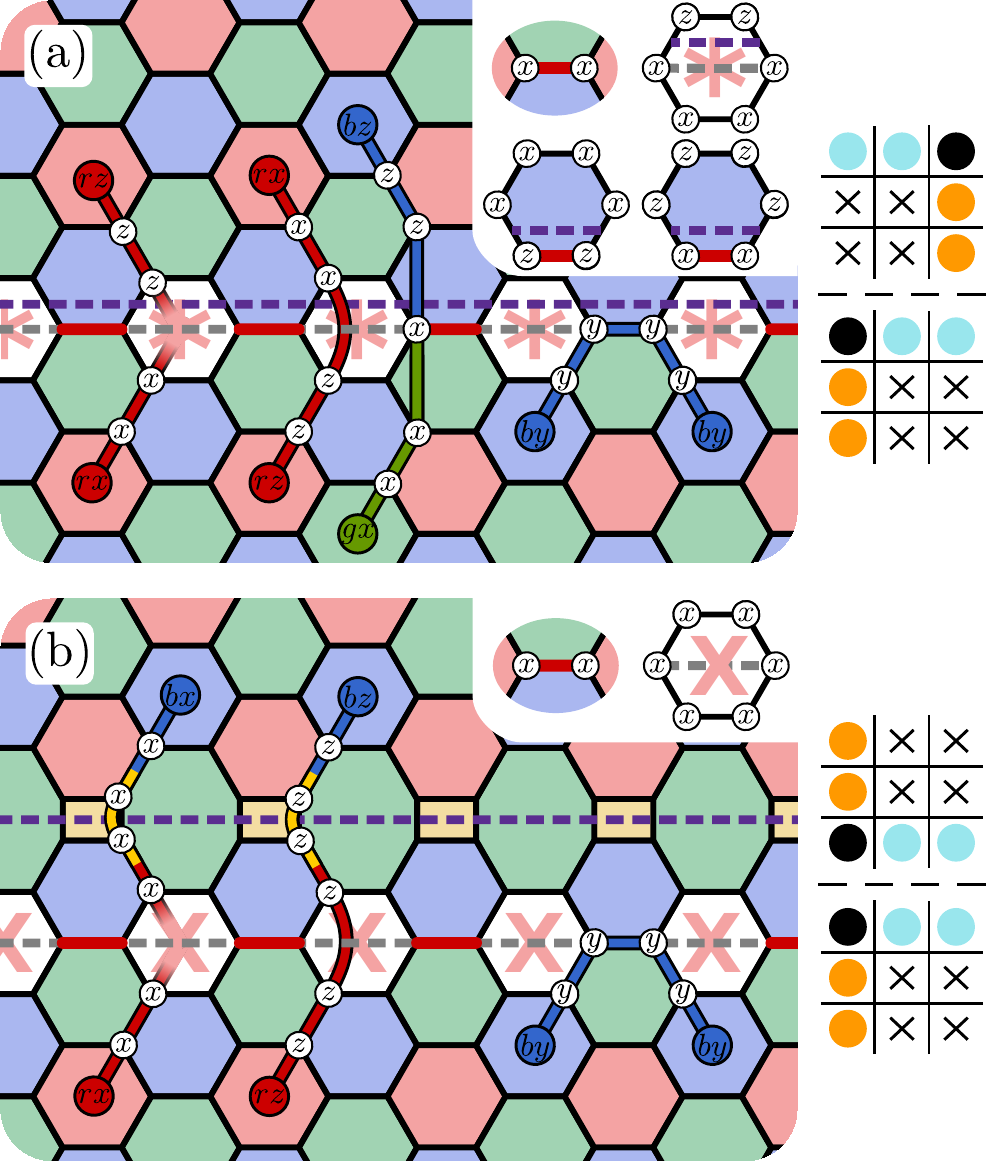}
	        \caption{
	            Semi-transparent domain walls change type when combined with an invertible color code domain wall.
	            Here, we show a domain wall in class $1A$ (as depicted in Fig.~\ref{fig:CC-CC-DW}~(a), marked by the grey dashed line, and how it transforms when an invertible domain wall (dashed purple line) is introduces next to it.
	            In \textbf{(a)} a Pauli-label permuting domain wall, analogue to the one shown in Fig.~\ref{fig:InvertDWsAndTwists}~(b), is introduced next to it, transforming it to a domain wall in class $3A$.
	            In \textbf{(b)} we introduce a colour permuting domain wall, as depicted in Fig.~\ref{fig:InvertDWsAndTwists}~(a), to obtain a class $2A$ domain wall.
            }
	        \label{fig:CC-CC-DW-with-invert}
	    \end{figure}

        The subfigures (a), (b), (c) and (d) in Fig.~\ref{fig:CC-CC-DW} correspond to the classes $1A$, $4B$, $4A$ and $1B$, respectively.
        Examples of members of classes 2 and 3 are obtained by adding a colour or Pauli permuting invertible domain wall to either side of one of the shown examples.
        This is shown in Fig.~\ref{fig:CC-CC-DW-with-invert},
        where a semi-transparent domain wall in class $1A$ gets transformed to a class $3A$ ($2A$) domain wall by complementing it with a Pauli(colour)-permuting invertible domain wall.

        Semi-transparent domain walls have previously been described abstractly in Ref.~\cite{KitaevKong12}.
        In general, a domain wall is described by two ``tunnelling maps" describing how the bulk excitations get transformed when approaching the domain wall from either side.
        In this picture, non-transparent domain walls correspond to non-invertible tunnelling maps.
        In App.~\ref{app:DWsStackedTC}, we describe how to explicitly calculate the tunnelling map in the sector of trivial wall excitations for phases equivalent to stacks of toric code, based on Ref.~\cite{Beigi11}.
        Calculating the full tunnelling map for arbitrary phases goes beyond the scope of this work, but will be covered in future work, see Ref.~\cite{Magdalena2022bulktoboundary}.

        Let us terminate this section by remarking that semi-transparent domain walls appear in color code lattice surgery protocols~\cite{Landahl14,Thomsen22}.
        In fact, lattice surgery with the color code makes use of all domain walls presented in this work so far, opaque, semi-transparent and invertible ones in both the temporal and spatial orientations.
        This is the subject of the following section.
	   
	\subsection{Lattice surgery}
        \label{sec:Cond1dLS}
        
        Lattice surgery is a protocol to make fault-tolerant joint measurements of Pauli observables between multiple logical qubits~\cite{HorsemanSurgery, Brown17, Kesselring18, Litinski19Game}.  
        These fault-tolerant operations are carried out by merging and subsequently splitting disjoint code patches which we achieve by changing the measured stabiliser terms.
        A sufficiently large set of lattice surgery operations can implement the Clifford group by measurement.
        Together with the preparation of noisy magic states and distillation protocols, we recover a universal set of fault-tolerant logic gates.
        
        Lattice surgery methods give rise to very resource efficient proposals for implementing fault-tolerant logical gates in topological error-correcting codes~\cite{HorsemanSurgery, Litinski19Game, Landahl14, Thomsen22}.
        The color code has been shown to have an advantage over other topological codes in terms of the resource cost of its implementations in Ref.~\cite{Thomsen22}.
        Achieving this advantage requires the use of all different types of color-code boundaries and domain walls.
        In what follows, we aim to tie together the above discussion of anyon condensation in the color code in order to understand the role of these boundaries and domain walls appearing in overhead efficient color code lattice surgery based quantum computation.
        
\begin{figure}[tb]
            \centering
            \includegraphics[width=1.00\linewidth]{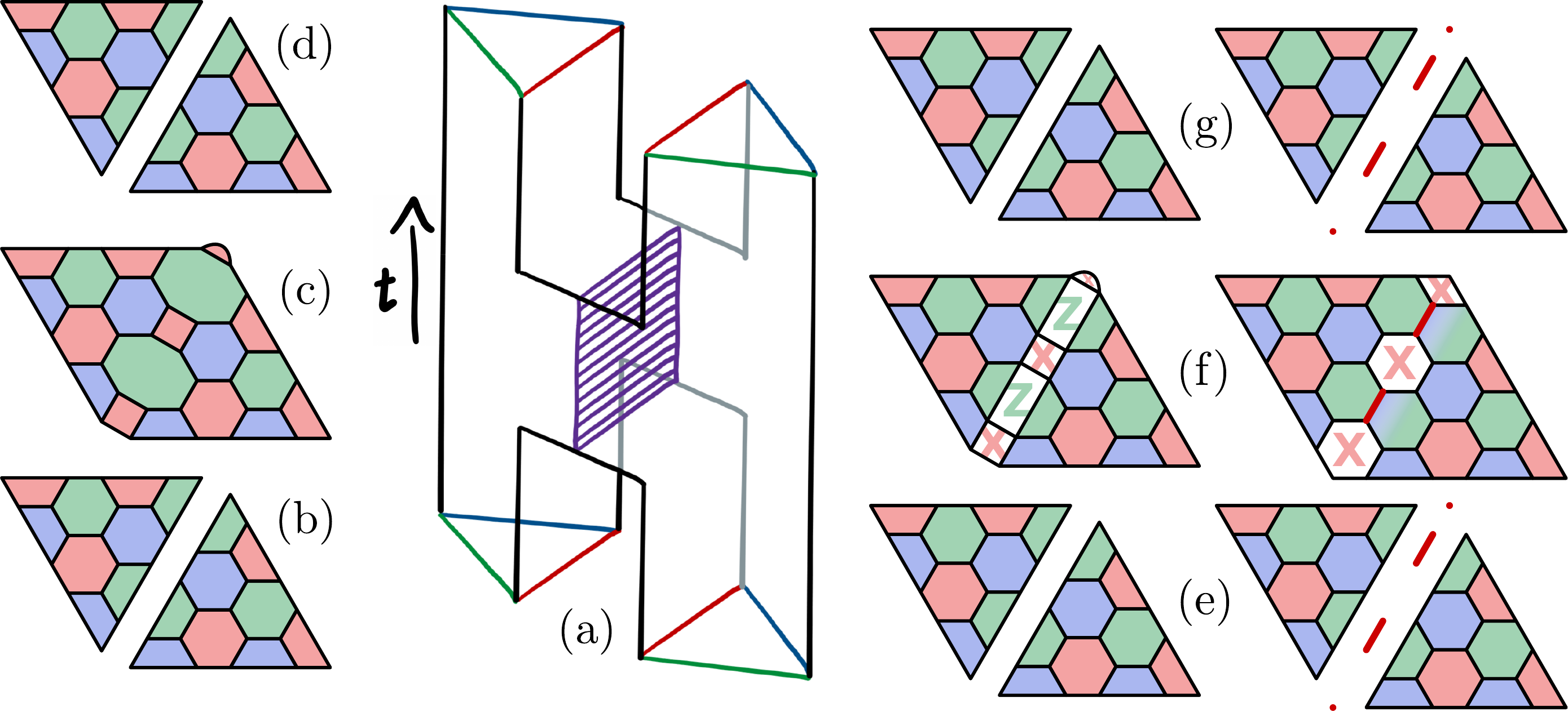}
            \caption{
                Lattice surgery between two triangular color codes.
                \textbf{(a)}~shows the space-time diagram of the operation, where two distinct triangular color code patches are merged and then split again.
                The purple region depicts a domain wall.
                (b), (c), and (d) show time-slices of the process.
                \textbf{(b)} depicts the initial configuration where we start with two logical qubits encoded in two disjoint triangular color codes.
                In \textbf{(c)}, we measure additional stabilisers supported on qubits on the red boundaries of both original triangular color codes.
                In doing so, we obtain two bits of classical information from reading out the $\overline{X}_1\overline{X}_2$ and $\overline{Z}_1\overline{Z}_2$ logical parities.
                Merging the codes as shown introduces a trivial domain wall in the purple region.
                In \textbf{(d)} we revert back to the initial configuration by measuring the initial stabiliser generators to split the big code patch.
                \textbf{(e)}-\textbf{(g)} show two possible ways of reading out only $\overline{X}_1\overline{X}_2$.
                In this case, a semi-transparent domain wall is introduced in the purple region.
                Two different microscopic realisations are shown.
                The left one does not require the use of additional auxiliary  qubits, see Fig.~\ref{fig:LatticeSurgeryDomainWalls}~(2) for the microscopic details of the stabilisers.
                The one on the right does require auxiliary qubits but features only stabiliser measurements of weight $6$ or lower.
                Note that the obtained semi-transparent domain wall in this case is the one introduced earlier in Fig.~\ref{fig:CC-CC-DW}~(a).
            }
            \label{fig:LatticeSurgeryspace-time}
        \end{figure}

        Let us begin by considering a simple example of a lattice surgery operation which captures many elements of the physics of more complex merging and splitting operations.
        In its simplest form, lattice surgery merges two disjoint color-code lattices along their adjacent boundaries before they are subsequently split~\cite{Landahl14}.
        This is shown in Fig.~\ref{fig:LatticeSurgeryspace-time}.
        In the initial configuration we start with two disjoint triangular color codes.
        Next, the two codes are merged.
        To achieve this merging operation we measure stabilisers between the two red boundaries of the two code patches to create one large code patch.
        Finally, we split the two codes again by measuring the initial stabilisers to return to the original code space.

        Looking at the stabilisers measured during the lattice surgery protocol depicted in Fig.~\ref{fig:LatticeSurgeryspace-time}~(c) and~(f), we can see that the product of the red $X$-type ($Z$-type) stabilisers in the seam is the product of two logical operators of the original code patches, namely $\overline{X}_1\overline{X}_2$ ($\overline{Z}_1\overline{Z}_2$).
        Hence, by measuring these additional stabilisers, we obtain the logical readings for the logical operators $\overline{X}_1\overline{X}_2$ ($\overline{Z}_1\overline{Z}_2$).
        Finally, we split the code patch to obtain the two initial triangular color codes again. Their logical qubits are now prepared in a Bell state, with the explicit state depending on the outcome of the logical parity measurements.
        
        Domain walls play a central role in lattice surgery operations.
        Let us study the domain walls we obtain along the seam when performing lattice surgery in our example of two triangular color codes.
        In the process depicted in the left of Fig.~\ref{fig:LatticeSurgeryspace-time}, we converted an opaque domain wall (b), which consisted of two red boundaries and did not let any anyons pass from one code to the other, into a fully transparent domain wall (c), effectively joining the two codes together.
        
        Performing a different measurement leads, in general, to a different domain wall.
        For instance, consider the case depicted on the right of Fig.~\ref{fig:LatticeSurgeryspace-time}.
        Here, we measure only $\overline{X}_1\overline{X}_2$.
        This requires us to find a set of commuting stabilisers which multiply to $\overline{X}_1\overline{X}_2$ while leading to a logical code patch which still encodes one logical qubit.
        We show a valid solution in Fig.~\ref{fig:LatticeSurgeryspace-time}~(f).
        Note that this measurement results in the semi-transparent domain wall shown in Fig.~\ref{fig:CC-CC-DW}.
        Similar protocols have been considered in Refs.~\cite{Landahl14,Litinski17}.
        
\begin{figure}[tb]
            \centering
            \includegraphics[width=1.00\linewidth]{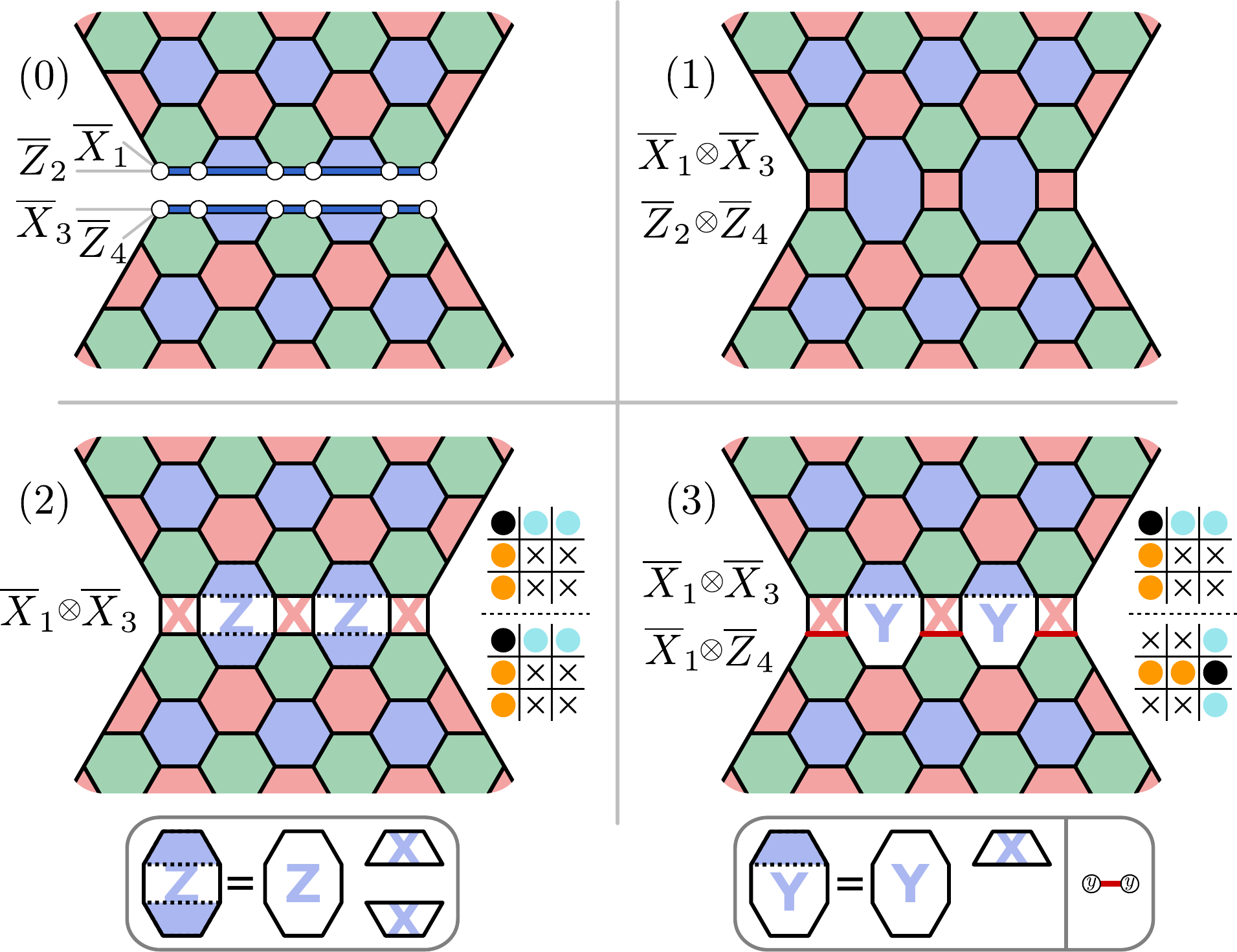}
            \caption{
                The four cases of possible measurements for two-patch color code lattice surgery.
                The logical operators supported on the interfacing boundaries of the code patches are shown in the top left.
                \textbf{(0)}~the trivial case, where no parity measurement is performed, leads to an opaque domain wall between the two patches.
                \textbf{(1)}~an invertible domain wall is obtained if two different logical operators on each boundary are addressed.
                A semi-transparent domain wall which is obtained when only one of the logical operators is addressed in both code patches \textbf{(2)} or in just one of the code patches \textbf{(3)}.
            }
            \label{fig:LatticeSurgeryDomainWalls}
        \end{figure}

        Let us generalise the observed behaviour and investigate the connection between the different types of domain walls and the types of Pauli-word(s) being measured.
        Microscopic examples for the case of two code patches are given in Fig.~\ref{fig:LatticeSurgeryDomainWalls}. 
        In this encoding we show two code blocks where both the top and bottom code block encode two logical qubits over a red boundary with blue boundaries on either side. The red boundary of the upper patch support the logical operators $\overline{X}_1$ and $\overline{Z}_2$ and the red boundary of the lower patch support $\overline{X}_3$ and $\overline{Z}_4$.
        
        (0) The trivial case where no measurement is performed results in an opaque domain wall.
        This occurs when logical qubits are left idling.
        
        (1) Invertible domain walls are obtained if two commuting Pauli words which address two different logical degrees of freedom on each boundary are measured.
        In the example presented in Fig.~\ref{fig:LatticeSurgeryDomainWalls}~(1), we measure $\overline{X}_1\overline{X}_3$ and $\overline{Z}_2\overline{Z}_4$.
        
        (2) As discussed before, a semi-transparent domain wall is obtained if only one of the two degrees of freedom supported on each boundary of a code patch is addressed.
        As an example, a measurement of $\overline{X}_1\overline{X}_3$, as in Fig.~\ref{fig:LatticeSurgeryDomainWalls}~(2), results in a semi-transparent domain wall.
        
        (3) If one of the boundaries involved in the lattice surgery has support on an even number of qubits, as is the case for the rectangular color code, it supports two commuting logical operators.
        In this case, it is possible to perform two commuting measurements involving the same logical operator on the other code patch.
        For instance, as in Fig.~\ref{fig:LatticeSurgeryDomainWalls}~(3), we might measure $\overline{X}_1\overline{X}_3$ and $\overline{X}_1\overline{Z}_4$.
        Such a measurement also leads to a semi-transparent domain wall.
        
        \begin{figure}[tb]
            \centering
            \includegraphics[width=.85\linewidth]{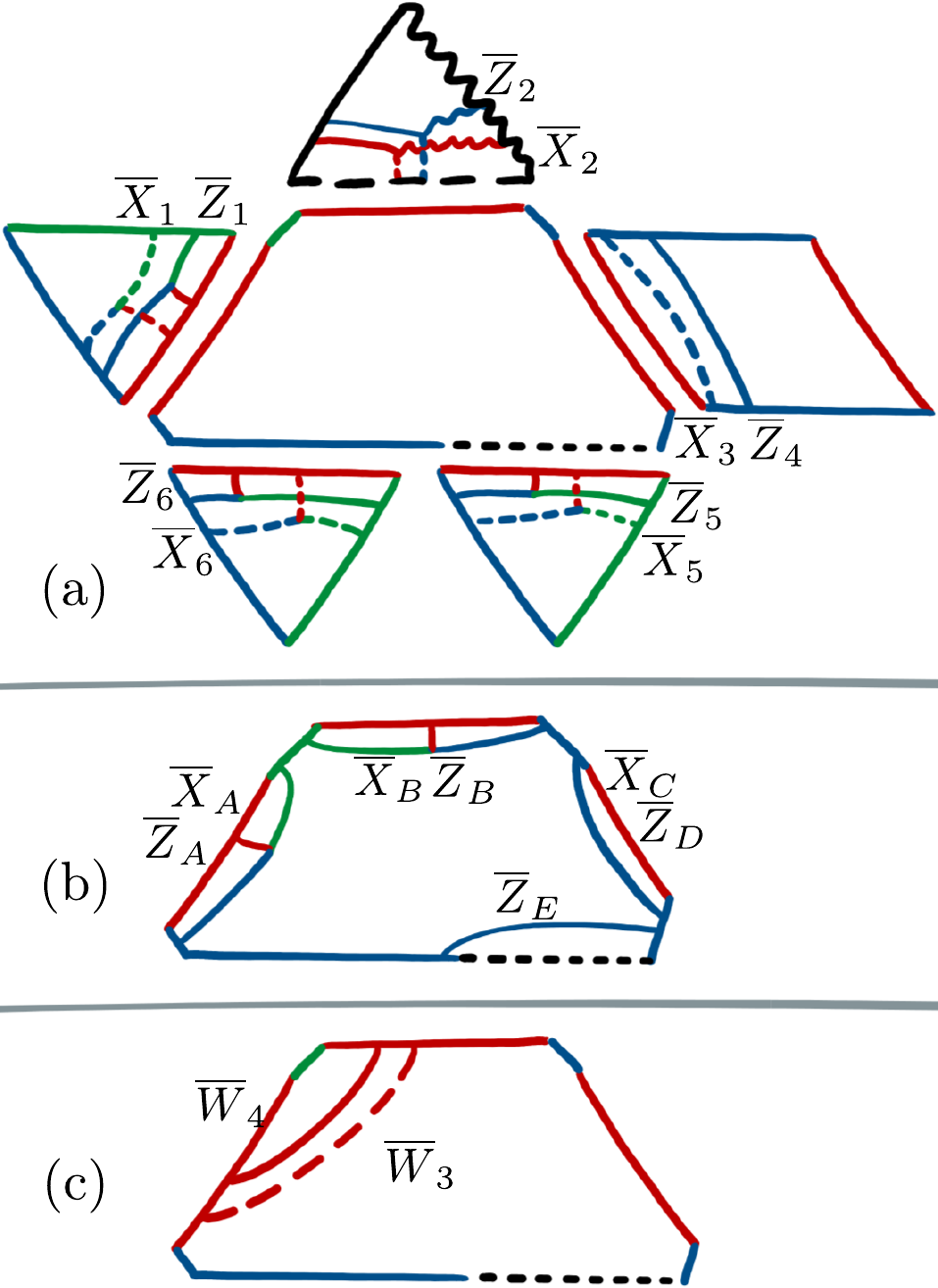}
            \caption{
                An example of two length-$6$ Pauli words being measured in parallel using color code lattice surgery.
                In the shown example, we measure $W_1 = \overline{X}_1 \overline{X}_2 \overline{X}_3 \overline{1}_4 \overline{1}_5 \overline{1}_6$ and $W_2 = \overline{Z}_1 \overline{Z}_2 \overline{X}_3 \overline{1}_4 \overline{Z}_5 \overline{1}_6$.
                The boundary configuration on the data patches (shown in \textbf{(a)}) and the logical Pauli words we are reading out, $W_1$ and $W_2$, determine the boundary configuration on the auxiliary patch.
                For instance, the boundary interfacing with the fifth qubit is a Pauli-$X$ boundary, as it is only involved in one of the two Pauli word measurements we perform.
                The auxiliary patch is initialised and read out using a red temporal boundary. This ensures that it is not initialised in an eigenstate of the logical operators depicted in \textbf{(b)} and we therefore avoid obtaining information about individual logical Pauli observables during the merging step. Instead, we obtain information about the products $W_1 W_3$ and $W_2 W_4$, where $W_3=\overline{X}_A\overline{X}_B\overline{X}_C$ and $W_4=\overline{Z}_A\overline{Z}_B\overline{Z}_D\overline{Z}_E$, shown in \textbf{(c)}, are the products of the logical operators in \textbf{(b)}. During the splitting step where we measure the red edge terms, we measure the values of $W_3$ and $W_4$, thereby enabling us to recover the values of the Pauli words $W_1$ and $W_2$
            }
            \label{fig:LSBigExample}
        \end{figure}
        Having discussed the different cases arising in two-patch color code lattice surgery, let us now move onto measurements of Pauli codewords with support on a larger number of patches. General lattice surgery operations are discussed in more detail in Ref.~\cite{Thomsen22}.
        We consider the example shown in Fig.~\ref{fig:LSBigExample}.
        Here, we aim to measure the two Pauli words $W_1 = \overline{X}_1 \overline{X}_2 \overline{X}_3 \overline{1}_4 \overline{1}_5 \overline{1}_6$ and $W_2 = \overline{Z}_1 \overline{Z}_2 \overline{X}_3 \overline{1}_4 \overline{Z}_5 \overline{1}_6$ which are encoded on five distinct patches of color code.
        Note, the two Pauli words commute, $\left[ W_1, W_2 \right] = 0$.
        This implies we are able to measure them at the same time using color code based lattice surgery \cite{Thomsen22}.
        In the example we find triangular code patches (see Fig.~\ref{fig:TriangularCCs}) as well as a rectangular code patch (see Fig.~\ref{fig:ReadOut}).
        Note how the rectangular patch encodes two logical qubits, indexed $4$ and $5$, leading to Pauli words of length $n=6$.
        To perform the measurement, we introduce an auxiliary  patch in the centre, such that it neighbours a boundary of each of the five code patches surrounding it.
        Furthermore, we choose the boundaries interfacing with code patches that are acted on non-trivially by both $W_1$ and $W_2$ to be red, and of Pauli type if they are acted on non-trivially only by one of $W_1$ and $W_2$, such as the Pauli-$X$ boundary interfacing with code patch number $5$.
        The exception is the boundary interfacing the code patch encoding qubit number $6$, which only gets acted on trivially by both $W_1$ and $W_2$.
        
        Let us study the auxiliary central patch in its own right.
        It encodes five logical qubits, which we label based on the logical operators that are supported on the boundaries interfacing the code patches, see Fig.~\ref{fig:LSBigExample}~(b).
        In the lattice surgery protocol, we measure the parity of the depicted logical operators on the auxiliary code patch and the logical operators on the surrounding code patches.
        In order to measure only $W_1$ and $W_2$, and no additional information, we initialise the auxiliary code patch using a red temporal boundary, ensuring that it is not initialised in an eigenstate of the depicted logical operators.
        As shown in Fig.~\ref{fig:LSBigExample}~(c), the products of the logical operators in (b) are red strings, namely $W_3=\overline{X}_A\overline{X}_B\overline{X}_C$ and $W_4=\overline{Z}_A\overline{Z}_B\overline{Z}_D\overline{Z}_E$.
        Thus, initialising with a red temporal boundary initialises the auxiliary code patch in an eigenstate of $W_3$ and $W_4$.
        Now, by performing the merging step in the lattice surgery protocol, we measure the products $W_1W_3$ and $W_2W_4$.
        Importantly, the measurements we have performed do not commute with $W_3$ and $W_4$.
        Thus, in order to infer $W_1$ and $W_2$, the values of $W_3$ and $W_4$ need to be obtained during the splitting step~\cite{HorsemanSurgery}.
        This is achieved by reading out the auxiliary code patch using a red temporal boundary.
        
        In the example we have examined, we encounter all types of domain walls.
        The opaque domain wall between the auxiliary code patch and the triangular code patch encoding logical qubit number $6$ is maintained throughout the protocol.
        The domain walls between the auxiliary patch and the triangular code patches number $1$ and $2$ are invertible or fully transparent.
        Note, the domain wall to the patch number $2$ is non-trivial, as it joins a red boundary and a Pauli-$X$ boundary together.
        The two domain walls between the auxiliary patch and the patch encoding qubits $3$ and $4$ as well as the one encoding qubit number $5$ are semi-transparent. 
        Here, the former is of type (3) in the above discussion, while the latter is of type (2).
        
        With this example we hope to have elucidated how color code lattice surgery can be used to efficiently perform logical gates on a wide range of logical encodings by making use of its rich set of domain walls.
        This should aid the design of further schemes using unconventional encodings such as the thin color code proposed in Ref.~\cite{Thomsen22}, which might be preferable in architectures with biased noise.
        
        \begin{figure}[tb]
            \centering
            \includegraphics[width=1\linewidth]{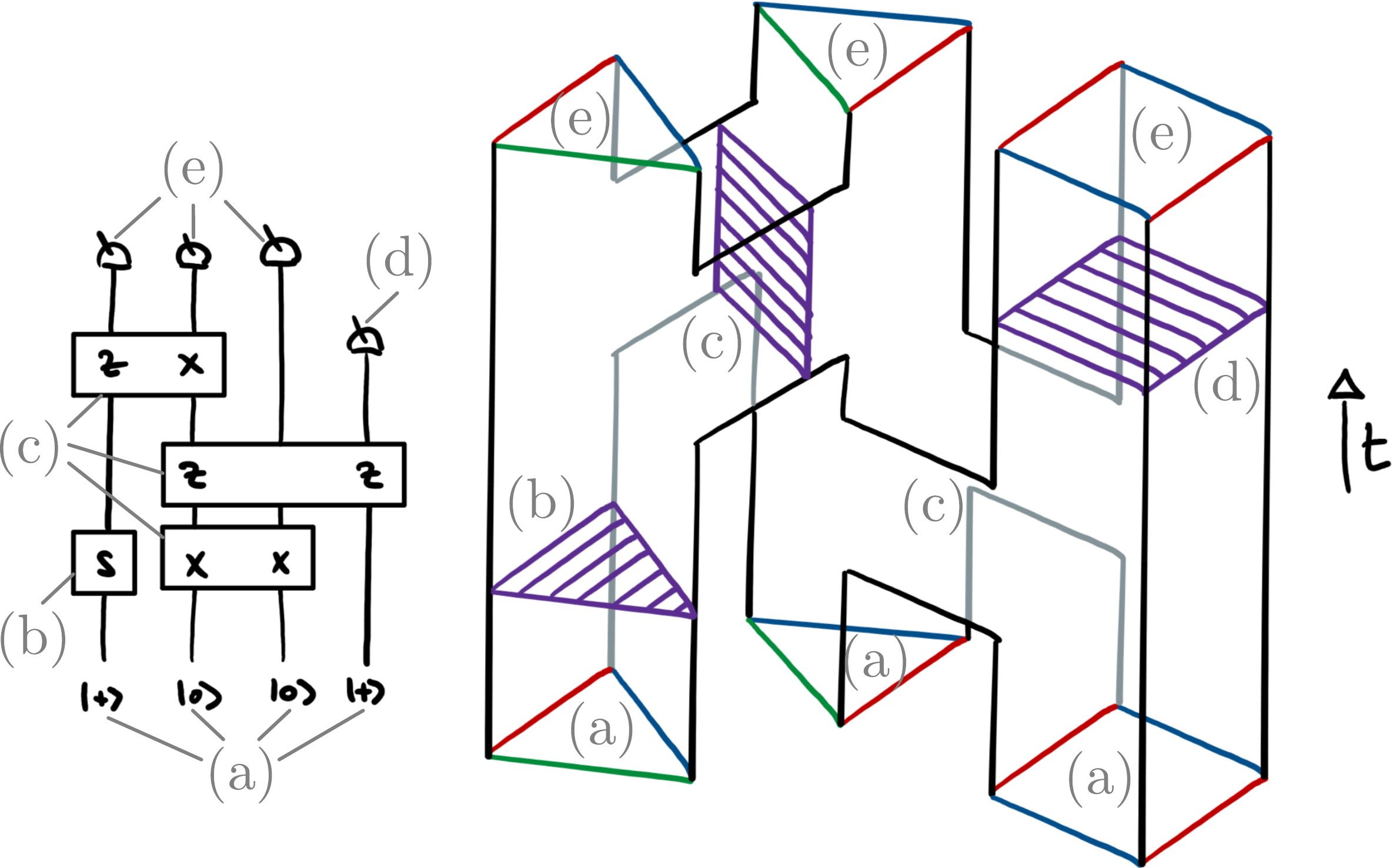}
            \caption{
                A circuit (left) is translated to a fault tolerant computation in the color code (right).
                All fault-tolerant logic gates are described with the condensation processes we present in Secs.~\ref{sec:Cond2d}-\ref{sec:Cond1d}.
                The spatial boundaries dictate the number of encoded qubits per code patch.
                Logical qubits are initialised \textbf{(a)} and read out \textbf{(e)} with temporal boundaries.
                A partial readout \textbf{(d)} corresponds to a domain wall obtained by partial condensation.
                Transversal gates \textbf{(b)} are applied using invertible domain walls corresponding to trivial condensation.
                To perform parity measurements between logical qubits, we deform the spatial boundaries of the code patches and introduce domain walls between them in lattice surgery protocols \textbf{(c)}.
            }
            \label{fig:SummaryCircuit}
        \end{figure}
        To conclude this part of the work, let us summarise how to translate from a circuit describing a quantum computation to a fault-tolerant implementation using the color code.
        We make our summary using the example presented in Fig.~\ref{fig:SummaryCircuit}, where we make use of all of the features discussed in Secs.~\ref{sec:Cond2d}-\ref{sec:Cond1d}.
        The operations we present give rise to a universal set of fault-tolerant logical operations for topological stabiliser codes, where non-Clifford gates are performed with magic state distillation.
        As we have now elaborated, all of these operations are described, both macroscopically and microscopically, with the unifying language of anyon condensation.

\section{Dynamically driven codes} 
    \label{sec:CondDynamic}
    
Dynamically driven `Floquet' codes~\cite{HastingsHaah21a} are a generalisation of subsystem codes where a time-ordered sequence of gauge operator measurements, or ``checks'', is specified to measure the syndrome data. This generalises subsystem codes~\cite{Poulin05} that are described by a generating set of all check measurements with no explicit time ordering. The honeycomb code~\cite{HastingsHaah21a} is an example of a Floquet code that has received considerable attention~\cite{Vuillot21, Gidney21, HaahHastings21b, Gidney22, Paetznick22} following its recent discovery, due to its practical implementation. Specifically, the honeycomb code has weight-six stabilisers that are inferred using only weight-two parity measurements. Furthermore, the honeycomb code can be realised on a planar lattice with boundaries~\cite{HaahHastings21b}.

The honeycomb code demonstrates the significance of specifying the order in which check operators are measured~\cite{HastingsHaah21a}. If we describe the honeycomb code as a subsystem code, where its gauge group is generated by its full set of check measurements, we arrive at a subsystem code that encodes no logical qubits. See Ref.~\cite{Poulin05} for an introduction to subsystem codes. Nevertheless, by specifying a constrained sequence of measurements where only a subset of the generators of the gauge group are measured at each time step, we find that each round of check measurements projects the system onto a new instantaneous code, such that logical qubits with an arbitrarily high distance are encoded in the system.

\begin{figure}[tb]
    \centering
    \includegraphics[width=0.4\linewidth]{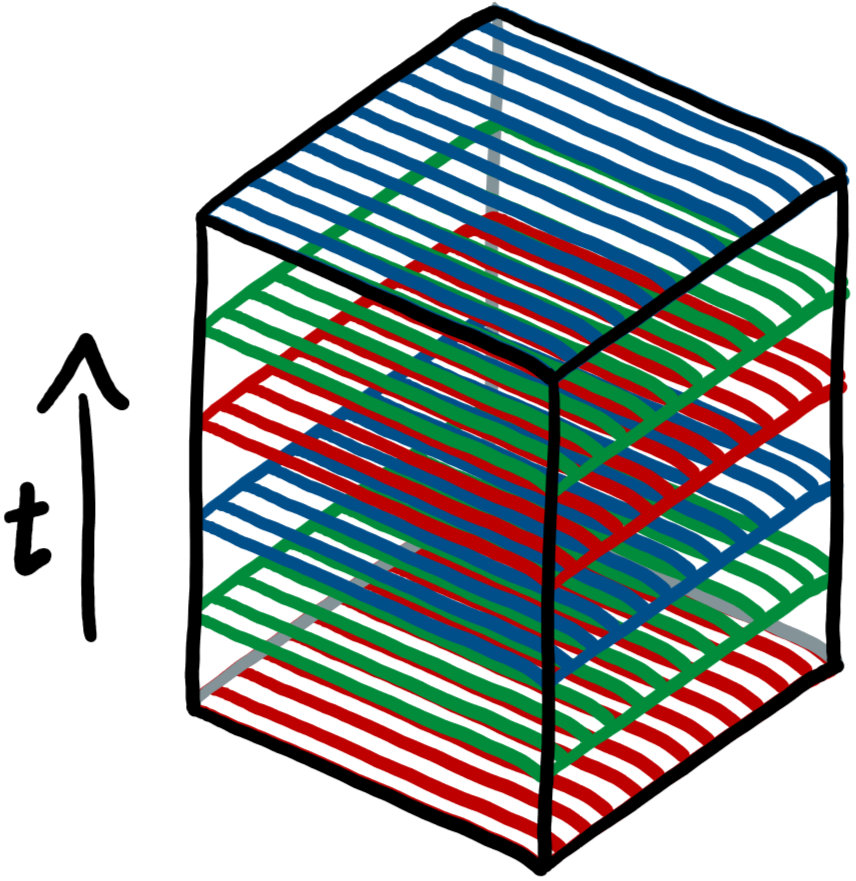}
    \caption{
        In Sec.~\ref{sec:CondDynamic} we derive dynamically driven codes by condensing the confined charges of different instances of the partially condensed color code in an ongoing sequence. The figure shows condensation operations in the space-time picture as coloured planes that lie orthogonal to the time-like direction.
    }
    \label{fig:CondensationInstantaneous}
\end{figure}

Here, we find that anyon condensation gives us a complementary perspective of dynamically driven codes. To this end we present a general construction for new types of such codes that we call dynamically condensed color codes; among which the honeycomb code is included. We obtain this generalisation from the observation that all of the instantaneous codes of the honeycomb code are examples of partially condensed color codes. By regarding the color code as a parent theory from which the instantaneous toric code states of a dynamically driven code can be derived, we can identify new transitions between different instances of the toric code via sets of weight-two projective measurements that act on the edges of the color-code lattice. Specifically, we find that we can reproduce the transformations of dynamically driven codes by condensing color codes anyons that have been confined by a previous condensation operation. Using other choices of edge measurements to make checks on the color-code lattice gives us additional freedom to design new dynamically driven codes. A sketch of our framework is shown in Fig.~\ref{fig:CondensationInstantaneous}.

We use our construction to introduce one specific example of a Floquet code that is of \emph{Calderbank-Shor Steane} (CSS) type. We find that, by measuring only weight-two parity checks on the edges of a three-colourable lattice, we can infer the values of both Pauli-$X$ and Pauli-$Z$ type stabilisers for each plaquette. We can use the outcomes of these stabilisers to detect the occurrence of errors. As the stabilisers we measure are precisely those of the color code we coin the code the \emph{Floquet color code}. We note that this code defined with periodic boundary conditions was discovered independently in Ref.~\cite{Davydova22, Bombin22PsiQ}. Let us remark that one might regard this choice of name as a misnomer. Although we measure check operators that infer the values of color-code stabilisers, the Floquet color code emulates the ground space of the toric-code phase.

The Floquet color code represents a generalisation beyond other examples of known dynamically driven codes. When expressed as a subsystem code, where the check operators are the generators of a gauge group, the Floquet color code has no geometrically local stabiliser operators. In contrast, the honeycomb code, for example, maintains a constant set of stabiliser operators whose cardinality is extensive in the system size. This is related to the fact that the Floquet color code emulates the toric-code phase. Although we measure all of the color-code stabilisers over a period of the Floquet color code, we never produce a simultaneous eigenstate of all of the color-code stabilisers at a single instant of the period of the Floquet color code. Specifically, this is because each time we perform a set of check measurements, we kick our system out of eigenstates of stabilisers that do not commute with the check operators that are measured.

One might be troubled that our new dynamically driven code does not maintain a constant stabiliser group. However, we present numerical results showing our code demonstrates a threshold comparable to the honeycomb code~\cite{Gidney22}. Our example therefore shows that no terms in the centraliser of a subsystem code need to remain sacred in order to encode and protect logical quantum information. We note that the observation that the Floquet color code does not have a constant stabiliser group has been made in independent work~\cite{Davydova22, Bombin22PsiQ}. Another Floquet code with this property, the automorphism code, is also presented in Ref.~\cite{aasen22automorphism}.

Two-dimensional codes can be realised on a planar qubit array by introducing boundaries. We therefore require suitable transformations for the boundary stabilisers of dynamically driven codes as we make deformations between instantaneous code states.
We find that regarding the color code as a parent theory for dynamically condensed color codes reveals a general rule to find appropriate boundary conditions as we perform the code deformations of a dynamically driven code. We complete this section by explaining boundary transformations from the perspective of bulk-condensed color codes before offering some concluding remarks on dynamically driven codes.

\subsection{Dynamically condensed color codes}
\label{SubSec:DCCC}

In what follows, we will describe the instantaneous stabiliser group for dynamically condensed color codes. We will explain how the instantaneous stabiliser group is transformed as we make different choices of weight-two edge measurements on the three-colourable lattice.

Dynamically driven codes are described by a series of instantaneous stabiliser groups. Each round of check measurements that are made in the sequence projects the system onto a new instantaneous stabiliser group. Each of the instantaneous stabiliser groups of dynamically condensed color codes $\mathcal{S}_\a$ are obtained by condensing boson $\a$ of the color code. The honeycomb code~\cite{HastingsHaah21a} for example moves, up to a local basis change, through a series of three instantaneous stabiliser groups $\mathcal{S}_{\rx}$, $\mathcal{S}_{\gy}$ and $\mathcal{S}_{\bz}$. In practice we find that we can transform between any two anyon-condensed color codes $\mathcal{S}_\a$ and $\mathcal{S}_\b$ provided $\a$ and $\b$ correspond to bosons of the color code that share neither the same colour or Pauli label.

The deconfined charges of the condensed color code give rise to logical operators of an instantaneous stabiliser group. Specifically, we regard physical string operators that transport deconfined charges over large non-trivial cycles of the lattice as the extensive logical operators. In the case of \( \mathcal{S}_\rx \) the logical operators are generated by the string operators for \( \ry \), \( \rz\), \( \gx \) and \( \bx \) charges. 

The edge terms for an instantaneous stabiliser group are stabilisers for the condensed code. They correspond to the string operators that transport the condensed charge. For \(\mathcal{S}_\rx \) this is a weight-two string operator that transports the \( \rx \) charges in the parent color code model. Indeed, we can generate longer string operators that transport these charges by taking the products of red Pauli-$X$ edge terms that are included in \( \mathcal{S}_\rx\).

\begin{figure}[tb]
    \centering
    \includegraphics[width=1.00\linewidth]{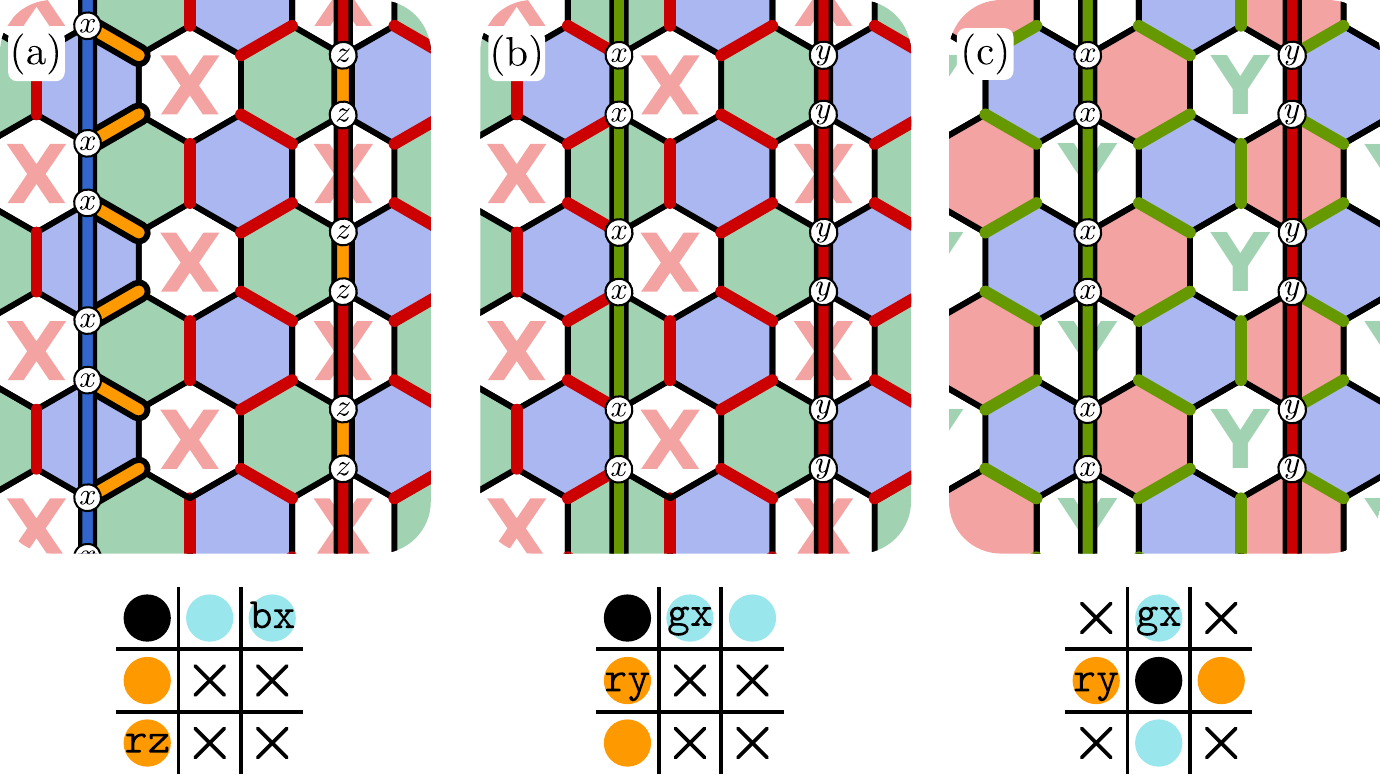}
    \caption{
        Logical operators of dynamically condensed color codes. The logical operators corresponding to string operators for \bx (\rz) charges,~\textbf{(a)}, are equivalent to the string operators for \bx (\ry) charges,~\textbf{(b)}.
        They differ by red $XX$ edge terms, the hopping terms of the condensed \rx anyon, which are stabiliser terms.
        We highlight the red edge checks we use to identify this equivalence in \textbf{(a)}.
        The logical operators obtained in \textbf{(b)} commute with both instantaneous stabiliser groups $\mathcal{S}_\rx$ and $\mathcal{S}_\gy$.
        Therefore, these logical operators are maintained as we transform between $\mathcal{S}_\rx$ and $\mathcal{S}_\gy$.
        We show the stabilisers of $\mathcal{S}_{\gy}$ instantaneous stabiliser group together with the same logical operators corresponding to \gx and \ry string terms in \textbf{(c)}.
    }
    \label{fig:FloquetCodeAnyonStrings}
\end{figure}

We can check that the deconfined charges of the condensed code are identified by fusion with a condensed charge at the microscopic level (see Sec.~\ref{sec:CondTheory}). We find that logical operators that are identified by condensation are equivalent up to multiplication by edge operators. We depict this equivalence for the case of the condensed code \( \mathcal{S}_\rx \) in Figs.~\ref{fig:FloquetCodeAnyonStrings}~(a) and~(b), where we show the equivalence between \( \ry \) and \( \rz \) operators, as well as equivalence between \(\gx\) and \(\bx\) operators, by multiplication with the edge operators that transport the condensed \( \rx \) charges.
We can therefore regard string operators \( \ry \) and \( \rz \) equivalently as Pauli-$Z$ logical operators, and \(\gx\) and \(\bx\) equivalently define Pauli-$X$ logical operators. Likewise, as discussed previously in Sec.~\ref{sec:CondStabs}, we can identify the deconfined red bosons with, say, an electric charge of the toric code model, $\ry \simeq \rz \equiv \e$, and, similarly, the two deconfined bosons that have a Pauli-$X$ label with a magnetic flux of the toric code; $\gx \simeq \bx \equiv \m$.

Let us now look at how we transform between different instantaneous stabiliser groups in dynamically condensed color codes. We will concentrate on a single transformation to explain how the logical operators are modified under a transformation process, but we note that we can compile a long sequence of transformations of this type. Additionally, we will show how we detect error events from the weight-two check measurements.

As we will see, we can transform between any condensed color codes $\mathcal{S}_\a$ and $\mathcal{S}_\b$ provided $\a$ and $\b$ are bosons that share neither their colour label nor their Pauli label.
In other words, we require \b to be a confined charge in $\mathcal{S}_\a$.
It is this observation that gives us a generalised construction for dynamically driven codes. As such, without loss of generality, we concentrate on one projection from the initial instantaneous code $\mathcal{S}_{\rx}$ onto $\mathcal{S}_{\gy}$. We complete this transformation by measuring all of the Pauli-$Y$ edge operators on green edges. It will be helpful to tabulate the bosonic charges of $\mathcal{S}_{\rx}$ as follows:
\begin{equation}
    \begin{tabular}{ c c | c | c }
        & \r & \g & \b \\
        \x ~ & \blackcirc & \cyancirc & \cyancirc \\ \hline    
        \y ~ & \orangecirc & \cross & \cross \\ \hline    
        \z ~ & \orangecirc & \cross & \cross 
    \end{tabular} ~ . \label{Eqn:Inst1}
\end{equation}
 Measuring the green Pauli-$Y$ edge operators to condense the confined $\gy$ charges then maps us onto a code with the following bosonic charges
\begin{equation}
    \begin{tabular}{ c c | c | c }
        & \r & \g & \b \\
        \x ~ &  \cross & \cyancirc & \cross \\ \hline \y ~ & \orangecirc  & \blackcirc &  \orangecirc \\ \hline
        \z ~ & \cross & \cyancirc &  \cross 
    \end{tabular} ~ . \label{Eqn:Inst2}
\end{equation}
In the former, $\mathcal{S}_{\rx}$, the $\rx$ charges are condensed such that, up to exchange of \e and \m labels, we can regard the deconfined $\ry$ and $\rz$ anyons equivalently as electric charges $\e$ of the toric code ($\ry \simeq \rz \equiv \e$), and similarly $\gx$ and $\bx$ can both be regarded as deconfined magnetic charges $\m$ of the toric code ($\gx \simeq \bx \equiv \m$). All other charges are confined. In the latter, the $\gy$ anyons are condensed and we have $\gx \simeq \gz \equiv \m$ and $\ry \simeq \by \equiv \e$.
Fig.~\ref{fig:FloquetCodeAnyonStrings} shows the microscopic details of the logical operators as the transformation is made.

Importantly, both $\mathcal{S}_\rx$ and $\mathcal{S}_\gy$ share a pair of logical operators.
These are string operators for $\gx$ and $\ry$ charges.
This can be read directly by comparing the two boson tables in Eqs.~\eqref{Eqn:Inst1} and~\eqref{Eqn:Inst2} where we use different colours to correspond to different species of toric-code bosons.
One can readily check that both of these boson tables have a common pair of deconfined bosons, that correspond to these anti-commuting logical operators and braid non-trivially with one another.
Importantly, this means that logical operators are maintained throughout the transformation and, as such, logical information is preserved.
It follows that measuring the $\gy$ edge operators does not reveal any logical information from the corresponding string-like logical operators of the deconfined charges.
See Fig.~\ref{fig:FloquetCodeAnyonStrings} for a microscopic illustration.

Let us further emphasise this point that the transformation determines the charge labelling convention for the new code, as this will be important in Sec.~\ref{subsubsec:Bdrytransforms} where we consider the transformation of boundary stabilisers for dynamically driven codes.
Specifically, we find that the particle identification across the transformation can be read directly from the boson tables of the condensed color codes before and after the transformation. By identifying \ry and its corresponding logical operators with, say, the electric charge \e in $\mathcal{S}_\rx$, it follows that the same logical operators must also transport \e charges in the transformed code $\mathcal{S}_\gy$. As such, the \ry charge of the transformed code, together with charges that are identified with it under the condensation operation, must also be identified with the \e particle. Likewise, by identifying \gx with the magnetic particle \m in the original code $\mathcal{S}_\rx$, it follows that \gx, and its corresponding string terms that give rise to logical operators, must also be identified with the \m particle in the transformed code $\mathcal{S}_\gy$.
We reflect this identification of charges across the transformation using a consistent colouring convention for the two types of deconfined charges in Eqs.~\eqref{Eqn:Inst1} and~\eqref{Eqn:Inst2}. Specifically, we use different colours to correspond to different species of toric-code bosons. In both boson tables in Eqs.~\eqref{Eqn:Inst1} and~\eqref{Eqn:Inst2}, \ry is coloured orange in both cases, denoting an \e charge, and \gx is coloured blue, indicating the identification of this particle with \m.

        Let us recall the above argument for a specific transformation holds in general for any transformation where a charge that was previously deconfined becomes condensed.
        If we were to measure the hopping terms of a deconfined charge, such as the red Pauli-$Y$ or Pauli-$Z$ edge terms, or the green or blue edge terms with a Pauli-$X$ label, we would perform a readout.
        As we have discussed in Sec.~\ref{sec:Cond2dMax} and Sec.~\ref{sec:Cond2dPart}, condensing a deconfined charge results in the readout of logical information.
        To summarise, it is essential that the ongoing condensation operation condenses a confined charge of $\mathcal{S}_\a$ in our construction for dynamically condensed color codes, in order to maintain coherent logical information encoded in the dynamically driven code.
       
       \begin{figure}[tb]
            \centering
            \includegraphics[width=.85\linewidth]{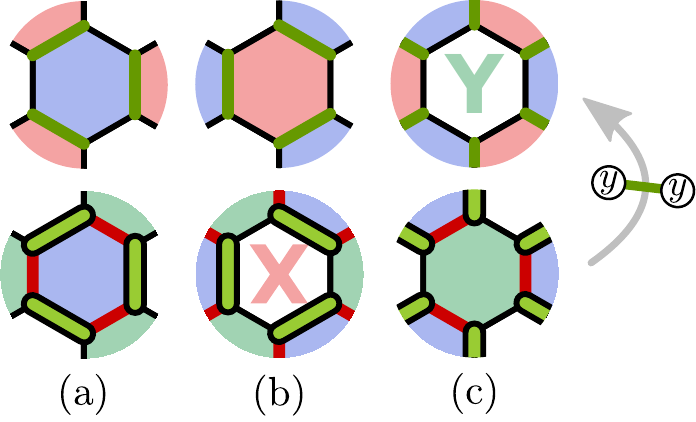}
            \caption{
                    A blue~\textbf{(a)}, red~\textbf{(b)} and green~\textbf{(c)}, plaquette of the $\mathcal{S}_\rx$ and $\mathcal{S}_\gy$ instantaneous stabiliser groups shown at the bottom and top, respectively. The $\mathcal{S}_\rx$ instantaneous stabiliser group is overlaid with the green $YY$ edge checks that are used to make the transformation. During the transformation: The blue stabiliser~\textbf{(a)} begins in an eigenstate of the Pauli-$Y$ stabiliser. We therefore read its known eigenstate by measuring the green $YY$ edge checks; The red plaquette~\textbf{(b)} is not in an eigenstate of the Pauli-Y stabiliser, we therefore project the system onto a random eigenstate of this stabiliser; Both the Pauli-$X$ and Pauli-$Z$ stabilisers on the green plaquette~\textbf{(c)} anti commute with the green $YY$ edge checks. We therefore kick the system out of eigenstates of these stabilisers, leaving the green plaquettes in an eigenstate of the Pauli-$Y$ plaquette stabiliser only.
            }
            \label{fig:FloquetStabTransfMicro}
        \end{figure}
        In addition to transforming the logical operators, measuring the green Pauli-$Y$ edge terms also serves to both infer some stabiliser data, and to reinitialise new stabilisers.
        Measurement of the green edge terms also means that certain stabilisers of $\mathcal{S}_\rx$ are removed from the system.
        We will discuss this in more detail for some specific examples in the following section (Sec.\ref{sec:FloqCC}), but let us provide an overview of the mechanics of stabiliser readout with dynamically condensed color codes here. In what follows we explain how the stabiliser group is transformed under a single condensation process, see also Fig.~\ref{fig:FloquetStabTransfMicro}.

       We are considering the transformation where we project the stabiliser group $\mathcal{S}_\rx$ onto $\mathcal{S}_\gy$.
       Here, we perform green edge measurements that commute with stabilisers on the red and blue plaquettes.
       As such, the system remains in an eigenstate of all stabiliser generators on red and blue plaquettes. We can also use the green edge measurements to infer the values of the $S^\ry_f$ and $S^\by_f$ operators.
       Given that we began in an eigenstate of $S^{\by}_f$ we learn its value for a second time. Comparing its new value to its original value allows us to identify errors that have occurred during the interim period, see Fig.~\ref{fig:FloquetStabTransfMicro}~{(a)}. This gives rise to a detection cell, introduced in see Sec.~\ref{sec:Prelimspace-time} and discussed further in Sec.~\ref{SubSubSec:DetectionCells}. In contrast, the stabiliser group $\mathcal{S}_\rx$ did not include $S^\ry_f$ terms, see Fig.~\ref{fig:FloquetStabTransfMicro}~{(b)}. Measuring the green Pauli-$Y$ edge terms therefore initialises the system in eigenstates of these stabilisers, and in turn the $S^\rz_f$ stabiliser given that we maintain an eigenstate of the $S^\rx_f$ stabiliser throughout the transformation. However, given the outcome of this inferred stabiliser measurement is random, these measurements provide no new syndrome data with respect to the red plaquettes.

       The transformation also removes stabilisers from the system. The green Pauli-$Y$ edge measurements anti-commute with the $S^\gx_f$ and $S^\gz_f$ stabilisers. These stabilisers are therefore not included in $\mathcal{S}_\gy$. Indeed, the new stabiliser group only includes $S^\gy_f$ on the green lattice faces. We depict this in Fig.~\ref{fig:FloquetStabTransfMicro}~{(c)}.

        \subsection{The Floquet color code}
        \label{sec:FloqCC}

        With the discovery that we can transform between any pair of anyon-condensed color codes from $\mathcal{S}_\a$ to $\mathcal{S}_\b$ provided bosons $\a$ and $\b$ share neither colour nor Pauli labels, we find a new degree of freedom that we can use to design new dynamically driven codes that follow different sequences of check measurements. To this end, we introduce the Floquet color code; a specific example of a dynamically condensed color code (see also~\cite{Davydova22, Bombin22PsiQ} where this code has recently been introduced in independent work on a lattice with periodic boundary conditions).
        \begin{figure}[tb]
            \centering
                \includegraphics[width=0.75\linewidth]{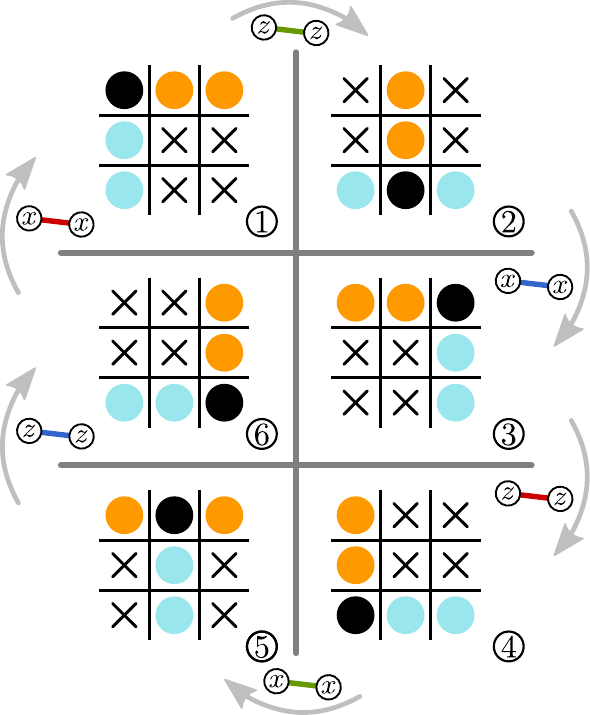}
            \caption{
                The boson tables of the different condensed color codes that are produced at each stage of the Floquet color code measurement sequence.
                }
            \label{fig:FloquetColorCodeMeasurementPattern}
        \end{figure}
        The Floquet color code follows a sequence of six instantaneous stabiliser groups
        \begin{equation}
            \dots \rightarrow    \mathcal{S}_\rx \rightarrow     \mathcal{S}_\gz \rightarrow    \mathcal{S}_\bx \rightarrow    \mathcal{S}_\rz \rightarrow    \mathcal{S}_\gx \rightarrow    \mathcal{S}_\bz \rightarrow \dots,
        \end{equation}
        see Fig.~\ref{fig:FloquetColorCodeMeasurementPattern}, where we use boson tables to show how the electric and magnetic charges are transformed as the Floquet color code undergoes code deformations.
 
Let us now describe quantum error correction using the Floquet color code before presenting the numerical results from our threshold simulation. As we have discussed, dynamically condensed color codes maintain canonical pairs of anti-commuting logical operators provided the transformations we use respects the rules that are detailed in the previous subsection (Sec.~\ref{SubSec:DCCC}).
One can check that our sequence respects these results, so let us concentrate on stabiliser measurements.

\subsubsection{Detection cells}

\label{SubSubSec:DetectionCells}

Fault-tolerant error correction requires that we identify all types of errors over time. In addition to measuring the occurrence of physical errors that act on the qubits of the system, we must also identify errors where measurement apparatus return the incorrect results. In the previous section we described how we transform between different instantaneous stabiliser groups as we perform check measurements. However, we use more general objects to identify the occurrence of errors in practice. We therefore define detection cells, see also Sec.~\ref{sec:Prelimspace-time}, that we use to identify measurement errors, as well as physical errors.

Detection cells compare the value of some Pauli check that we measure to the value of the same check that was measured at some earlier time. If the final reading deviates from the initial reading then we declare that an error event has been detected. These events can be thought of analogously with point-like anyonic excitations in the space-time picture.
Error events can be caused by physical errors that occur between its initial and final reading, or by measurement errors that change the value of either reading of the check measurement.

\begin{figure}[tb]
    \centering
    \includegraphics[width=0.85\linewidth]{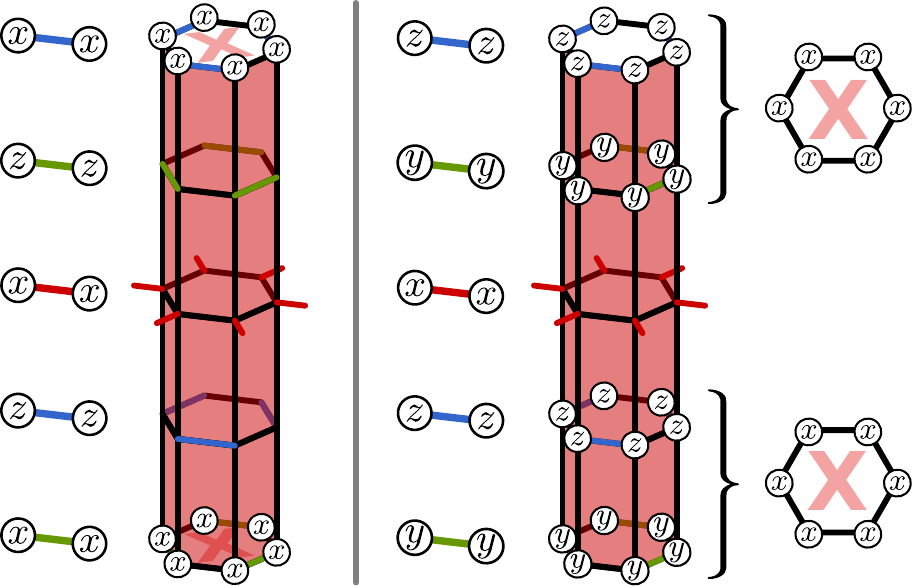}
    \caption{
       Detection cells for the Floquet color code (left) and the honeycomb code (right). For the Floquet color code, we measure the red Pauli-$X$ stabiliser at the initial and final time in the figure. At the initial time we infer its value by measuring the green Pauli-$X$ edge checks, and we compare its value to the reading of the same stabiliser at the final time in the figure, where we infer its value from measuring the blue Pauli-$X$ edge checks. One can check that all of the check measurements that are made in between the two readings of the red Pauli-$X$ stabiliser commute with the stabiliser of interest. As such we obtain a detection cell by comparing the first and last reading of the stabiliser. In the case of the honeycomb code we infer the value of the red Pauli-$X$ stabiliser by taking the product of all of the check measurements during the first two time steps; both the green Pauli-$Y$ checks and the blue Pauli-$Z$ checks, and we compare the value of this stabiliser to the value inferred from the green and blue check measurements that are collected at the final two time steps. The Pauli-$X$ stabiliser commutes with the red Pauli-$X$ edge checks that are performed in between the two readings of the stabiliser.
    }
    \label{fig:FloquetStabsComparison}
\end{figure}

Let us first consider the example of a detection cell corresponding to a Pauli-$X$ stabiliser on a red plaquette, see Fig.~\ref{fig:FloquetStabsComparison}(left), although we remark that no generality is lost here, as our periodic measurement sequence is invariant under cyclic permutations of the arbitrary color labels, or exchange of the Pauli-$X$ and Pauli-$Z$ labels. We initialise the red Pauli-$X$ stabiliser at the earliest time, shown at the bottom of the figure, when we measure the green Pauli-$X$ edge checks. We compare the stabiliser measurement to the measurement of the same stabiliser, performed at the final time at the top of the figure, where we infer its value from measuring the blue Pauli-$X$ edge checks. Assuming no errors occur, we expect the reading of this stabiliser to be the same at both the first and last time.

It is important that all the check measurements that are made in the interim period between the initial and final measurements of the detection cell commute with the red Pauli-$X$ stabiliser. This will mean that both measurements of the stabiliser will have the same value, provided no errors occur. In between the initial and final readout of the red Pauli-$X$ stabiliser, we measure blue and green Pauli-$Z$ edge checks, and red Pauli-$X$ edge checks. One can readily verify that all of these check measurements commute with the red Pauli-$X$ stabiliser. As such, we obtain a detection cell that identifies errors that occur in between the initial and final readout of the stabiliser of interest, as well as any measurement errors that may occur during either reading of the stabiliser.

\begin{figure}[tb]
    \centering
    \includegraphics[width=0.8\linewidth]{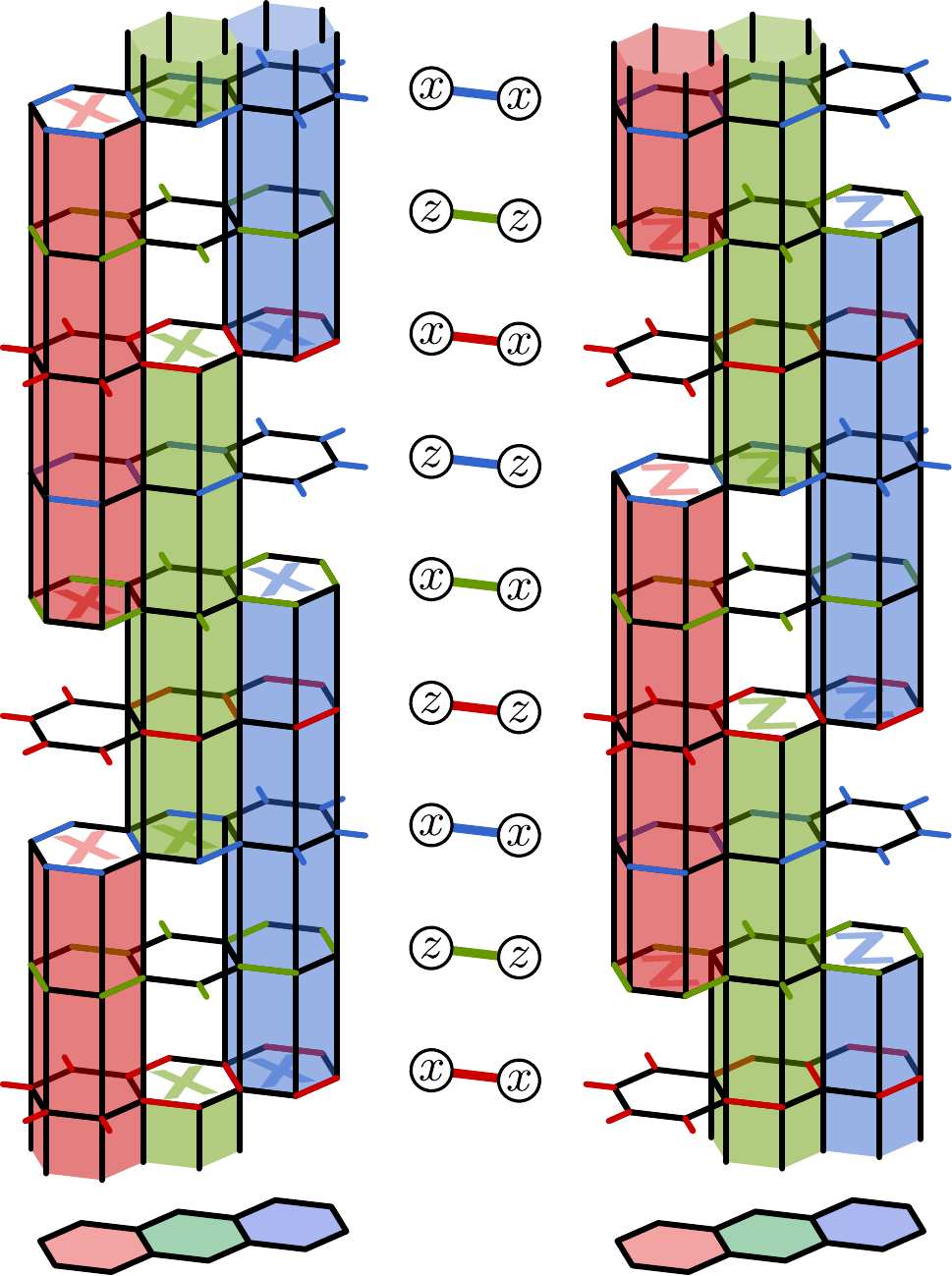}
    \caption{The space-time diagram showing detection cells corresponding to Pauli-$X$ type stabilisers and Pauli-$Z$ type stabilisers shown, respectively on the left and right where time runs upwards along the page. Time runs upwards, and is marked by the points in time at which the edge measurements are made. We show the edge measurements in the centre of the figure. 
    Red Pauli-$X$(Pauli-$Z$) type cells are initialised when the green Pauli-$X$(Pauli-$Z$) edge checks are measured, and they are read out a second time when we measure the blue Pauli-$X$(Pauli-$Z$) edge checks, thereby completing the detection cell. Likewise, green (blue) detection cells are initialised when we measure the blue (red) edge checks, and are read out a second time when we measure the red (green) edge checks. As the stabilisers of the Floquet color code do not commute with all the edge checks, we observe short intervals where a qubit does not support some given cell. For example, the red Pauli-$X$ type detection cell is not supported over the interval where the Pauli-$Z$ edge checks are measured. As such, we find a temporal gap between detection cells of the same type.
    }
    \label{fig:FloquetCodeStabVolumes}
\end{figure}

Let us now look at how all of the detection cells are supported. We find that all qubits support four detection cells at any given instant; two Pauli-$X$ type cells and two Pauli-$Z$ type cells. In Fig.~\ref{fig:FloquetCodeStabVolumes} we show red, green and blue detection cells in space- time with the Pauli-$X$ type cells on the left of the figure and the Pauli-$Z$ type cells on the right.
Note how the left and right diagrams are equivalent up to conjugation by a Hadamard and a shift by three time-steps.
Let us thus without loss of generality concentrate on the Pauli-$X$ detection cells.
As we have already discussed, the red Pauli-$X$ detection cells are initialised when we measure the green Pauli-$X$ edge checks, and are read out a second time when we measure the blue Pauli-$X$ edge checks. Similarly, the detection cells corresponding to green (blue) stabilisers are initialised when we measure the blue (red) edge checks, and they are read out at the final instant when we measure the red (green) edge checks. By construction, every qubit supports one stabiliser of each color. We can therefore see in the diagram that at any given instant, every qubit supports two Pauli-$X$ type detection cells.

An unusual feature of the Floquet color code is that we do not maintain a constant group of stabilisers. Rather, stabilisers are constantly reinitialised and later checked to obtain detection cells that identify error events. For example, the red Pauli-$X$ stabilisers do not commute with red Pauli-$Z$ edge check measurements, we cannot maintain a red Pauli-$X$ detection cell as we measure Pauli-$Z$ edge checks. We must therefore reinitialise the red Pauli-$X$ stabiliser after we measure the red Pauli-$Z$ edge checks. Likewise, we do not maintain a green(blue) Pauli-$X$ detection cell over the interval where we measure green(blue) Pauli-$Z$ edge checks. This is represented in the figure by the temporal gap between two detection cells. Nevertheless, we see that every qubit always supports two detection cells at any given time.

\subsubsection{The error syndrome of the Floquet color code}
\label{SubSubSec:FCCsyndrome}

    \begin{figure}[tb]
        \centering
        \includegraphics[width=0.85\linewidth]{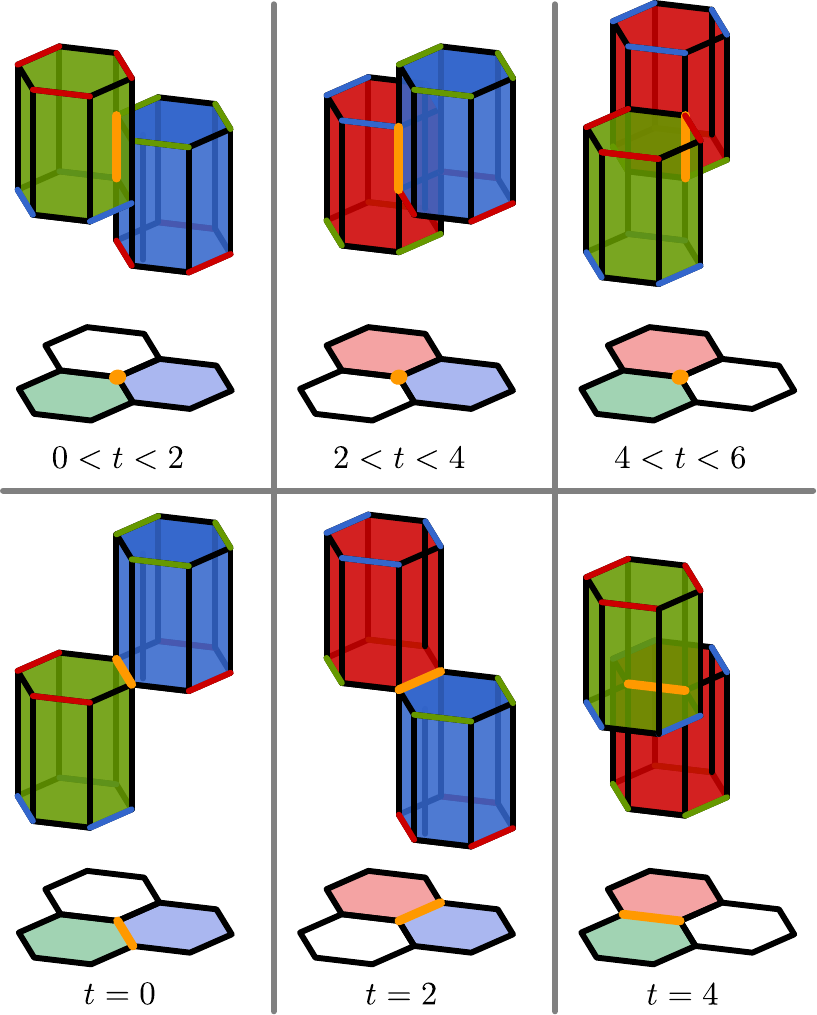}
        \caption{
            The error syndrome of physical errors (top) and measurement errors (bottom). 
            Let us consider the physical error shown in the top left: A qubit supports a green and a blue Pauli-$Z$ type detection cell after the blue Pauli-$Z$ edge checks are measured, up until the instant when the proceeding green Pauli-$Z$ edge checks are measured. Therefore, a physical error that occurs on a qubit during the interval $0 < t < 2$ will produce two detection events on the green and the blue detection cell it supports.
            To understand how measurement errors manifest, let us focus on the case of a measurement error on a red edge (bottom left): The red edge checks read out the green detection cell, and simultaneously reinitialise the stabiliser used for a blue detection cell. Therefore a measurement error on a red edge check will violate a green and a blue detection cell.
        }
        \label{fig:FloquetErrorSyndrome}
    \end{figure}

Let us now look at how the detection cells respond to errors. As the Floquet color code is a CSS code, we concentrate only on bit-flip errors, but remark that an equivalent discussion will hold for Pauli-$Z$ type errors acting on Pauli-$X$ type detection cells. Pauli-$Y$ errors can be regarded as the product of a Pauli-$X$ and Pauli-$Z$ type error. In Fig.~\ref{fig:FloquetErrorSyndrome} we show the occurrence of physical errors at different time intervals over a period, as well as a measurement error on different types of edge check. We will concentrate our discussion in the main text on errors that create a pair of detection events on red and blue detection cells, but remark an equivalent discussion will hold for any pair of colours, up to a cyclic permutation of colour labels. We show examples of all types of errors in Fig.~\ref{fig:FloquetErrorSyndrome}.

Let us first look at a bit-flip error. A qubit supports a red and a blue Pauli-$Z$ type detection cell after the instant the red Pauli-$Z$ type checks are measured, up to the instant where the proceeding blue Pauli-$Z$ type checks are measured. A bit-flip error that occurs in this interval will therefore violate the two detection cells the qubit is supporting at this time, thereby identifying a pair of error events. We show the two violated detection cells in Fig.~\ref{fig:FloquetErrorSyndrome}~(top middle).

Measurement errors also create a pair of detection events in the error syndrome. Each edge check that is measured performs two tasks; it contributes to the read out of a stabiliser thereby completing a detection cell, and it is also used to reinitialise a stabiliser to produce a new detection cell. A measurement error on a given edge check will create detection events on both of its associated detection cells, as shown in Fig.~\ref{fig:FloquetErrorSyndrome}~(bottom middle). The figure shows a single measurement error on a green edge check that reads out a blue detection cell and reinitialises a red detection cell.

Given that all types of errors, both bit flips on physical qubits and measurement errors, create detection events in pairs in the space-time bulk, we have a conservation law among error detection events~\cite{Brown20parallelized, Brown22conservation} that enables us to employ standard decoding methods such as minimum-weight perfect matching~\cite{Dennis02, Wang2003confinement, Brown22conservation} or union find~\cite{Delfosse17}. In what follows we describe simulations to evaluate the threshold using a minimum-weight perfect matching decoder implemented with $\mathtt{PyMatching}$~\cite{Higgott21}.

\subsubsection{Numeric simulations for the Floquet color code and comparison with the honeycomb code}
\label{SubSubSec:FloquetComparison}

\begin{figure}[tb]
    \centering
    \includegraphics[width=\columnwidth]{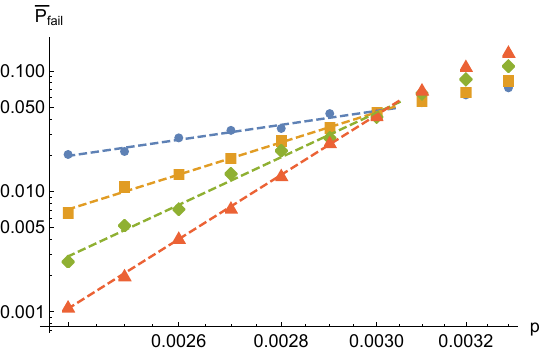}
    \caption{
        Threshold plot for the Floquet color code with periodic boundary conditions using the standard depolarising noise model~\cite{Gidney22}. Source code for the simulations is made available in an online repository~\cite{FCCsourceCode}. Data for codes of $L \times L$ red hexagons with $L = 4,\,8,\,12,\,16$ are shown in blue, yellow, green and red respectively. We obtain a threshold of $\sim 0.3\%$, which is approximately the same as that obtained for the equivalent noise model~\cite{Gidney2021stim}. Simulations have been conducted using $\mathtt{Stim}$, and decoding has been performed using a minimum-weight perfect matching decoder, implemented using $\mathtt{PyMatching}$~\cite{Higgott21}, that concentrated only on stabiliser measurements obtained from Pauli-$Z$ edge checks. Further improvements might be obtained by exploiting correlations between the detection events that are measured by the Pauli-$X$ and Pauli-$Z$ type stabilisers under the circuit-level depolarising noise model.
    }
    \label{fig:threshold}
\end{figure}

To assess its performance, we simulate fault-tolerant quantum error correction with the Floquet color code under circuit level noise (see the caption of Fig.~\ref{fig:threshold} for details).
For the standard depolarising circuit noise model used in Ref.~\cite{Gidney22} we obtain a threshold of $\sim 0.3\%$ using a matching decoder. The decoder matches detection events from the Pauli-$Z$ type detection cells independent of the syndrome information that is measured by the Pauli-$X$ type detection cells. To contrast with other work, Ref.~\cite{Gidney22} reports a threshold of $0.2-0.3\%$ with the honeycomb code under the same noise model using a minimum-weight perfect-matching decoder that accounts for correlations. 

Using a basic decoding algorithm, we have shown that the Floquet color code has a threshold that is competitive with the honeycomb code. This may be surprising, given that we are simulating a code that does not have a constant stabiliser group. Rather, local stabilisers are constantly being removed from the system and reinitialised with different transformations between instantaneous stabiliser groups. In spite of this, the Floquet color code 
obtains syndrome data at the same rate as the honeycomb code.

In what follows, we discuss three factors to explain the similar thresholds obtained for the honeycomb code and the Floquet color code. These factors are: 
\begin{enumerate}[(i)]
    \item the rate at which detection cells are evaluated,
    \item the stabiliser operators of the two Floquet codes,
    \item similar syndrome data structure for decoding.
\end{enumerate}

To review, up to a local change of basis, two periods of the honeycomb code check measurements are as follows:
\begin{equation}
\dots \rightarrow    \mathcal{S}_\rx \rightarrow     \mathcal{S}_\gy \rightarrow    \mathcal{S}_\bz \rightarrow    \mathcal{S}_\rx \rightarrow    \mathcal{S}_\gy \rightarrow    \mathcal{S}_\bz \rightarrow \dots. 
\end{equation}
We also show a corresponding detection cell for the honeycomb code in Fig.~\ref{fig:FloquetStabsComparison} (right).

Let us summarise the difference in how detection cells are measured for each of the Floquet codes in regards to point~(i).

The Floquet color code measures two distinct stabilisers per plaquette whereas the honeycomb code only measures one stabiliser per plaquette. On the other hand, the honeycomb code obtains new detection cells for each of its corresponding stabilisers at double the rate of each of the stabilisers for the Floquet color code. Overall, detection events are measured at an equivalent rate for each Floquet code.

In regards to point~(ii), both codes have very similar structure in the sense that detection cells correspond to weight-six stabilisers that are obtained over five rounds of weight-two check measurements. And given that both codes produce syndrome data at the same rate, as discussed in point~(i), it is perhaps unsurprising that the Floquet color code and the honeycomb code demonstrate comparable threshold error rates.

Let us finally discuss decoding using the minimum-weight perfect-matching decoder for the two Floquet codes; point~(iii). To summarise the remaining discussion in this section, we find that both decoders have similar decoding graphs in the sense that errors give rise to pairs of events at detection cells in very similar configurations in space-time. We can exploit this structure for both codes to design a minimum-weight perfect-matching algorithm.

Let us discuss how we can obtain a matching decoder for the honeycomb code. We find the honeycomb code demonstrates a parity conservation law among its detection events~\cite{Brown20parallelized, Brown22conservation} over two subsets of its detection cells. We obtain these two conservation laws by dividing the cells according to the time the cell is initialised. One subset includes all the detection cells that begin initialisation by check measurements made at odd time steps and the other subset begin their initialisation by checks made at even time steps. We leave it to the reader to verify this fact. Nevertheless, the detection cells of some given symmetry for the honeycomb code share the same structure as those of, say, the Pauli-$X$ type detection cells of the Floquet color code, see Fig.~\ref{fig:FloquetCodeStabVolumes}. Given a detection-event parity symmetry, one can exploit its corresponding conservation law to design a matching decoder~\cite{Brown20parallelized,Brown22conservation}.

Likewise, we obtain two subsets of detection-event parity conservation laws with the Floquet color code. This has already been mentioned implicitly in Sec.~\ref{SubSubSec:FCCsyndrome}. Using that the code is a CSS code, we can trivially separate the decoding problem for the Pauli-$X$ type detection cells and the Pauli-$Z$ type detection cells. Indeed, as we have discussed in detail, errors give rise to defects in pairs in the Floquet color code if we subdivide the results from the detection cells in this way. Given that the Pauli-$X$ type detection cells are initialised at odd time steps and Pauli-$Z$ type detection cells are initialised at even time steps, we can readily see an equivalence in the structure of the decoding graph for the two different Floquet codes.

\subsection{The boundaries of Floquet codes}
\label{SubSec:FloquetBoundaries}
    
Dynamically driven codes with a local group of stabiliser generators can be realised on a planar array of qubits by encoding logical information on a lattice with boundaries~\cite{HastingsHaah21a, HaahHastings21b}. In order to design a dynamically driven code with boundaries, we must also find suitable transformations between the boundary stabiliser operators as we perform deformations between instantaneous stabiliser groups. Here we appeal to the physics of the underlying parent phase to find a systematic way of obtaining suitable boundary transformations for examples of dynamically driven codes. In what follows, we will show how the boundaries of dynamically condensed color codes can be derived using the structure of the parent anyon theory of the color-code model. We will go on to demonstrate our general theory by producing microscopic boundary conditions for the Floquet color code.

 \subsubsection{Boundaries of condensed color codes}
 \label{SubSubSec:BoundariesCCC}
 
    The toric-code phase that is realised by dynamically condensed color codes has a well-understood boundary theory~\cite{Levin13}. We encode logical qubits by introducing ``rough'' and ``smooth'' boundaries to the lattice~\cite{Bravyi98, Dennis02}, where a rough boundary condenses electric charges, labelled \(\e\), and a smooth boundary condenses magnetic charges, \( \m \). We must therefore look for boundaries for dynamically driven codes, together with their associated transformations, that reproduce the behaviour of the rough and smooth boundaries of the toric code. We find that we can derive the boundaries of dynamically condensed color codes from the parent color-code model. Let us review the boundaries of the color code before discussing how the boundaries of dynamically driven codes are obtained from the parent theory. We pay particular attention to the color-code boson table to help to elucidate our construction.

We describe six different boundaries for the color code in 
Sec.~\ref{sec:CondMax}. Three boundaries are associated to 
color labels of the color-code bosons and three boundaries are associated to Pauli labels. We call these color boundaries and Pauli boundaries, respectively. The boundary label denotes the types of boson the boundary condenses. 

We recall that the color boundaries correspond to the columns of the boson table of the color code and the Pauli boundaries correspond to the rows of the boson table. Indeed, the table is defined such that all the bosons 
in a column share a common color label and all of the bosons in a row share a common Pauli-label. Specifically, the boson table is designed such that the rows and columns represent distinct Lagrangian subgroups of the color-code anyon model, see Sec.~\ref{sec:CondMax}.
    
    Let us look once again at the charges of the boson table that remain deconfined as we condense a single charge. Upon condensing a single color-code boson, up to a symmetry in exchange of the $\e$ and $\m$ labels, we identify the charges that share a row with the condensed charge with the electric charges of the toric code, and we identify the charges that share a column with the condensed charge with the magnetic charges of the toric code. Correspondingly, we require suitable boundaries to condense these charge types at appropriate locations.
    
    We posit that we obtain appropriate rough and smooth boundary terms that respectively absorb \(\e\) and \(\m\) charges for dynamically condensed color codes by choosing appropriate boundaries in the corresponding color-code theory. Without loss of generality, let us take the color code where the red Pauli-$X$ charge is condensed, $\mathcal{S}_\rx$ as an example.
    In this condensed color code we have that the deconfined electric charges of the condensed theory correspond to the $\ry$ and $\rz$ charges of the parent theory. Both of these charges have a common red colour label, and are therefore both absorbed, uniquely, by a red colour boundary. As such, red colour boundaries of a color code become rough boundaries that absorb electric charges of the condensed theory $\mathcal{S}_\rx$.
    Similarly, the magnetic fluxes of the condensed theory correspond to the $\gx$ and $\bx$ charges of the parent theory. Both of these charges are uniquely absorbed by a Pauli-$X$ boundary, as they share an $X$ Pauli label. We therefore find that the Pauli-$X$ boundaries of a color code become the smooth boundaries that absorb the magnetic charges of $\mathcal{S}_\rx$.
    \begin{figure}[tb]
        \centering
        \includegraphics[width=0.95\linewidth]{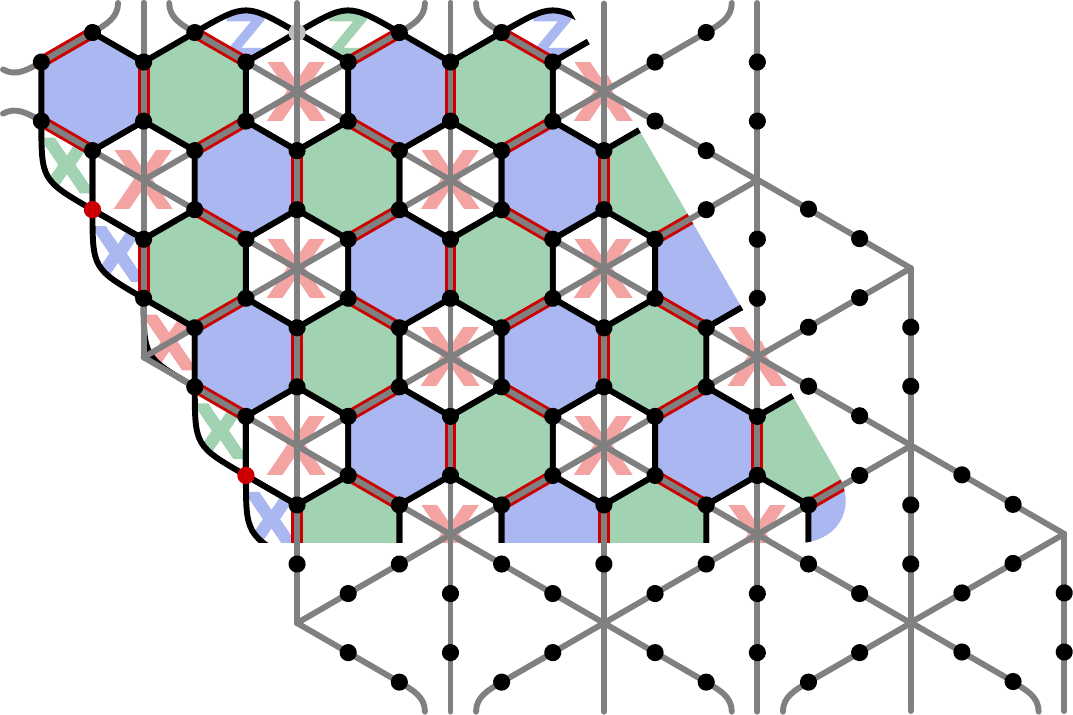}
        \caption{
            The boundaries of the Floquet color code at any point in time can be related to smooth and rough boundaries of the corresponding surface code.
            We demonstrate this here by drawing the surface code in grey on top of the Floquet color code with \rx being condensed.
        }
        \label{fig:FloquetCodeTCBoundaries}
    \end{figure}
    Indeed, in Fig.~\ref{fig:FloquetCodeTCBoundaries} we show a condensed color code $\mathcal{S}_\rx$ where the parent theory had two distinct red boundaries and two distinct Pauli boundaries. The figure shows the corresponding toric code lattice overlaid, with qubits on the edges. We observe that the Pauli-$X$ boundaries produce smooth boundaries in the condensed theory whereas the red boundaries produce rough boundaries.

\subsubsection{Boundary transformations}
\label{subsubsec:Bdrytransforms}
\begin{figure}[b]
    \centering
    \includegraphics[width=1.00\linewidth]{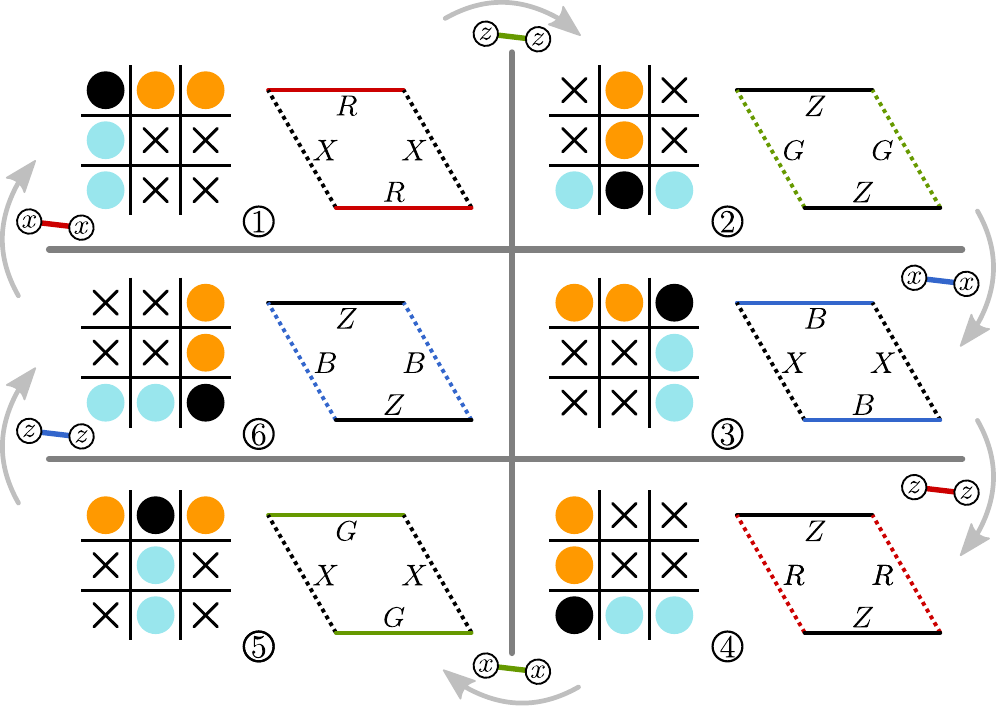}
    \caption{
        The boundaries of the Floquet color code as it changes in time.
        The rough boundaries are always located on the left and right boundaries of the code, the smooth boundaries on the top and bottom boundaries.
        The corresponding boundary of the parent color code, however, change over time as indicated by the capital letters where $R$, $G$ and $B$ are the red, green and blue coloured boundaries, and $X$ and $Z$ denote a parent color-code theory with an $X$-type and $Z$-type Pauli boundary, respectively.
    }
    \label{fig:FloquetCodeXZXZBoundaries}
\end{figure}

We have argued that the lattice geometry at the boundary of the instantaneous stabiliser groups of a dynamically condensed color codes can be determined by the corresponding excitations that are absorbed at the boundary for the parent color-code anyon theory. We can go on and follow our rule to its conclusion to learn how the boundaries must transform as we perform deformations between different instantaneous stabiliser groups of dynamically condensed color codes, see also Ref.~\cite{HaahHastings21b}.
As an explicit example, in Fig.~\ref{fig:FloquetCodeXZXZBoundaries} we show the boundary transformations through a single period of the Floquet color code in terms of the boundaries of its corresponding parent theory together with their corresponding boson tables.
We explain this choice of boundary transformations throughout this section.

Without loss of generality we will follow the details of the transformation between the first and second step of the period of the Floquet color code as an example, where we transform from $\mathcal{S}_\rx$ onto $\mathcal{S}_\gz$; see steps 1 and 2 in Fig.~\ref{fig:FloquetCodeXZXZBoundaries}, but we note that the following discussion will hold between any pair of condensed color-codes, $\mathcal{S}_\a$ and $\mathcal{S}_\b$, with bosons \(\a\) and \(\b\) condensed, provided \(\a\) and \(\b\) share neither a colour label nor a Pauli label.

Let us posit again that rough boundaries must remain rough after a Floquet code transformation and likewise a smooth boundary must remain smooth. More specifically, this means that the rough boundaries must condense the bosons of the parent theory that are identified with electric charges both before and after the transformation and, likewise, the smooth boundaries must condense the magnetic charges throughout the transformation. We must therefore follow how the electric and magnetic charges are maintained throughout the transformation at the level of the parent color-code anyon theory; see Sec.~\ref{SubSec:DCCC}. To restate our result briefly, if we choose a convention for \(\mathcal{S}_\rx\) where we have  identify red parent bosons with electric charges, and Pauli-$X$ parent bosons with magnetic charges, then it follows that, after the transformation from \(\mathcal{S}_\rx\) onto \(\mathcal{S}_\gz\), then Pauli-$Z$ parent bosons correspond to electric charges and green parent bosons correspond to magnetic charges. This convention is laid out explicitly in the boson tables shown in Fig.~\ref{fig:FloquetCodeXZXZBoundaries}.

\begin{figure}[tb]
    \centering
    \includegraphics[width=0.85\linewidth]{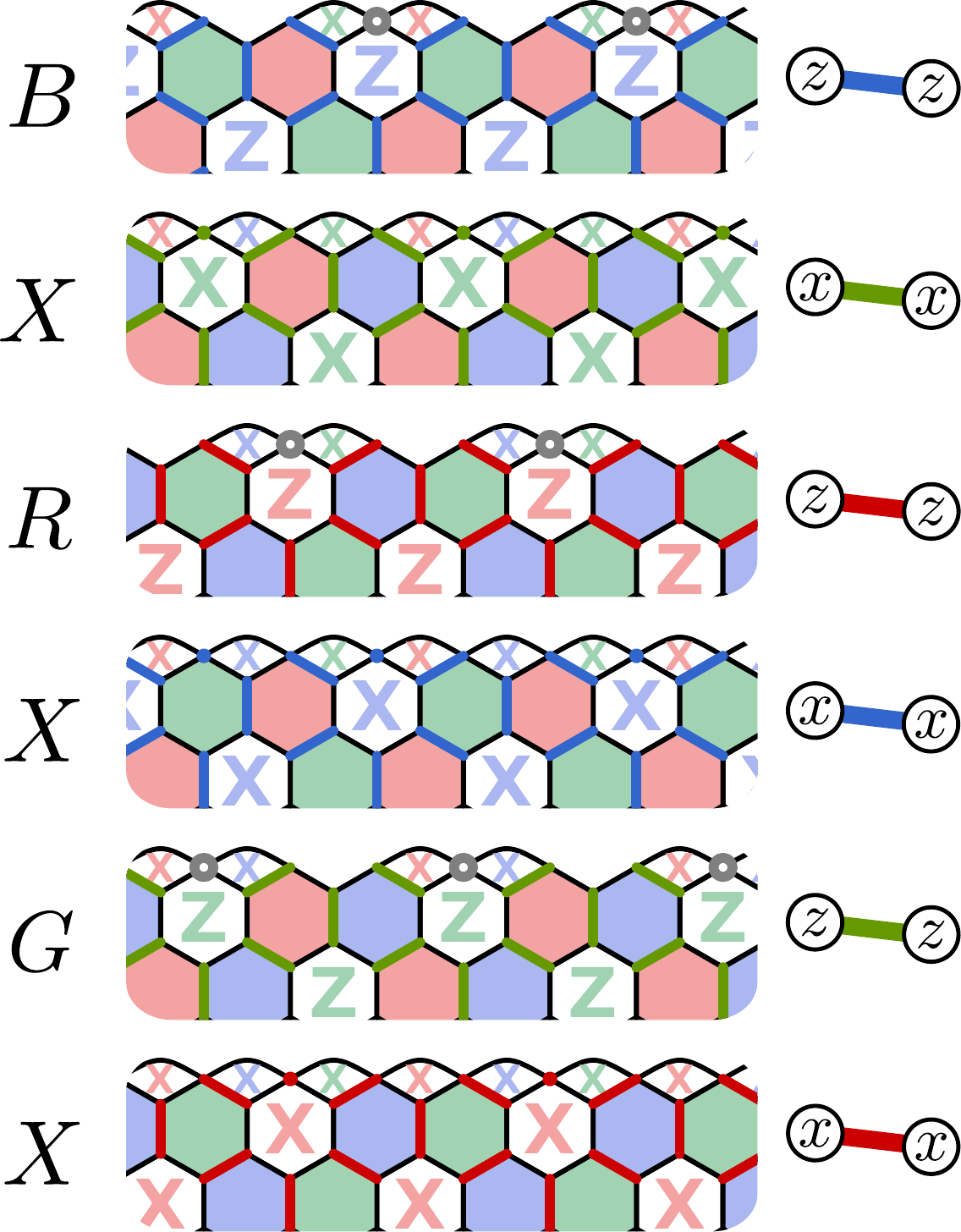}
    \caption{
    Microscopic realisations of the smooth boundaries of the Floquet color code as the code transforms between different instantaneous stabiliser groups over a single period of the code.
    The letters on the left indicate the boundary of the parent color code phase.
    We measure weight-two edge checks at each round, where the measurements are given by the key illustrated to the right of the figure. We also perform single-qubit measurements at certain time steps, shown by coloured spots. Furthermore, certain qubits are not measured at a given step. These are marked by grey circles.
    }
    \label{fig:FloquetBoundaries}
\end{figure}

We find that constraining the electric and magnetic charge labels as we transform \(\mathcal{S}_\rx\) onto \(\mathcal{S}_\gz\) also constrains the boundary transformation. With the convention used above, we have that the red (Pauli-$Z$) parent bosons of  \(\mathcal{S}_\rx\) (\(\mathcal{S}_\gz\)) correspond to electric charges and Pauli-$X$ (green) parent bosons of  \(\mathcal{S}_\rx\) (\(\mathcal{S}_\gz\)) correspond to magnetic charges. This fixes the boundaries of the code. We specifically have that the red (Pauli-$Z$) boundaries correspond to the rough boundaries of \(\mathcal{S}_\rx\) (\(\mathcal{S}_\gz\)) and that the Pauli-$X$ (green) boundaries correspond to the smooth boundaries of \(\mathcal{S}_\rx\) (\(\mathcal{S}_\gz\)). If follows that, at the level of the parent theory, red boundaries must transform onto Pauli-$Z$ boundaries as we transform from \(\mathcal{S}_\rx\) onto \(\mathcal{S}_\gz\), and likewise, Pauli-$X$ boundaries must transform onto green boundaries in the parent theory under the transformation of interest. This is the result shown in Fig.~\ref{fig:FloquetCodeXZXZBoundaries}. In the following section (Sec.~\ref{sec:MircoBoundariesFloquet}), we give microscopic details demonstrating the boundary transformation that we have described here at a macroscopic level. 

To conclude here, let us summarise the prescription we have given to determine boundary transformations for dynamically condensed color codes. As discussed in \ref{SubSec:DCCC}, the electric and magnetic charge labels are constrained throughout a transformation. The transformation can be displayed clearly in boson tables once a charge label convention is specified at the initial step. Furthermore, boundary stabiliser terms can be obtained at the level of the parent theory by reading the respective rows and columns of the boson table that support the deconfined charges of the condensed anyon theory, see Sec.~\ref{SubSubSec:BoundariesCCC}. Finally, we have argued that constraints among the charge labels under the transformation, together with a rule for finding check operators from the parent theory at the boundary of an instantaneous code, can determine how we transform the boundaries of dynamically condensed color codes.

\subsubsection{Microscopic details at the boundary of the Floquet color code}
\label{sec:MircoBoundariesFloquet}

Having given an overview for how the boundaries of the dynamically condensed color codes transform, we can also present the microscopic details for how the boundaries transform throughout a period of the Floquet color code. We show these details in Fig.~\ref{fig:FloquetBoundaries}. In the figure we concentrate on a smooth boundary that oscillates between color labels and Pauli-$X$ labels.
The case of a rough boundary transformation is obtained by replacing Pauli-$X$ and Pauli-$Z$ labels.
Details are given in the figure caption.

\begin{figure}[tb]
    \centering
    \includegraphics[width=0.8\linewidth]{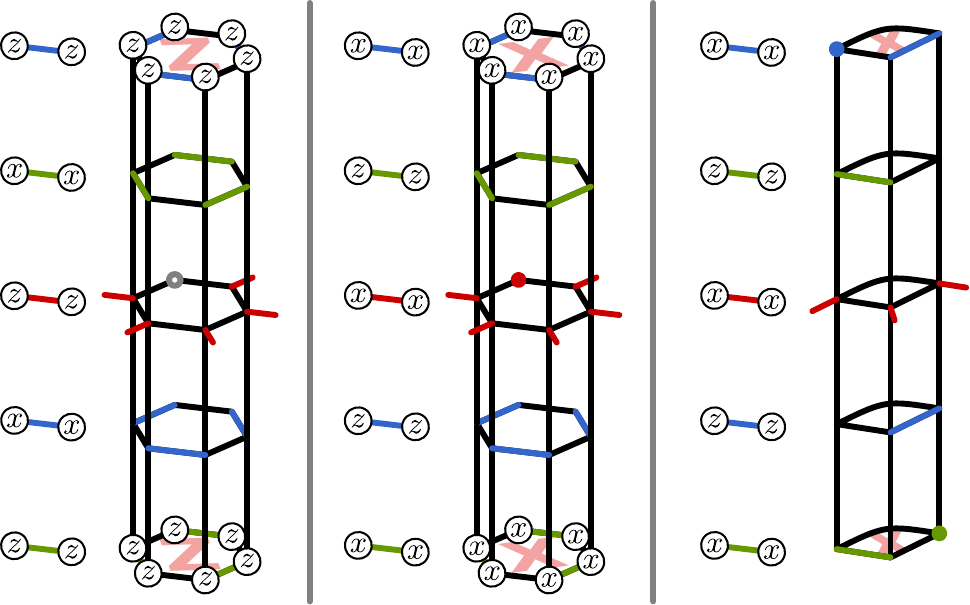}
    \caption{
        Detection cells at the smooth boundary of the Floquet color code. These measurements correspond to a single period with corresponding stabilisers as discussed in Fig.~\ref{fig:FloquetBoundaries}. We show detection cells for a red Pauli-$Z$ stabiliser (left), a red Pauli-$X$ stabiliser (middle) as well as the detection cell for a weight-three Pauli-$X$ stabiliser (right).
    }
    \label{fig:FloquetBoundariesStabs}
\end{figure}

We also show the detection cells at the boundary of the Floquet color code in Fig.~\ref{fig:FloquetBoundariesStabs}. Specifically, we show a boundary detection cell corresponding Pauli-$Z$ stabiliser (left), a weight-six boundary Pauli-$X$ stabiliser (middle), and a weight-three Pauli-$X$ stabiliser (right). In each figure one can check that all of the intermediate measurements in the schedule commute with the initial and final inference of the stabiliser at the first and last time step. One can also check that the detectors at the boundary are consistent with the behaviour of detectors at their respective rough and smooth boundaries, as we expect. We can check this by going through an equivalent analysis to that we have used in SubSubSec.~\ref{SubSubSec:FCCsyndrome}.

\begin{figure}[b]
    \centering
    \includegraphics[width=1.00\linewidth]{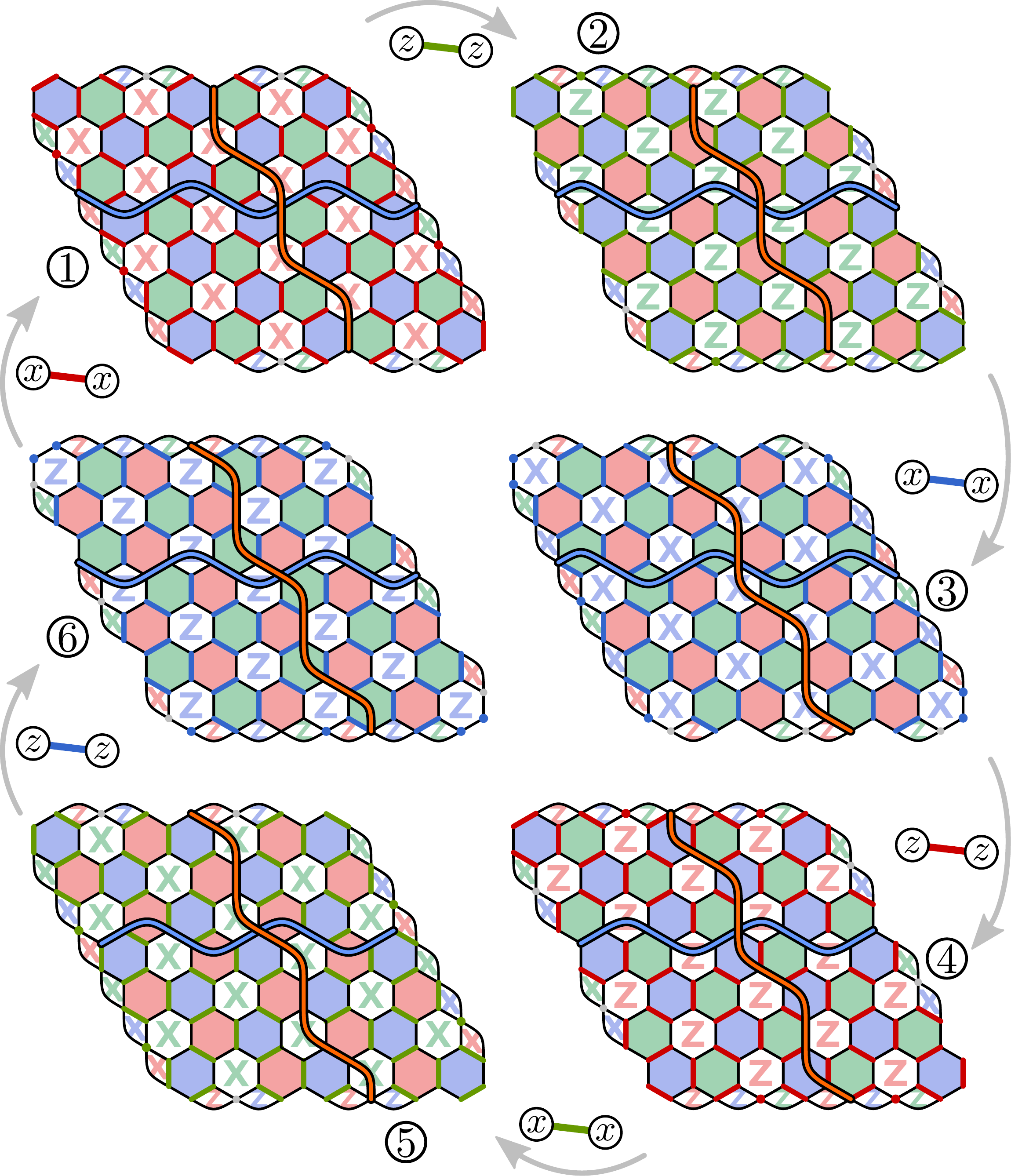}
    \caption{
        The Floquet color code as it evolves throughout the six steps of the measurement scheme.
        We show the logical $X$ ($Z$) operator in cyan (orange).
    }
    \label{fig:FloquetSurfaceCodes}
\end{figure}

We show a period of the Floquet color code with boundaries in Fig.~\ref{fig:FloquetSurfaceCodes}. We also show the support of its logical operators. One may worry that a Floquet code may drift over an array of qubits as transformations are performed. Using the lattice geometry and the measurement pattern we have specified, we find that the footprint of our code remains static.

\subsection{Remarks on Floquet codes}

We have presented a generalised construction for dynamically driven codes called dynamically condensed color codes. Our construction contains the honeycomb code. Moreover, it has enabled us to present a new dynamically driven code that we call the Floquet color code. We have also given a general boundary theory that allows us to write down a valid stabiliser group at the code boundary. Let us now discuss various aspects of dynamically driven codes that may suggest new research directions.

\subsubsection{Topological phases for Floquet codes}
\label{SubSubSec:FloquetPhases}

From the perspective of topological order, the Floquet color code presented in this work and its close cousins like the honeycomb code are in the toric code phase at any given time step. However, from one time step to the next its local representation, given by the instantaneous stabiliser group, changes. In a space-time picture, this corresponds to an invertible domain wall within the toric-code phase appearing periodically in time.

The here presented dynamically condensed color code inherit features from the well studied parent phase.
In our construction, we benefit from exploiting details of the color code, such as its symmetries and boundaries~\cite{Yoshida15, Kesselring18}. For example, we have used the color code boundaries to construct boundaries for dynamically driven codes systematically. 
Naturally, we inherit beneficial properties of the parent stabiliser group such as the geometric locality and bounded weight of the stabiliser checks.

Continuing in this spirit, in future work it will be interesting to show how we can produce other types of topological defects in dynamically driven codes, such as twist defects, that can be used to perform fault-tolerant gates. In addition to finding how they manifest in the microscopic details of an instantaneous stabiliser group, it is also important to find how they transform as they undergo a full period of check measurements for a Floquet code. We may discover an intuitive way of designing these objects by appealing to the details of the parent color-code theory.

On the other hand, we notice that not all sequences of condensation lead to a valid code. In certain sequences, the measured 2-body checks do not let us infer enough syndrome information to perform error correction. It will be interesting to understand the conditions on the validity of a measurement sequence in terms of the macroscopic parent anyon theory and its microscopic stabiliser realisation.

In recent work, see Ref.~\cite{aasen22automorphism}, a general construction of \textit{automorphism codes} based on string-net models was presented. They use an automorphism of a topological order and the associated domain wall in the microscopic description to define a Floquet code that implements the chosen automorphism in the time direction over a cycle of measurements. Given a phase and a representation of an automorphism, their construction gives rise to a unique Floquet code. 

Our construction offers a different perspective on dynamically driven codes.
At any time step, we condense a different boson in a parent theory to describe the instantaneous stabiliser group.
This allows to construct more measurement sequences to drive codes in the condensed phase.
We leave the connection of dynamically condensed to automorphism codes from Ref.~\cite{aasen22automorphism} for future work.
We expect that the automorphism formalism has to be extended to more general (invertible) bimodule categories in order to capture dynamically condensed codes.

The dynamically condensed color codes that we have studied provide us with an intuitive model that shows us that we can design dynamically driven codes by condensing bosons of a parent color-code theory. Taking the perspective of a parent anyon theory with a non-trivial condensate may illuminate a way for us to design a much broader landscape of dynamically condensed codes. A first step would be to find other topological stabiliser codes corresponding to richer topological phases with non-trivial condensates and perform a similar study as we have conducted for the color code. Going beyond stabiliser codes, we could consider taking a non-Abelian parent phase with a non-trivial condensate to construct a \textit{non-Abelian dynamically driven code}. One possibility to enforce non-trivial condensates is by doubling an anyon theory. One could take some number of copies of an anyon model with a gappable boundary and dynamically condense bosons in there.

The concept of condensing topological excitations can also be applied to higher-dimensional models. In Ref.~\cite{Aasen2020topologicalDefect} a network of (partly) condensing defects within a $(3+1)$-dimensional topological theory gave rise to various Fracton models. Using similar defects interleaved (periodically) in time can define dynamical condensation in three dimensions. Moreover, it would be interesting to see how known three-dimensional subsystem codes, like the gauge color code \cite{Bombin_2015}, relate to the concept of topological condensation. Gauge fixing could be related to picking a particular description of a condensate. Changing the gauge would then correspond to an invertible domain wall within the condensate. It may be valuable to find a unifying theory between gauge fixing and dynamically driven code measurement cycles, as such a theory may provide us with new ways of performing fault-tolerant logic gates with other types of topological codes.

\subsubsection{A unifying framework for quantum error-correcting codes}

Dynamically driven codes have provided us with a new way to read out syndrome data from topological phases using a certain sequence of check measurements.
The fact that we can obtain syndrome data using only weight-two measurements is particularly appealing from a practical perspective. 
A generalisation of this idea may lead us to ways of modifying more general quantum codes with constant encoding rate in the number of physical qubits~\cite{Breuckmann2021}.
As such, it is valuable to find a unifying picture to describe Floquet codes together with other types of codes evenhandedly.
We might expect any such framework to include different types of circuits that read out stabiliser operators.
Let us discuss here other syndrome readout circuits, and check measurement sequences for other subsystem codes.
We do this with the aim to identify some similarities between Floquet codes and subsystem codes, to attempt to demystify the physics of these new models.

It is a challenging exercise to point to the differences between dynamically driven codes and more conventional codes.
It is argued that choice of sequence is essential for a dynamically driven code to maintain logical information.
However, in the case of subsystem codes, given that we have a large set of non-commuting measurements that must be checked to obtain a complete set of syndrome data, a suitable sequence must also be chosen, even if the sequence is found trivially.
For instance, for CSS subsystem codes, one typically assumes a sequence where we measure all of the Pauli-$X$ checks simultaneously, followed by the Pauli-$Z$ checks.
Other subsystem codes require a non-trivial sequence of check measurements to read out stabilisers~\cite{Bombin10topological}.
On the other hand, one can easily conceive of poor choices of readout sequences where stabilisers are read out inefficiently.
Furthermore, at each step in the sequence, the subsystem code is projected onto a new instantaneous stabiliser group, where certain terms of the gauge group become fixed and join the stabiliser group of the code, while other terms of the gauge group are kicked out of the fixed subspace.
The authors of Ref.~\cite{Higgott21subsystem} even consider changing the sequence of check measurements to improve a subsystem code to correct biased noise. 

The key innovation that the examples of dynamically driven codes show us, beyond subsystem codes, is that the logical operators need not remain constant throughout a sequence of check measurements, but rather, they can be steadily deformed throughout a sequence of check measurements.
As we have shown here, it is not even necessary to maintain a static stabiliser group as we undergo code transformations.
Hopefully, these innovations will lead us to discover other new codes that are practical for experimental realisation.

Another way we might consider unifying Floquet codes with other codes is at the level of syndrome extraction circuits.
Fault-tolerant syndrome readout circuits have been proposed~\cite{Steane97, Knill2005} where a code is teleported onto a second auxiliary system, such that the teleportation operation also reveals syndrome data.
One could view this teleportation operation as a code deformation, where the stabilisers of a code are teleported onto a new set of qubits with new support.
One could imagine a strategy for syndrome readout where a code is periodically transferred between two subsystems.
We might regard this strategy for syndrome readout as a dynamically driven code.
A similar approach for syndrome readout has also been suggested using measurement-based quantum computation, where a three-dimensional cluster state is prepared over time using a two-dimensional qubit array~\cite{Raussendorf07fault}.
We discuss this point of view from the perspective of fault-tolerant logic gates in the following section, see Sec.~\ref{sec:FTgates}.

\subsubsection{Fault-tolerant logic gates}
\label{sec:FTgates}

The realisation of scalable quantum computation with dynamically driven codes will require the development of fault-tolerant logic gates.
Given that the dynamically condensed color codes realise the toric code phase, we might expect to be able to perform Clifford gates with dynamically driven codes using a generalisation of code deformation.
These operations might include braiding punctures~\cite{Raussendorf06} or braiding twist defects~\cite{Bombin10, Brown17}.
The ongoing development of quantum computation with dynamically driven codes will require a more general theory to implement the microscopic details of the measurements to realise fault-tolerant logic gates.
As we have already discussed in Sec.~\ref{SubSubSec:FloquetPhases}, we may find such constructions by appealing to the physics of the parent theory of a dynamically driven code that is created by condensation.
The work of Hastings and Haah~\cite{HaahHastings21b} has already considered some lattice surgery operations~\cite{HorsemanSurgery} with Floquet codes.
If we additionally show how to initialise Floquet codes in a magic state, we can realise universal quantum computation using magic-state distillation assuming we can perform the Clifford gates.

The development of Floquet codes may also give us new ways of realising fault-tolerant non-Clifford gates.
If one subscribes to measurement-based fault-tolerant quantum computation as an example of a Floquet-code encoding~\cite{Raussendorf07fault}, then we can regard the gate constructions given in Refs.~\cite{Bombin18, Brown2020a} as Floquet codes that realise non-Clifford operation on their dynamically changing code space.
Indeed, in both of these cases, a two-dimensional qubit array is driven between different codes using measurements.
Additionally, a just-in-time decoder is used to reproduce the physics of a three-dimensional code to perform non-Clifford operations in $(2+1)$-dimensional space.
It will be interesting to determine to what extent we need to periodically drive these two-dimensional systems between different stabiliser codes to realise non-Clifford gates.
These examples of two-dimensional non-Clifford gates are well understood by considering the error correction system in a static three-dimensional space-time.
It may be an instructive exercise to develop a space-time theory for dynamically driven codes.

\subsubsection{General noise models}

For more practical implementations of fault-tolerant quantum computing, we require robust codes that demonstrate a high threshold against the noise models that real qubits experience.
It has been shown that the threshold is highly sensitive to local modifications to codes undergoing noise models with biases towards particular types of error~\cite{Tuckett18, bonilla2021, Dua2022, Srivastava2022, SanMiguel2022}.
In addition to changes to the local bases of the physical qubits of the code, our general construction for designing dynamically condensed color codes also offers us other ways of designing dynamically driven codes.
We might, for instance, chose different condensation paths through the color-code boson table.
It may be that certain paths are better suited to obtain high thresholds for different choices of biased noise models.
Topological codes have also been optimised into rectangular configurations with different height and width in order to reduce overhead in the biased noise setting~\cite{bonilla2021, Chamberland22}.
One can observe that there are no logical Pauli-$X$ operators composed of physical Pauli-$Z$ terms at any instance of the Floquet color code (see Fig.~\ref{fig:FloquetSurfaceCodes}).
This code may therefore serve as a candidate for a thin Floquet code under noise biased to introduce dephasing errors to data qubits.

A key difference between dynamically driven codes and more conventional static codes is the action of measurement errors.
For stabiliser codes, high-performance fault-tolerant syndrome readout circuit can be found by studying an idealised situation where measurements are error-free~\cite{Tuckett20, bonilla2021}.
Then, the decoding strategy does not differ significantly when we extend to the case where measurements are noisy, up to the dimensionality of the decoding problem.
In contrast, evaluation of detection cells for dynamically driven codes depends on a large number of measurements. Moreover, a single measurement error may affect a large number of detection cells.
For instance a measurement error in the honeycomb code will trigger four detection events.
In the physically motivated limit that readout errors are very common, it may not be straight forward to determine the performance of a code tailored to correct for biased noise acting on data qubits by only considering the logical operators of the instantaneous stabiliser groups of a dynamically driven code.
Given the non-trivial role of measurements in dynamically driven codes, we might expect that a high rate of measurement error may significantly compromise the performance of a tailored error correction strategy.

Rather, given that readout is very noisy in many real architectures, it may be interesting to find dynamically driven codes that are particularly robust to measurement errors.
In spite of their many similarities with respect to error correction, see Sec.~\ref{SubSubSec:FloquetComparison}, a key difference between the honeycomb code and the Floquet color code is the number of measurements needed to evaluate a detection cell.
A detection cell of the honeycomb code depends on twelve measurement outcomes whereas a detection cell of the Floquet color code only depends on six measurements, see Fig.~\ref{fig:FloquetStabsComparison}.
This difference may lead to a significant contrast in their performance, particularly in the limit where measurement errors are dominant.

We will also need to consider more general types of errors in a real quantum system such as qubit leakage errors~\cite{Aliferis07} and fabrication defects~\cite{Strikis21}.
Solutions to these problems typically require the use of additional auxiliary qubits and modifications to the syndrome readout circuit.
Simple and elegant modifications have been proposed~\cite{suchara2014leakage, Ghosh15} to deal with leakage for the surface code.
It will be important for their practical development to find circuits that perform these same tasks for dynamically driven codes.
For the case of fabrication defects, Ref.~\cite{Strikis21} has shown a syndrome readout protocol that demonstrated a threshold with topological codes using a defective qubit array.
The protocol adds punctures to the topological code to isolate the defective components.
To obtain syndrome data needed to demonstrate a threshold, the protocol also requires code deformations to the boundaries of the punctures to transform the boundaries between different types; rough and smooth.
We expect that we can adapt this protocol for dynamically condensed color codes, employing the boundary theory that we presented in Sec.~\ref{SubSec:FloquetBoundaries} to transform between their rough and smooth boundaries.

\section{Discussion and outlook}
    \label{sec:Discussion}

    Many architectures for practical fault-tolerant quantum computation are based on the manipulation of topological phases of matter.
    In this work we have argued that anyon condensation gives us a framework to describe ways of encoding logical qubits with topological quantum error-correcting codes as well as many of the mechanisms for performing robust logical operations.
    We have demonstrated the application of this framework explicitly with the color code as our key example, expressing the many fault-tolerant encodings and implementations of logic gates in the color code as applications of anyon condensation.
    This framework has also shown us how to generalise many of these color-code encodings and logic gates using the notion of partial condensation, where a subset of bosons of the color code are condensed on a region of the lattice to manipulate topological degrees of freedom.
    Finally, we have reformulated the notion of a dynamically-driven Floquet code in terms of a condensate of a parent color code theory, and this new perspective has enabled us to generalise known Floquet codes as well as providing a constructive way to design the boundary stabilisers of our Floquet code construction.

    With our work, the catalogue of topological features available in the color code~\cite{Kesselring18} has been significantly extended to now include semi-punctures, semi-transparent domain walls and generalised twist defects that appear at their end points.
    Importantly, the language we use to describe these features is independent of their orientation in the $(2+1)$-dimensional space-time volume describing the computation.
    Because of this symmetry, we can make use of a given topological feature in many different ways to perform computational tasks.
    This perspective allows for these topological objects to be used for fault-tolerant state initialisation and readout, to apply logical gates, and to transform between inequivalent schemes of topological encoding.
    As a \emph{practical application} of this new framework, we expect that the large zoo of feasible operations available to us can be used to significantly decrease the physical resource overheads of fault-tolerant logical operations that plague existing constructions. Notions of lattice surgery as discussed in Ref.~\cite{Thomsen22} can be seen as first steps in this direction. The framework developed here is expected to be highly instrumental in developing schemes that are even more resource economical, 
    contributing to the quest of identifying 
    schemes of fault tolerant quantum computing with reasonable overheads.

    More broadly, we have shown that partial anyon condensation can provide us a number of novel operations within the framework of topological fault-tolerant quantum computing.
    These operations complement known logic operations that are completed with constant-depth unitary circuits such as transversal gates, as well as more conventional measurement-based operations that manipulate punctures and braiding operations for non-Abelian point-like objects such as twists.
    We have found that the color-code model to be particularly illustrative of this idea, due to its rich and numerous symmetries exhibited at the level of its low-energy excitations.
    Moving forward, it will be exciting to understand how similar mechanisms are manifest in more general topological phases.
    We may study, for instance, other $(2+1)$-dimensional topological quantum field theories~\cite{Cong2017universal}, including non-Abelian theories~\cite{laubscher2019universal}, as well as higher-dimensional topological phases and fracton phases.
    In order to make progress along these lines, one could seek other examples of condensation processes in other phases, and to develop a theory of anyon condensation for general classes of phases.

    Lastly, we have also shown that dynamically-prepared Floquet codes can be rederived, and generalised, within the framework of anyon condensation of color-code excitations.
    We were able to design key features of this novel type of code, such as the code boundaries, by appealing to the physics of the parent color-code model.
    Given the practicality of their realisation, the development of Floquet codes may provide better ways of realising different types of non-trivial topological phases.
    In future work, it will be interesting to discover to what extent our theory of Floquet codes from anyon condensation can be generalised to other types of topological phases.
    Further afield, it may also be that developments of the theory of dynamical codes may show us more practical ways of implementing quantum low-density parity-check codes~\cite{Breuckmann2021QLDPC}.

\begin{acknowledgements}
    The authors would like to thank D.~Barter and J.~Bridgeman for several discussions on semi-transparent domain walls that initiated this project, and we thank D.~Williamson for asking questions that led us to our construction for dynamically driven codes.
	The authors are also grateful for helpful discussions and comments from
	D.~Aasen,
	A.~Bauer, 
	A.~Cross, 
	M.~Davydova,
    P.J.~Derks, 
	T.~Ellison, 
	C.~Gidney, 
    O.~Higgott,
	M.~Newman, 
    A.~Townsend-Teague,
	Z.~Wang,
    and 
	J.~Wootton. BJB is also grateful for the hospitality of the Center for Quantum Devices at the University of Copenhagen where parts of this work were completed.
	This work has been supported by the 
	BMBF (RealistiQ, QSolid, MUNIQC-Atoms), 
	the DFG (CRC 183),
	and the Einstein Foundation 
    (Einstein Research Unit on quantum devices).
    This research is also part of the Munich Quantum Valley (K-8), which is supported by the Bavarian state government with funds from the Hightech Agenda Bayern Plus.
    This work is also supported by the Australian Research Council via the Centre of Excellence in Engineered Quantum Systems (EQUS) project number CE170100009, and by the ARO under Grant Number W911NF-21-1-0007.
    F.~T.~is supported by the Sydney Quantum Academy.
    B.~J.~B.~has changed affiliation to IBM Quantum during the preparation of this manuscript.
\end{acknowledgements}
	
\appendix

\section{Domain walls and boundaries in stacked toric code phases}\label{app:DWsStackedTC}

In this appendix, we elaborate how results on the classification and construction of boundaries and domain walls in (Abelian) quantum double models apply to a phase that is equivalent to $n$ layers of the toric code. We will see that in this case, many properties of a given (abstractly defined) boundary are easily computable. In particular, the color code is equivalent to 2 layers of the toric code (see Sec.~\ref{sec:PrelimUnfolding}), so $n=2$.

First, let us show why it is important to study boundaries, even if one is interested in domain walls.

\subsection{Folding trick}
 \label{sec:TheoryFoldingTrick}
 
    Until now we have seen how to understand domain walls in terms of condensation and symmetry.
    However, in practical scenarios, one might want to have a simpler picture to understand/construct domain walls.
    To this end, we can employ the \textit{folding trick}.
    It relates a domain wall to a boundary of a ``folded" phase, composed of the phases on both sides of the domain wall of interest.
    
\begin{figure}[tb]
        \centering
        \includegraphics[width=.75\linewidth]{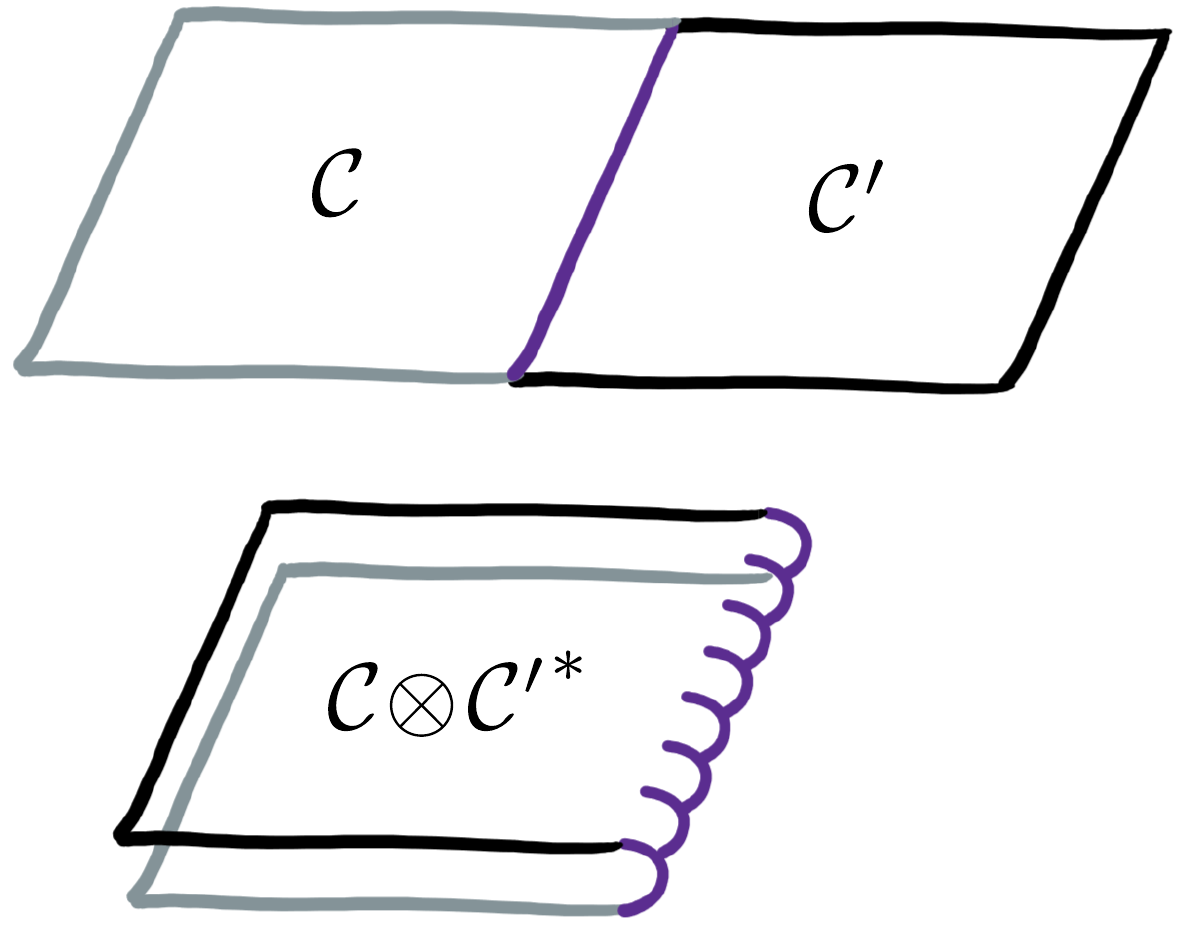}
        \caption{
        A $\cC-\cC'$ domain wall (top) is equivalent to a boundary of stacked phase $\cC\otimes\cC'^\ast$ (bottom).
        This equivalence is often referred to as the \textit{folding trick}.
        Note that $\cC'$ is inverted by the fold.
        In this work, however, since we consider qubit stabiliser models, $\cC'^\ast=\cC'$.
        }
        \label{fig:FoldingTrick}
    \end{figure}

    The folding trick allows us to think about any $\cC$-$\cC'$ domain wall as a boundary of the stacked theory $\cC\otimes\cC'$~\footnote{Technically, one of the phases needs to be inverted. But for stabiliser codes, as considered here, the inversion does not change the phase.}.
    The folding trick can be easily understood pictorially by folding a plane with a domain wall along it and identifying the cut with a boundary to a vacuum, see Fig.~\ref{fig:FoldingTrick}.
    Mathematically speaking, the folding trick is used to classify domain walls in terms of boundaries \cite{KitaevKong12, Beigi11, bridgeman2020}, which is a simpler task since a topological boundary (of an Abelian phase) is fully described by a Lagrangian subgroup.
    
    In order to think about domain walls as boundaries of a stacked phase, we establish what the structure of the Lagrangian subgroup (folded) tells us about the domain wall (unfolded).
    The fold gives a natural decomposition of the anyon fusion group $\cC\otimes\cC'$.
    A generator of $\cL$ of the form $\a\otimes\a'$, without $\a$ and $\a'$ appearing individually means that in the unfolded picture $\a\mapsto\a'$ when passing through the domain wall.
    For an invertible domain wall, every generator is of that form, for an opaque domain wall all generators can be brought to the form $\a\otimes\one$ or $\one\otimes\a'$.
    The semi-transparent domain walls are the ones with generators of both of the above types.
    
    The equivalence of domain walls and certain boundaries shows that, mainly for computational purposes, it suffices to consider how to construct condensable bosons for boundaries.
    In Sec.~\ref{sec:CondDWs}, we will show how the Lagrangian subgroup of any boundary of layers of toric codes can be directly calculated based on its classifying data.

\subsection{General structure of boundaries}
	In Sec.~\ref{sec:CondDWs} we showed that domain walls -- gapped interfaces of topological phases -- can be understood in terms of the condensate formed by the mobile anyons. This classifies all possible domain walls up to automorphisms on both sides of the interface. In this section, we will explicitly show how this categorises domain walls from $n$ to $m$ layers of the toric code. In particular, we will see how the condition that the mobile anyons have to form a condensate constrains the domain wall.

	As the general case, we consider a domain wall from $n$ layers of the toric code, $TC_n$, to $m$ layers $TC_m$. We show that any such domain wall is equivalent to an opaque domain wall on a subset of layers and a fully transparent domain wall on the remaining layers up to symmetries on both sides of the interface, $\Aut(TC_n)$ and $\Aut(TC_m)$.
	In the language of anyon condensation an equivalent statement is that any condensate of $n$ layers of toric codes is in the phase of $k$ layers of toric codes, where $k\leq n$. We will prove this in the remainder of this section using the tools laid out in the main text of this article.
	
	Imagine a stabiliser model for $n$ layers of toric codes, the quantum double model \cite{Kitaev03} for $G=\bZ_2^n$. By construction, it is a qubit Pauli stabiliser code. Any condensate can be modelled by adding the corresponding hopping terms to the stabiliser group generators (see Sec.~\ref{sec:CondStabs}). In toric codes, any such string operator is a (multi-)qubit Pauli word. Hence, the stabiliser group in the condensed phase will be a qubit Pauli group. From Ref.~\cite{Bombin12}, we know that any such code is in the same phase as $k$ layers of toric codes. Counting the cosets making up the condensate given a set of condensable bosons and the conditions on them shows that $k\leq n$. This completes one proof of the above statement.
	
	Albeit the simplicity of the stabiliser based proof, the microscopic description of the condensate is not necessary.
	We have given a simple proof that demonstrating how condensates are constrained by appealing to the microscopic details of the $G=\bZ_2^n$ toric code at the level of the stabiliser group. We find that it is also instructive to re-derive the result at the level of the low-energy particle theory of the quantum double model. Specifically, one can also work out the condensed phase explicitly by looking at the conditions defining the condensed phase when an arbitrary boson in the $n$ layer toric code phase is condensed. Checking that the fusion group and the modular data of the resulting anyons coincide with $n-1$ layers of toric codes is straight forward using $\bZ_2$ arithmetic. The proof for any number of condensed bosons then follows by induction.
	
\subsection{Calculating Lagrangian subgroups}
It is a well established fact in the mathematical condensed-matter literature that the boundaries of a quantum double model of a finite group $G$ are classified by a subgroup $N\subset G$ and a so-called \emph{2-cocycle class} $[\psi]\in H^2(N,U(1))$ \cite{davydov2017149, Beigi11}. Moreover, 
Ref.~\cite{Beigi11} has established a formula to calculate the Lagrangian subgroup of a boundary $(N,\psi)$ of a generic quantum double model, Abelian or Non-Abelian. In this section, we explicitly show the formula in the case of $n$ layers of the toric code and give some intuition for the quantities appearing in the expressions. We end the section with some simple examples to see how familiar boundaries of the toric and color code come out of this description.

The topological phase of $n$ layers of the toric code is modelled by a quantum double model with gauge group $G=\bZ_2^n$. For the remainder of this section we represent its elements by binary vectors of length $n$ and the group multiplication by entry-wise addition modulo 2,
\begin{align}
    \vb{a}\oplus\vb{b} := \mqty(a_1+b_1 \mod 2\\ a_2+ b_2 \mod 2\\ \vdots\\ a_n+b_n\mod 2).
\end{align}
Any subgroup is of the form $\bZ_2^k$ for $k\leq n$ and generated by $k$ independent generators. Any such subgroup, together with a 2-cocycle on it, defines a boundary of the quantum double model.
A 2-cocycle on the subgroup is a function $\psi: N\times N\to U(1)$ with the property
\begin{align}
    \psi(\vb{b},\vb{c})\psi(\vb{a},\vb{b}\oplus \vb{c}) = \psi(\vb{a}\oplus \vb{b},\vb{c})\psi(\vb{a},\vb{b})
\end{align}
for all $\vb{a},\vb{b},\vb{c}\in N$. 2-cocycles are sorted into \textit{equivalence classes} where two 2-cocycles $\psi$ and $\Tilde{\psi}$ are equivalent if there exists a function $\beta: N\to U(1)$ such that
\begin{align}
    \widetilde{\psi}(\vb{a},\vb{b}) = \psi(\vb{a},\vb{b})\frac{\beta(\vb{a})\beta(\vb{b})}{\beta(\vb{a}\oplus \vb{b})}.
\end{align}
The set of these equivalence classes is called \textit{second cohomology group} and denoted by $H^2(N,U(1))$. For more details on the origin of the this equivalence relation and group cohomology, we refer to the appendix of Ref.~\cite{chen2013}. In fact, the type of boundary defined by a pair of subgroup and 2-cocycle only depends on the equivalence class that contains the 2-cocycle. In every 2-cocycle class there is a special representative that satisfies
\begin{align}
    \psi(\vb{0},\vb{a})=\psi(\vb{a},\vb{0})=1\qcomma\forall \vb{a},
\end{align}
where $\vb{0}$ denotes the identity element in $N$. We call such a 2-cocycle \textit{normalised}. Let us elaborate on the form of inequivalent normalised 2-cocycles for $N=\bZ_2^k$. In fact, for $k=1$, there is only a single equivalence class, represented by the trivial 2-cocycle
\begin{align}
    \psi_0(\vb{a},\vb{b}) = 1\quad\forall \vb{a},\vb{b}.
\end{align}
For $k=2$, however, there exist one nontrivial 2-coycleclass 
which is represented by
\begin{align}\label{eq:2cocycle_Z2Z2}
    \psi_1(\vb{a},\vb{b}) = (-1)^{a_1b_2}.
\end{align}
Since it is of order two, i.e., $(\psi_1)^2 = \psi_0\equiv 1$, we write $H^2(\bZ_2\times \bZ_2,U(1)) = \bZ_2$. Luckily, we can construct a representative of any non-trivial 2-cocycle class on $\bZ_2^k$ for any $k$ from the non-trivial 2-cocycle $\psi_1$.
Using the Künneth formula for group cohomology \cite{chen2013} and the fact that $\bZ_2$ only has trivial 2-cocycles, one can show that for $k>1$ the 2-cocycles  over $\bZ_2^k$ decompose into 2-cocycles defined on pairs of tensor factors. Abstracly,
\begin{align}
    H^2(\bZ_2^k,U(1)) =& \bigotimes_{(i,j)\text{ pairs}} H^2((\bZ_2)_i\times(\bZ_2)_j,U(1))
    \nonumber\\
    =& \bigotimes_{(i,j)\text{ pairs}} \bZ_2
    =\bZ_2^{\binom{k}{2}},
\end{align}
where $(\bZ_2)_i$ refers to the $i$th tensor factor of $\bZ_2^k$ and $\binom{k}{2}=k(k-1)/2$ is the number of pairs in the $k$ factors. For the 2-cocycles themselves, this means that we can write the (normalised) representative of any 2-cocycle class as a product of (normalised) 2-cocycles each of which only acts non-trivially on a pair of factors $(\bZ_2)_i\times (\bZ_2)_j$ in $\bZ_2^k$,
\begin{align}
\begin{split}
    \psi_{\{n_{i,j}\}}(\vb{a},\vb{b}) =& \prod_{(i,j)\text{ pairs}} \psi_1(\vb{a}_{i,j},\vb{b}_{i,j})^{n_{i,j}}\\
    =& (-1)^{\sum_{i<j}n_{i,j}a_ib_j},
\end{split}
\end{align}
where $\vb{a}_{i,j}$ denotes the restriction of $\vb{a}$ on $(\bZ_2)_i\times(\bZ_2)_j\subset\bZ_2^k$ and the set of $n_{i,j}=0,1$ labels the 2-cocycle classes and indicates if the 2-cocycle is non-trivial on pair $(i,j)$.
Taken together, inequivalent boundaries of $n$ layers of the toric code are 
classified by a subgroup $\bZ_2^k\subset\bZ_2^n$ and a set of $\bZ_2$ numbers $\{n_{i,j}\,|\, (i,j)\text{ pairs of factors in }\bZ_2^k\}$.

Now that we have introduced the quantities that define the bulk model (finite group, $G=\bZ_2^n$) and a boundary (subgroup $N\simeq\bZ_2^k$ and a 2-cocycle class in $H^2(\bZ_2^k,U(1))$), we can turn our attention on how to describe the bulk anyons and their condensation at the boundary in terms of this defining data. The bulk anyons are labelled by a pair of group elements, in our case $\bZ_2^n\times\bZ_2^n$, where the first factor represents the electric part and the other one the magnetic part. For example, $e_1=(0,\dots|1,\dots)$, $m_1=(1,\dots|0\dots)$ are, respectively, the electric and magnetic anyon on the first layer. Their composite particle, the fermion on the first layer, is represented by their sum $f_1 = e_1\oplus m_1 = (1,0,\dots|1,0,\dots)$. The modular data (monodromy and topological spin) can all be computed by the so-called \textit{character function} $\chi_{\bullet}(\bullet):\bZ_2^n \times \bZ_2^n \to \{\pm1\}$,
\begin{align}
    \chi_{\vb{a}}(\vb{b}) = (-1)^{\sum_{i=1}^n a_ib_i}.
\end{align}
Note that this function couples the first and the second argument on every ``layer" individually. This is related to the fact that in layers of toric code an anyon has non-trivial topological spin (of $-1$) if it has both an electric and a magnetic component on the same layer. For the exact formulas we refer to Ref.~\cite{coste2000}.

The way we calculate the Lagrangian subgroup corresponding to a boundary labelled by a subgroup $N$ and a 2-cocycle $\psi_{n_{i,j}}$ is to construct an indicator function $m^{N,\{n_{i,j}\}}: \bZ_2^n\times \bZ_2^n \to \{0,1\}$, which evaluates to 1 if anyon $(\vb{g}|\vb{h})$ can condense, i.e., is part of the Lagrangian subgroup, and 0 if not. Applying the construction outlined in Ref.~\cite{Beigi11} to our case (where $G=\bZ_2^n$, $N\simeq\bZ_2^k$ and the 2-cocycles described above), this function reads
\begin{subequations}\label{eq:nlayers_indicator_function}
\begin{align}
\begin{split}
    &m^{N,\{n_{i,j}\}}(\vb{g}|\vb{h}) = \\
    &\delta_{\vb{g}\in N}\frac{1}{\abs{N}}\sum_{\vb{l}\in N} \chi_{\vb{h}}(\vb{l})\psi_{\{n_{i,j}\}}(\vb{l},\vb{g})\psi_{\{n_{i,j}\}}(\vb{g},\vb{l})
\end{split}\\
\begin{split}
    = \delta_{\vb{g}\in N}\frac{1}{2^k}\sum_{\vb{l}\in N}(-1)^{\sum_{i=1}^n l_ih_i}\\
    \times (-1)^{\sum_{m=0}^k\sum_{j>m=0}^k \vb{l}|_{m} n_{mj}\vb{g}|_m + \vb{l}|_{j}n_{mj}\vb{g}|_m)},
\end{split}
\end{align}
\end{subequations}
where $|_i$ denotes the projection onto the $i$th tensor factor of $N$. We see that the $g$-part (magnetic component) is purely determined by the subgroup $N$. For the origin of this function, we refer to Refs.~\cite{Beigi11, Magdalena2022bulktoboundary}.
By which $h$-part (electric component) it is accompanied is determined by the 2-cocycle characterising the boundary. To understand the formula better, let us look at some examples.

\textbf{Example 1: surface code boundaries, $n=1$:}\\
The simplest example are the toric code boundaries, where $G_1=\bZ_2$. The 4 bulk anyons are labelled by $\{(0|0), (0|1), (1|0), (1|1)\}$, corresponding to the trivial, electric, magnetic and fermionic charge. The group $\bZ_2$ has two trivial subgroups, $N_1=\{0\}$ and $N_2=G$. Both have only trivial 2-cocycles, so the indicator function reduces to
\begin{align}
    m^{N_i}(g|h) =& \delta_{g\in N_i}\frac{1}{\abs{N_i}}\sum_{l\in N_i} (-1)^{hl}\\
    =& \delta_{g\in N_i}\prod_{l\in N_i}\delta_{hl=0},\nonumber
\end{align}
where we have identified that the second part of the right hand side is only non-zero if $hl=0$ for all $l\in N_i$. This yields the Lagrangian subgroups
\begin{subequations}
\begin{align}
    \cL_{N_1} =& \{(0|0),(0|1)\}\qq{and}\\
    \cL_{N_2} =& \{(0|0),(1|0)\},
\end{align}
\end{subequations}
which are exactly the pure electric anyons for $N_1$ and the pure magnetic anyons for $N_2$ as expected for a single layer of toric code.

\textbf{Example 2: color code boundaries, $n=2$:}\\
The color code can be unfolded to two layers of toric code (see Sec.~\ref{sec:PrelimUnfolding}),
i.e., a 
\emph{quantum double model} with $G_2=\bZ_2\times\bZ_2$. Besides the two trivial subgroups, $N_1=\{(0,0)\}$ and $N_2=G_2$, we find that $G_2$ has three other subgroups, $N_3 = \langle (0,1)\rangle$, $N_4=\langle (1,0)\rangle$ and $N_5=\langle(1,1)\rangle$, each of which is isomorphic to $\bZ_2$. As discussed above, $N_2$ has two 2-cocycle classes, whereas the other subgroups only host a trivial one. In total, this gives 6 boundaries corresponding to the six Lagrangian subgroups in Eqs.~\eqref{eq:CC_Lagrangian}. In the following, we will see how they can be recovered with formula in Eq.~\eqref{eq:nlayers_indicator_function}.

Let us first consider the trivial 2-cocycle on any of the subgroups. This gives a similar formula to the one in the previous example,
\begin{align}
    m^{N_i}(g_1g_2|h_1h_2) =& \delta_{(g_1g_2)\in N_i}\frac{1}{\abs{N_i}}\sum_{l\in N_i} (-1)^{h_1l_1+h_2l_2} \nonumber\\
    =& \delta_{(g_1g_2)\in N_i}\prod_{l\in N_i}\delta_{\vb{l}\cdot\vb{h}=0\text{ mod }2},
\end{align}
which yields the Lagrangian subgroups
\begin{subequations}
\begin{align}
    \cL_{N_1} =& \{(00|ab)\,|\,a,b=0,1\},\\
    \cL_{N_2} =& \{(ab|00)\,|\,a,b=0,1\},\\
    \cL_{N_3} =& \{(0a|b0)\,|\,a,b=0,1\},\\
    \cL_{N_4} =& \{(a0|0b)\,|\,a,b=0,1\},\\
    \cL_{N_5} =& \{(aa|bb)\,|\,a,b=0,1\}.
\end{align}
\end{subequations}
Again, we recover the pure $e$,
respectively $m$ condensing boundaries with the two trivial subgroups $N_1$ and $N_2$ (with trivial 2-cocycle).
There is an additional Lagrangian subgroup for the non-trivial 2-cocycle $\psi_1$ on $N_2=G$, see Eq.~\eqref{eq:2cocycle_Z2Z2}. Plugging this into the indicator function gives
\begin{subequations}
\begin{align}
    m^{N_2,\psi_1}&(g_1g_2|h_1h_2) =\\
    &\delta_{(g_1g_2)\in G} \frac{1}{4} \sum_{a_1,a_2=0,1}(-1)^{a_1(h_1+g_2)+a_2(h_2+g_1)}\nonumber \\
    =& \delta_{h_1+g_2 = 0\text{ mod }2}\delta_{h_2+g_1 = 0\text{ mod }2},
\end{align}
\end{subequations}
Evaluating the function for every anyon label, we get the sixth Lagrangian subgroup
\begin{align}
    \cL_{N_2,\psi_1} = \{(ab|ba)\,|\,a,b=0,1\}.
\end{align}
Depending on the unfolding one chooses to map the color code to two toric code layers the above Lagrangian subgroups correspond to different boundaries . For example, choosing the standard mapping \eqref{eq:AnyonUnfolding}, $\cL_{N_1}$ corresponds to the $x$ boundary, $\cL_{N_4}$ to the red boundary, $\cL_{N_2,\psi_1}$ to the $z$ boundary, etc.\,.

\textbf{Example 3: color code domain walls, $n=4$:}\\
Via the folding trick, domain walls in the color code are in one-to-one correspondence to boundaries of two layers of color code. Each color code anyon is labelled by a $\bZ_2^{\times 4}$ element so in the folded picture (see Sec.~\ref{sec:TheoryFoldingTrick}), they are labelled by $\bZ_2^{\times 4}\times\bZ_2^{\times 4}$ elements, each factor corresponding to the anyons on either side of the domain wall (before folding). We label the magnetic fluxes on the left(right) side of the domain wall by $m_i^{l(r)}$ and the electric charges by $e_i^{l(r)}$, respectively. For our purposes in $m^{N,n_{ij}}(\bullet)$, we represent them by binary strings
\begin{subequations}
\begin{align}
    (\vb{g}|\vb{h}) =& (g_1g_2g_3g_4|h_1h_2h_3h_4)\\
    \simeq &
    (m^l_1)^{g_1}(m^l_2)^{g_2}(m^r_1)_{g_3} (m^r_2)^{g_4}\nonumber \\
    &\times(e^l_1)^{h_1} (e^l_2)^{h_2} (e^r_1)^{h_3} (e^r_2)^{h_4}.
\end{align}
\end{subequations}
Note that we associated the left side with the first two bits in $\vb{g}$, respectively $\vb{h}$, and the last two with the right side of the domain wall.

There are 270 different domain walls in the color code corresponding to the different subgroups of $\bZ_2^4$ and potential non-trivial 2-cocycles on them. We will consider an exemplary subset thereof in this section, the three subgroups $N_1=\langle (1000)\rangle\simeq\bZ_2$, $N_2 = \langle (1010)\rangle\simeq\bZ_2$ and $N_3 = \langle (1000), (0010)\rangle\simeq\bZ_2\times\bZ_2$, where the latter can have a non-trivial 2-cocycle.

Plugging in $N_1$ with a trivial 2-cocycle into Eq.~\eqref{eq:nlayers_indicator_function} gives the indicator function for the corresponding Lagrangian subgroup
\begin{align}
    m^{N_1}(\vb{g}|\vb{h}) =& \delta_{\vb{g}\in\{(0000),(1000)\}} \frac{1}{2}\sum_{l=0}^1(-1)^{h_1l_1}\\
    =& \delta_{\vb{g}\in\{(0000),(1000)\}}\delta_{h_1,0}.
    \nonumber 
\end{align}
The associated Lagrangian subgroup is given by
\begin{align}
\begin{split}
    \cL_{N_1} &= \{(a_1000|0a_2a_3a_4)\,|\,a_i=0,1 \}\\
    &\simeq \langle m_1^l,e_2^l,e_1^r,e_2^r\rangle.
    \nonumber 
\end{split}
\end{align}
We see that the Lagrangian subgroup is generated by anyons supported only on the left or the right side of the domain wall so it corresponds to a fully opaque wall when unfolded (see Sec.~\ref{sec:TheoryFoldingTrick}). This can be traced back to the fact that $N_1$ is only supported on a single tensor fact of the input group $\bZ_2^{\times 4}$. This generalises to all boundaries corresponding to any subgroup that factorises over the input tensor factors and a trivial 2-cocycle.

$N_2$ also has a single generator, $(1010)$. This generator, however, is supported on two tensor factors each of which was associated with a different side of the domain wall. For this subgroup the indicator function reads
\begin{align}
\begin{split}
    m^{N_2}(\vb{g}|\vb{h}) =& \delta_{\vb{g}\in\{(0000),(1010)\}}\frac{1}{2}\sum_{l=0}^1 (-1)^{l(h_1+h_3)}\\
    =& \delta_{\vb{g}\in\{(0000),(1010)\}}\delta_{h_1+h_3 = 0 \text{ mod } 2}
\end{split}
\end{align}
giving rise to the Lagrangian subgroup
\begin{align}
\begin{split}
    \cL_{N_2} &= \{(a_10a_10|a_2a_3a_2a_4\,|\,a_i=0,1)\}\\
    &\simeq \langle m_1^lm_1^r,e_1^le_1^r,e_2^l,e_2^r \rangle.
\end{split}
\end{align}
When unfolded to a domain wall, this boundary corresponds to a semi-transparent domain wall where the anyons $m_1$ and $e_1$ can freely pass through and the $e_2$ particles condense individually from both sides.

For $N_3$, there are two associated boundaries, one for each 2-cocycle class. For the trivial 2-cocycle the calculation goes through similarly to before. The indicator function reads
\begin{align}
\begin{split}
    m^{N_3}(\vb{g}|\vb{h}) =& \delta_{\vb{g}\in\langle (1000),(0010)\rangle}\frac{1}{4}\sum_{l_1,l_2=0}^1(-1)^{l_1h_1+l_2h_3}\\
    =& \delta_{\vb{g}\in\langle (1000),(0010)\rangle} \delta_{h_1=0}\delta_{h_3=0}
\end{split}
\end{align}
and the associated Lagrangian subgroup is
\begin{align}
\begin{split}
    \cL_{N_3} =& \{(a_10a_20|0a_30a_4)\,|\, a_i=0,1\}\\
    \simeq& \langle m_1^l,m_1^r,e_2^l,e_2^r\rangle.
\end{split}
\end{align}
Again, since the Lagrangian subgroup is generated by anyons supported on a single side of the domain wall, it corresponds to an opaque one. On each side, $m_1$ and $e_2$ can condense individually. The fact that it is fully opaque can again be traced back to the fact that the chosen subgroup factorises over the fold, i.e., is generated by generators solely supported on the first two or last two factors of $\bZ_2^{\times 4}$. This argument does not hold anymore for the boundary associated with the same subgroup but a non-trivial 2-cocycle of the form
\begin{align}
    \psi^{N_3}(\vb{a},\vb{b}) = (-1)^{a_1b_3}.
\end{align}
Plugging this into the indicator function gives
\begin{align}
\begin{split}
    m^{N_3,\psi}(\vb{g}|\vb{h}) =& \delta_{\vb{g}\in N_3}\frac{1}{4}\sum_{l_1,l_2=0}^1(-1)^{l_1(h_1+g_3)+l_2(h_3+g_1)}\\
    =& \delta_{\vb{g}\in N_3}\delta_{h_1+g_3 = 0\text{ mod }2}\delta_{h_3+g_1 = 0\text{ mod }2}.
\end{split}
\end{align}
Note that the non-trivial 2-coycle couples the contraints on $\vb{h}$ and $\vb{g}$.
The associated Lagrangian subgroup reads
\begin{align}
\begin{split}
    \cL_{N_3,\psi} =& \{(a_10a_20|a_2a_3a_1a_4)\,|\,a_i=0,1\}\\
    \simeq& \langle m_1^le_1^r,m_1^re_1^l,e_2^l,e_2^r \rangle.
\end{split}
\end{align}
Even though the defining group factorises over the individual tensor factors of the input group the domain wall associated to the Lagrangian subgroup above is semi-transparent. The 2-cocycle couples the constraints on $\vb{g}$ and $\vb{h}$ in the indicator function in a way that $m_1^l$ only condenses when accompanied by $e_1^r$ and vice versa. As a domain wall this means that $m_1$ can pass through from either side but gets transformed into $e_1$, 
i.e., the non-trivial automorphism of the upper layer toric code gets applied. On the lower layer $e_2^l$ and $e_2^r$ condense individually, making the domain wall semi-transparent. When comparing $\cL_{N_3}$ with $\cL_{N_3,\psi}$ we see that the non-trivial 2-cocycle effectively ``twists" the magnetic part in $\cL_{N_3}$ by appending electric charges on a different layer. This generalises to any non-trivial 2-cocycle on any subgroup isomorphic to $\bZ_2\times\bZ_2$. It appends the $m$-anyons on either of the two factors (of the subgroup) with electric charges on the other one.

\textbf{Example 4: Trivial 2-coycle:}\\
As seen in the first three examples, Eq.~\eqref{eq:nlayers_indicator_function} 
takes a particularly simple form for a boundary associated to a 
trivial 2-cocycle on the chosen subgroup $N$,
\begin{subequations}\label{eq:nlayers_indicator_trivialcocycle}
\begin{align}
m^N(\vb{g}|\vb{h}) =& \delta_{\vb{g}\in N}\frac{1}{2^k} \sum_{\vb{l}\in N}(-1)^{\sum_i l_ih_i}\\
=& \delta_{\vb{g}\in N}\prod_{\vb{l}\in N}\delta_{\vb{h}\cdot \vb{l}=0\text{ mod } 2}.
\end{align}
\end{subequations}
We see that the constraints on $\vb{g}$ and $\vb{h}$ decouple. The Lagrangian subgroup consists of anyons with $g\in N$ accompanied with a charge that has even overlap with any element in $N$, i.e for which $\sum_{i}l_ih_i=0 \mod 2 \,\forall l\in N$. This agrees with our understanding that in order for two 
anyons braiding trivially in layers of toric code the $e$ and the $m$ part has to overlap on an even number of layers. As a consistency check, we can count all $\vb{g},\vb{h}$ for which $m^N(\vb{g}|\vb{k})=1$. In any Abelian anyon model, the order of a Lagrangian subgroup is the square root the total number of anyons in the bulk. In our case, where the bulk has $2^{2n}$ anyons, we hence expect $2^n$ different $(\vb{g}|\vb{k})$ that condense. The $\vb{g}$ label has to be in $N$, giving rise to $2^k$ different values. Additionally, $\vb{h}\in\bZ_2^n$ is constrained by $k$ independent constraints (one for each independent generator of $N$) of the form $\vb{h}\cdot \vb{l}=0\mod 2$, each of which reduces the number of valid values by a factor of 2. Put together, $\vb{h}$ can take $2^{n-k}$ different values, independent of $\vb{g}$, and we get $2^k 2^{n-k}=2^n$ anyons for which $m^N(\vb{g}|\vb{k})=1$.

\input{main.bbl}
\end{document}

%% file: main.bbl
%